\documentclass [PhD] {uclathes}
\newcommand{ \be }{\begin{equation}}
\newcommand{ \ee }{\end{equation}}
\newcommand{ \bea }{\begin{eqnarray}}
\newcommand{ \eea }{\end{eqnarray}}
\newcommand{ \bf }{\begin{figure}[htpb]}
\newcommand{ \ef }{\end{figure}}
\newcommand{ \bmn }{\begin{minipage}}
\newcommand{ \emn }{\end{minipage}}
\newcommand{ \bt }{\begin{table}[htpb]}
\newcommand{ \et }{\\end{table}}
\newcommand{ \la }{\langle}
\newcommand{ \ra }{\rangle}

\newcommand{ \rar }{\rightarrow}

\newcommand{ \as }{$\alpha_{s}$}
\newcommand{ \pt }{$p_{T}$}
\newcommand{ \dzero }{$D^{0}$}
\newcommand{ \vv }{$v_{2}$}
\newcommand{\sqrts}{\mbox{$\sqrt{s}$}}
\newcommand{\sNN}{\mbox{$\sqrt{s_{_{\mathrm{NN}}}}$}}
\newcommand{\dAu}{\textit{d}+Au}

\newcommand{\AuAu}{Au+Au}

\newcommand{\pp}{\mbox{\textit{p}+\textit{p}}}

\newcommand{\mev}{\mbox{$\mathrm{MeV}$}}
\newcommand{\gev}{\mbox{$\mathrm{GeV}$}}
\newcommand{\gevcc}{\mbox{$\mathrm{GeV/}c^2$}}

\newcommand{\gevc}{\mbox{${\mathrm{GeV/}}c$}}

\newcommand{\raa}{\mbox{$R_{AA}$}}
\newcommand{\RAA}{\mbox{$R_{AuAu/dAu}$}}

\newcommand{\dedx}{\mbox{$dE/dx$}}

\newcommand{\nbin}{\mbox{$N_{\mathrm{bin}}$}}

\usepackage{amssymb}
\usepackage{amsthm}
\usepackage{revsymb}
\usepackage{rotate}
\usepackage{epsfig}
\usepackage{eepic}
\usepackage{epic}
\usepackage{graphics}
\usepackage[center]{caption2}
\usepackage{wrapfig}
\usepackage{doublespace}
\usepackage{multicols}
\usepackage{multirow}
\usepackage{amsmath}
\usepackage{indentfirst}

\begin{document}
\ifx\href\undefined\else\hypersetup{linktocpage=true}\fi



\title{
Measurement of charm production cross-section and leptons from its
semileptonic decay at RHIC }

\author         {Yifei Zhang}
\department {Physics}
\supervisor{Ziping Zhang, Nu Xu}
\degreeyear     {2007}
\copyrightyear{2007}
\nosignature


\dedication{ \textsl{Dedicated to my dear family} \vskip 1 in
}


\acknowledgments{ I would never forget people whom I was/am
working with. And I can not finish my Ph.D project without their
support and help. This thesis includes many efforts and
contributions from them.

Firstly, I would thank Prof. Ziping Zhang and Prof. Hongfang Chen
for guiding me into the heavy ion collision physics field and
offering me the great opportunity to participate the STAR data
analysis. I appreciate their continuous supervision and support in
the last six years. I thank Dr. Hans-Georg Ritter and Dr. Nu Xu
who provide me the environment and opportunity to work with the
great RNC group in LBNL. I thank Dr. Nu Xu for his guidance and
fruitful discussions on many aspects of my research in the last
two years. I greatly benefited from Dr. Nu Xu's profound and broad
understanding of physics. I am deeply inspired by his assuredness
and encouragement. I would especially thank Dr. Zhangbu Xu from
BNL for his rich experience in data analysis and patience in
instructing me all the analysis details throughout the thesis.

I would like to thank Dr. Haibin Zhang and Dr. Chen Zhong for
their helpful cooperation in BNL and plentiful discussions. Thanks
also go to Dr. Xin Dong, Dr. Haidong Liu, Dr. Yan Lu and Dr.
Lijuan Ruan for their amount of discussions on work and for the
joyful time we had together.

I am very lucky to work with the RNC group in LBNL. I am impressed
by the folksy and enthusiastic people in the group and their open
mind and meticulosity on research. I learned a lot from the
fruitful discussion with them. Especial thanks to Dr. Jim Thomas,
Dr. Ernst Sichtermann and Dr. Andrew Rose who gave me lots of help
and useful discussions on the HFT software. I also appreciate Dr.
Marco van Leeuwen and Dr. Vi Nham Tram for their valuable
discussion on charm physics and help on the correction of charm
paper and my thesis. I also thank other members in RNC group for
their kind assistances.

I would thank all the members of spectra and heavy flavor working
group. Especial thanks to Dr. Bedanga Mohanty for the plenty of
discussions on many aspects of physics. I would thank to Prof.
Huan.Z. Huang, Dr. Xiaoyan Lin from UCLA for many helpful
discussions on heavy flavor physics. I also thank Dr. Thomas
Ullrich, Dr. Manuel, Dr. Weijiang Dong and other people who gave
me lots of useful suggestions and discussions on the detailed data
analysis of heavy flavor.

I would especially thank Prof. Charles Whitten from UCLA for his
elegant wording and grammatical corrections on my thesis.

Many thanks to our high energy physics group (HEPG) in USTC.
Especial thanks to Prof. Xiaolian Wang, Prof. Zizong Xu, Prof.
Jian Wu and Dr. Ming Shao for their continuous discussions and
support. I would thank Dr. Yane Zhao and Dr. Yi Zhou for their
hospitable help on my living. And many thanks to my classmate Dr.
Qing Shan, Zebo Tang and other students in our group for their
hardwork on the MRPC production and testing.

Thank the PDSF and RCF software group for their offered resources
and continuous development and support on software. Thank the
whole STAR Collaboration for building such a beautiful detector
and provide the enormous amount of data with high quality. I would
particularly thank STAR TOF group and HFT group for the hardwork
on detector upgrades and brightening the future.

Finally, what I will remember to the end of my life is the eternal
love, the tolerant consideration and selfless dedication from my
family. Their continuous support and understanding have always
been my motivity of never giving up.

 }




\abstract {\par The strong interaction, one of the four
fundamental forces of nature, is confining: there is no single
quark as a color-triplet state observed experimentally. {\em
Quantum Chromodynamics} (QCD) is believed to be a correct basic
gauge field theory of strong interactions. Lattice QCD
calculations predict a phase transition from hadronic gas to a new
matter, {\em Quark-Gluon Plasma} (QGP), at high energy
nuclear-nuclear collisions. In this new form of matter, quarks are
deconfined and approach local thermalization. One of the ultimate
goals of the heavy ion collision experiments is to search for the
QGP matter and study its properties. The Relativistic Heavy Ion
Collider (RHIC) located at Brookhaven National Laboratory (BNL)
provides a high energy density environment to create and search
for the QGP matter by colliding ions like Au at energies up to
\sNN=200 \gev. Recent experimental studies at RHIC have given
strong evidences that the nuclear matter created in \AuAu\
collisions at \sNN=200 \gev\ has surprisingly large collectivity
and opacity as reflected by its hydrodynamic behavior at low \pt\
and its particle suppression behavior at high \pt.

Charm quarks provide a unique tool to probe the partonic matter
created in relativistic heavy-ion collisions at RHIC energies. Due
to their large quark mass ($\simeq 1.3$ GeV/$c^{2}$), charm quarks
are predicted to lose less energy than light quarks via only gluon
radiation. A measurement of the nuclear modification factor for
the charmed hadrons semileptonic decayed single electrons compared
to light hadrons is valuably important to complete the picture of
the observed jet-quenching phenomenon and help us better
understand the energy-loss mechanisms at parton stage in \AuAu\
collisions at RHIC.

Furthermore, the interactions between charm quarks and the medium
could boost the radial and elliptic flow resulting in a different
charm \pt\ spectrum shape. Due to its large mass, a charm quark
could acquire flow from the sufficient interactions with the
surrounding partons in the dense medium. The measurement of charm
flow and freeze-out properties is vital to test light flavor
thermalization and the partonic density in the early stage of
heavy ion collisions.

Charm quarks are believed to be produced only at early stages via
initial gluon fusions and its cross section can be evaluated by
perturbative QCD calculations. Thus study of the binary collision
(\nbin) scaling properties for the total charm cross-section from
different collision systems can test the theoretical assumptions
and determine if charm quarks are indeed good probes to the
partonic matter created in high energy heavy ion collisions. Charm
total cross-section measurement is also essential for the
separation of bottom contribution in non-photonic electron
measurement, and for the model calculations, which tries to
explain the observed similar suppression pattern of $J/\Psi$ at
RHIC and SPS.

In this thesis, we present the measurements of $D^{0}\rightarrow
K\pi$ at low \pt\ ($p_T\leq2$ \gevc) and non-photonic electron
spectra ($0.9\leq p_T\leq5$ \gevc) from $D^0$ semi-leptonic decay.
In addition, we use a newly proposed technique to identify muons
from charm decays at low \pt. The combination of all these three
measurements stringently constrains the total charm production
cross-section at mid-rapidity at RHIC. They also allow the
extraction of the charmed hadron spectral shape and a study of
possible charm radial flow in \AuAu\ collisions.

$D^0$ mesons were reconstructed from hadronic decay
$D^0\rightarrow K^-\pi^+$ ($\bar{D^0}\rightarrow K^+\pi^-$) with a
branching ratio of 3.83\% in minbias \AuAu\ collisions. The $D^0$
yields were obtained from fitting a Gaussian plus a linear (or a
second-order polynomial function) for residual background to the
invariant mass distributions of kaon-pion pairs after mixed-event
combinatorial background subtraction.

The inclusive muons at $0.17\leq p_T\leq0.25$ \gevc\ in minbias
\AuAu\ and central \AuAu\ collisions were analyzed by combining
the ionization energy loss (\dedx) measured in the Time Projection
Chamber (TPC) and the mass calculated from the Time Of Flight
(TOF) detector at STAR after the residual pion contamination
subtracted. The dominant background muons from pion/kaon weak
decays were statistically subtracted using the distribution of the
distance of closest approach to the collision vertex (DCA).

Inclusive electrons up to $p_T=5$ \gevc\ are identified by using a
combination of velocity ($\beta$) measurements from the TOF
detector and \dedx\ measured in the TPC. Photonic background
electrons are subtracted statistically by reconstructing the
invariant mass of the tagged $e^{\pm}$ and every other partner
candidate $e^{\mp}$ using a 2-dimensional invariant mass method.
Partner track finding efficiency is estimated from STAR Geant +
Monte Carlo embedding data. More than $\sim 95\%$ of photonic
background (photon conversion and $\pi^0$ Dalitz decay) can be
subtracted through this method. The \pt\ spectra for non-photonic
electrons in central \AuAu\, minbias \AuAu\ collisions and its
subdivided centralities will be presented. In addition, the method
to extract the elliptic flow \vv\ of non-photonic electron is also
developed in the thesis.

By combining these three independent measurements: $D\rightarrow
K\pi$ , muons and electrons from charm semileptonic decays in
minbias and central \AuAu\ collisions at RHIC, we observed that:
The transverse momentum spectra from non-photonic electrons are
strongly suppressed at $0.9\leq p_T\leq5$ \gevc\ in central \AuAu\
collisions relative to \dAu\ collisions. For electrons with
$p_T\gtrsim2$ \gevc, corresponding to charmed hadrons with
$p_T\gtrsim4$ \gevc, the suppression is similar to that of light
baryons and mesons. The blast wave fit to the electron spectra
with $p_T\lesssim2$ \gevc\ indicates that charmed hadrons may
interact and decouple from the system differently from
multi-strange hadrons and light hadrons. Future upgrades with a
direct reconstruction of charmed hadrons are crucial for more
quantitative answers. Charm differential cross-sections at
mid-rapidity ($d\sigma_{c\bar{c}}^{NN}/dy$) are extracted from a
combination of the three measurements covering $\sim90\%$ of the
kinematics. The total charm cross-section per nucleon-nucleon
collision ($\sigma_{c\bar{c}}^{NN}$) is reported as
$1.40\pm0.11(stat.)\pm0.39(sys.)$ mb in $0-12$\% central \AuAu\
and $1.29\pm0.12\pm0.36$ mb in minbias \AuAu\ collisions at
\sNN=200 \gev. The charm production cross sections are found to
follow the number of binary collisions scaling. This supports the
assumption that hard processes scale with binary interactions
among initial nucleons and charm quarks can be used as a probe
sensitive to the early dynamical stage of the system.

In the above measurements, the bottom contribution to the
non-photonic electron spectrum is neglected. The separation of
bottom and charm contributions in current non-photonic electron
measurements is very difficult. There are large uncertainties in
the model predictions for charm and bottom production in
high-energy nuclear collisions. Thus identification of bottom from
the non-photonic electron measurements is crucial to better
understand charm physics. In the discussion section of this
thesis, we will try a fit to non-photonic electron spectrum and
estimate the bottom contributions. We will also compare the $v_2$
distribution from simulation to the experimental data and estimate
the possible charm $v_2$. }



\makeintropages

\chapter{Introductions}

\section{Quantum ChromoDynamics}

\subsection{Confinement and asymptotic freedom}

{\em Quantum ChromoDynamics} (QCD)~\cite{QCDbook} is thought to be
a correct theory of the strong nuclear force, one of the four
fundamental forces of nature. It describe the strong interactions
among quarks, which are regarded as fundamental constituents of
matter, via their color quantum numbers. The strong interactions
among quarks are mediated by a set of force particles known as
gluons. Different from {\em Quantum ElectroDynamics} (QED) - the
gauge theory describing electromagnetic interaction, QCD is based
on the non-Abelian gauge group $SU(3)$, with gauge bosons (color
octet gluons), and hence the gluons could have self-interacting.
This results in a negative $\beta$-function and {\em asymptotic
freedom} at high energies and strong interactions at low energies.

These strong interactions are confining: the self-coupled gluons
strongly restrain the isolation of the quarks at large distance.
There is no single quark as a color-triplet state observed
experimentally. Only color-singlet bound states can propagate over
macroscopic distances. The only stable color-singlets with size of
the order of $1 fm$ are quark$-$antiquark pairs, mesons, and
three-quark states, baryons. At high energy reactions, like deep
inelastic scattering, the quark and gluon constituents of hadrons
act as quasi-free particles, partons. Such reactions can be
factorized into the convolution of non-perturbative parton
distribution functions, which cannot be calculated from first
principles directly. But with process-dependent functions ({\em
i.e. hard processes involving large momentum transfers}), the
reactions can be calculated as perturbative expansions in the
coupling constant \as.

In QED, the electrodynamic coupling constant $\alpha =
\frac{1}{137}$. However, due to the gluons self-interactions, the
renormalized QCD coupling shows renormalization scale ($\mu$)
dependence~\cite{alphas}. The running coupling $\alpha_{s}(\mu)$
can be written as: \be \label{equas}
\alpha_s(\mu)\equiv\frac{g_s^2(\mu)}{4\pi} \approx
\frac{4\pi}{\beta_{0}\ln(\mu^2/\Lambda_{QCD}^2)},
\ee where $g_s$, which is strong charge in the gauge group, is the
only parameter in the QCD Lagrangian besides the quark masses.
$\beta_0$ ($>$0) is the first coefficient of the $beta$-function
(renormalization neglects the higher orders). The strong force of
the gluon-gluon self-coupling becomes smaller at shorter distance
or with larger momentum transfers (\as$\rar$ 0 as
$\mu\rar\infty$), which is known as {\em asymptotic freedom}. In
this case, QCD can be calculated perturbatively. Many experiments
measured \as\ at different scales. Since some of the precise
measurements come from $Z^0$ decays, it has become universal to
use $\alpha_{s}(M_z)$ as the label. The
$\alpha_{s}(M_{Z})=0.1176\pm0.002$~\cite{PDG} comes from a fit to
the experimental data, and the QCD scale $\Lambda_{QCD}\sim200$
MeV. Fig.~\ref{alphas} shows the measured $\alpha_s$ at different
momentum transfer scale $\mu$ compared with Lattice QCD
calculations.

\bf \centering\mbox{
\includegraphics[width=0.60\textwidth]{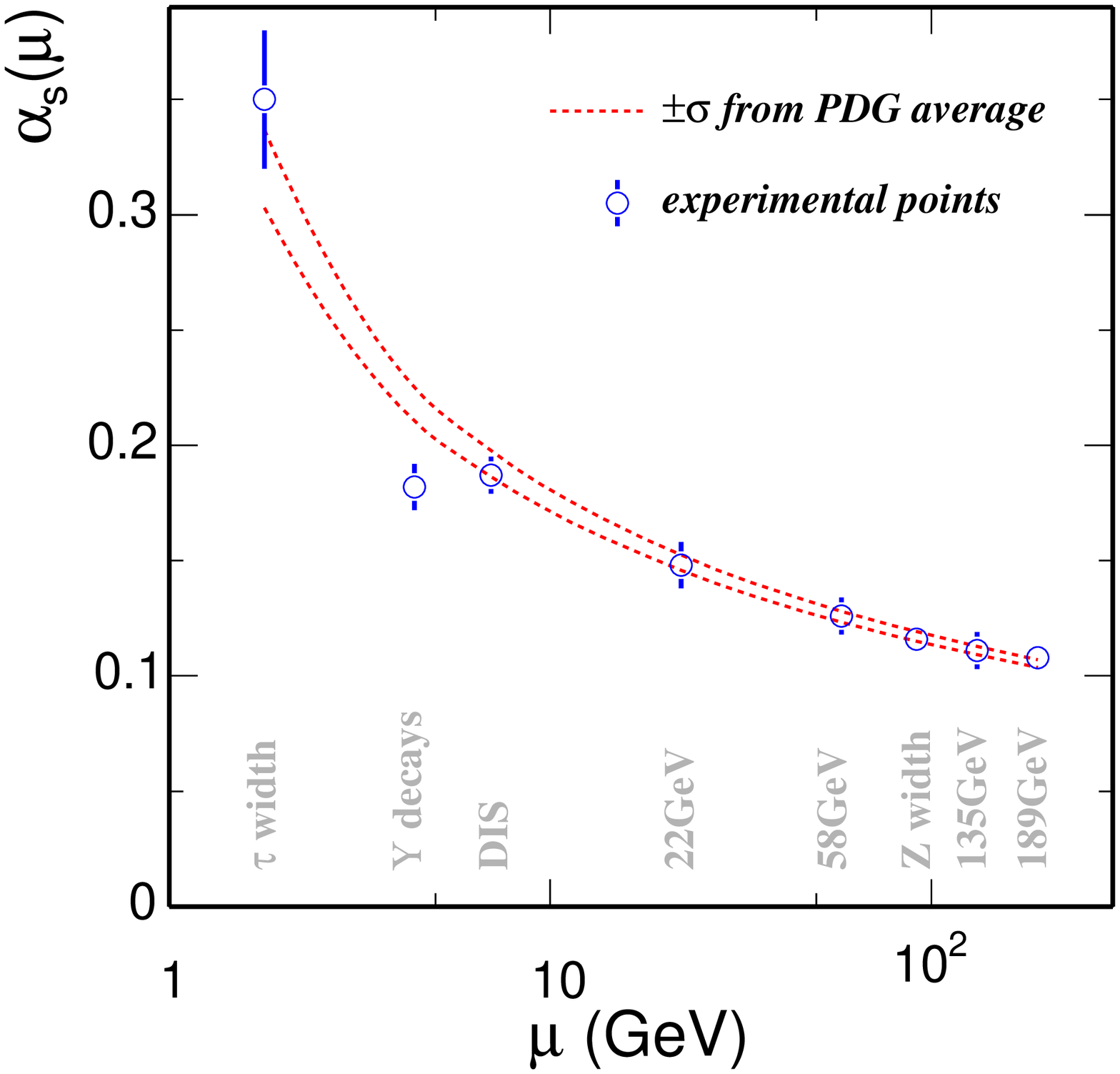}}
\caption[QCD running coupling constant \as]{Measured QCD running
coupling constant \as\ from different experiments compared with
Lattice QCD calculations.} \label{alphas} \ef

\subsection{Nonperturbative QCD}

Corresponding to the asymptotic freedom of QCD at high momentum
scales (short distances), the running coupling becomes larger at
lower momenta (longer distances). This effect is called {\em
infrared slavery}. Eq.~\ref{equas} shows strong coupling at
$\mu\sim\Lambda_{QCD}$. Perturbation theory is not expected to be
a reliable solution when \as\ becomes large. Thus nonperturbative
methods must be used to study the low-momentum, long-distance
behavior of QCD.

A closely related phenomenon is {\em hadronization}. This term
refers to the dynamical process in which a set of partons produced
in a short-distance interaction reorganize themselves, possibly
with the production of additional partons, to make the hadrons
that are observed in the final state. This again is expected to be
a low momentum transfer process, corresponding to the relatively
long timescale over which hadron formation takes place.

A nonperturbative approach that is widely used to study
confinement is {\em lattice QCD}~\cite{LQCDintro}. The continuum
field theory is replaced by one defined on a space-time lattice.
As long as the size of the lattice elements inside a hadron is
sufficiently small, hadronic properties and matrix elements should
be reliably calculable on massively parallel computers.

\subsection{Heavy quark masses and perturbative QCD (pQCD)}

The masses of the heavy c quark and b quark ($m_{c}\sim 1.3$
\gevcc, $m_{b}\sim 4.75$ \gevcc) are almost exclusively generated
through their coupling to the Higgs field in the electro-weak
sector, while the masses of light quarks (u, d, s) are dominated
by spontaneous breaking of chiral symmetry in QCD. This means that
in a {\em Quark-Gluon Plasma} (QGP), where chiral symmetry is
expected to be restored, light quarks are left with their bare
current masses while heavy-flavor quarks remain heavy. The heavy
quark masses are not modified by the QCD vacuum~\cite{PDG,Qmass}.
Hence once heavy quarks are pair-produced in early stage via
initial gluon fusions~\cite{LinPRC} at high energy collision,
their total number is conserved because the typical temperature of
the subsequently evolutional medium is much smaller than the heavy
quark masses. Therefore, heavy-flavor quarks are an ideal probe to
study early dynamics in high-energy nuclear collisions, and their
cross section can be evaluated by pQCD calculations.

\bf \centering\mbox{
\includegraphics[width=0.60\textwidth]{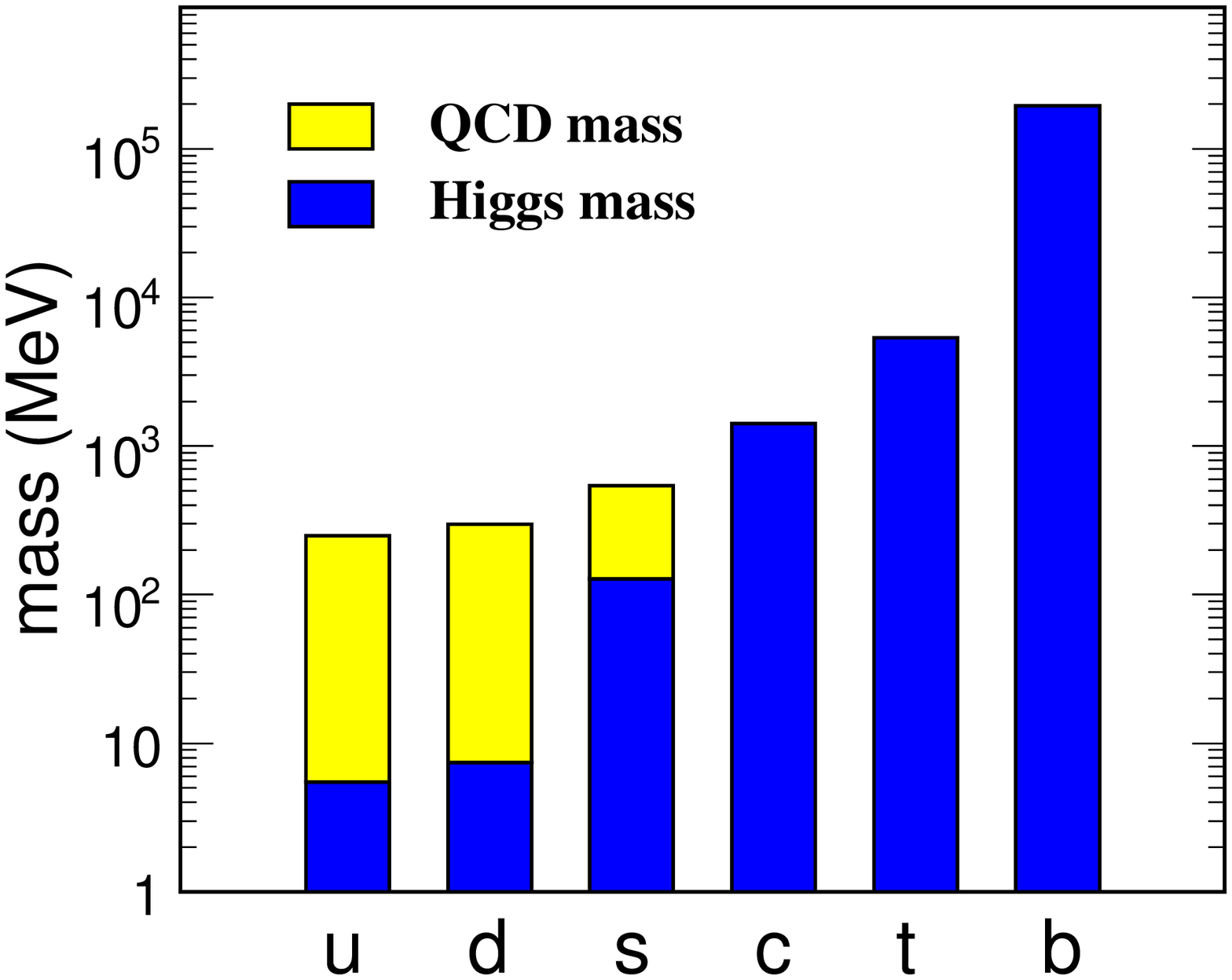}}
\caption[QCD running coupling constant \as]{Masses of the six
flavors. The masses generated by electroweak symmetry breaking
(current quark masses) are shown as dark boxes; A large fraction
of the light quark masses is due to the chiral symmetry breaking
in the QCD vacuum (constituent quark masses), shown as light
boxes. The numerical values were taken from Ref.~\cite{PDG}.}
\label{Qmass} \ef

QCD is formulated in terms of quarks and gluons while the
experimentally observed states are hadrons. As previously
mentioned, in principal the method of perturbation theory is
appropriate in the high-momentum scale, short-distance regime. It
is known as the QCD factorization theorem~\cite{CSS,Collins} and
states that for processes that have initial and/or observed final
state hadrons the differential cross-section has the following
form: \be \label{fac}
d\sigma(x,Q^2,m^2)=\prod_{h,h'}\sum_{i,f}f_{i/h}(x,\mu^2)\otimes
d\hat\sigma_{i\to f}(x,Q^2,m^2,\mu_r^2,\mu_F^2)\otimes
D_{h'/f}(x,\mu^2) + {\mathcal{O}}(\Lambda/Q), \ee where the factor
$f_{i/h}$ stands for the {\em Parton Distribution Function} (PDF)
of the parton $i$ inside the hadron $h$ present in the initial
state, and $Q$ and $x$ represent the hard scale and some
kinematical variable respectively. The parton distributions depend
on the factorization/renormalization scale $\mu^2$. The second
factor $d\hat\sigma_{i\to f}$, also known as the (Wilson)
coefficient function, represents the partonic hard scattering
cross section for the reaction $i\to f$ that depends on the
unphysical renormalization and factorization scales $\mu_r^2$ and
$\mu_F^2$ and on the masses of the heavy quarks $m^2$. The last
factor $D$ is the so called {\em Fragmentation Function} (FF). It
contains the information for the hadronization of the hard parton
$f$ (that is produced in the hard process described by the
partonic cross-section $d\hat\sigma$) into an observed hadron
$h'$. Since the concepts of parton distributions and hadronization
are only for the initial and final states of hadrons,
lepton-nucleon {\em Deep Inelastic Scattering} (DIS) and high
energy $e^+e^-$ collisions are performed to measure PDFs and FFs,
respectively.

The large masses ($\gg\Lambda_{QCD}$) of the heavy quarks, which
are 'hard' scales, make perturbative QCD
applicable~\cite{vogtXsec}. Asymptotic freedom is assumed in
calculating the interactions between two heavy hadrons on the
quark/gluon level but the confinement scale determines the
probability of finding the interacting parton in the initial
hadron. Factorization assumes that between the perturbative hard
part and the full expansion, non-perturbative parton distribution
functions are applied. The integral hadronic cross section in an
$AB$ collision where $AB=pp, pA$ or nucleus-nucleus can be written
as: \be
\sigma_{AB}(S,m^2)=\sum_{i,j=q,\bar{q},g}\int_{4m_Q^2/s}^{1}
\frac{d\tau}{\tau}\int
dx_idx_j\delta(x_ix_j-\tau)f_{i}^{A}(x_i,\mu_{F}^2)f_{j}^{B}(x_j,\mu_{F}^2)
\hat{\sigma}_{ij}(s,m^2,\mu_{F}^2,\mu_{R}^2), \ee where
$f_{i}^{A}$ and $f_{i}^{B}$ are the nonperturbative PDF determined
from experiments, $x_i$ and $x_j$ are the fractional momentum of
hadrons $A$ and $B$ carried by partons $i$ and $j$, $\tau=s/S$,
$s$ is partonic center of mass energy squared,
$\hat{\sigma}_{ij}(s,m^2,\mu_{F}^2,\mu_{R}^2)$ is hard partonic
cross section, which only depends on quark mass $m$, not kinematic
quantities, can be calculated in QCD in powers of
$\alpha_{s}^{2+n}$, known as {\em leading order} (LO), $n=0$; {\em
next-to-leading order} (NLO), $n=1$ .... Eq.~\ref{qxsecnlo}
express the partonic cross-section to NLO: \be
\hat{\sigma}_{ij}(s,m^2,\mu_{F}^2,\mu_{R}^2)=\frac{\alpha_s^2(\mu_R^2)}{m^2}\{f_{ij}^{(0,0)}(\rho)+4\pi\alpha_s(\mu_R^2)[f_{ij}^{(1,0)}(\rho)+f_{ij}^{(1,1)}(\rho)\ln(\mu_F^2/m^2)]+{\mathcal{O}}(\alpha_s^2)\},
\label{qxsecnlo} \ee where $\rho=4m^2/s$, $\mu_F$ is factorization
scale which separates hard part from nonperturbative part, $\mu_R$
is renormalization scale, at which the strong coupling constant
\as\ is evaluated. $\mu_F=\mu_R$ is assumed in evaluations of
parton densities, and $f_{ij}^{(a,b)}$ are dimensionless,
$\mu-$independent scaling functions. For LO, $a=b=0$ and
$ij=q\bar{q}$, $gg$. For NLO, $a=1$ ,$b=0,1$ and $ij=q\bar{q},gg$
and $qg,\bar{q}g$. $f_{ij}^{(0,0)}$ are always positive,
$f_{ij}^{(1,b)}$ can be negative. If $\mu_F^2=m^2$,
$f_{ij}^{(1,1)}$ does not contribute. Results of theoretical
calculations strongly depend on quark mass, m, factorization
scale, $\mu_F$, in the parton densities and renormalization scale,
$\mu_R$, in \as, {\em
etc.}~\cite{manganoXsec,vogtXsec,raufeisenXsec}.

\subsection{Deconfinement, equation of state and QCD phase}
Currently, the isolated quark has never been observed. Quarks and
gluons are confined in QCD matter as hadron gas. If the
temperature and energy density of the matter are sufficiently
high, the strong force among quarks and gluons may be greatly
reduced. Thus quarks and gluons could be deconfined from hadrons
and become disengaged with large distance among them in that hot
and dense condition. This phenomenon is called {\em
deconfinement}. The deconfinement indicates a phase transition
from a hadron gas to a new deconfined form of matter -- the QGP,
which is formed by these dissociative quarks and gluons with new
(color) {\em degree of freedom} (d.o.f). In the QGP phase, the
broken chiral symmetry will be restored. The evolution of energy
density and pressure with temperature from Lattice QCD
calculations illustrate the phase transition~\cite{LQCDpressure},
see Fig.~\ref{lQCDphase}.

\bf \centering \bmn[c]{0.5\textwidth} \centering
\includegraphics[width=0.95\textwidth]{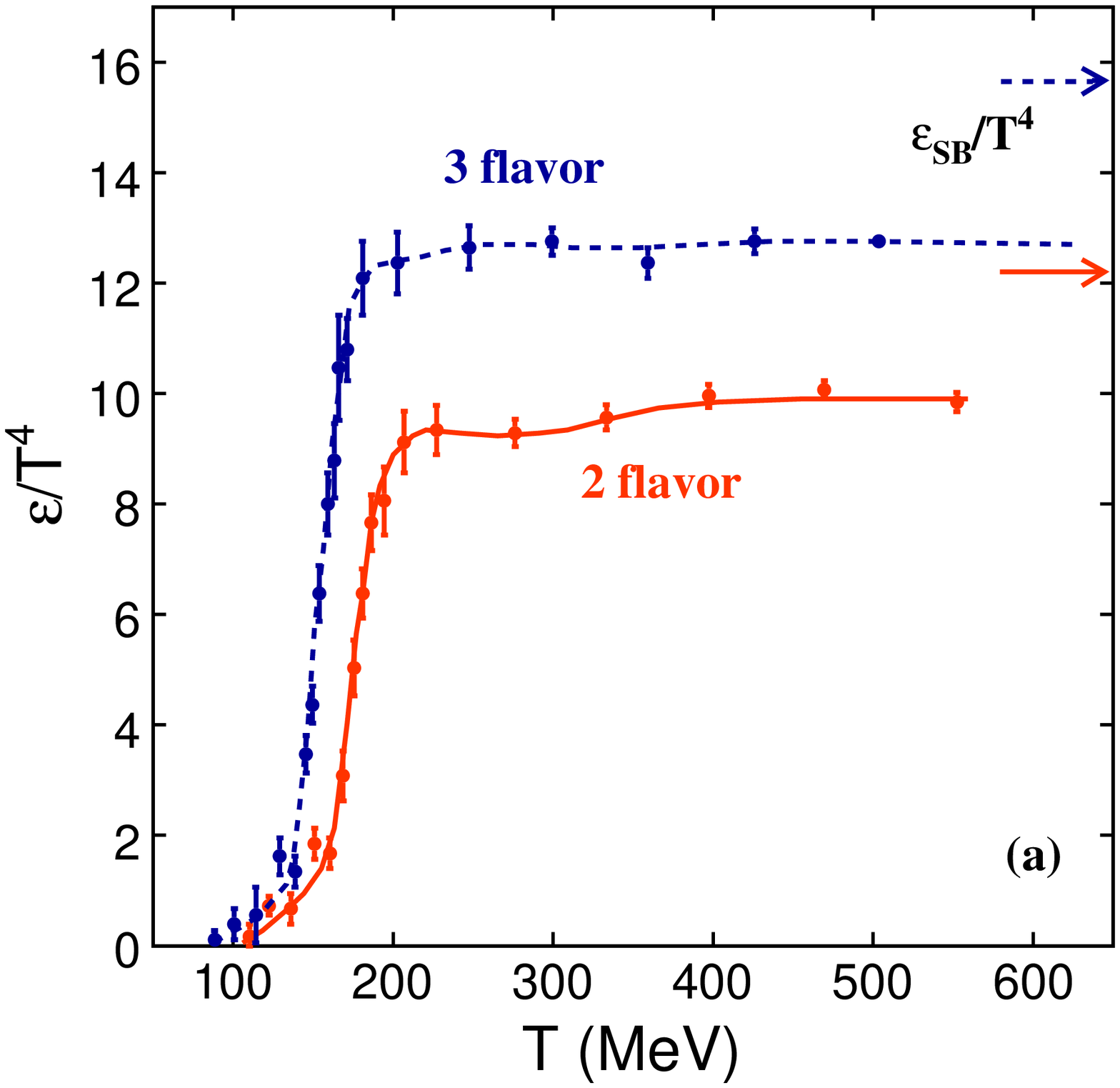}
\emn%
\bmn[c]{0.5\textwidth} \centering
\includegraphics[width=0.95\textwidth]{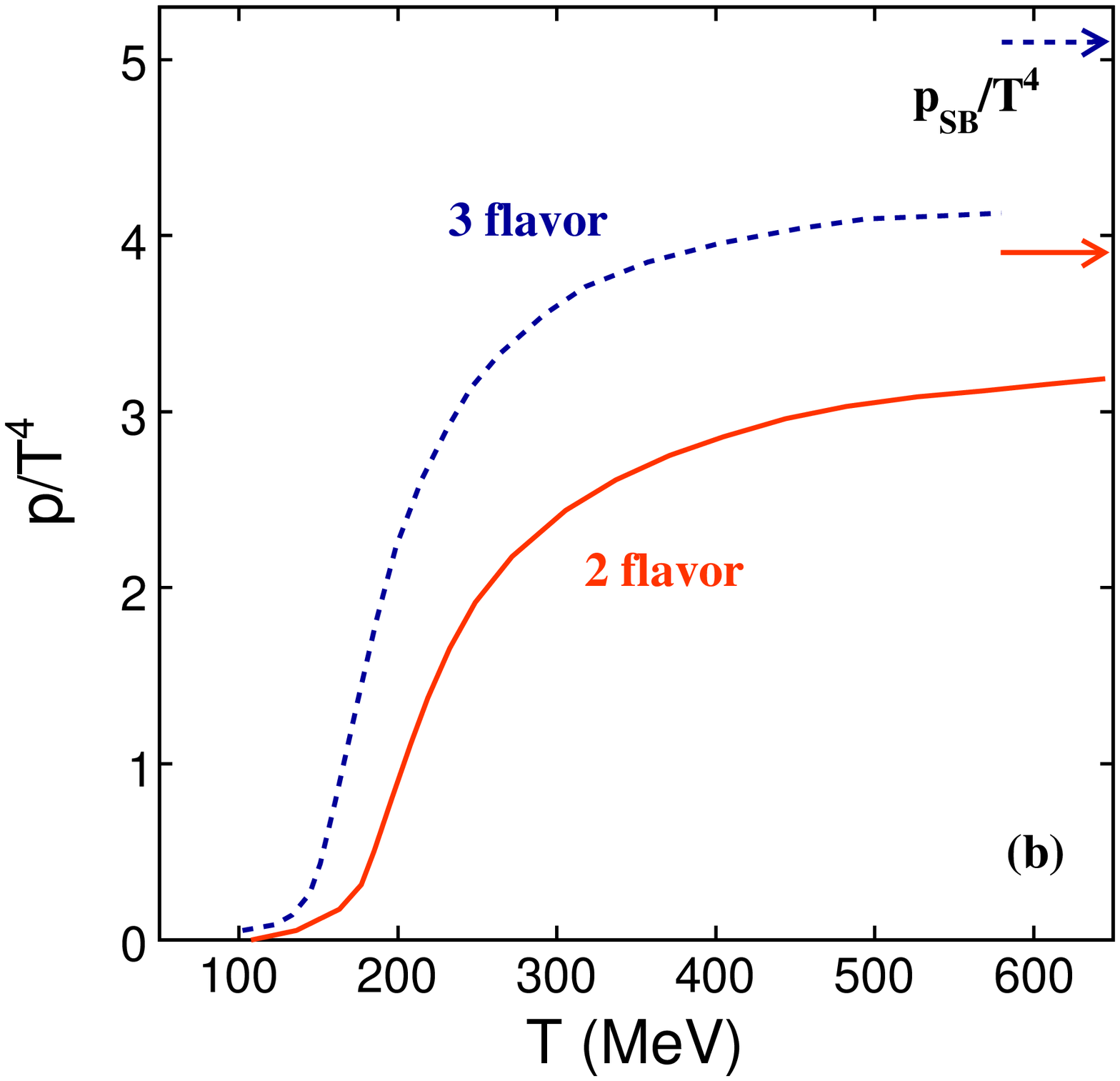}
\emn%
\caption[LQCD calculation for energy density and pressure]{The
evolution of $\varepsilon/T^4$ (a) and $p/T^4$ (b) with the
increase of temperature $T$ for 3 flavor and 2 flavor
configurations. The arrows indicate the SB limit for each case.}
\label{lQCDphase} \ef

Large increase of degrees of freedom at $T_c$ is seen in the rapid
change in energy density and pressure. The critical temperature is
predicted to be $T_c\sim150-180$ MeV, and the energy density at
the critical point is $\varepsilon_c(T_c)\sim1-3$ GeV/fm$^3$
($\sim0.17$ GeV/fm$^3$ for nuclear matter)~\cite{LQCDreview}. The
arrows indicate the {\em Stefan-Boltzmann} (SB) limits, applicable
for the systems with massless, non-interacting quarks and gluons.
The similar deviation of the curves from the SB limits indicates
that besides the effect of quark masses, the quasiparticles, which
form the new matter, must interact with one another. The various
heavy-ion experiments devoted to the creation and detection of new
forms of highly excited matter also indicate that, at the
reachable energy scales, the produced phase exhibits a strong
collective behavior (see next section). All these theoretical
predictions and experimental observations are incompatible with a
weakly interacting QGP.

Based on local 4-momenta conservation and other conserved currents
(e.g. baryon number), \be
\partial_{\mu}T^{\mu\nu}=0, ~~~~~~~ \partial_{\mu}j^{\mu}=0, \label{epconserv} \ee
from the {\em Hydrodynamic} model, the equations of relativistic
ideal hydrodynamics are \be j^{\mu}(x)=n(x)u^{\mu}(x), \ee \be
T^{\mu\nu}=[\varepsilon(x)+p(x)]u^{\mu}u^{\nu}-g^{\mu\nu}p(x), \ee
where $\varepsilon(x)$ is the energy density, $p(x)$ is the
pressure, $n(x)$ is the conserved number density, and
$u^{\mu}(x)=\gamma(1,v_x,v_y,v_z)$ with
$\gamma{\,=\,}1/{\textstyle\sqrt{1{-}v_x^2{-}v_y^2{-}v_z^2}}$ is
the local four velocity of the system.

Eq.~\ref{epconserv} contains 5 equations for 6 unknown fields
$\varepsilon,n,p,v_x,v_y,v_z$. To fully describe the system one
needs an {\em equation of state} (EOS), which relates pressure,
energy and baryon density. The thermal condition of the system can
be fixed by pressure, $p$ and temperature, $T(\varepsilon)$. The
EOS, which describes the system response to changes of the thermal
condition, is separately constructed for a relativistic ideal QGP
phase (dashed line in Fig.~\ref{EOSfig}), $\partial
p/\partial\varepsilon=\frac{1}{3}$, and a hadron resonances gas
(dotted line), $\partial p/\partial\varepsilon\sim\frac{1}{6}$. By
combining these two EOSs, the EOS indicates the phase transition
(solid line) can be obtained from Maxwell construction with a bag
constant $B$. Here $B$ is tuned to the desired phase transition
temperature: $B^{1/4}=230$ MeV gives $T_c(n=0) = 164$ MeV at
vanishing net baryon density.

\bf \centering\mbox{
\includegraphics[width=0.60\textwidth]{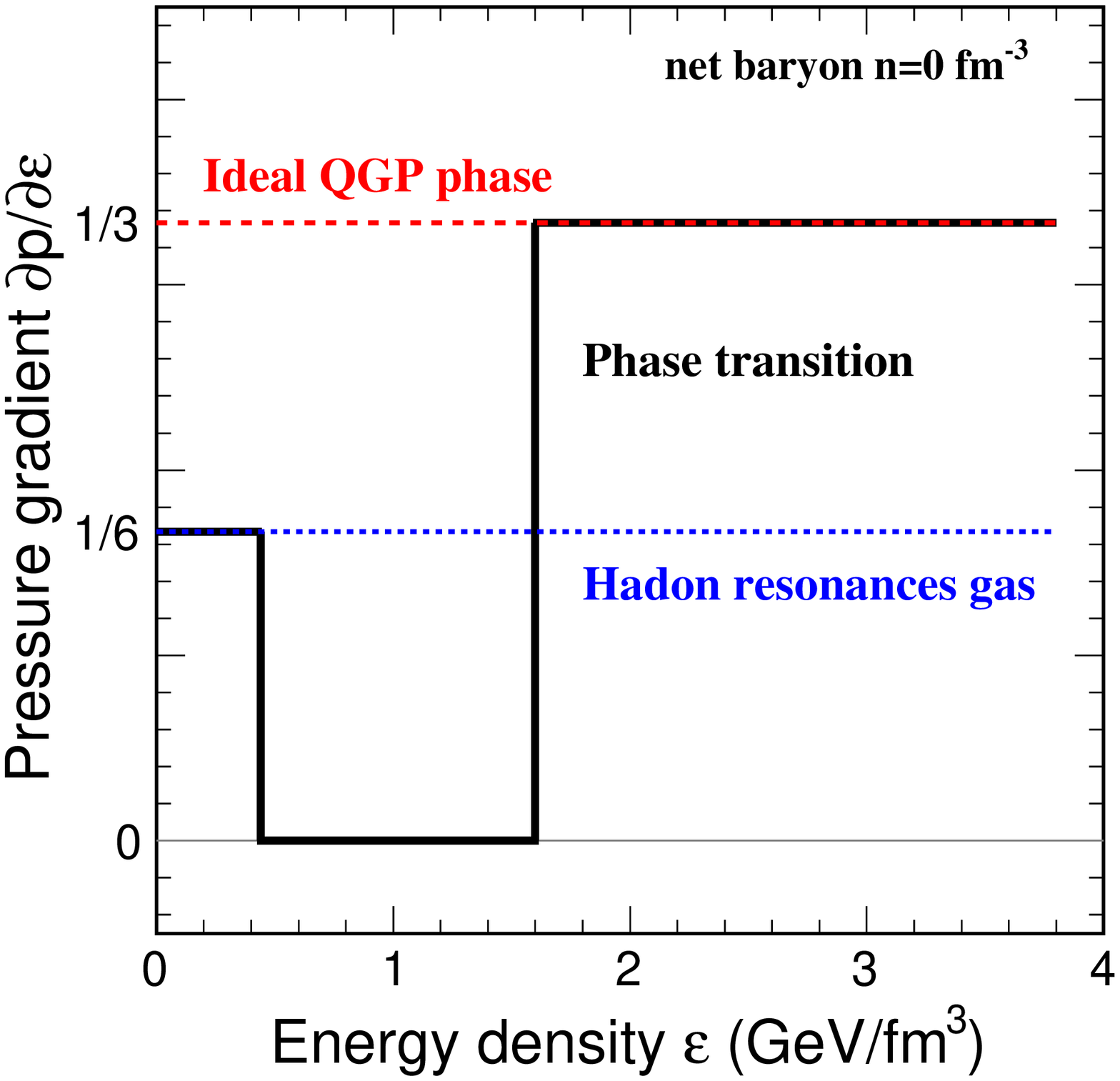}}
\caption[Equation of state]{Equations of state for ideal QGP
phase, hadron resonances gas and phase transition at vanishing net
baryon density.} \label{EOSfig} \ef


\section{Relativistic Heavy Ion Collisions}

With very high energy density and temperature in a bulk system,
the new QGP phase of matter is predicted to be formed by the
deconfined quarks and gluons. In order to search for the
production of QGP matter and to study its properties, relativistic
heavy-ion collisions have been performed to provide the hot and
dense environment under laboratory conditions. The Relativistic
Heavy Ion Collider (RHIC) at Brookhaven National Laboratory is
built to create and search for this novel form of matter by
colliding Au ions at energies up to \sNN=200 \gev.

In relativistic heavy-ion collisions, due to the Lorentz
contraction in the moving direction, the collision geometry
between the two nuclei behaves as two thin disks approaching each
other nearly with speed of light. The heavy-ion collisions can be
approximated as the interpenetrating collisions of their
constituent nucleons with partonic interactions at high energy.
The number of participating nucleons and the produced particle
multiplicities in the final hadron state can be calculated by the
Glauber model which relates these quantities to the size of the
impact parameter, $b$, which is defined as the distance between
the perpendicular bisectors along the colliding direction of the
two ions. The impact parameter is large for peripheral collisions,
consisting of a small number of participants and small
multiplicities. A central collision occurs when the impact
parameter is small. Only central collisions of two heavy nuclei at
high energies are expected to provide a sufficiently hot and dense
environment to produce the QGP. In central heavy-ion collisions,
with sufficient interactions of the partons, the chemical and
local thermal equilibrium of the quark-gluon system could be
reached and thus the QGP forms. At the initial stage of the
collision with the primary hard scattering, high \pt\ jets, heavy
quark pairs, direct photons, {\em etc.} are created due to large
momentum transfers. Due to high energy densities and pressure
gradients, the QCD system expands and cools down. The partons can
hadronize to mesons and baryons via inelastic interactions. When
the chemical freeze-out point (temperature) is reached, the
inelastic interactions will stop, and the relative particle yields
will not change. After that, the elastic interactions
re-distribute the transverse momenta among hadrons. The particle
kinetic freeze-out point (temperature) will occur after the
elastic interactions stop.

\subsection{Energy loss and jet quenching}

Particle yields will change due to energy loss in the strong
interacting medium created in nuclear-nuclear collisions. These
modifications of high \pt\ particle yields can be used as unique
tools to study the medium properties. Experimentally, the widely
used observable quantity of particle energy loss is called the
nuclear modification factor $R_{AB}$, which is defined as the
ratio of the spectra in $A+B$ collisions and those in \pp\
collisions, scaled by the number of binary nuclear-nuclear
collisions~\cite{highpt130}: \be \label{nmfeq}
R_{AB}(p_T)=\frac{d^2N^{AB}/dp_Td\eta}{T_{AB}d^2\sigma^{pp}/dp_Td\eta},
\ee where $T_{AB}=\la N_{bin}\ra/\sigma_{inel.}^{pp}$ accounts for
the nuclear overlap geometry, averaged over the measured
centrality class~\cite{HICintro}. $\la N_{bin}\ra$ is the
equivalent number of binary collisions calculated from the Glauber
model.

As presented in Fig.~\ref{starraa} (a), high \pt\ suppression of
\nbin\ scaled hadron production in central \AuAu\ collisions
relative to \pp\ collisions has been observed in RHIC experiments
at \sNN=200 \gev. This is considered as evidence for the energy
loss of the energetic partons, high \pt\ jets interacting with the
hot and dense medium created in central \AuAu\ collisions at
RHIC~\cite{dAuhighpt}. Due to the absence of nuclear effects such
as shadowing~\cite{eks98,deFlorian03,Kopeliovich03,vogt04}, the
Cronin effect~\cite{accardi04,cattaruzza04,barna04}, {\em etc.},
hard processes are expected to scale with the number of binary
collisions ($R_{AB}=1$). Thus the nuclear suppression phenomenon
was not seen in \dAu\ collisions. The enhancement in the
intermediate \pt\ region for $R_{dAu}$ is due to the Cronin
effect. The consistent picture of away-side jet quenching has also
been observed in dihadron azimuthal angle correlations.
Fig.~\ref{starraa} (b) shows the associated hadron
(2$<$\pt$<$$p_{T}(trig)$ \gevc) azimuthal distribution relative to
a triggered hadron (4$<$\pt$<$6 \gevc) after subtraction of the
elliptic flow and pedestal contributions. The enhanced correlation
at near-side ($\Delta\phi\sim0$), which means the pair is from a
single jet, was observed in \pp, \dAu\ and \AuAu\ collisions. The
pair from the back-to-back jet correlation at away-side
($\Delta\phi\sim\pi$) in central \AuAu\ collisions shows a
dramatic suppression relative to those in \pp\ and \dAu\
collisions~\cite{dAuhighpt,jetspectra}.

\bf \centering \bmn[c]{0.5\textwidth} \centering
\includegraphics[width=1.\textwidth]{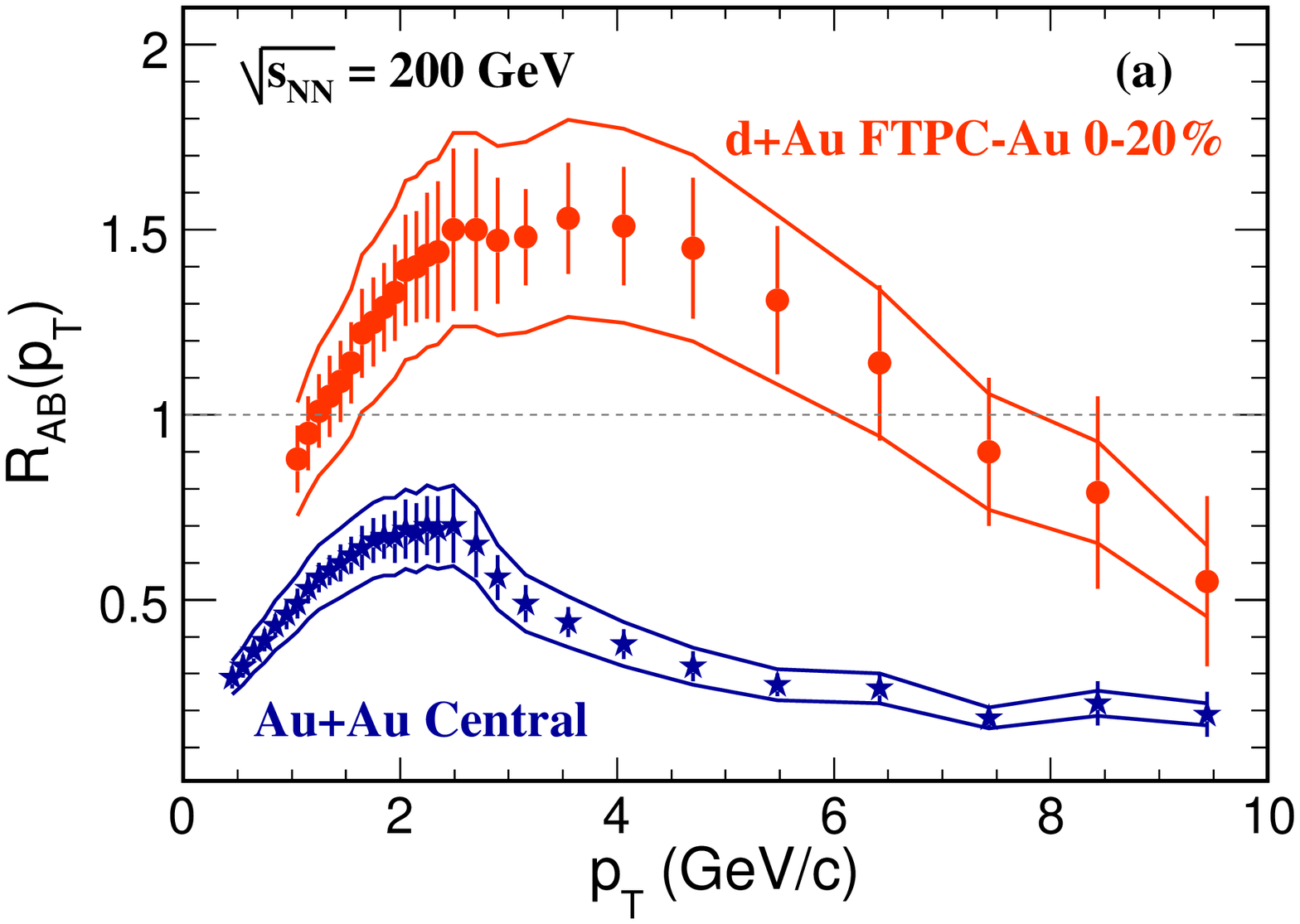}
\emn%
\bmn[c]{0.5\textwidth} \centering
\includegraphics[width=1.\textwidth]{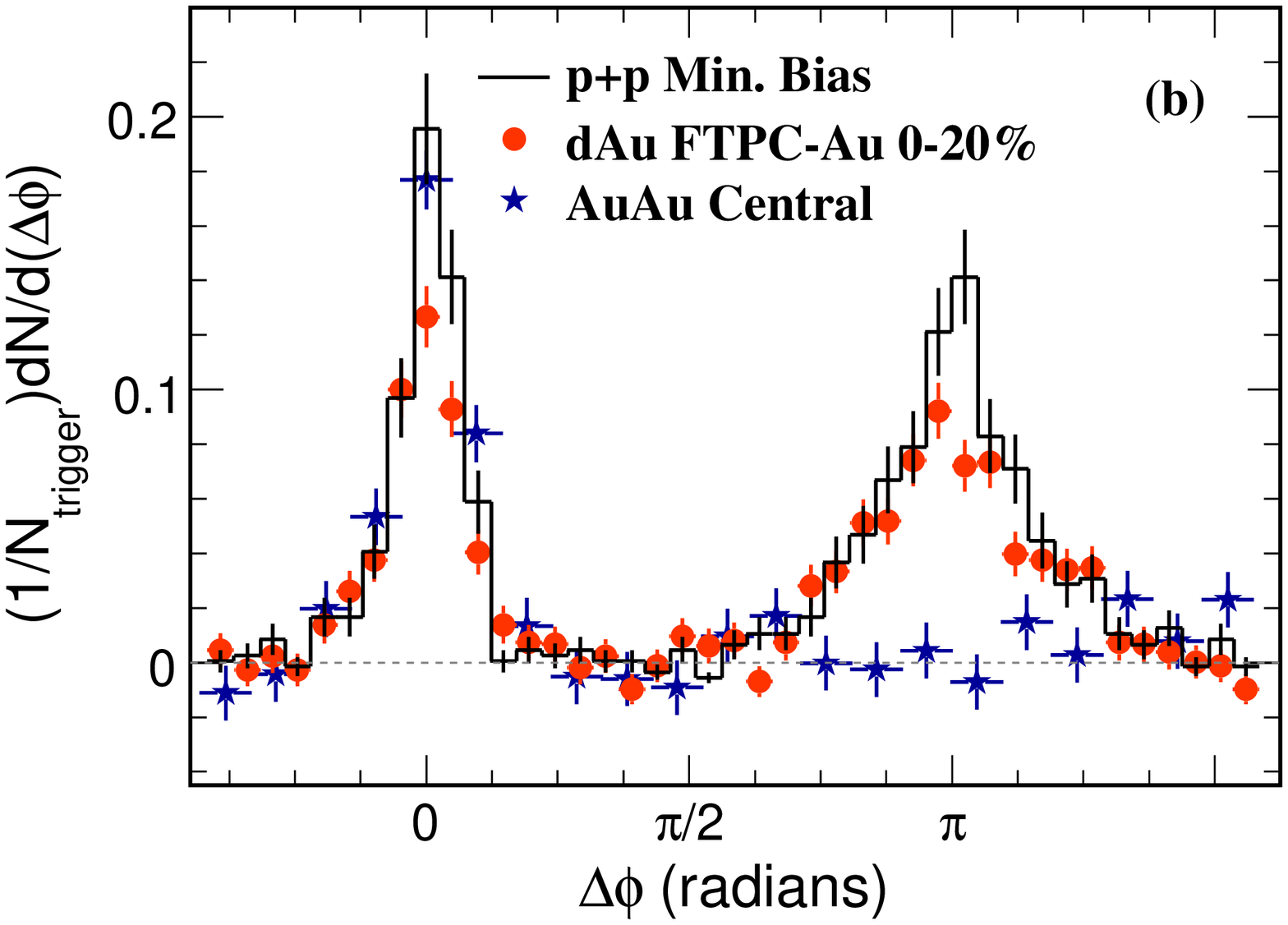}
\emn%
\caption[Nuclear modification factor and dihadron azimuthal
correlation]{Panel (a): $R_{AB}$(\pt) for minimum bias and central
\dAu\ collisions and central \AuAu\ collisions. The bands show the
normalization uncertainties. Panel (b): Two particle azimuthal
distributions in \pp, \dAu\ and central \AuAu\ collisions.}
\label{starraa} \ef

All these high \pt\ suppression or away-side jet quenching
observations indicate that light quarks strongly lose energy in
the interacting dense medium. However, due to their large quark
mass and small gluon radiative angle, heavy quarks are predicted
to lose less energy compared to light quarks since pQCD
energy-loss calculation assumes only gluon
radiation~\cite{deadcone,heavyDMPRL,armesto}. The finite mass
generalization of the small $x$ (soft radiation $x\ll$1) invariant
DGLAP radiation spectrum is given by \be \label{dceq}
\omega\frac{dN_g^{(0)}}{d^3\vec{k}}\simeq\frac{C_R\alpha_s}{\pi^2}\frac{k^2}{(k^2+m_g^2+x^2M^2)^2},
\ee where $\omega(k)$ is the energy carried by radiated gluons in
the medium with momentum $k$, and \as\ is the strong coupling
constant. The color charge factor
$C_R=\frac{N_c^2-1}{2N_c}=\frac{4}{3}$ with the number of color
flavor $N_c=3$ in this case. The gluon radiation angle is defined
as $\theta_c\equiv\sqrt{m_g^2+x^2M^2}/(xE)$. Due to large quark
mass $M$, the gluon radiation is suppressed at smaller angles
$\theta$$<$$\theta_c$. This effect is known as the "dead cone"
phenomenon.

\bf \centering \bmn[c]{0.5\textwidth} \centering
\includegraphics[width=1.\textwidth]{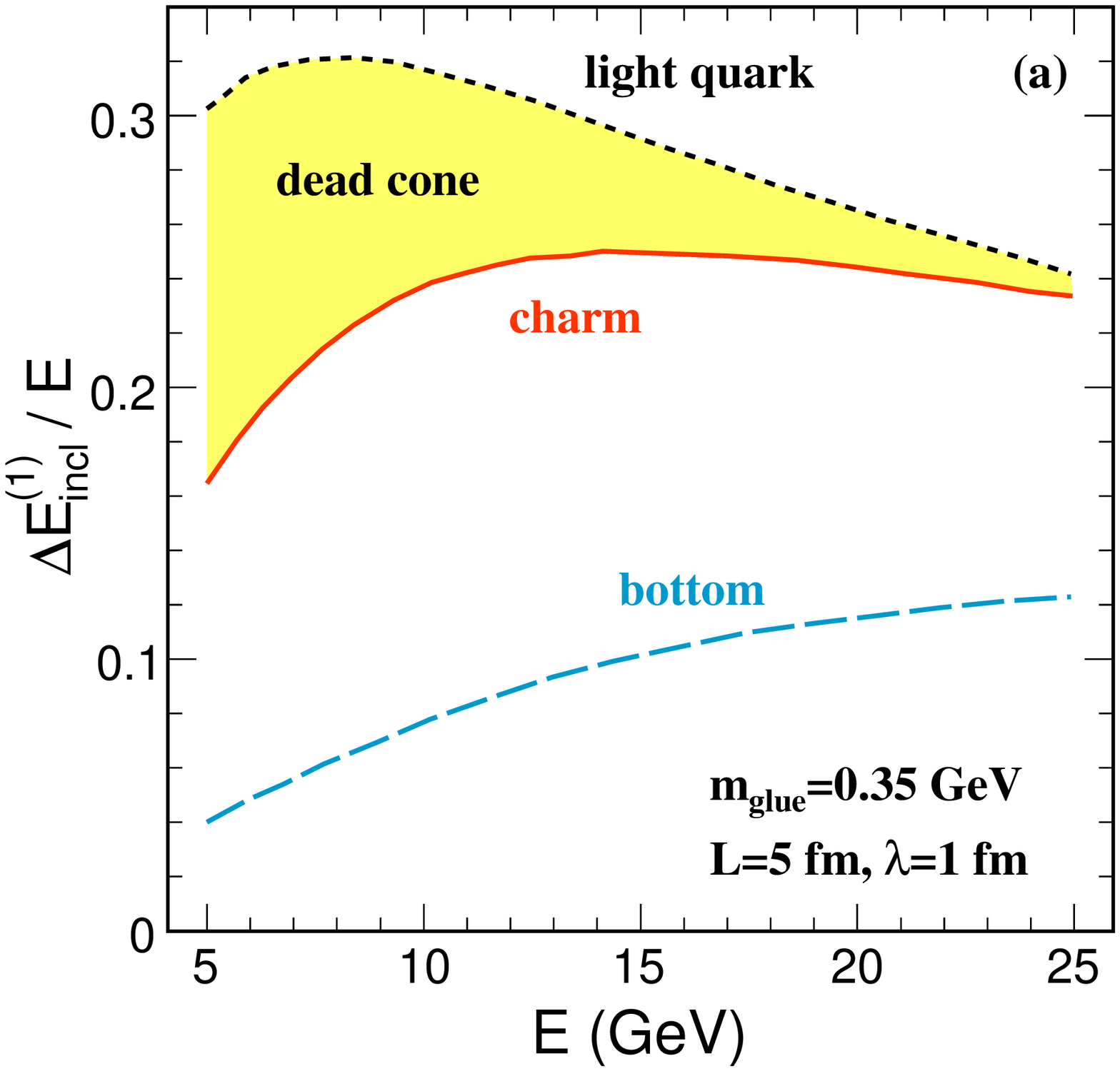}
\emn%
\bmn[c]{0.5\textwidth} \centering
\includegraphics[width=1.\textwidth]{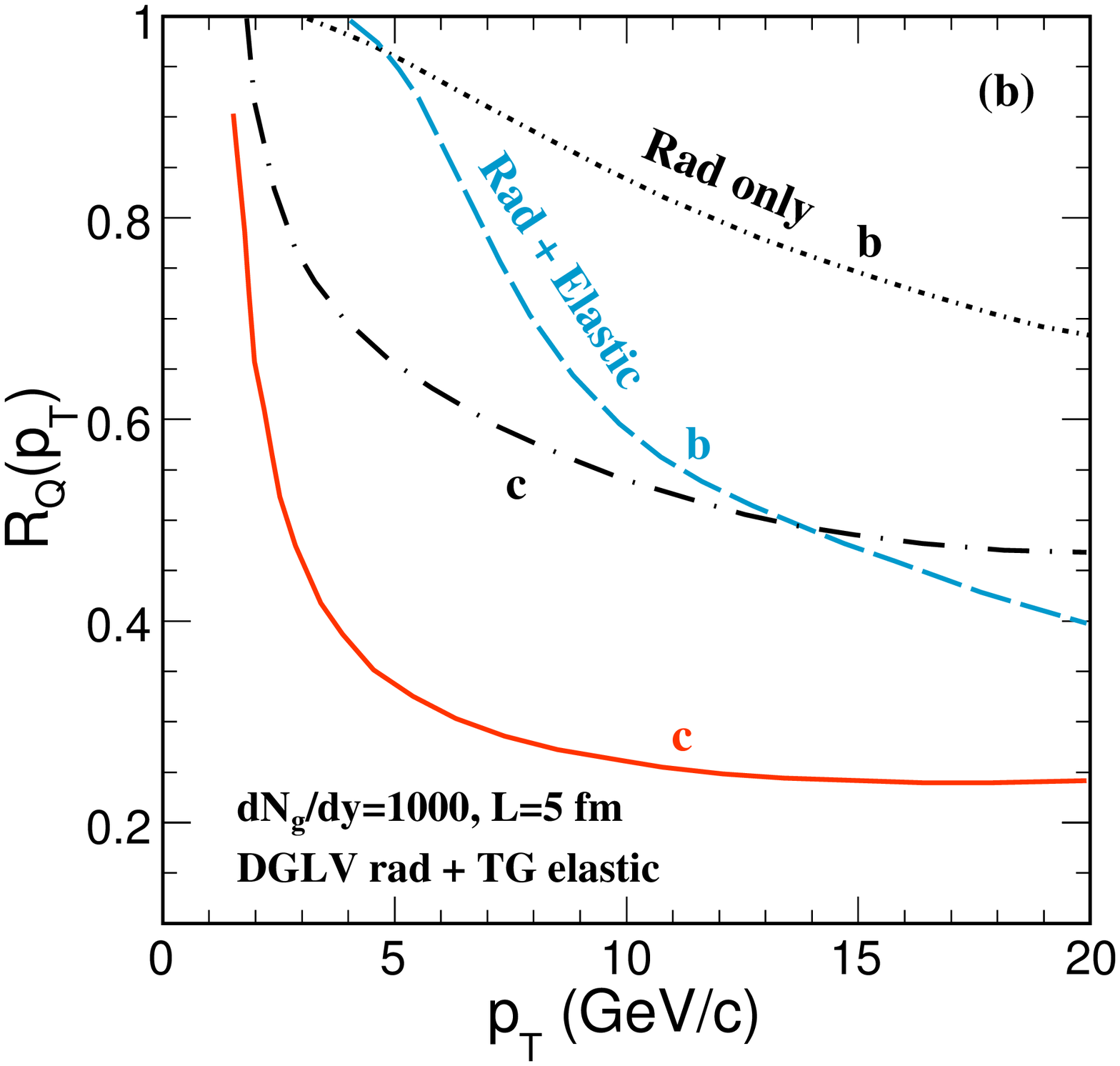}
\emn%
\caption[Theoretical calculations of heavy quark energy
loss]{Panel (a): The $1^{st}$ order in opacity fractional energy
loss for heavy quarks is shown as a function of their energy in a
plasma characterized by \as=0.3, Debye mass $\mu$=0.5 \gev, and
$L=5\lambda=5$ fm. Panel (b): Heavy quark nuclear modifications as
a function of \pt\ for fixed $L=5$ fm and $dN_g/dy=1000$. Short-
(bottom) and long- (charm) dot-dashed curve include only radiative
energy loss, while dashed (bottom) and solid (charm) curve include
elastic energy loss as well.} \label{hqeloss1} \ef

From study on the non-abelian analog of the {\em Ter-Mikayelian}
plasmon effect for gluons and the extended GLV energy loss method,
the heavy quark energy loss to first order in the opacity,
$L/\lambda_g$ ($L$ is the path length, $\lambda_g$ is the gluon
mean free path), are numerically estimated as the solid curves
(charm) and long-dashed curve (bottom) shown in
Fig.~\ref{hqeloss1} (a), which is much less than light quarks
(short-dashed curve). The shaded area indicates the so called
"dead cone" effect.

\bf \centering\mbox{
\includegraphics[width=0.7\textwidth]{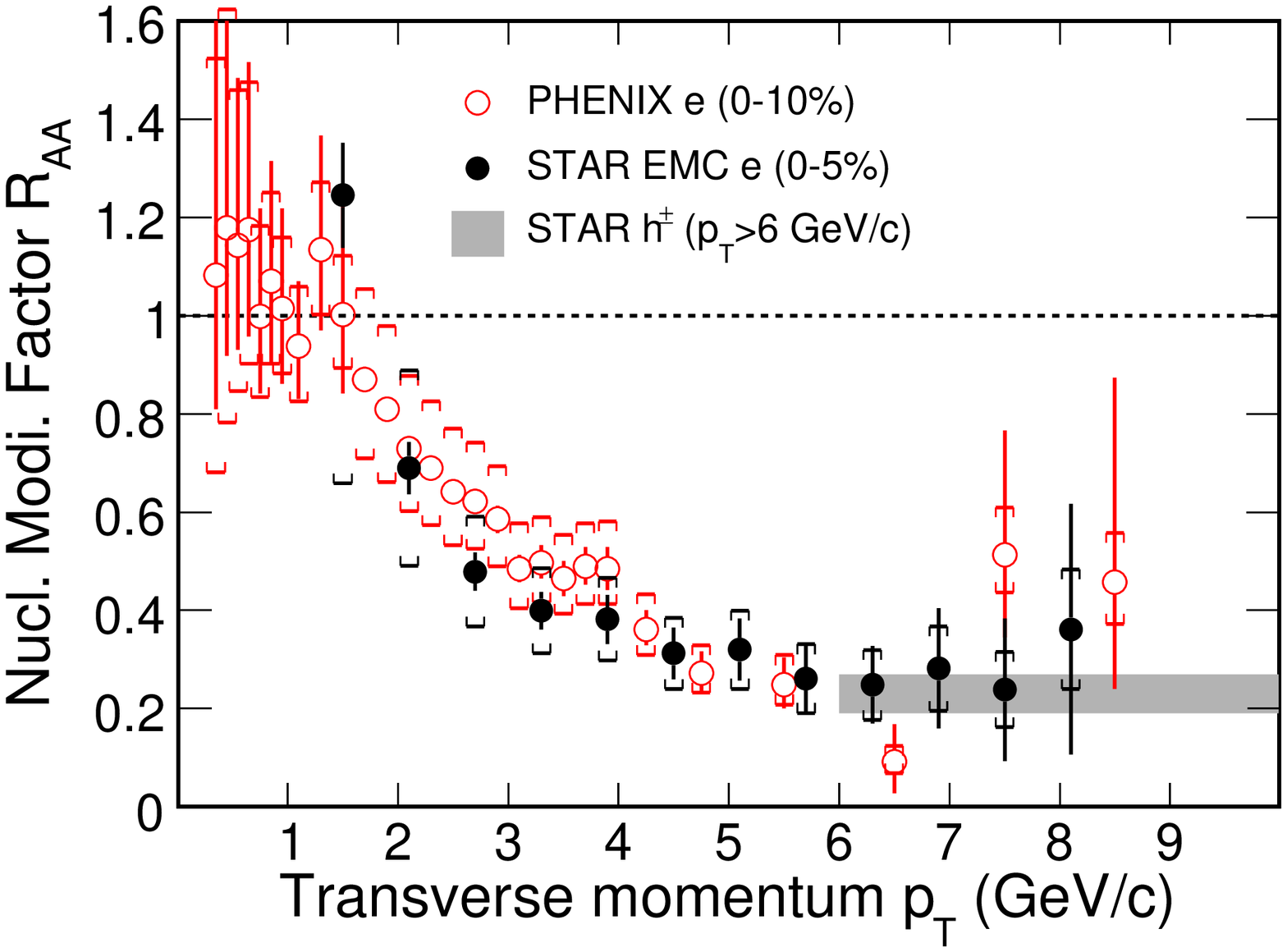}}
\caption[Non-photonic electron high \pt\ suppression]{Nuclear
modification factors of non-photonic electron from PHENIX (open
circle) and STAR (solid circle) are suppressed in similar
magnitude with light flavor hadrons (grey band).} \label{neraa}
\ef

However, recent measurements of the \pt\ distributions and nuclear
modification factor for the non-photonic electrons (presumably
from heavy quark decays) at high \pt\ show similar suppression to
those from light hadrons~\cite{starcraa,phenixcraa}, see
Fig.~\ref{neraa}. This renews the interest in understanding the
heavy quark energy loss mechanism in central \AuAu\ collisions at
RHIC. From theoretical calculations, which take elastic
(collisional) energy loss into account, the nuclear modifications
of heavy quarks are expected to have larger suppression compared
to those with only radiative energy loss~\cite{Wicks05}, see
Fig.~\ref{hqeloss1} (b). Accounting for elastic energy loss seems
to produce good agreement with experiments for heavy flavor energy
loss, but this prescription fails in the case of light flavors.
These will be discussed in more detail in chapter 5.

\subsection{Collective motion and freeze-out}
The measured hadron spectra, especially in the soft sector at
transverse momenta ${p_{T}}_{\sim}^<1.5$ \gevc, reflect the
properties of the bulk of the matter produced in heavy-ion
collisions after elastic interactions have stopped among the
hadrons at kinetic freeze-out. At this stage the system is already
relatively dilute and "cold". However, from the final state hadron
spectra at kinetic freeze-out, one can obtain the information
about the earlier hotter and denser stage. Since different hadrons
have different production (hadronization) mechanisms, the
characteristics of the different stages of the collision systems
can be explored by analysis of the transverse momentum
distributions for various hadron species. Fig.~\ref{mtspecbw} (a)
shows the measured $m_{T}$ ($\equiv\sqrt{p_{T}^2+mass^2}$) spectra
for light hadrons ($\pi, K, p$), $\Lambda$, $\Xi$ and
multi-strange hadrons ($\Phi$, $\Omega$) in 200 \gev\ central
\AuAu\ collisions~\cite{pikpspectra,k0slambda,phispectra,xiomega},
and charmed hadron ($D^0$)~\cite{YFSQM06} in 200 \gev\ minimum
bias \AuAu\ collisions (will be discussed in Chapter 4).

\bf \centering \bmn[c]{0.48\textwidth} \centering
\includegraphics[width=1.\textwidth]{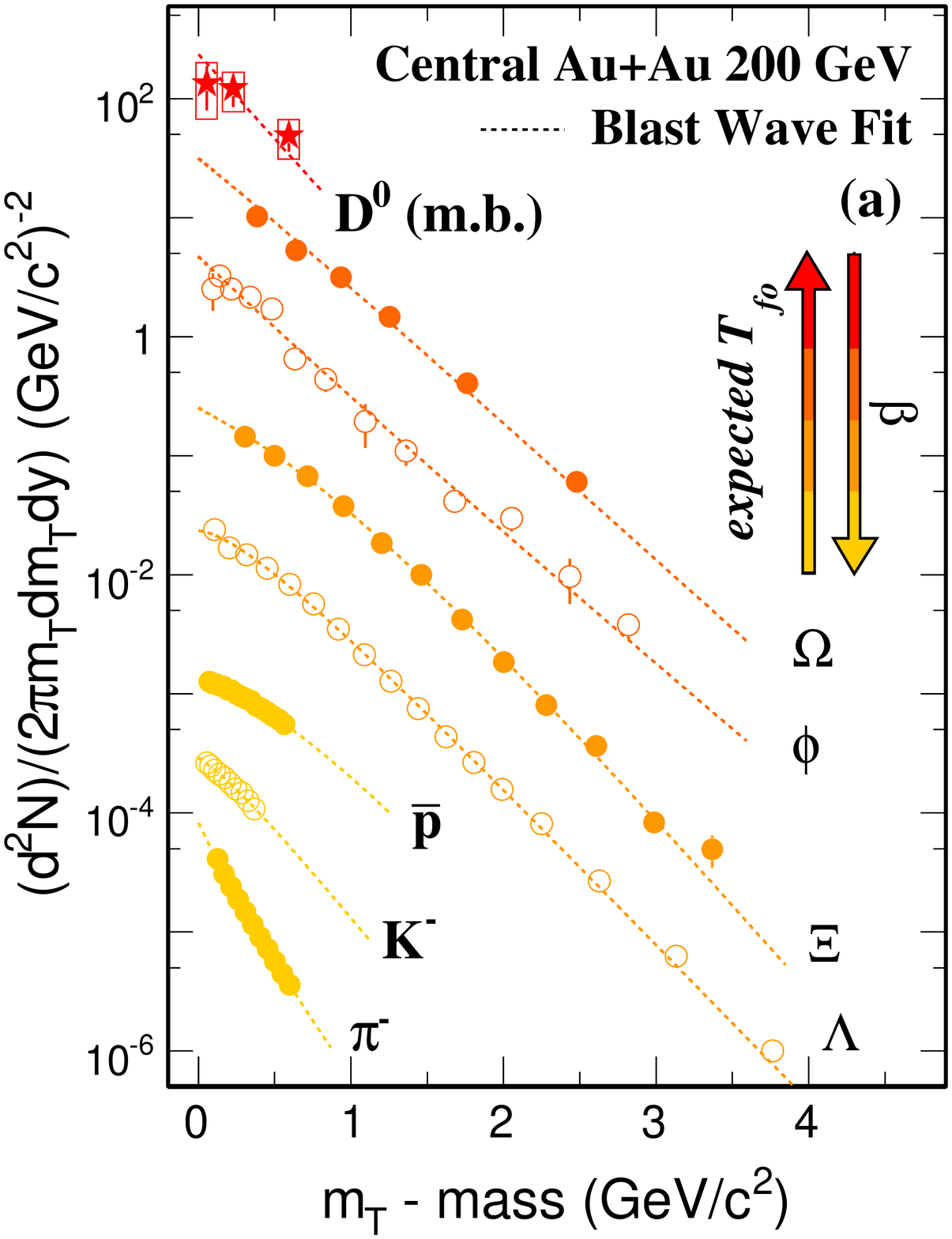}
\emn%
\bmn[c]{0.48\textwidth} \centering
\includegraphics[width=1.\textwidth]{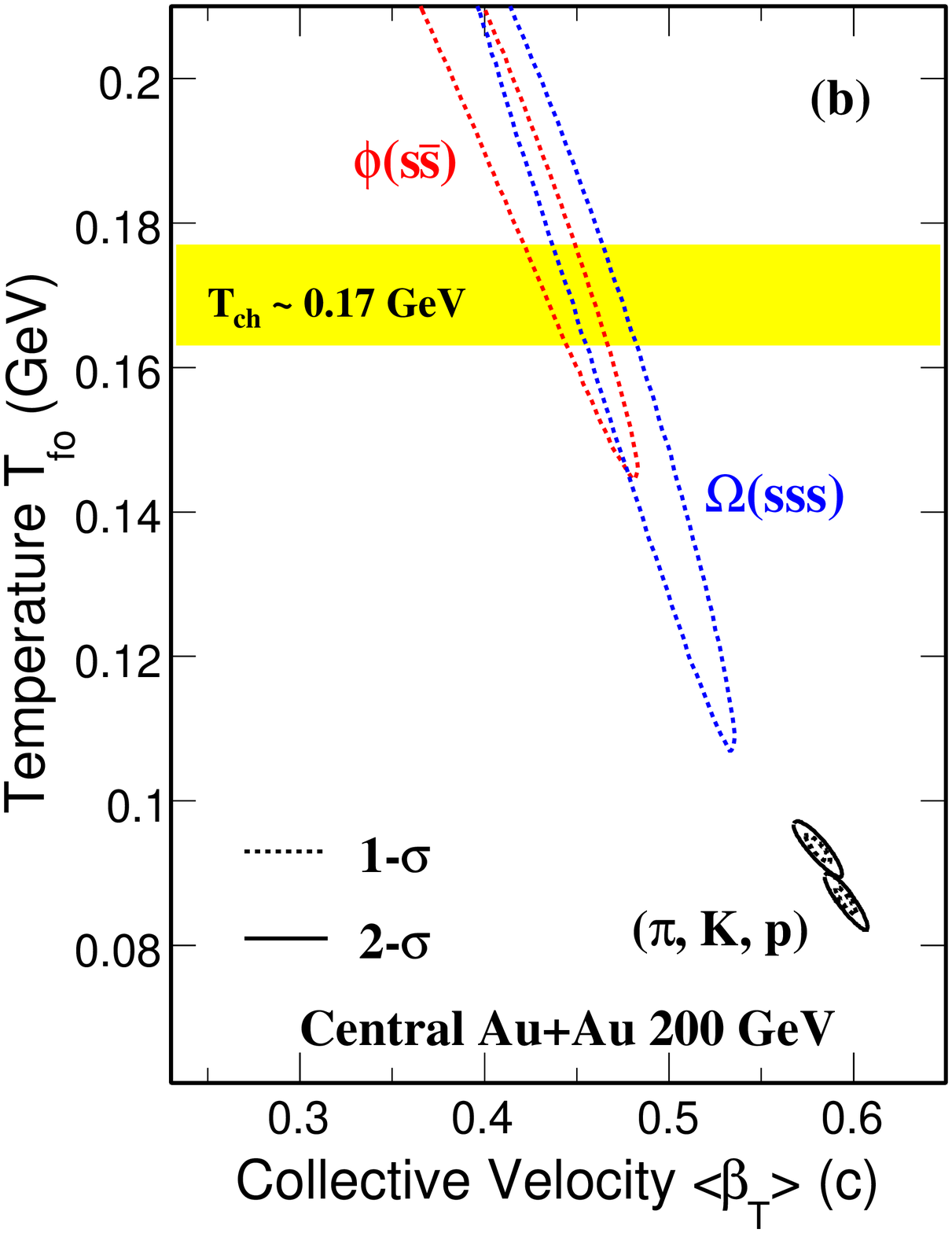}
\emn%
\caption[Blast-wave fit for PID spectra and fit parameters]{Panel
(a): The $m_{T}$ spectra for light hadrons ($\pi, K, p$),
$\Lambda$, $\Xi$ and multi-strange hadrons ($\Phi$, $\Omega$) in
200 \gev\ central \AuAu\ collisions, and charmed hadron ($D^0$) in
200 \gev\ minimum bias \AuAu\ collisions are shown in symbols. The
Blast Wave fit results are shown in curves. The BW fit for $D^0$
was done by combining $D^0$ spectrum and the measured charm decay
leptons spectra below 2 \gevc. The BW fits were done for $\pi^-$,
$K^-$, $\bar{p}$ simultaneously and for other particles
separately. The arrows show the expected increasing freeze-out
temperature and decreasing collective velocity from bottom to top.
Panel (b): Blast-wave parameters $T_{fo}$ vs. $\la\beta_T\ra$
contour plot from the simultaneous fits to light hadrons
($\pi,K,p$) spectra in 5-10\% and 0-5\% \AuAu\ collisions and
separate fits to multi-strange hadrons $\phi(s\bar{s})$,
$\Omega(sss)$ spectra in 0-10\% central \AuAu\ collisions.}
\label{mtspecbw} \ef

In the Blast Wave thermal model~\cite{blastwave}, local thermal
equilibrium is assumed and only particle mass and temperature of
the system modify particle transverse momentum. After the
collision, the bulk system expands and becomes more and more
dilute and cold, while the particle collective velocity develops
larger and larger. Extraction of model parameters characterize the
random, generally interpreted as a kinetic freeze-out temperature
$T_{fo}$ and collective, radial flow velocity $\la\beta\ra$,
aspects. Fig.~\ref{mtspecbw} (b) shows the fit parameters $T_{fo}$
versus $\la\beta\ra$ for different particles in central \AuAu\
collisions. Simultaneous fit results for light hadrons ($\pi, K,
p$) give large flow velocity and small freeze-out temperature in
central collisions, indicating that light hadrons kinetically
freeze out late with strong collectivity when system evolutes a
more rapid expansion after chemical freeze-out. Compared to light
hadrons, the larger temperature and smaller flow velocity obtained
from the fits indicates that multi-strange hadrons freeze out at
earlier stage during the evolution of the bulk system. Since
multi-strange hadrons are expected to have smaller hadronic
scattering cross sections and their transverse momentum
distributions will not change significantly after chemical
freeze-out, the kinetic freeze-out temperature from the fit to
those particles is consistent with the chemical freeze-out
temperature $T_{ch}$, albeit with still large uncertainties. The
chemical freeze-out temperature $T_{ch}$ is close to the critical
temperature $T_{c}$, indicating the temperature of the system
created in the collisions is greater than $T_{c}$ and hence the
phase transition may take place and QGP may form at RHIC energy of
\sNN=200 \gev.

Due to relatively heavy quark mass and much smaller hadronic
scattering cross section, heavy flavor hadrons are expected to
freeze-out early and difficultly participate in collective motion.
Thus the larger freeze-out temperature and smaller flow velocity
are expected for heavy flavor hadrons. But currently the
experimental statistics is not good enough to distinguish the
freeze-out properties between D-meson and multi-strange hadrons,
this will be discussed in chapter 5.

\subsection{anisotropic flow}

Recent heavy-ion collisions experiments at RHIC have successfully
generated a lot of exciting results. Besides strong suppression of
particles with high momentum created in the collisions - jet
quenching, the large azimuthal anisotropy - elliptic flow is one
of the most remarkable discoveries. All observations indicate that
the extremely opaque quark-gluon matter formed in the collisions
exhibits highly collective, near-hydrodynamic
behavior~\cite{hydrointro}. In non-central heavy-ion collisions,
the initial coordinate spacial anisotropy, which is formed by the
geometrical overlap of the two nuclei, will produce anisotropic
pressure gradients in the high dense matter which then
subsequently produce a final-state momentum space anisotropy of
the produced particles via rescattering. Because of the
self-quenching of coordinate space anisotropy, the measurement of
the particle azimuthal anisotropy distributions with respect to
(w.r.t) the reaction plane can reveal the information on the
dynamics at the early stage of the collision.

The initial space anisotropy in the overlapping region (x,y),
which is the plane transverse to the beam axis z, is characterized
by the eccentricity: \be \varepsilon=\frac{\la y^2-x^2\ra}{\la
y^2+x^2\ra} \ee

The final particle spectrum in momentum space can be expanded into
a Fourier expression in terms of particle azimuthal $\phi$
distributions w.r.t the reaction plane $\Psi_r$: \be
\label{floweq} E\frac{d^{3}N}{dp^{3}}=\frac{d^{2}N}{2\pi
p_{T}dp_{T}dy} (1+\sum_{n=1}^{\infty}2v_{n}cos[n(\phi-\Psi_{r})]),
\ee The Fourier coefficients represent anisotropy parameters and
can be extracted as \be v_n = \la cos[n(\phi-\Psi_{r})]\ra, \ee
where the first and second coefficients $v_1$, $v_2$ are called
the directed and elliptic flow.

The identified particle elliptic flow $v_2$ shows strong
transverse momentum dependence. In the low \pt\ region, particle
collective motion shows hydrodynamical
behavior~\cite{hydrointro,starwhitepaper}. As is well known,
hydrodynamic models, which assume ideal relativistic fluid flow
and negligible relaxation time compared to the time scale of the
equilibrated system, successfully reproduce the experimental data
and reasonably describe the mass ordering for $v_2$ at low \pt.
Thus the particle $v_2$ follows $m_{T}-m$
scaling~\cite{Yanthesis}, as shown in Fig.~\ref{v2mt} (a). Here
$m_T\equiv\sqrt{p_T^2+m^2}$ and $m$ is the particle mass. On a
$m_{T}-m$ scale, at intermediate \pt\ (1.5 GeV/$c<p_T<4$ \gevc),
the particle $v_2$ is split into meson and baryon groupings, and
this phenomenon may be described by coalescence/recombination
models~\cite{coalKo,coalMolnar}. In these models, mesons and
baryons are hadronized by coalescing two and three co-moving
quarks, respectively. The simple picture is that the hadron
$v_2^h$ can be described by its constituent quark $v_2^q$ as: \be
v_2^h(p_T)\simeq n_qv_2^q(p_T/n_q), ~~~~n_q(meson)=2,
~~~~n_q(baryon)=3, \ee where $n_q$ is the number of constituent
quarks. Thus, $v_2$ may follow the {\em Number-of-Quark} (NQ)
scaling. This scaling behavior is observed in \AuAu\ collisions at
RHIC. Fig.~\ref{v2mt} (b) shows the scaling of identified particle
$v_2$ divided by $n_q$ as a function of $(m_T-m)/n_q$. These
exciting results are considered as indicators for the observation
of the partonic level collectivity (deconfinement) in heavy-ion
collisions at RHIC energies.

\bf \centering \bmn[c]{0.5\textwidth} \centering
\includegraphics[width=1.\textwidth]{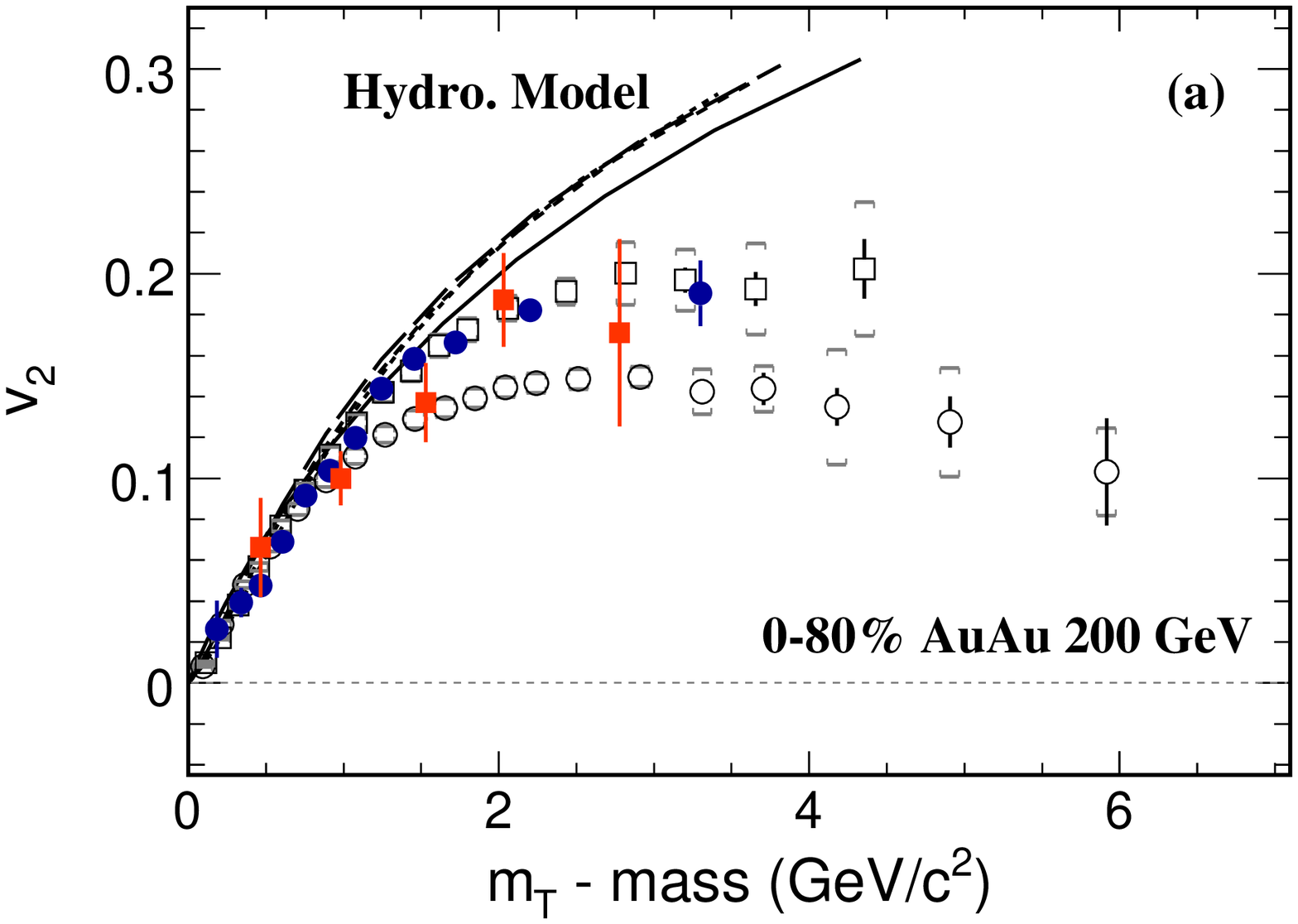}
\emn%
\bmn[c]{0.5\textwidth} \centering
\includegraphics[width=1.\textwidth]{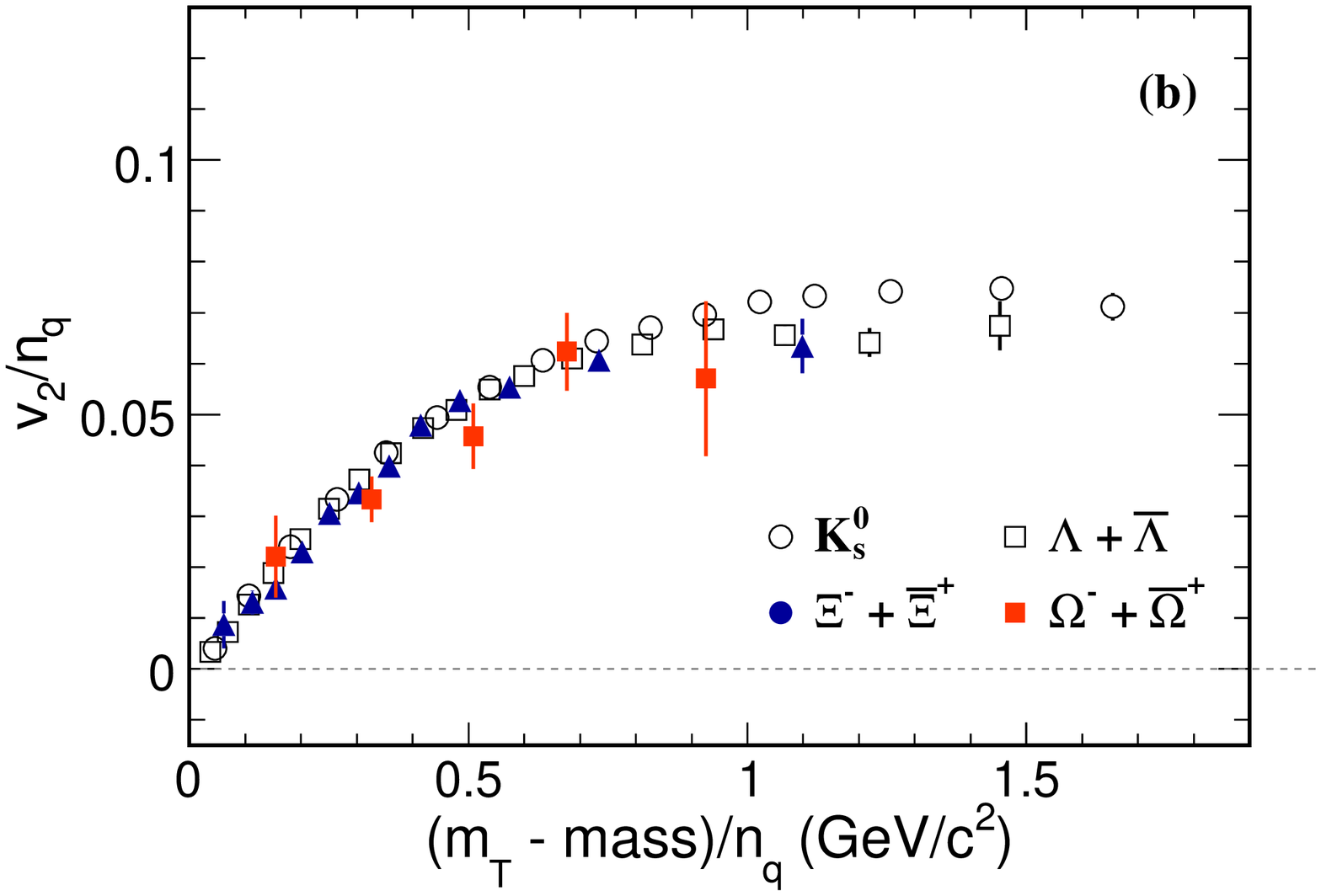}
\emn%
\caption[Identified particle elliptic flow $v_2$ and its $m_T-m$
scaling]{Panel (a): Identified particle $v_2$ as a function of
$m_T-m$ in 0-80\% \AuAu\ collisions at \sNN=200 GeV. Open circles,
open squares, solid triangles and solid circles represent $K_s^0$,
$\Lambda+\bar{\Lambda}$, $\Xi+\bar{\Xi}$ and
$\Omega+\bar{\Omega}$, respectively. Hydrodynamic calculations are
shown as curves for comparison. Panel (b): Identified particle
$v_2$ scaled by the number of constituent quarks ($n_q$) versus
$(m_T-m)/n_q$.} \label{v2mt} \ef

\subsection{Charm production at RHIC}

Charm quarks are a unique tool to probe the partonic matter
created in relativistic heavy-ion collisions at RHIC energies.
Because of their large quark mass ($\simeq 1.3$ GeV/$c^{2}$),
charm quarks are predicted to lose less energy compared to light
quarks when pQCD energy-loss calculation assumes only gluon
radiation~\cite{deadcone,heavyDMPRL,armesto}. Recent measurements
of the \pt\ distributions and nuclear modification factor for the
non-photonic electrons (presumably from heavy quark decays) at
high \pt\ show similar suppression to those from light
hadrons~\cite{starcraa,phenixcraa}. This renews the interest in
understanding the picture of the observed jet-quenching phenomena
at the partonic stage in \AuAu\ collisions at RHIC. Furthermore,
theoretical calculations have shown that interactions between the
surrounding partons in the medium and charm quarks could change
the measurable kinematics\cite{teaney,vanHCharmflow,dd1}, and
could boost the radial and elliptic flow resulting in a different
charm \pt\ spectrum shape. Due to its heavy mass, a charm quark
can acquire flow from the interactions with the constituents of a
dense medium in analog to Brownian motion. Its decoupling from
medium also depends on the cross-section of the interaction and
medium density. Therefore, the measurement of charm flow and
freeze-out properties is vital to test light flavor thermalization
and the partonic density in the early stage of heavy ion
collisions
\cite{xu2flow,linCharmflow,kocharmflow,teaney,jamiecharmflow}. The
charmed hadron kinetic freeze-out temperature $T_{fo}$ and the
radial flow velocity $\beta$ can be derived from blast-wave
model~\cite{blastwave} fits to data.

These postulations and predictions are based on the assumptions
that charm quarks are produced only at early stages via initial
gluon fusion and that their production cross-section can be
calculated by perturbative QCD \cite{LinPRC,cacciari}. Study of
the binary collision (\nbin) scaling properties for the total
charm cross-section from \dAu\ to \AuAu\ collisions can test these
assumptions and determine if charm quarks are indeed good probes
with well-defined initial and final states. Charm total
cross-section measurement is also essential for model
calculations, which tries to explain observed similar suppression
pattern of $J/\Psi$ at RHIC and
SPS~\cite{pbm,pbmjpsi,luicjpsi,Kosjpsi,kinematicjpsi,mclerranjpsi}.
It has been shown that low \pt\ muons and intermediate \pt\
electrons from charmed hadron semileptonic decays can stringently
constrain the charm total cross-section and are sensitive to the
radial flow and the freeze-out condition~\cite{ffcharm}.

\subsection{Bottom contribution in non-photonic electron measurements}

Due to the absence of the measurement of B-mesons and precise
measurement of D-mesons, it is difficult to separate bottom and
charm contributions experimentally in current non-photonic
electron measurements for both spectra and elliptic flow $v_2$. As
discussed previously, the suppression behavior of heavy quarks is
quite different from light quarks due to the "dead cone"
effect~\cite{deadcone}, and this is especially true for the bottom
quark. Even when the elastic energy loss is included, the bottom
quark still loses much less energy. The bottom contribution may
reduce the energy loss of non-photonic electrons from heavy flavor
decays. But as shown in Fig.~\ref{eraav2theory} (a), the
suppression of the non-photonic electron \raa\ is as large as
light hadrons. Both the theoretical result with charm energy loss
only and the theoretical calculations with charm+bottom energy
loss by assuming large $\hat{q}$ or counting elastic energy loss
can describe the data within
errors~\cite{DGLV06,Wicks05,RappRaa,Ivancoll,armeloss}.

\bf \centering \bmn[c]{0.48\textwidth} \centering
\includegraphics[width=1.\textwidth]{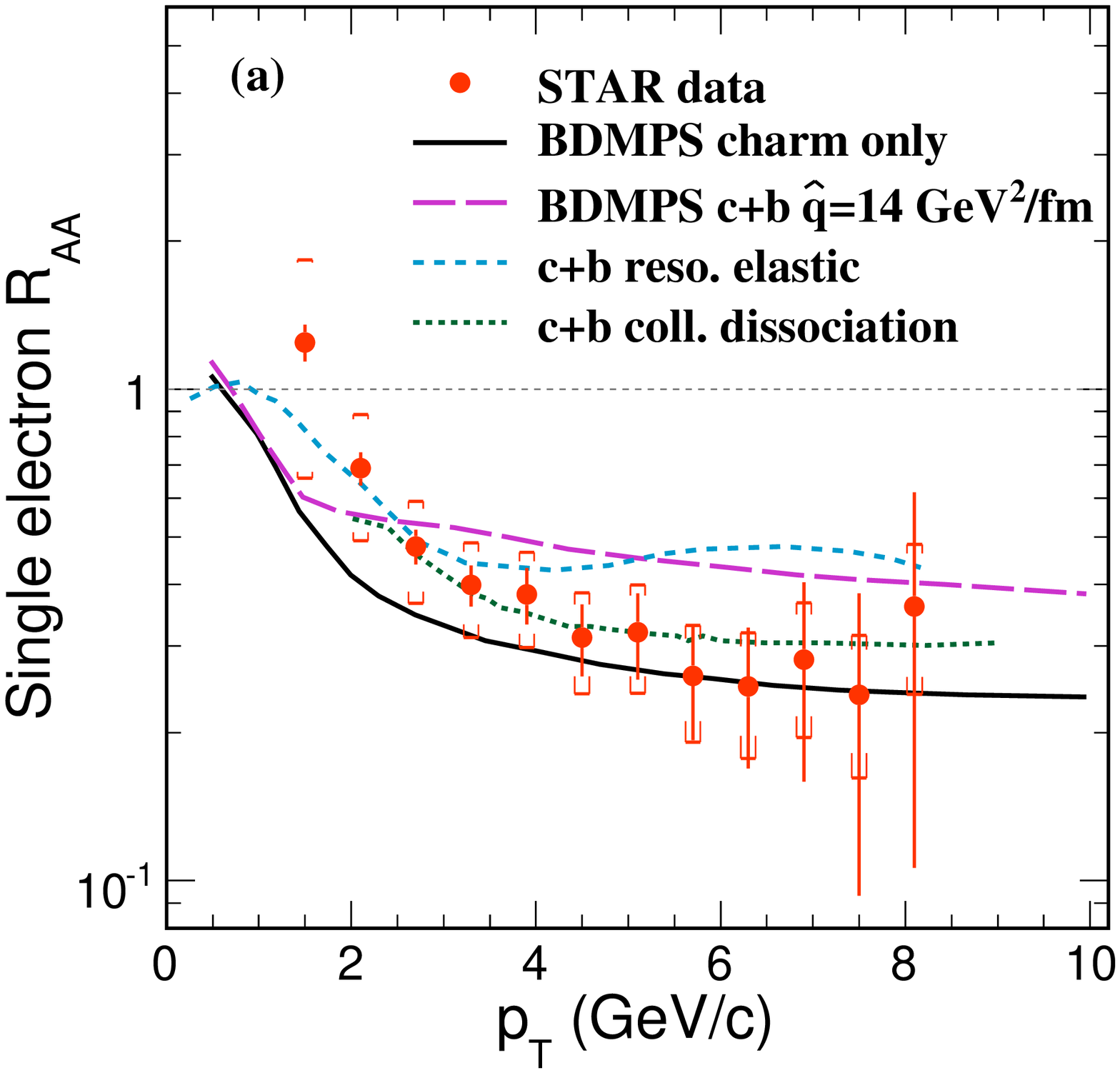}
\emn%
\bmn[c]{0.48\textwidth} \centering
\includegraphics[width=1.\textwidth]{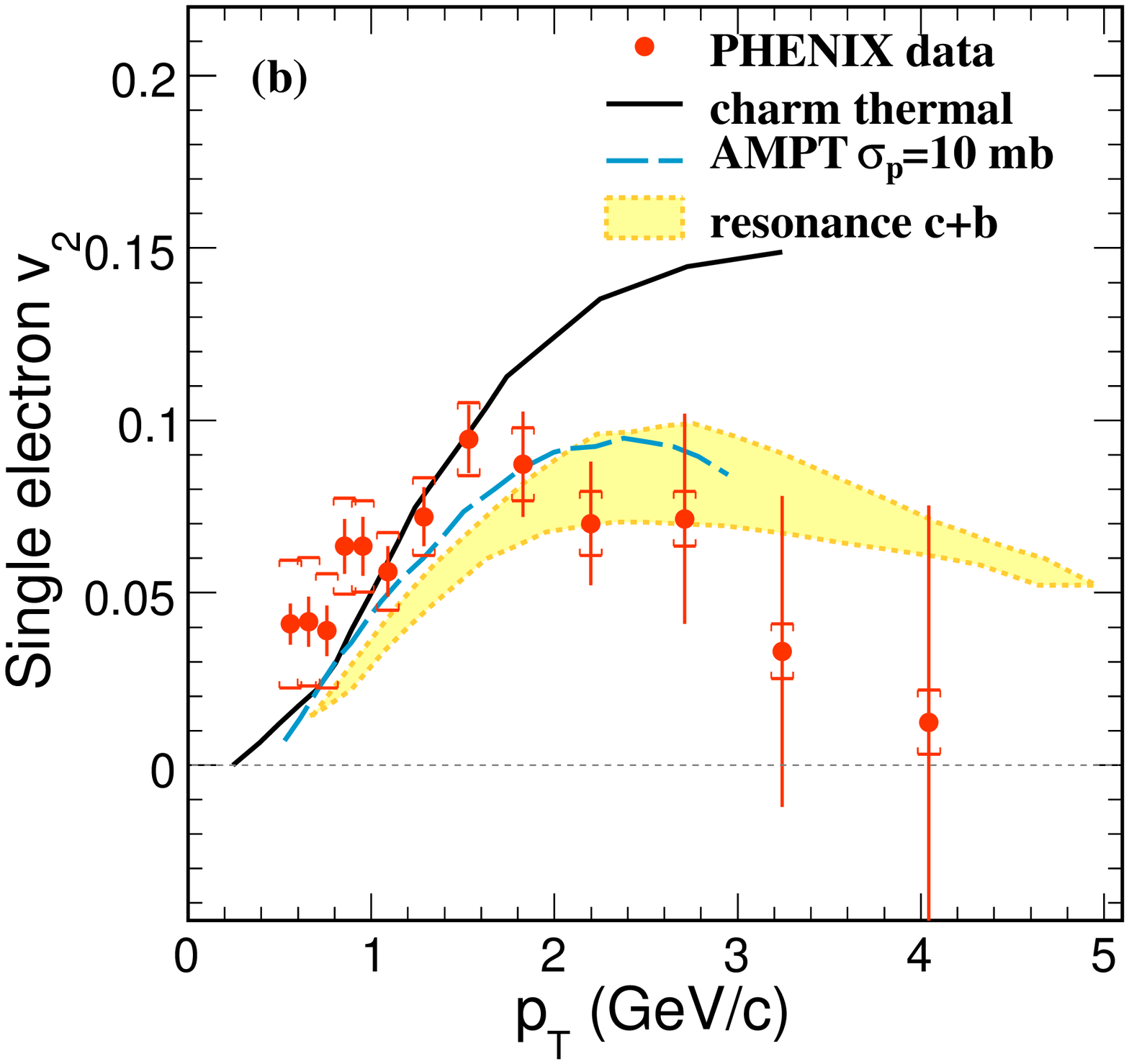}
\emn%
\caption[Comparisons of non-photonic electron \raa\ and $v_2$ with
theories]{Panel (a): Non-photonic electron \raa\ measured from
STAR (circle) compared to theories. Panel (b): Non-photonic
electron $v_2$ measured from PHENIX (circle) compared to
theories.} \label{eraav2theory} \ef

Recently, PHENIX has measured the non-photonic electron
$v_2$~\cite{Phenixv2}, shown in Fig.~\ref{eraav2theory} (b). The
observed large elliptic flow of the non-photonic electron may
indicate strong coupling of heavy quarks with medium. There are
many theoretical calculations for the non-photonic electron $v_2$,
such as charm thermal+flow model (solid curve)~\cite{kocharmflow},
A multi-phase transition (AMPT) model assume cross section
$\sigma_{p}$=10 mb (dashed curve)~\cite{AmptCharmflow}, resonance
states of D-/B- mesons (band)~\cite{vanHCharmflow}, {\em etc.} The
comparison with theories also showes that both the model results
with charm only and the results with charm+bottom have good
agreement with data within errors.

Thus, the puzzle of the bottom contributions in non-photonic
electron spectra and $v_2$ still remains. We will have a detail
discussion about this issue in chapter 5.

\chapter{Experimental Set-up}

\section{The RHIC accelerator}

The Relativistic Heavy Ion Collider (RHIC) located at Brookhaven
National Laboratory (BNL) is built to accelerate and collide heavy
ions and polarized protons with high luminosity. The RHIC
accelerator facility consists of two concentric storage rings,
called blue and yellow rings, sharing a common horizontal plane in
the tunnel. The super-conducting magnets in each ring with a
circumference of 3.8 km are designed to bend and focus the ion
beams. The counter-rotating beams can collide with one another at
six location along their 3.8 km circumference. The top
center-of-mass collision energy for heavy ion beams is 200 GeV per
nucleon pair. The operational momentum increases with the
charge-to-mass ratio, resulting in the top energy of 125 GeV/u for
lighter ion beams and up to 250 GeV/u for polarized proton beams.
The average luminosity for gold-on-gold collisions at \sNN = 200
GeV is $8\times10^{26} cm^{-2}s^{-1}$ without electron cooling and
$7\times10^{27} cm^{-2}s^{-1}$ with electron cooling.

Fig.~\ref{rhic} shows the RHIC accelerator complex including the
accessorial accelerators used to bring the ions beams up to RHIC
injection energy, and strip electrons from the atoms. Negatively
charged gold ions are partially stripped of their electrons and
then accelerated to 15 MeV/u in the Tandem Van de Graaff facility.
After a charge selection by bending magnets, beams of gold ions
are transferred to the Booster Synchrotron and accelerated to 95
MeV/u through the Tandem-to-Booster line. Then the gold ions are
injected into the Alternating Gradient Synchrotron (AGS) and
accelerated to 10.8 GeV/u. The beams are transferred to RHIC
through the AGS-to-RHIC Beam Transfer Line. Finally, beams are
injected to RHIC and fully accelerated to the top collision energy
100 GeV/nucleon. For \pp\ collisions, proton beams are injected
from the 200 MeV Linac into the booster, followed by acceleration
in the AGS and injection into RHIC.

\bf \centering\mbox{
\includegraphics[width=0.8\textwidth]{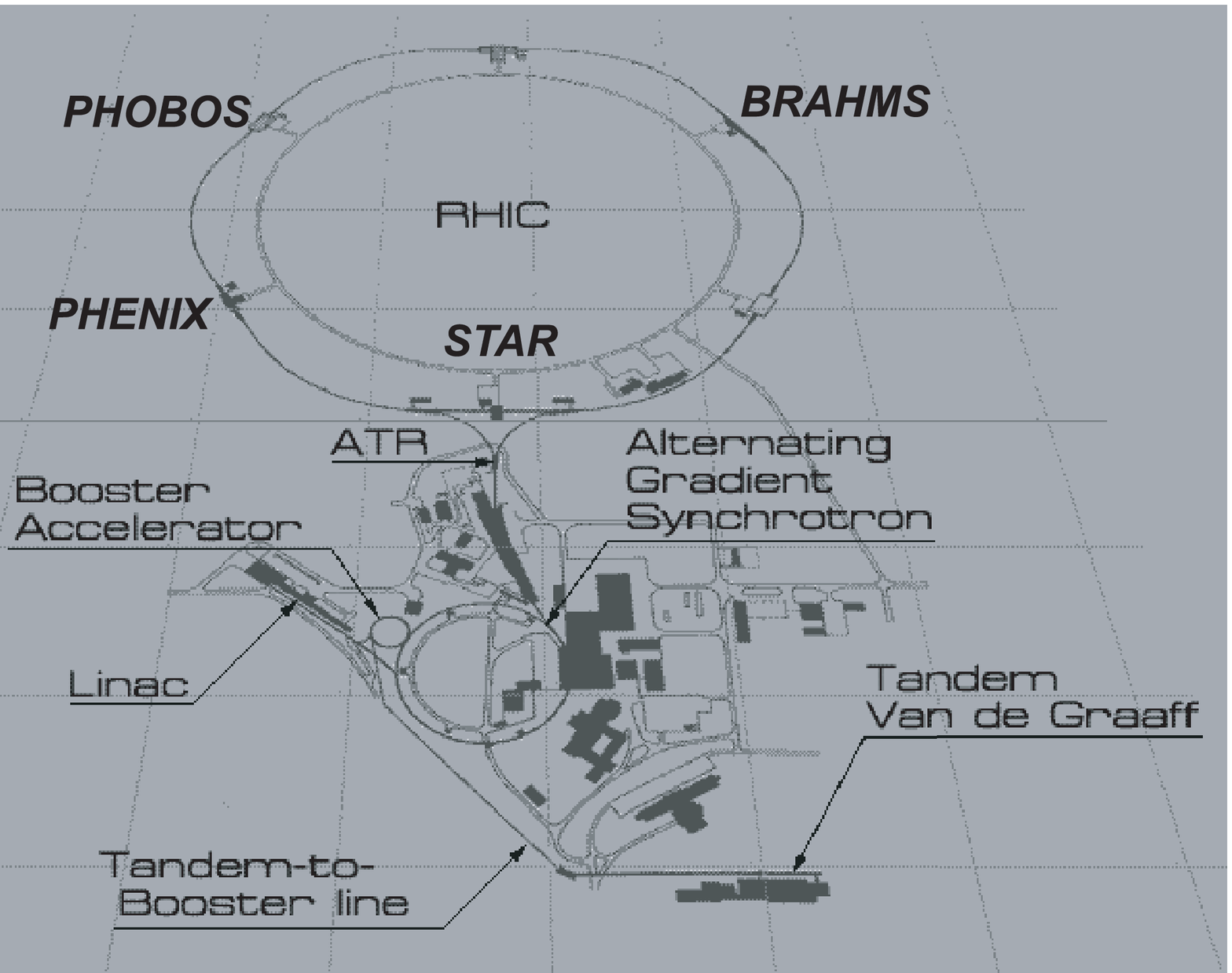}}
\caption[The RHIC complex]{Schematic of the RHIC complex. RHIC's
two 3.8-kilometer rings collide relativistic heavy ions and
polarized protons at six intersection points.} \label{rhic} \ef

There are six interaction points along the rings, and 4 of them
are equipped with detectors. They are two large experiments STAR
(6 o'clock), PHENIX (8 o'clock) and two small ones PHOBOS (10
o'clock) and BRAHMS (2 o'clock), respectively.

\section{The STAR detector}

To investigate the behavior of strongly interacting matter at high
energy density and to search for signatures of the new matter form
of QGP, The Solenoidal Tracker at RHIC (STAR) was specially
constructed for measurements of hadron production over a large
solid angle with high precision tracking, high quality momentum,
and good particle identification capability. It has an azimuthal
symmetric acceptance and covers large range around mid-rapidity.
STAR consists of several subsystems and a main tracker - the {\em
Time Projection Chamber} (TPC) located in a homogenous solenoidal
analyzing magnet.

\bf \centering\mbox{
\includegraphics[width=0.7\textwidth]{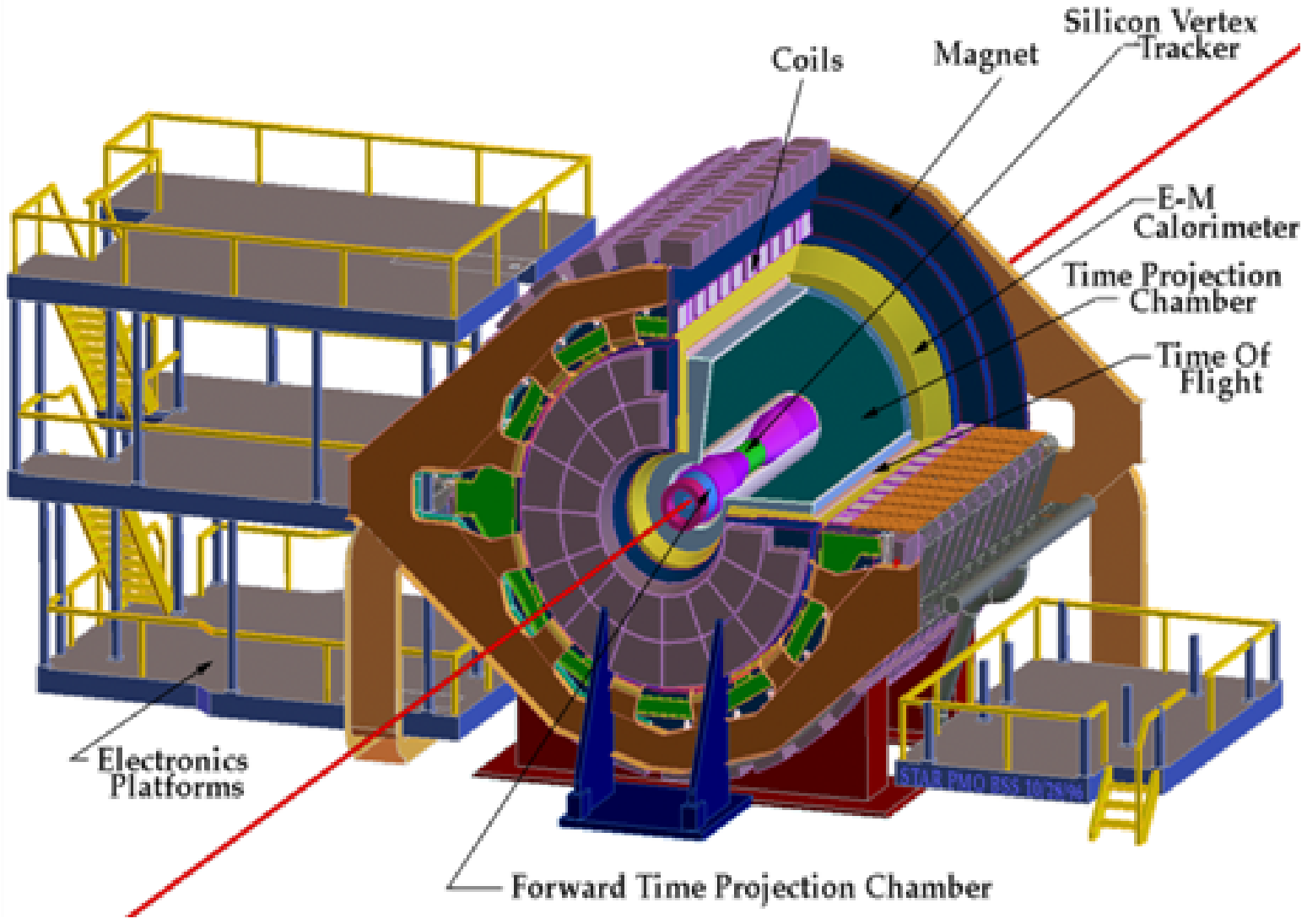}}
\caption[Perspective view of STAR detector]{Perspective view of
STAR detector, with a cutaway for viewing inner detector systems.}
\label{starlo} \ef

The layout of the STAR experiment is shown in Fig.~\ref{starlo}.
Fig.~\ref{star} shows the cutaway view of the STAR detector. The
STAR magnet is cylindrical in design with a length of 6.85 m and
has inner and outer diameters of 5.27 m and 7.32 m, respectively.
A uniform magnetic field of 0.5 T (full field or reversed full
field) or 0.25 T (half field) is provided for the tracking of
charged particles. The main tracker - TPC for charged particle
tracking and identification is 4 meters long and located at a
radial distance from 50 to 200 cm from the beam axis including 45
layers. It covers a pseudo-rapidity range $|\eta|<1.5$ and in full
azimuth ($2\pi$). There are inner detectors {\em Silicon Vertex
Tracker} (SVT) and {\em Silicon Strip Detector} (SSD) close to the
interaction vertex, which provides additional high precision space
points on track so that it improves the position resolution and
allows us to reconstruct the secondary vertex of weak decay
particles. Both the TPC and SVT contribute to particle
identification using ionization energy loss (\dedx), with an
anticipated combined energy loss resolution of $\sigma=7\%$. The
momentum resolution reaches a value of $\delta p/p=0.02$ for a
majority of the tracks in the TPC. The momentum resolution
improves as the number of hit points (nhitpts) along the track
increases and as the particle¡¯s momentum decreases, as expected.

\bf \centering\mbox{
\includegraphics[width=0.8\textwidth]{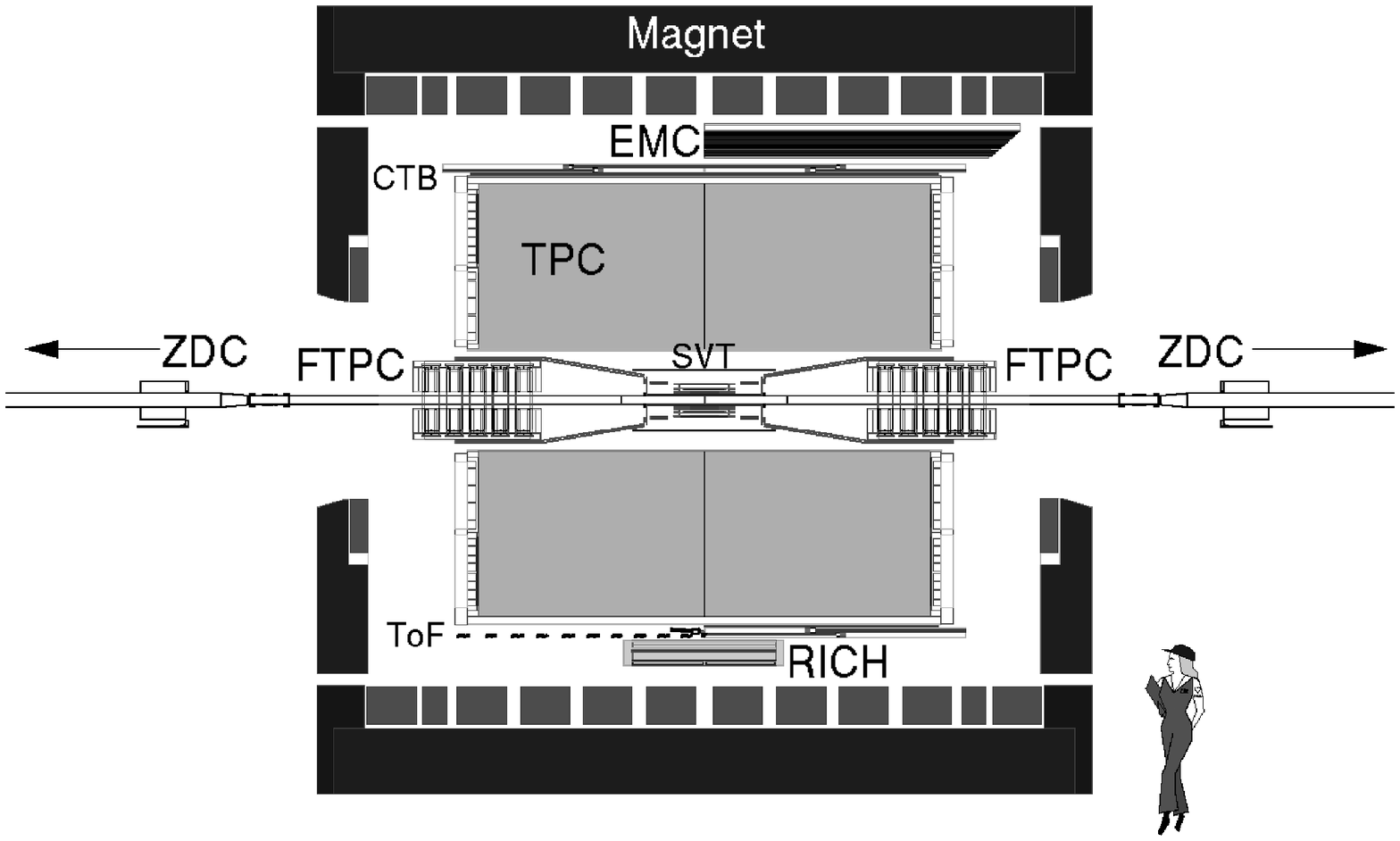}}
\caption[Cutaway view of STAR]{Cutaway view of the STAR detector.
It includes the partial installed ElectroMagnetic Calorimeter
(EMC) and two prototypes Time-of-Flight (TOF) detectors.}
\label{star} \ef

In the extended large rapidity region, there are two {\em Forward
TPC} (FTPC) detectors installed covering $2.8<|\eta|<3.8$ with
complete azimuth to track particles at forward and backward
rapidity. One prototype tray of {\em Time-Of-Flight} (TOF)
detector using scintillator materials (TOFp) was installed since
Run II and another prototype tray of TOF detector using {\em
Multigap Resistive Plate Chamber} (MRPC) technology (TOFr) was
installed since Run III, covering $-1<|\eta|<0$ and $1\over120$ of
full azimuth. In future run, TOFp will be removed and a full
barrel TOF detector based on MRPC technique will be installed,
which is expected to extend the PID capability of STAR greatly.
Part of barrel {\em ElectronMagnetic Calorimeter} (BEMC) and
endcap EMC (EEMC) were also installed since Run II. They are used
to measure the electromagnetic probes - electrons and photons.

The fast detectors are designed for providing inputs of the
trigger system, such as {\em Zero-degree Calorimeter} (ZDC), {\em
Central Trigger Barrel} (CTB) and {\em Beam-Beam Counters} (BBC).
Two ZDCs locates on each side $\sim 18$ m away from the collision
points. Each is centered at $0^\text{o}$ and covers $\sim 2.5$
mrad. In the collision, usually just the overlap parts of the two
colliding ions interact with each other. The left fragments are
continuously going forward. The charged fragments are bent by the
dipole magnets and the outgoing neutrons can be detected by the
ZDCs. The ZDC signals are used for monitoring the heavy ion beam
luminosity and for the experiments triggers. A minimum bias
trigger was obtained by selecting events with a pulse height
larger than that of one neutron in each of the forward ZDCs, which
corresponds to 95 percent of the geometrical cross section. The
CTB is a collection of scintillating tiles surrounding the outer
cylinder of the TPC. The CTB will be mostly used to select central
triggered events in heavy ion collisions by measuring the
occupancy of those CTB slats. 
The BBC subsystem covers $3.3<|\eta|<5.0$, measuring the
"beam-jets" at high rapidity from {\em Non-Singly Diffractive}
(NSD) inelastic \pp\ interactions. It consists of two disk shaped
scintillating detectors, with one placed at each endcap of the TPC
(3.5 m from TPC center). Each BBC disk is composed of
scintillating tiles that are arranged in a hexagonal closest
packing. The \pp NSD trigger sums the output of all tiles on each
BBC and requires a coincidence of both BBC's firing above noise
threshold within a time window.

\section{The Time Projection Chamber - TPC}

TPC is the primary tracking device of the STAR
detector~\cite{tpctech}. The TPC is designed to record the tracks
of particles, provide the information of their momenta, and
identify the particles by measuring their ionization energy loss
(\dedx). Charged particles are identified over a large momentum
range from 0.15 \gevc\ to $\sim30$ \gevc. For half field, the
lower limit $\sim0.075$ \gevc\ can be achieved. The TPC consists
of a 4.2 m long cylinder with 4.0 m in diameter. The cylinder is
concentric with the beam pipe, and the inner and outer radii of
the active volume are 0.5 m and 2.0 m, respectively. The TPC
covers the full region of azimuth ($0<\phi<2\pi$) and covers the
pseudo-rapidity range of $|\eta|<2$ for inner radius and
$|\eta|<1$ for outer radius. Fig.~\ref{tpc} shows a cutaway view
of the TPC structure.

\bf \centering\mbox{
\includegraphics[width=0.8\textwidth]{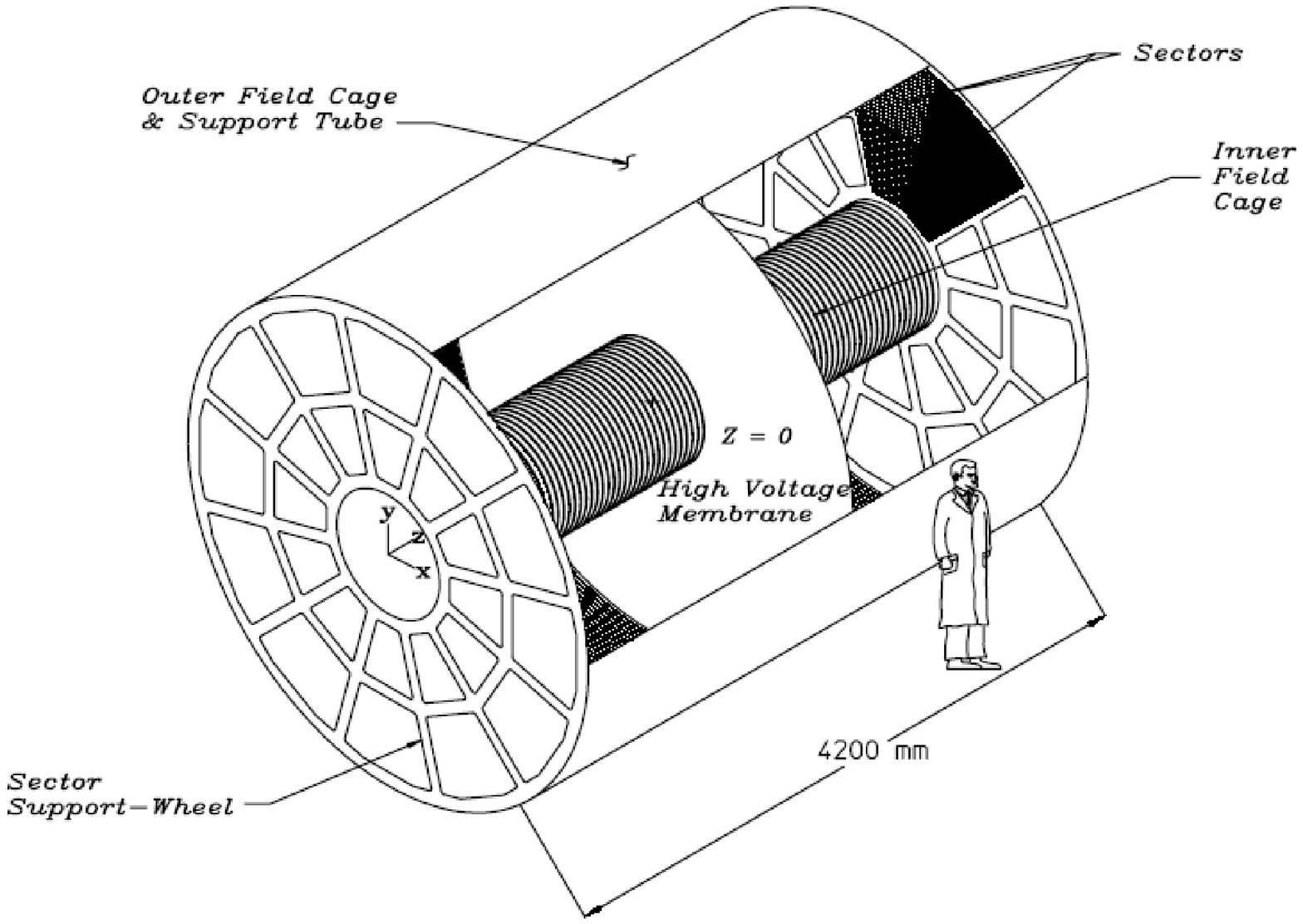}}
\caption[Cutaway view of the TPC detector]{Cutaway view of the TPC
detector at STAR.} \label{tpc} \ef

The TPC is divided into two parts by the central membrane. It is
typically held at 28 kV high voltage. A chain of 183 resistors and
equipotential rings along the inner and outer field cage create a
uniform drift filed ($\sim 135$ V/cm) from the central membrane to
the ground planes where anode wires and pad planes are organized
into 12 sectors for each sub-volume of the TPC. The working gas of
the TPC is P10, the mixture of $90\%$ Ar and $10\%$ CH$_4$,
regulated at 2 mbar above the atmospheric pressure. The electron
drift velocity in P10 is relatively fast, $\sim 5.45$ cm/$\mu$s at
$130$ V/cm drift field. The gas mixture must satisfy multiple
requirements and the gas gains are $~\sim 3770$ and $\sim 1230$
for the inner and outer sectors working at normal anode voltages
(1170 V for inner and 1390 V for outer), respectively. The readout
system is based on the {\em Multi-Wire Proportional Chamber}
(MWPC) with readout pads. Each readout pad is also divided into
inner and outer sub-sectors, while the inner sub-sector is
designed to handle high track density near collision vertex.
136,608 readout pads provide $(x,y)$ coordinate information, while
$z$ coordinate is provided by 512 time buckets and the drift
velocity. Typical resolution is $\sim 0.5-1.0$ mm.

When charged particles traverse the TPC, they liberate the
electrons from the TPC gas due to the \dedx. These electrons are
drifted towards the end cap planes of the TPC and collected by a
readout pad. The signal is amplified and integrated by a circuit
containing a pre-amplifier and a shaper. Then it is digitalized
and then transmitted over a set of optical fibers to STAR {\em
Data AcQuisition system} (DAQ).

At the DAQ stage, raw events containing millions of ADC values and
TDC values were recorded. Raw data were then reconstructed into
hits, tracks, vertices, the collision position through the
reconstruction chain of TPC by Kalman method. The TPC
reconstruction process begins by the 3D coordinate space points
finding. This step results in a collection of points reported in
global Cartesian coordinates. The {\em Timing Projection chamber
Tracker} (TPT) algorithm is then used to reconstruct tracks by
helical trajectory fit. The resulted track collection from the TPC
is combined with any other available tracking detector
reconstruction results and then refit by application of a Kalman
filter routine $-$ a complete and robust statistical treatment.
The primary collision vertex is then reconstructed from these
global tracks and a refit on these tracks with the {\em distance
of closest approach} ($dca$) less the 3 cm is preformed by a
constrained Kalman fit that forces the track to originate from the
primary vertex. As expected, the vertex resolution decreases as
the square root of the number of tracks used in the calculation.
The primary vertex resolution is $\sim 350$ $\mu$m with a track
multiplicity above 1000. The reconstruction efficiency including
the detector acceptance for primary tracks depends on the particle
type, track quality cuts, \pt, track multiplicity {\em etc}. The
typical value for the primary pions with $N_{fit}\ge25$ and
$|\eta|<0.7$, $dca<3.0$ cm is approximate constant at $p_T>0.4$
\gevc: $>\sim 90\%$ for \AuAu\ peripheral collisions and $\sim
80\%$ for central collisions, respectively.

The TPC provide the track momentum and the \dedx\ information for
charged particles identification. For a particle with charge z (in
units of e) and speed $\beta=v/c$ passing through The mean rate of
\dedx\ is given by the Bethe-Bloch
equation~\ref{dEdxBB}~\cite{PDG}:

\be -\frac{dE}{dx} =
Kz^2\frac{Z}{A}\frac{1}{\beta^2}\left[\frac{1}{2}\ln\frac{2m_ec^2\beta^2\gamma^2T_{max}}{I^2}-\beta^2-\frac{\delta}{2}\right]
\label{dEdxBB} \ee

The meaning of each symbol can be referred to~\cite{PDG}.
Different types of particles (different rest masses) with the same
momentum have different kinematic variables $\beta$ ($\gamma$),
which may result in distinguishable \dedx. The typical resolution
of \dedx\ in \AuAu\ collisions is $\sim 8\%$, which makes the
$\pi$/$K$ separation up to $p\sim 0.7$ \gevc\ and proton/meson
separation up to $p\sim 1.1$ \gevc.

Combined the TPC with other detectors, such as the TOF detector
and the BEMC, {\em etc.}, the capability of particle
identification can be greatly improved, particle separation can be
extended to higher \pt\ region.

\section{The prototype TOF detector based on MRPC technology}

A full barrel TOF detector, based on the MRPC technology, has been
proposed in the STAR detector upgrade programs. There will be 120
trays with 60 on east side and 60 on west side. For each tray,
there will be 33 MRPCs. The TOFp detector (a prototype based on
scintillator technology) was installed since Run
II~\cite{tofpBill}. It replaced one of CTB trays, covering
$-1<\eta<0$, and $\pi/60$ in azimuth. It contains 41 scintillator
slats with the signal read out by {\em Photo Multiplier Tubes}
(PMTs). The resolution of TOFp is $\sim 85$ ps in \AuAu\
collisions. However, due to the significant higher cost by the
PMTs, this design will not be used in the full TOF upgrade.

On the other hand, in Run III and Run IV, new prototypes of TOF
detector based on MRPC (TOFr) were installed also covering
$-1<\eta<0$ and $\pi/60$ in azimuth. In Run III, 28 MRPC modules
were installed in the tray and 12 of them were equipped with
electronics, corresponding to $\sim 0.3\%$ of the TPC
acceptance~\cite{tofpikp}. In Run IV, a new tray with 24 modules
were installed at the same place as Run III. But only 12 modules
were equipped with valid electronics, which means the acceptance
in Run IV was roughly similar to that in Run III.

The trigger system of the TOF detector is the two pVPDs, each
staying 5.4 m away from the TPC center along the beam
line~\cite{tofpBill}. They provide a starting timing information
for TOF detectors. Each pVPD consists of three detecting element
tubes covering $\sim 19\%$ of the total solid angle in
$4.43<|\eta|<4.94$. Due to different multiplicities, the effective
timing resolution of total starting time is 25 ps, 85 ps and 140
ps for 200 GeV \AuAu, \dAu\ and \pp\ collisions, respectively.


\bf \centering\mbox{
\includegraphics[width=0.8\textwidth]{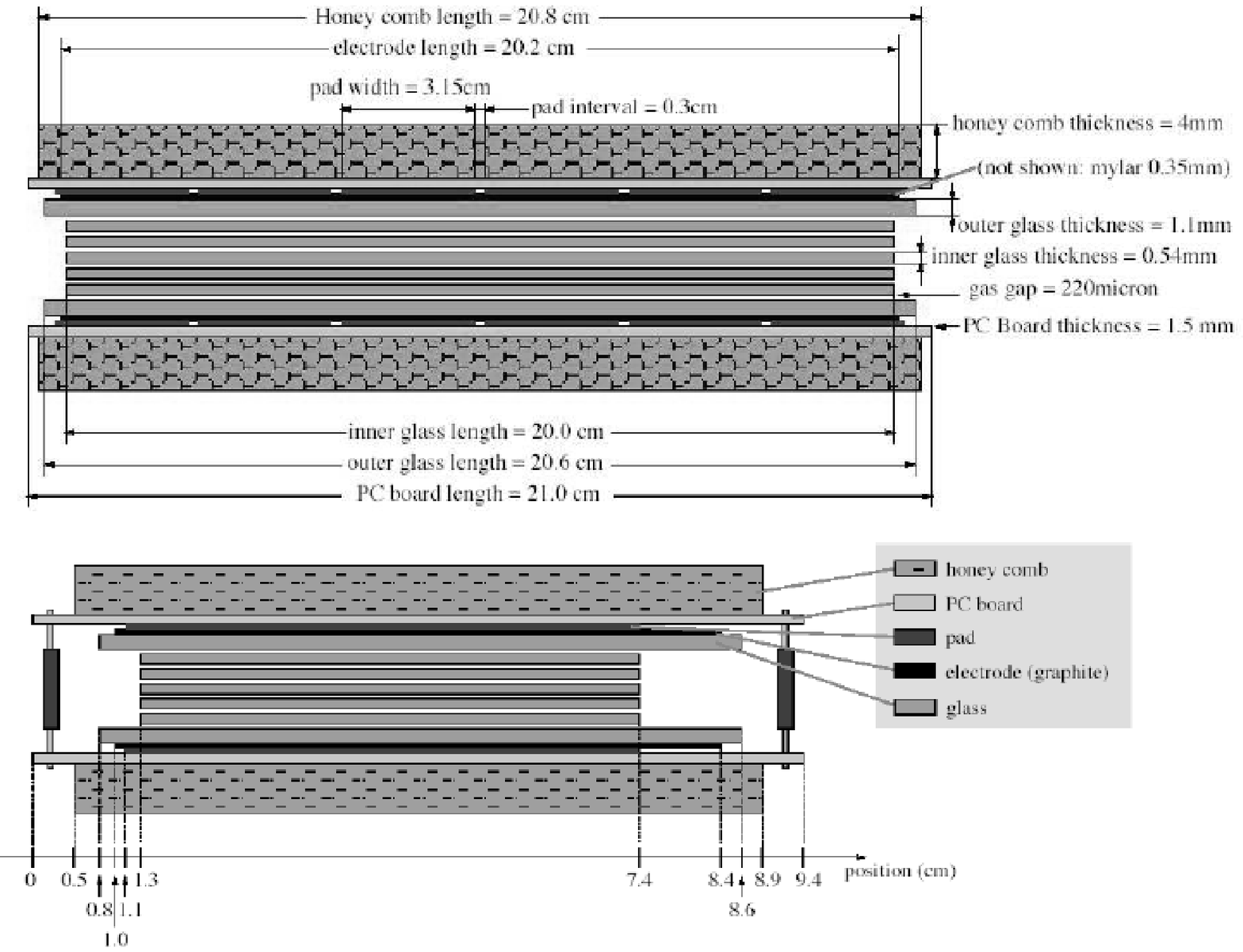}}
\caption[Two-side view of a MRPC module]{Two-side view of a MRPC
module~\cite{mrpctest}.} \label{mrpc} \ef

{\em Resistive Plate Chambers} (RPCs) were developed in 1980s.
MRPC technology was first developed by the CERN ALICE group less
than 10 years ago. Fig.~\ref{mrpc} shows the two side views (long
edge view on top and short edge view on bottom) of an MRPC module
appropriate for STAR~\cite{mrpctest}. An MRPC basically consists a
stack of resistive plates, spaced one from the other with equal
sized spacers (such as fish line) creating a series of gas gaps.
It works in avalanch mode. Electrodes are connected to the outer
surfaces of the stack of resistive plates while all the internal
plates are left electrically floating. Initially the voltage on
these internal plates is given by electrostatics, but they are
kept at the correct voltage due to the flow of electrons and ions
created in the avalanches. There are six read-out strips on each
module in this design. The first beam test for 6-gap MRPCs at CERN
PS-T10 facility with $p_{lab} = 7$ GeV/c pions beam resulted in a
$\sim 65$ ps timing resolution with more than $~95\%$ detecting
efficiency and the module is capable of working at high event rate
(500 Hz/cm$^2$)~\cite{mrpctest}. These modules were then assembled
in a prototype TOF tray and tested in the AGS radiation area.
Similar resolution was obtained. In RHIC Run III and Run IV, the
MRPC modules in TOFr trays installed in the STAR detector were
applied on the high voltage of 14 kV and with the working gas of
$95\%$ freon and $5\%$ isobutane. The charged particle detecting
efficiency is $>95\%$ at high voltage plateau.

TOF system calibrations include the start time calibration from
pVPDs and TOFr/TOFp flight time calibration. The main sources need
to be considered are global time offset due to different
electronics delays, the correlation between the amplitude and the
timing signals, the correlation between the hit position and the
timing signals {\em etc}. Detailed calibrations on TOF systems can
be found in ~\cite{LijuanThesis,tofpikp} (TOFr) and
~\cite{tofpBill} (TOFp).

\chapter{Non-photonic electron measurement in 200 GeV Au+Au collisions}
Due to the large combinatorial background, direct reconstruction
of heavy-flavor mesons via hadronic decay channels is difficult in
current high energy nuclear collisions at
RHIC~\cite{stardAucharm}. The indirect measurement of single
leptons through heavy-flavor semileptonic decays has been used as
an efficient way to study heavy flavor production. The transverse
momentum (\pt) distributions of single electron and the extracted
total charm cross-section have been measured in 200 GeV \dAu\
collisions~\cite{stardAucharm,Xinthesis}. Study of the binary
collisions (\nbin) scaling properties for the charm total
cross-section among \dAu\ to \AuAu\ collisions can test if
heavy-flavor quarks are produced exclusively at initial impact.
Modification of the single lepton production in nuclear collisions
can reveal the heavy-flavor energy-loss. Thermal parameters
extracted from the \pt\ distributions of single electron are
useful for us to understand the heavy-flavor hadrons freeze-out
and flow properties in medium in nuclear-nuclear collisions.
Therefore, it is very important to measure single electron \pt\
distributions in \AuAu\ collisions.

In this chapter, analysis details of single electron \pt\
distributions in 200 GeV \AuAu\ collisions will be presented. In
addition, due to large photonic background, it is difficult to
measure single electron \vv\ in current experimental environment,
but a method to extract single electron \vv\ will be proposed.

\section{Non-photonic electron transverse momentum
distributions}

\subsection{Data sets and cuts}
In RHIC Run IV (year 2004), STAR experiment collected abundant
triggered minimum bias (minbias) events and central events from
$\sqrt{s_{NN}}$=200 GeV \AuAu\ collisions. A 0-80\% minbias \AuAu\
collision centrality is defined by the charged particle reference
multiplicity (refmult) measured at mid-rapidity ($|\eta|<0.5$)
requiring number of TPC fit points (nFitPts) $>9$ and the
distance-of-closest-approach (dca) $<3$ cm. The sub-centralities
for the data analysis are selected by applying the refmult cuts
from the calculation of Glauber model~\cite{Glauber}, which gives
the average percentage of the number of events in each centrality
bin, see Fig.~\ref{multcen}. For the centrality dependence study
of the non-photonic electrons, the minbias event sample was
subdivided into three centrality bins: 0-20\%, 20-40\% and
40-80\%. Table~\ref{dataset1} lists the data sets under the
trigger selections used in this analysis from Run IV 200 GeV
\AuAu\ collisions. The right column lists two numbers for the size
of event samples after the location of collision vertex along beam
axis ($V_z$) cuts, one is the number of TPC triggered events used
for the photonic background electron analysis, the other is the
number of TOF triggered events used for inclusive electron
identification.

\begin{table}[hbt]
\caption[Data sets list]{Data sets from Run IV used in this
analysis} \label{dataset1} \vskip 0.1 in
\centering\begin{tabular}{|c|c|c|c|} \hline \hline
Trigger & Centrality(refmult cut) & Vertex Z cut (cm) & Events Size (TPC, TOF) \\
\hline
        & 0-20\% [319,1000) & $|V_{Z}|<30$ & 3.28M, 1.93M \\
        \cline{2-4}
minbias & 20-40\% [150,319) & $|V_{Z}|<30$ & 3.36M, 1.97M
\\ \cline{2-4}
        & 40-80\% [14,150) & $|V_{Z}|<30$ & 6.55M, 3.85M
\\ \cline{2-4}
        & 0-80\% [14,1000) & $|V_{Z}|<30$ & 13.2M, 7.8M \\ \hline
central & 0-12\% & $|V_{Z}|<50$ & 21.7M, 15.5M \\
     \hline \hline
\end{tabular}
\end{table}

The online centrality selection in \AuAu\ minbias collisions was
based on the energy deposited in the two Zero-Degree Calorimeters
(ZDC)~\cite{STAR,starwhitepaper}. Since the 5-10\% refmult cuts in
central triggers will give some combination of ZDC- and TPC-based
centrality, which is not 5-10\% actually. But in the analysis, due
to the limited statistics, all the central triggers are accepted,
the centrality is 0-12\%, the bias from realistic 0-10\% is about
5\%. The TOF trigger set up in Run IV was to select events with a
valid pVPD coincidence. The trigger efficiency for the TOF events
is $\sim$60\%. The tracks matched to the TOF should have good ADC
signal ($>30$) and at least one TOF hit. In Run IV, one TOFr tray,
which is based on MRPC technology, and one TOFp tray, which uses
RPC modules, were installed outside the TPC and calorimeters. The
TOF detector covers $1/120$ full barrel and negative
pseudorapidity ($-1<\eta<0$).

\bf \centering\mbox{
\includegraphics[width=0.6\textwidth]{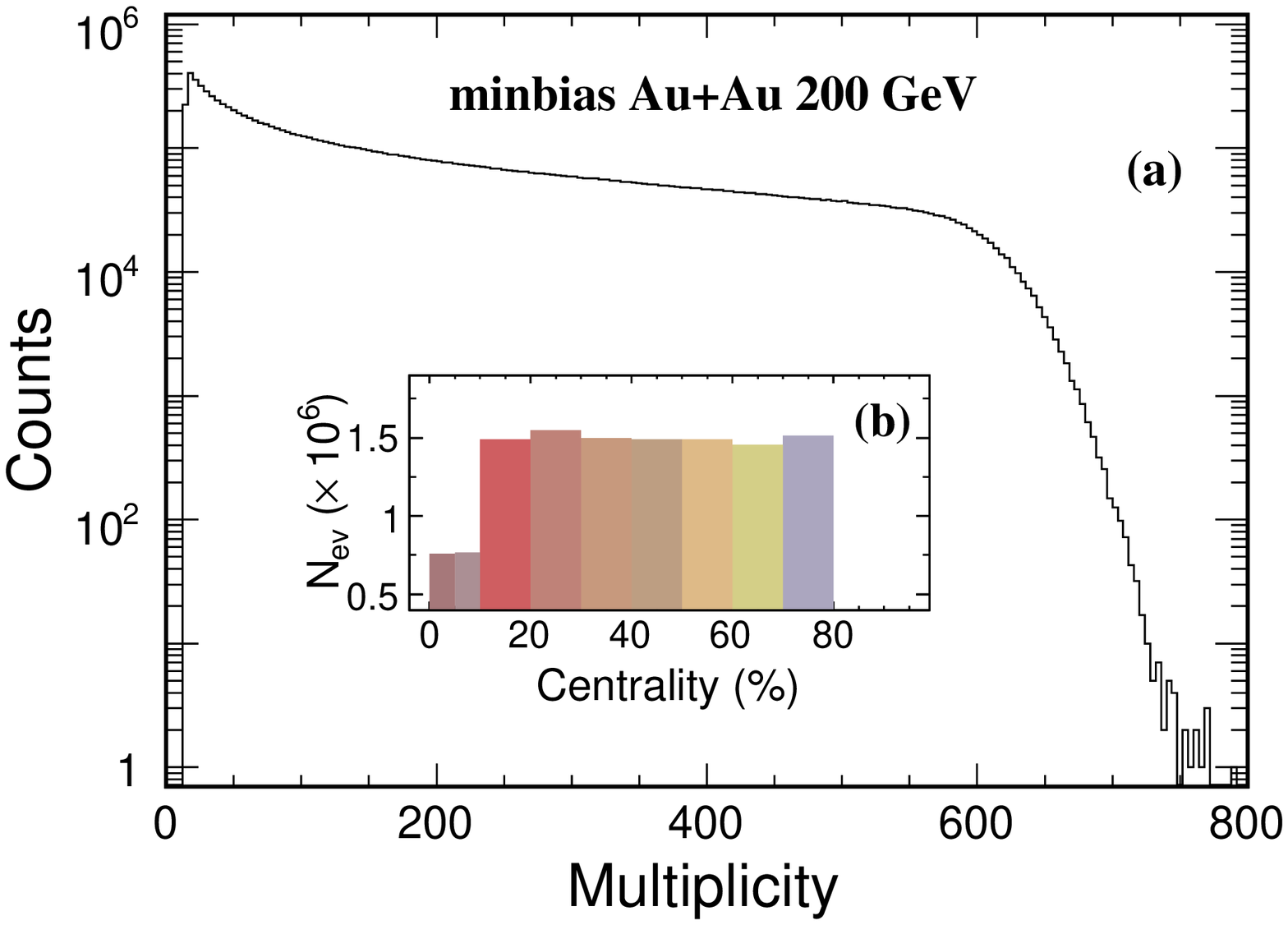}}
\caption[Multiplicity and centrality distributions]{Panel (a)
shows the multiplicity for minbias triggered events, which follows
Glauber distribution. Panel (b) shows the centralities by cutting
the multiplicity calculated from Glauber model.} \label{multcen}
\ef

\subsection{Electron identification}
The charged particle is usually identified by the ionization
energy loss (\dedx) measured in the STAR main detector
TPC~\cite{tpctech}. With TPC \dedx\ only, electron can be
identified only up to \pt$\sim0.8$ GeV/c~\cite{IanPRC,IanThesis}
due to large hadron contaminations. But the very small electron
mass gives it the ability to be separated from other hadrons by
measuring the flight timing information. The TOF detector was
designed to measure the velocity ($\beta$) of electrons and
hadrons~\cite{startof1,stardAucharm,pidNIMA,Xinthesis}.

By using a combination of velocity measured from the TOF and
\dedx\ measured in the TPC, inclusive electrons are identified up
to \pt$\sim4$ \gevc\ in minbias \AuAu\ collisions and up to
\pt$\sim5$ \gevc\ in central \AuAu\ collisions. Panel (a) of
Fig.~\ref{ePID} shows the 2-D scattering plot of \dedx\ measured
from TPC vs. momentum ($p$) for the charged particles with good
TOF hits matched in \AuAu\ collisions. Panel (c) shows the
$1/\beta$ vs. $p$. Due to the different mass, charged particles
can be separated by measuring their velocity. Electron band is
around unity and merged with pion band. Panel (b) shows the \dedx\
vs. $p$ after the particle velocity cut ($|1/\beta-1|<0.03$),
after most of the charged hadrons rejected, the pure \dedx\ bands
for electron and pion are left. And they are separated well.

\bf \centering \bmn[c]{0.5\textwidth} \centering
\includegraphics[width=1.1\textwidth]{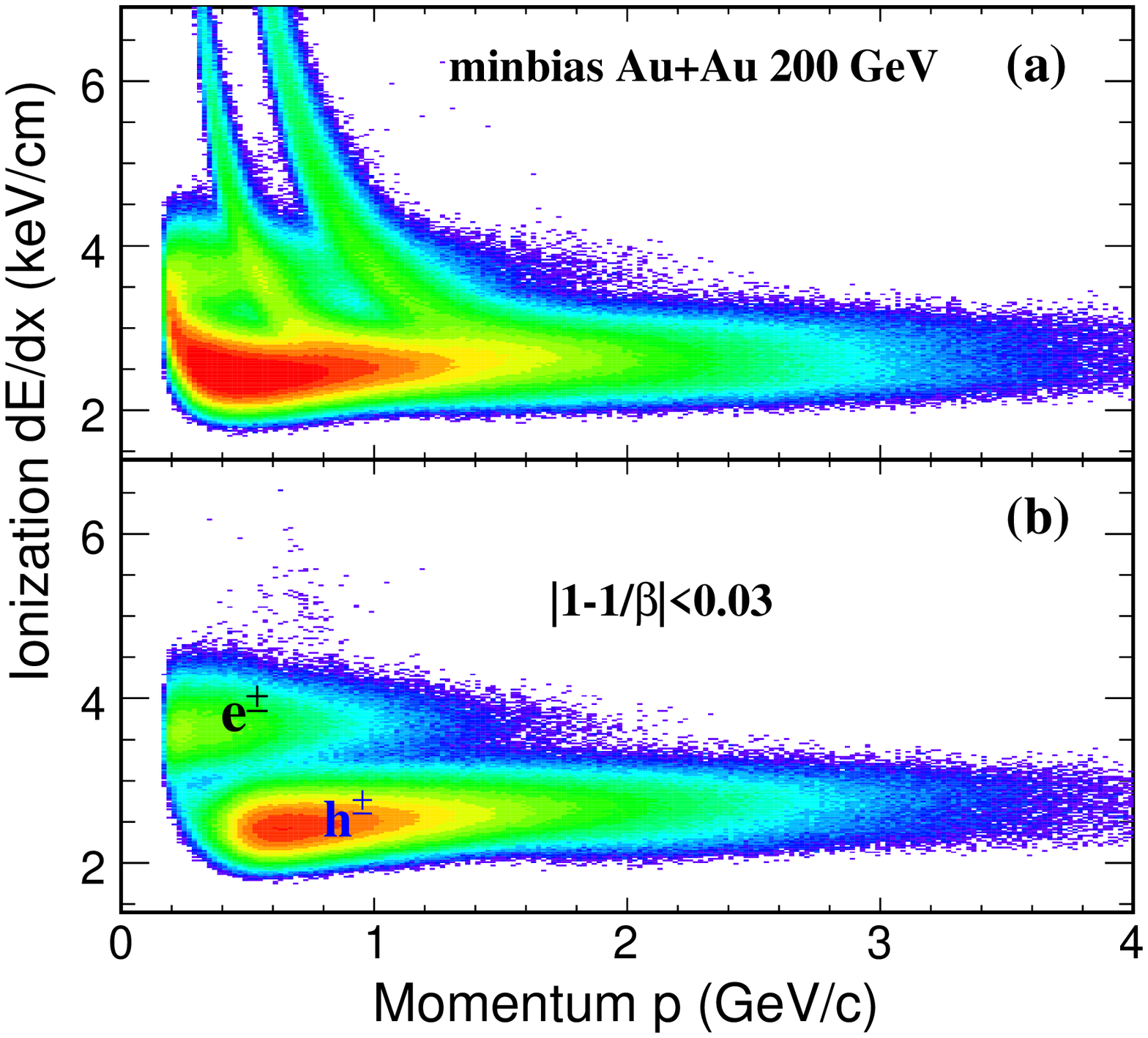}
\emn%
\bmn[c]{0.5\textwidth} \centering
\includegraphics[width=0.95\textwidth]{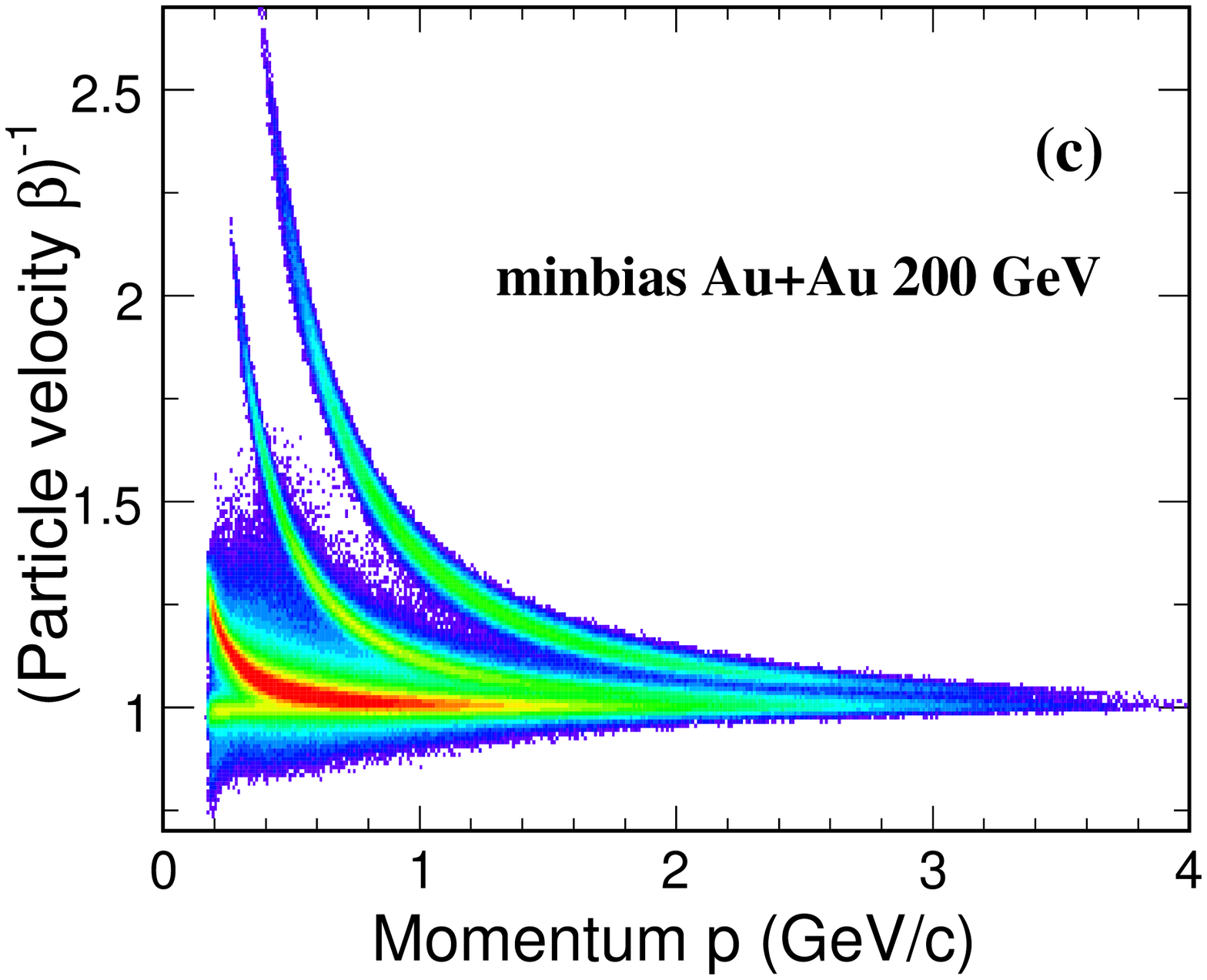}
\emn \caption[Ionization energy loss and particle velocity vs.
momentum]{Panel (a): Ionization energy loss measured by TPC vs.
momentum. Panel (b): Ionization energy loss measured vs. momentum
after TOF velocity cut. Panel (c): Particle velocity measured by
TOF vs. momentum.} \label{ePID} \ef

Table~\ref{inclecut} lists the cuts for inclusive electron
selection.

\begin{table}[hbt]
\caption[Inclusive electron cuts list]{Cuts for inclusive electron
selection} \label{inclecut} \vskip 0.1 in
\centering\begin{tabular}{|c|c|} \hline \hline
Primary track? & yes\\
\hline TOF hits matched? & yes\\
\hline Global dca & $<3.0$ cm\\
\hline nFitPts & $\geq$ 25\\
\hline nFitdedxPts & $\geq$ 15\\
\hline nFitPts/nMax & $>0.52$\\
\hline $1/\beta-1$ & (-0.03, 0.03)\\
\hline pseudorapidity & (-1.0, 0)\\
\hline TOFr ADC & $>30$\\
\hline \hline
\end{tabular}
\end{table}

The \dedx\ distributions are not exactly gaussian. The
$log_{10}((\dedx)/(\dedx_{Bichsel}))$ (abbreviated as \dedx),
which follows gaussian distribution, was defined as the
logarithmic ratio of measured \dedx\ and predicted \dedx\ from
Bichsel model~\cite{Bichsel}. After the velocity cut, the
2-dimension histogram of \dedx\ as a function of \pt\ was
projected into several \pt\ bins. The 2-Gaussian function cannot
describe the shoulder region between hadron peak and electron peak
well at lower \pt, a function of exponential+gaussian was
performed instead. At $p_T {}^{>}_{\sim}
 1.5$ \gevc, the other hadrons, such as protons, contaminate the
low \dedx\ window and statistics cannot enable us to distinguish
the difference of these two fits, so 3-Gaussian fit was used in
this \pt\ region. Then the inclusive electron raw yields were
obtained from these fits in each \pt\ bin. Fig.~\ref{dEdxFit}
shows the \dedx\ distributions and fitting results in several
typical \pt\ bins after TOF velocity cut. The integral yields by
applying a cut on $3\sigma>\dedx>\dedx_{mean}$ are used to
estimate the hadron contamination compared to the fits,
$\dedx_{mean}$ is the central value of the \dedx\ distribution.
During the energy deposition in the TPC, the merged tracks are
measured as larger energy loss, which dominate the tails shown on
the right of electron gaussian peak, see Fig.~\ref{dEdxFit}. It is
more obvious in central collisions than peripheral. The higher the
multiplicity is, the larger the possibility of the track mergence.
There are two small gaussian contributions to the tail: one is due
to two merged pions, the other is considered as a mergence of one
electron and one pion. The contamination of the tail to the
electron is quite small, less than percent, even it looks obvious
in the log scale. The tails fade at higher \pt.

\bf \centering \bmn[b]{0.5\textwidth} \centering
\includegraphics[width=1.0\textwidth]{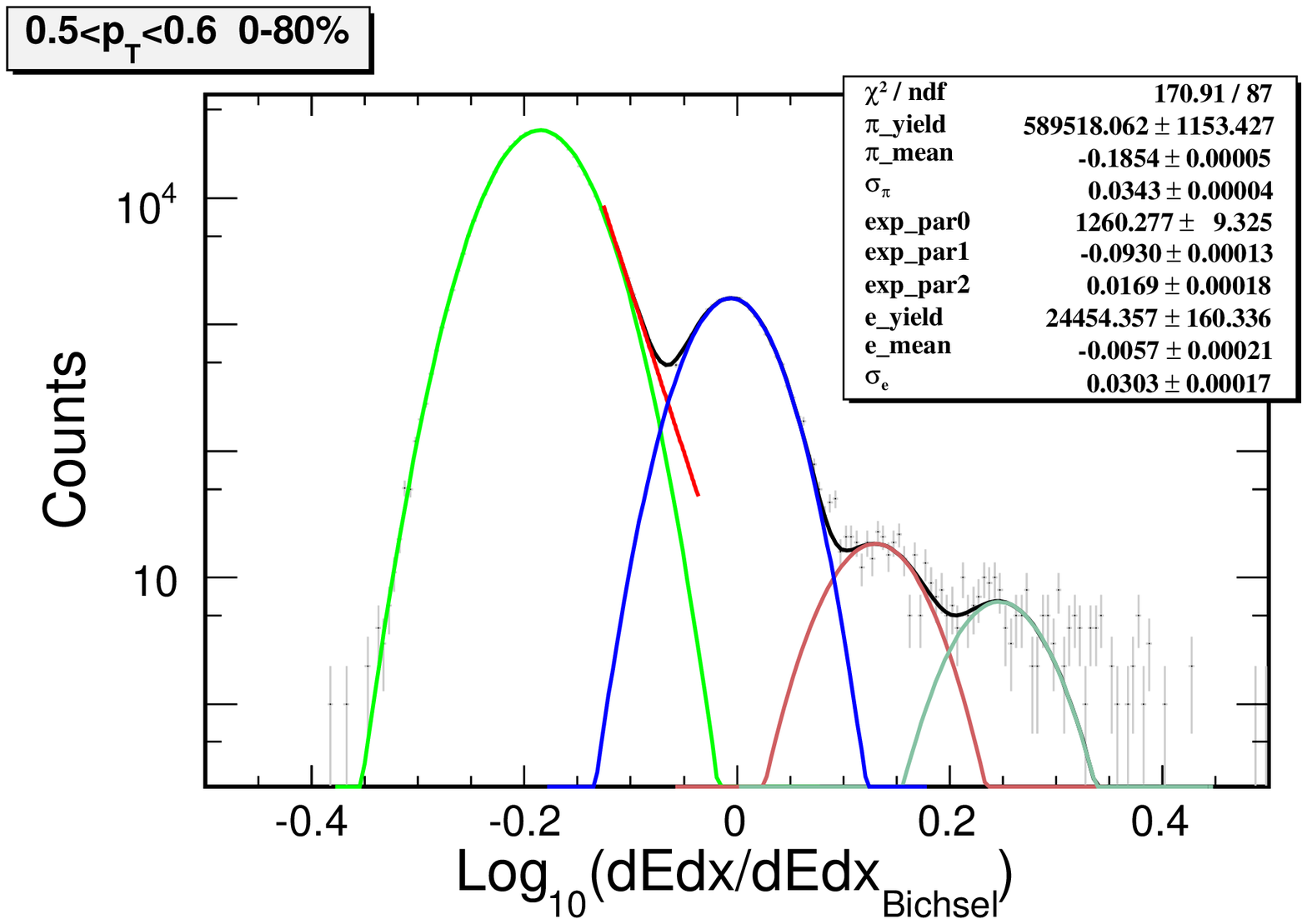}
\emn%
\bmn[b]{0.5\textwidth} \centering
\includegraphics[width=1.0\textwidth]{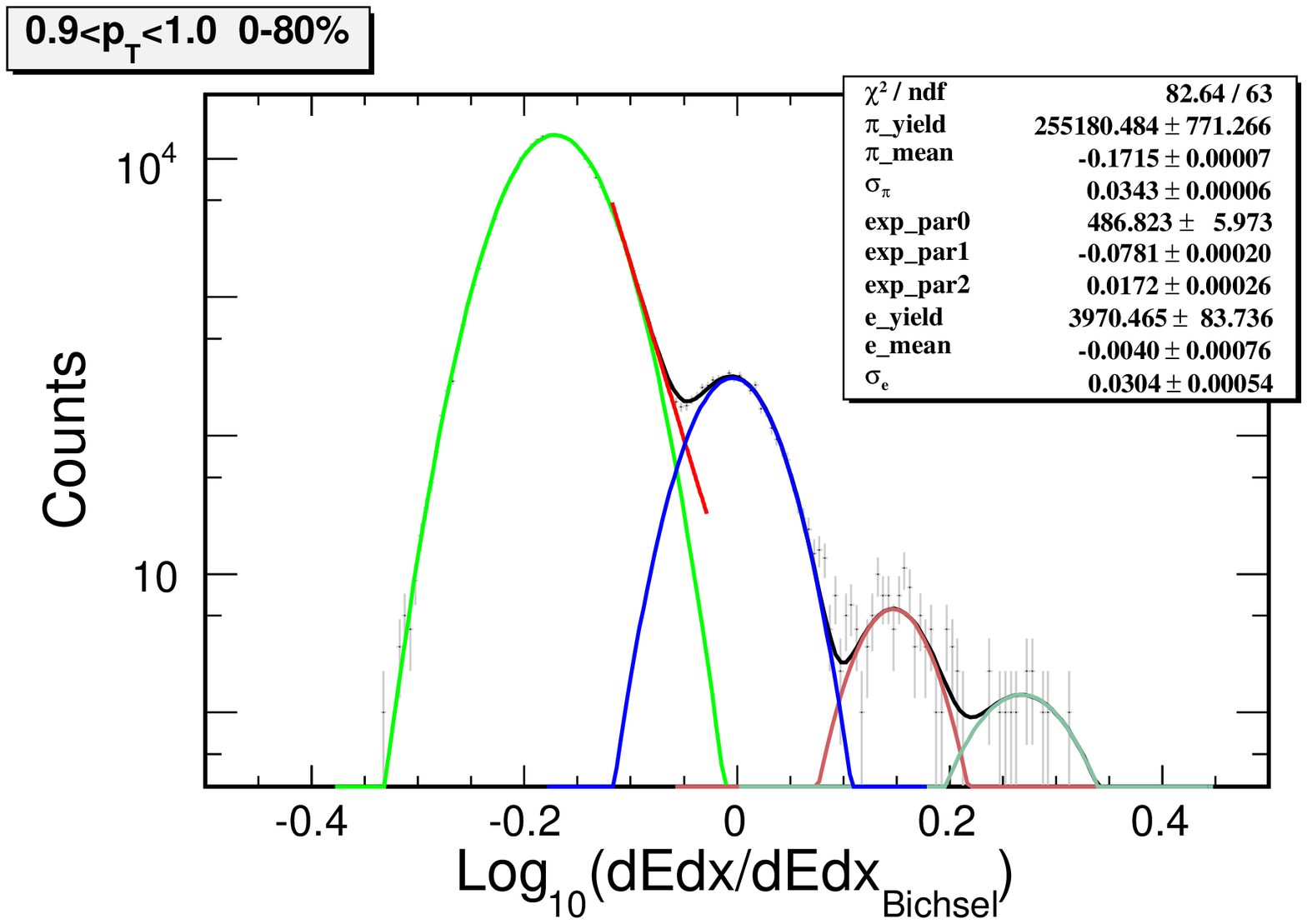}
\emn\\[10pt]
\bmn[b]{0.5\textwidth} \centering
\includegraphics[width=1.0\textwidth]{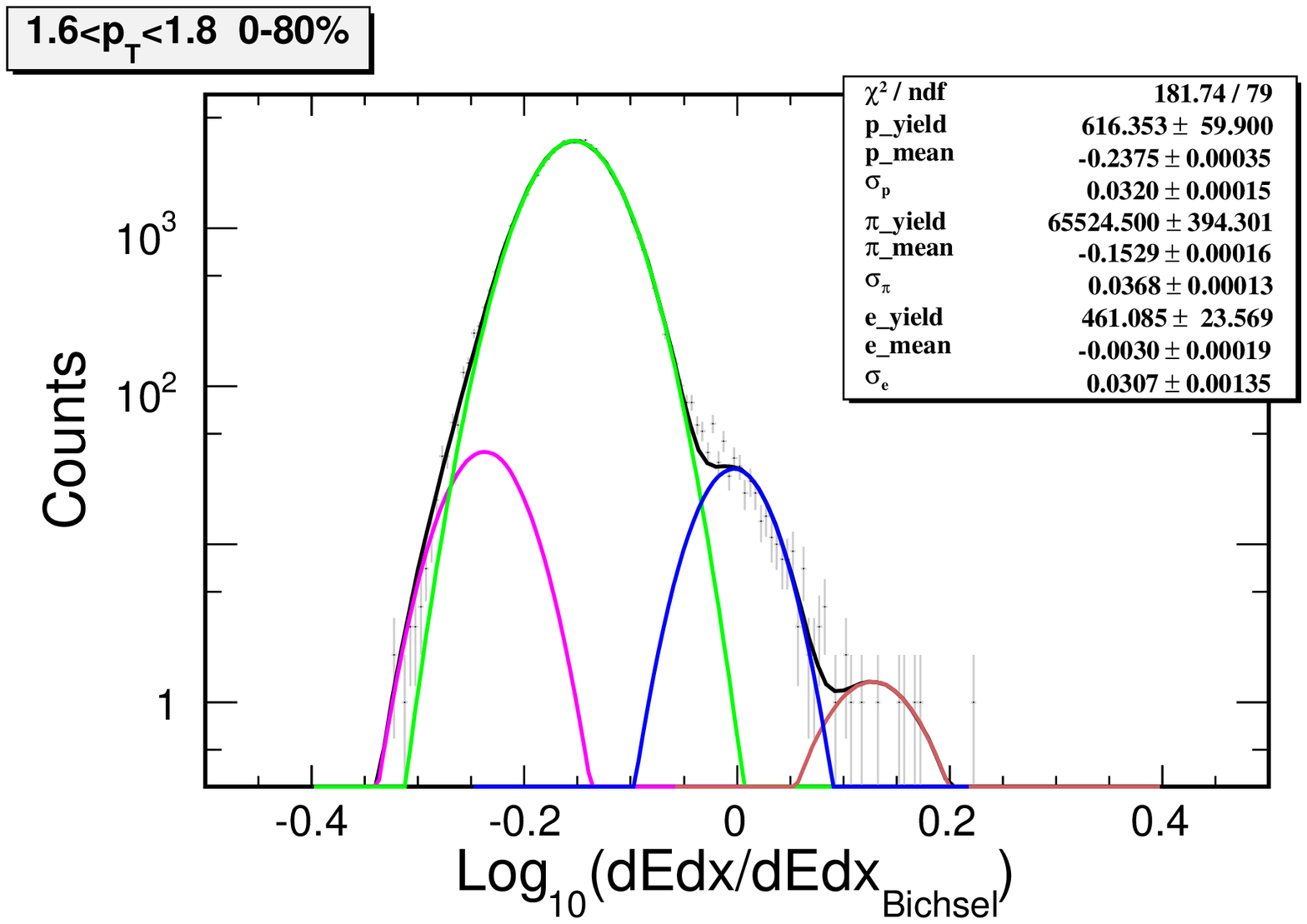}
\emn%
\bmn[b]{0.5\textwidth} \centering
\includegraphics[width=1.0\textwidth]{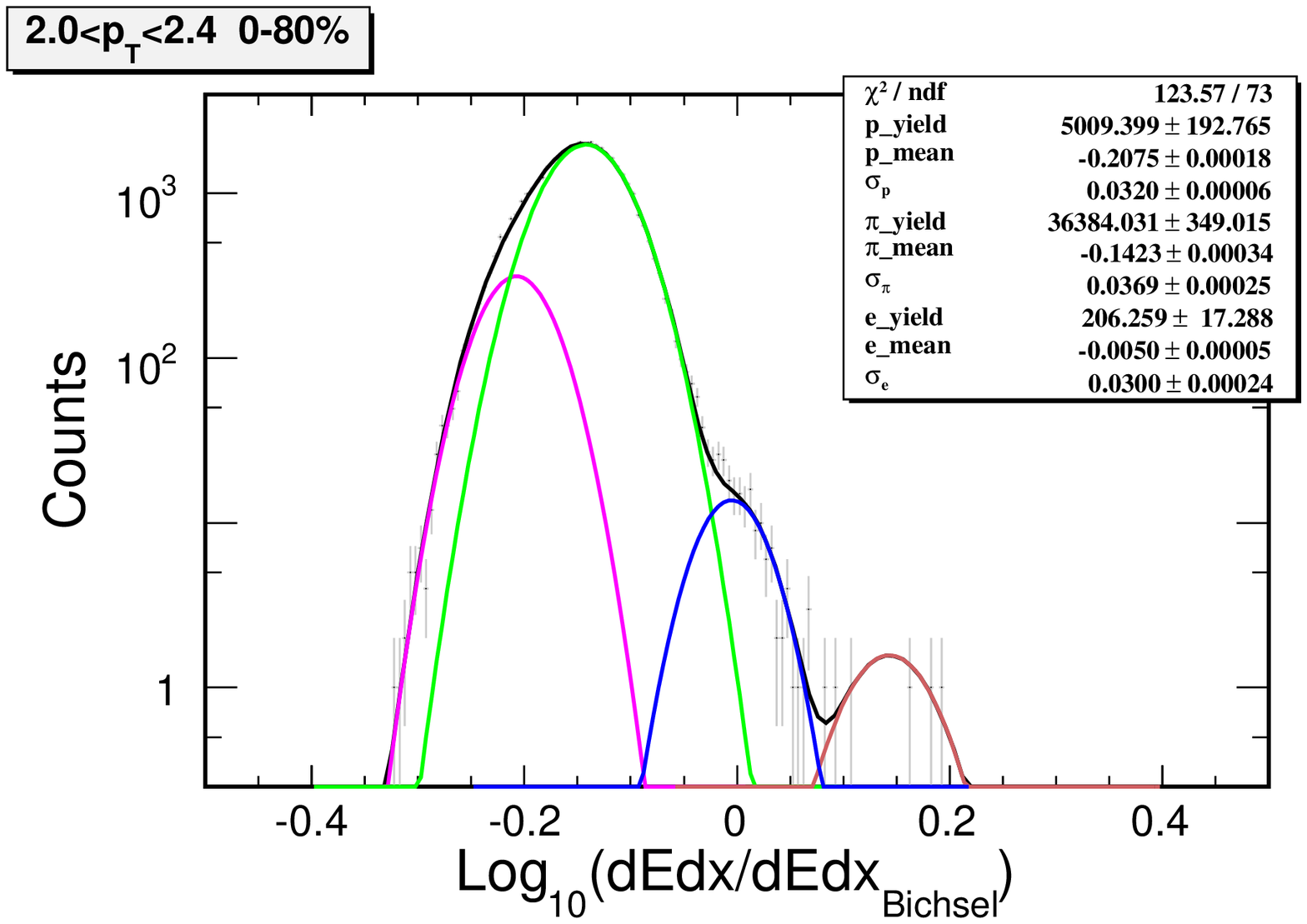}
\emn\\[10pt]
\bmn[b]{0.5\textwidth} \centering
\includegraphics[width=1.0\textwidth]{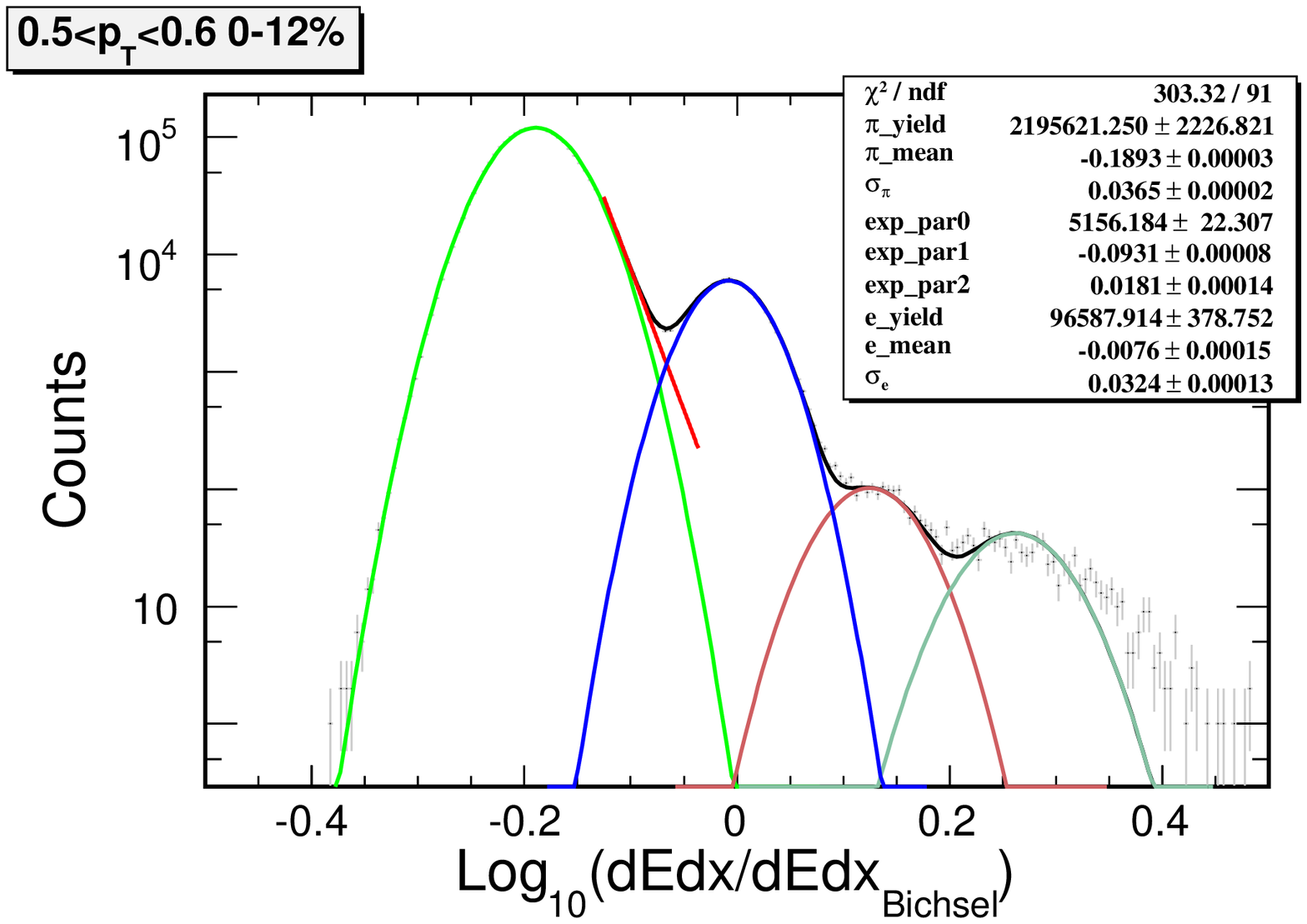}
\emn%
\bmn[b]{0.5\textwidth} \centering
\includegraphics[width=1.0\textwidth]{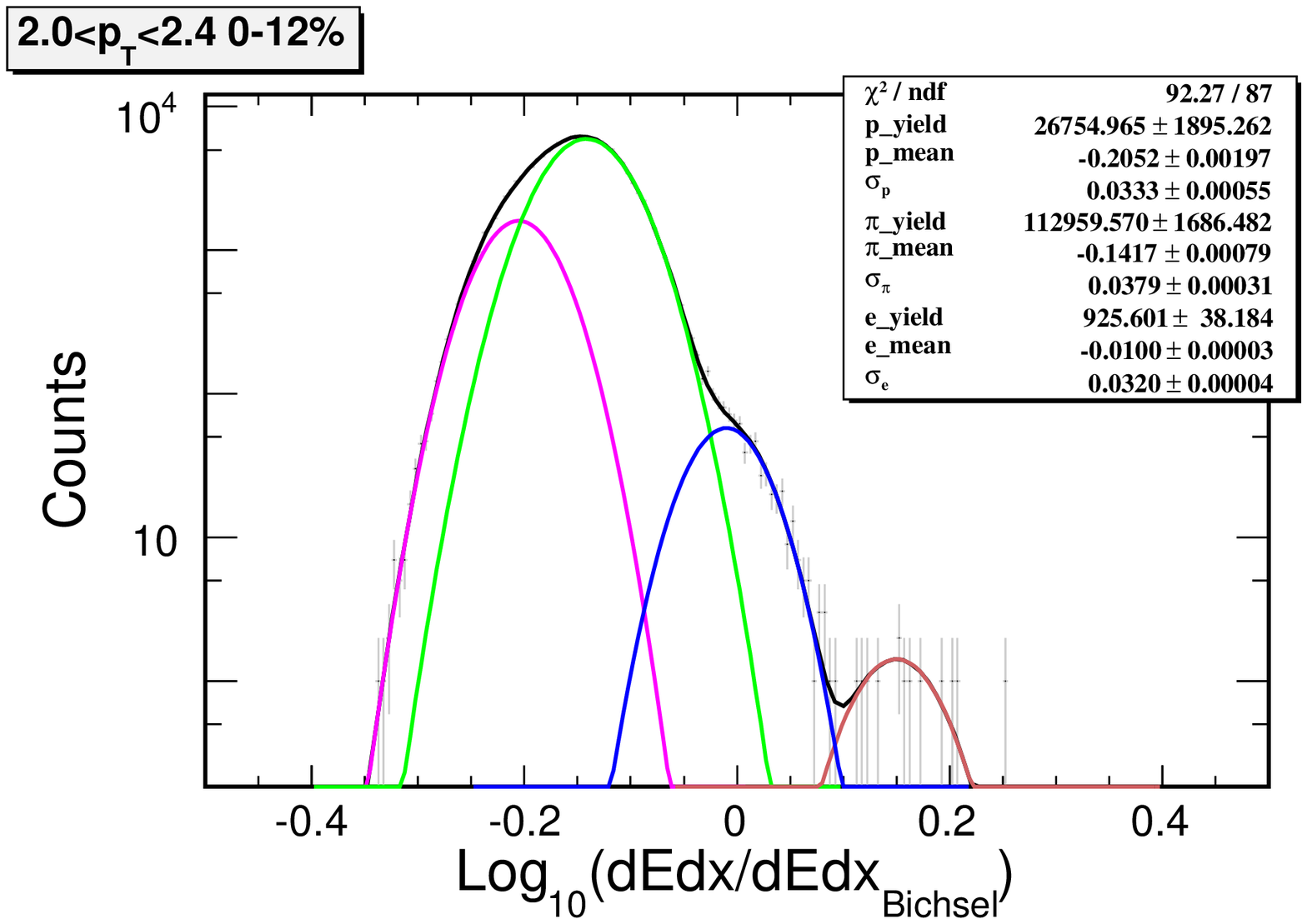}
\emn%
\caption[\dedx\ fit for inclusive electron raw yield extraction]
{\dedx\ projections in several \pt\ bins after the TOF velocity
cut. The electron raw yields were extracted from the fit.
Exponential (red curve) + gaussian (blue curve) function is used
at low \pt. Multi-gaussian function is performed at higher \pt.
The tails due to merged tracks are described by gaussian
functions, they fade at higher \pt. The \dedx\ distributions in
minbias collisions (first 4 panels) and central collisions (last 2
panels) are similar.} \label{dEdxFit} \ef

Fig.~\ref{dEdxtail} show the difference of the fit to the tails by
fixing width and opening width. The yields of the inclusive
electron change less than 1\%.

\bf \centering \bmn[b]{0.5\textwidth} \centering
\includegraphics[width=1.0\textwidth]{plots/fitdedx0_12_3.eps}
\emn%
\bmn[b]{0.5\textwidth} \centering
\includegraphics[width=1.0\textwidth]{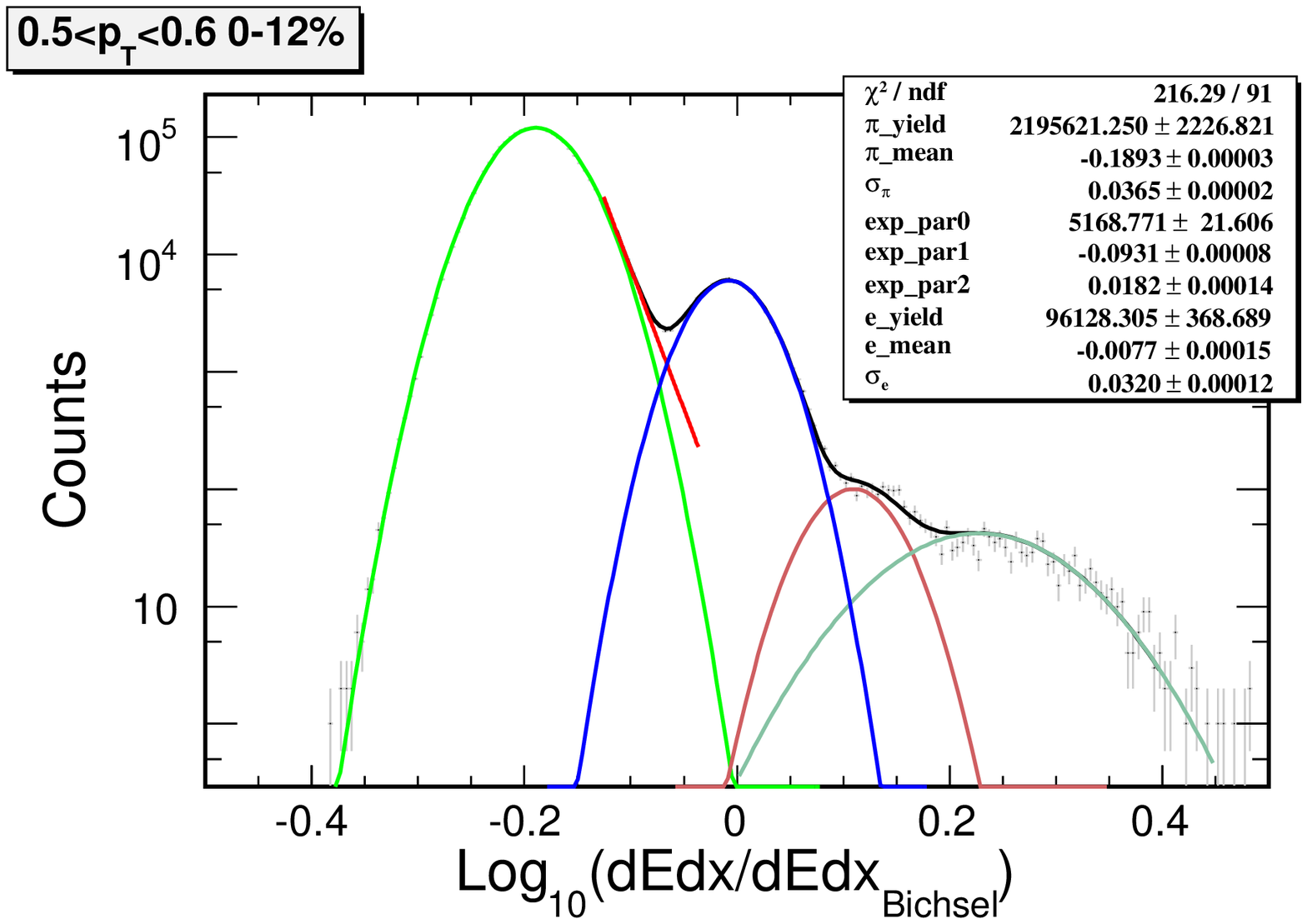}
\emn\\[10pt]
\caption[Fit \dedx\ tail with different width]{Left panel shows
the gaussian fit to the tail with fixed width. Right panel shows
the fit with open width. The difference of electron raw yields is
$<1\%$.} \label{dEdxtail} \ef

Since \dedx\ projected in a momentum bin is more like gaussian
distribution than that projected in a \pt\ bin. The $n\sigma$(\pt)
is defined as \be n\sigma(p_{T}) \sim
A\times(\dedx-\dedx_{Bichsel})/(\dedx_{Bichsel}\times
\sqrt{ndEdxFitPts}) \label{tofCut} \ee

$A$ is a constant, which only depends on the detector environment.
$n\sigma$ is gaussian distribution in each \pt\ bin. $ndEdxFitPts$
is the number of fit points used for calculating the \dedx\ of a
charged particle during the tracking in the TPC. A 2-Gaussian
function was used to fit the $n\sigma_{e}$ shown as
Fig.~\ref{fitnsigma}. The difference between fit to $n\sigma_{e}$
and fit to log \dedx\ is less than 5\% point to point at lower
\pt. At higher \pt\ the difference is even smaller, due to the
smaller difference of momentum and \pt.

\bf \centering \bmn[b]{0.5\textwidth} \centering
\includegraphics[width=1.0\textwidth]{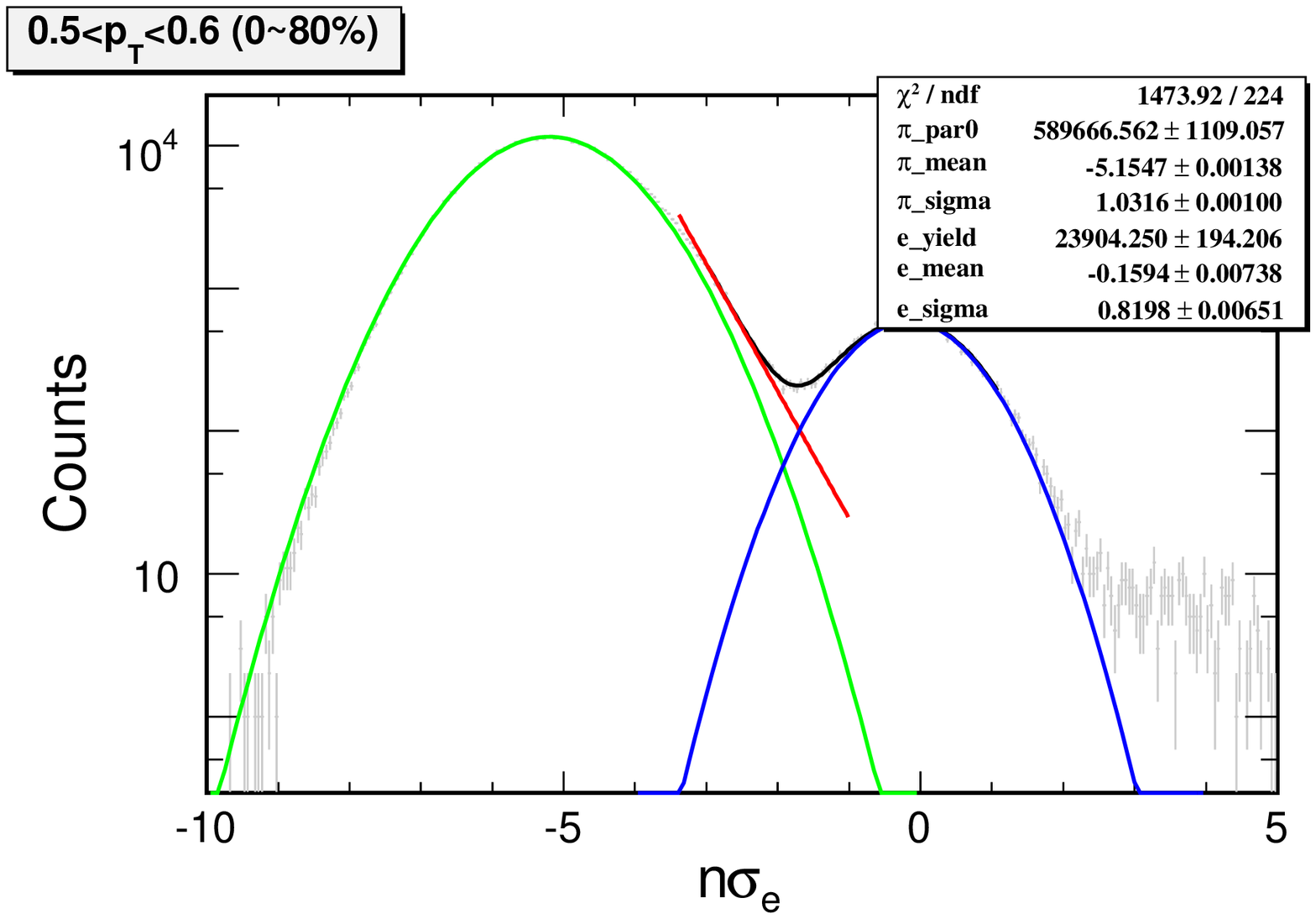}
\emn%
\bmn[b]{0.5\textwidth} \centering
\includegraphics[width=1.0\textwidth]{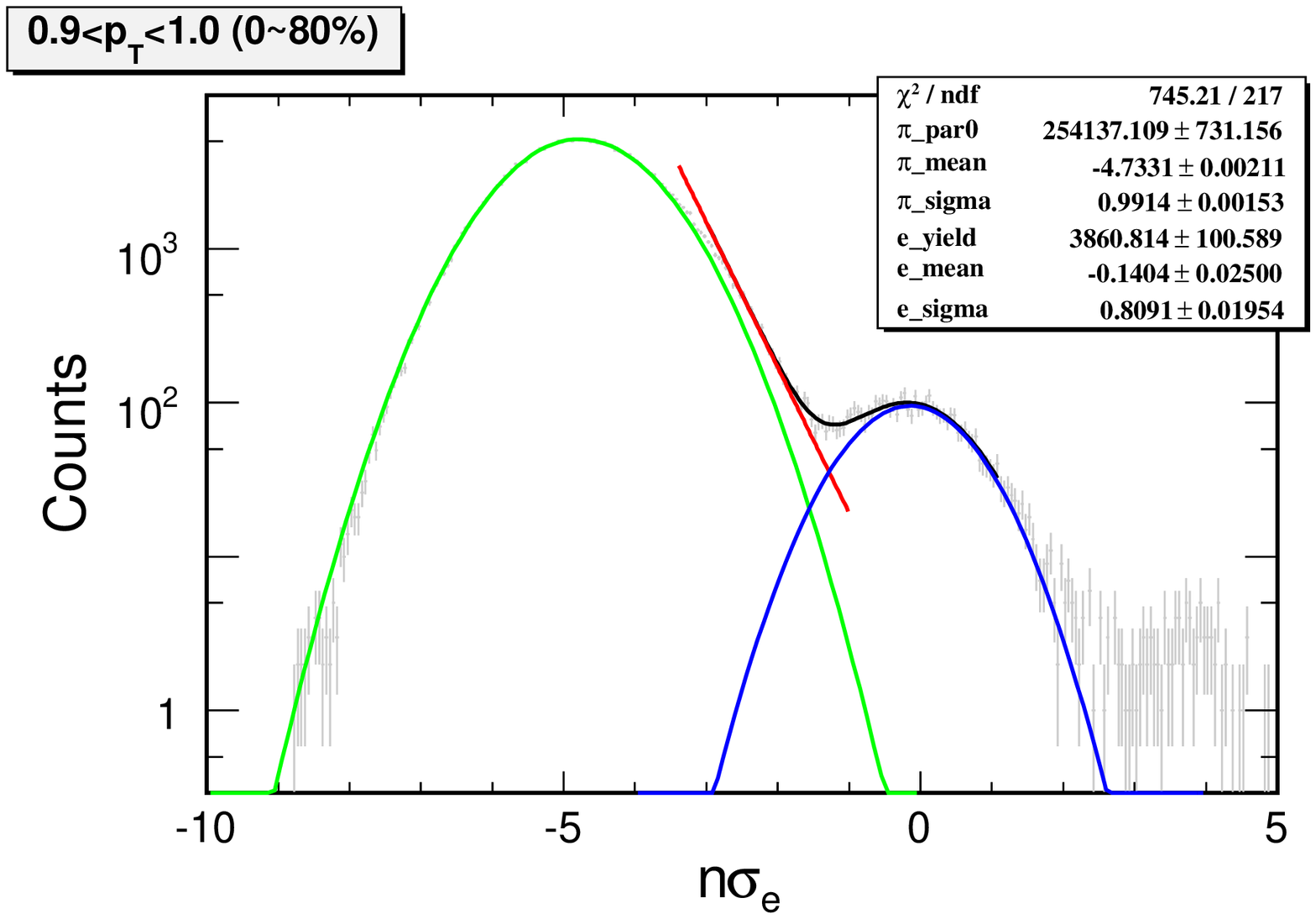}
\emn\\[10pt]
\caption[Fit to $n\sigma_{e}$ distributions]{The difference
between fit to $n\sigma_{e}$ and log \dedx\ is less than 5\% point
to point at lower \pt. This was used to check the difference of
the electron raw yields derived from \dedx\ vs. $p$ and \pt.}
\label{fitnsigma} \ef

\subsection{Acceptance and efficiencies}
During the detector tracking and the extraction of the inclusive
electron raw yields, a part of them were lost due to the detector
acceptance, the reconstruction and cut efficiency {\em etc.}. The
final yields of inclusive electron need to be corrected by the
acceptance and efficiencies:

\begin{itemize}
\item TPC tracking efficiency $-$ track reconstruction efficiency,
TPC acceptance. \item TOF matching efficiency $-$ TPC track with
good TOF hits matched, TOF acceptance. \item number of \dedx\ fit
points cut efficiency.
\end{itemize}

Due to the lower statistics of electrons, the TPC tracking
efficiency was studied from pion embedding for each centrality in
\AuAu\ collisions, shown in left panel of Fig.~\ref{eff1}. The
different vertex Z cuts and eta cuts do not effect the tracking
efficiency much. The TOF matching efficiencies used for electrons
were determined from real data as the number of pion tracks
matched to TOF divided by the number of pion tracks reconstructed
from TPC. The centrality dependence of the matching efficiency is
shown in right panel of Fig.~\ref{eff1}. But different from
electrons, $\sim$10\% of pions could be scattered or decay,
especially at low \pt, which was corrected in the total
efficiencies.

\bf \centering \bmn[b]{0.5\textwidth} \centering
\includegraphics[width=1.0\textwidth]{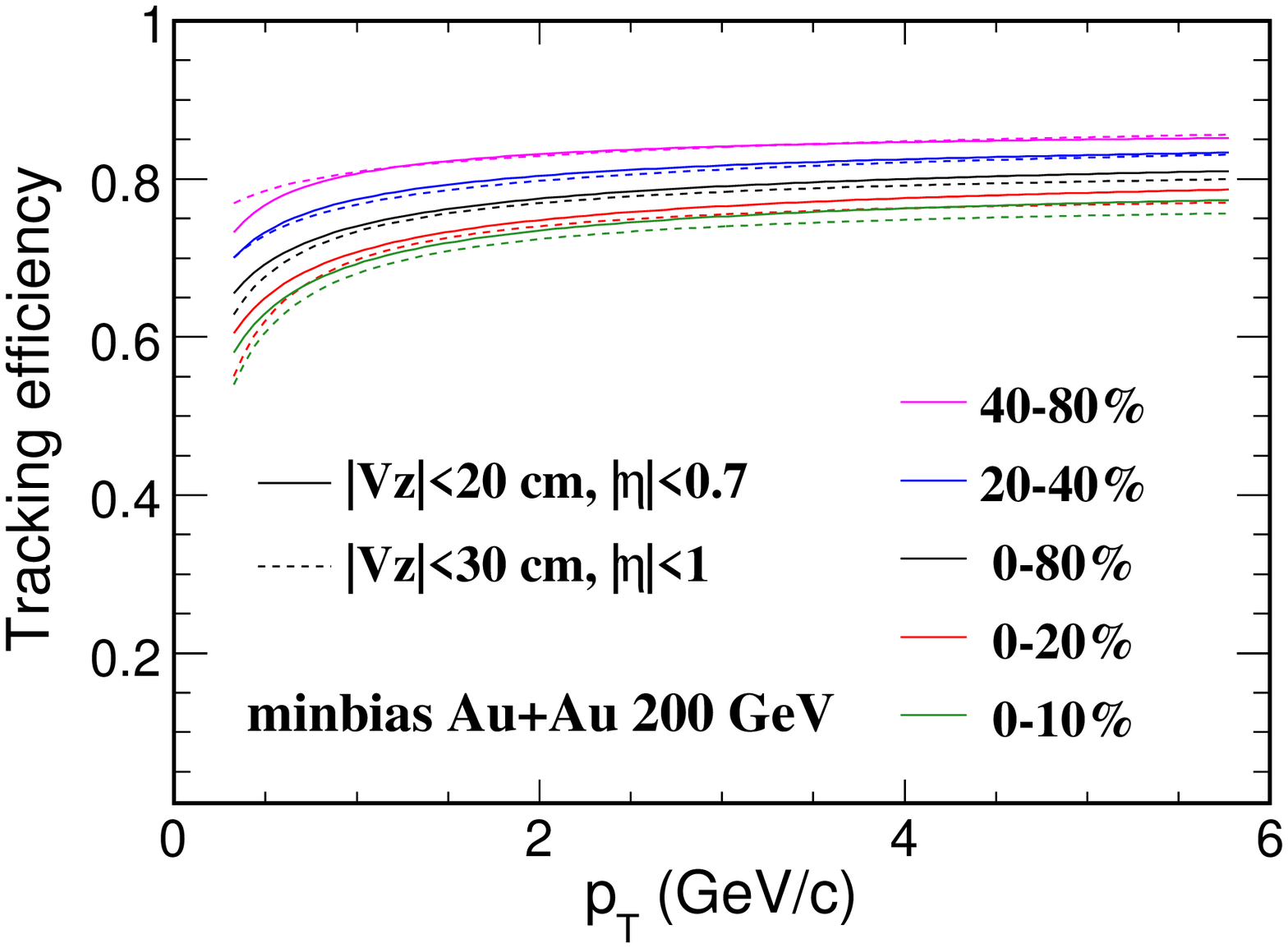}
\emn%
\bmn[b]{0.5\textwidth} \centering
\includegraphics[width=1.0\textwidth]{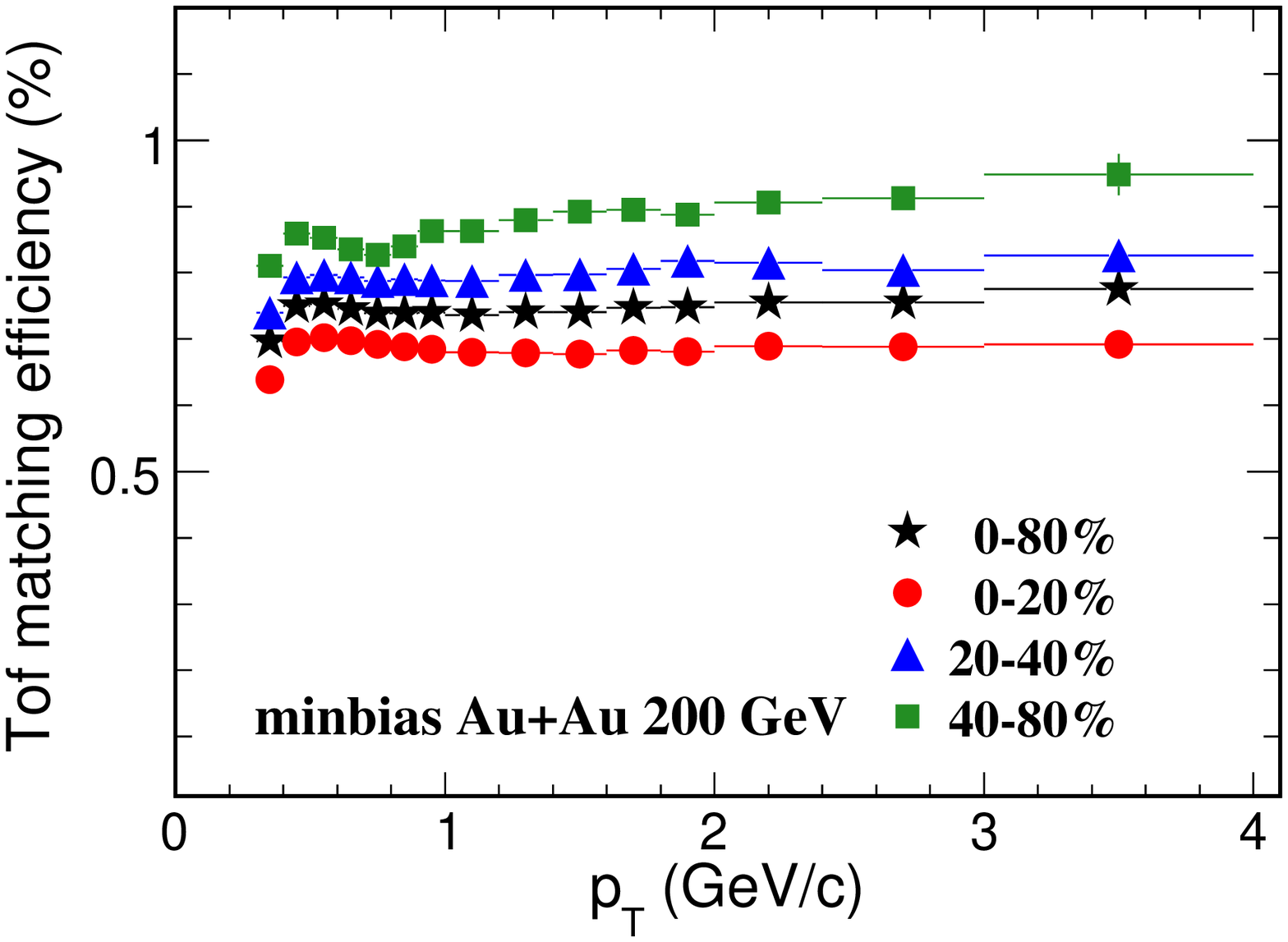}
\emn\\[10pt]
\caption[TPC tracking efficiency and TOF matching efficiency]{Left
panel: Centrality dependence of TPC tracking efficiency from
embedding data. Right panel: Centrality dependence of TOF matching
efficiency from real data.} \label{eff1} \ef

The TPC tracking efficiency does not include the ndEdxFitPts cut
efficiency, since it is technically difficult to study \dedx\ in
embedding data. So the ndEdxFitPts cut efficiency was determined
from real data as the number of tracks by cutting on
ndEdxFitPts$geq$15 divided by the number of total tracks, shown in
Fig.~\ref{eff2}.

\bf \centering \bmn[b]{0.6\textwidth} \centering
\includegraphics[width=1.0\textwidth]{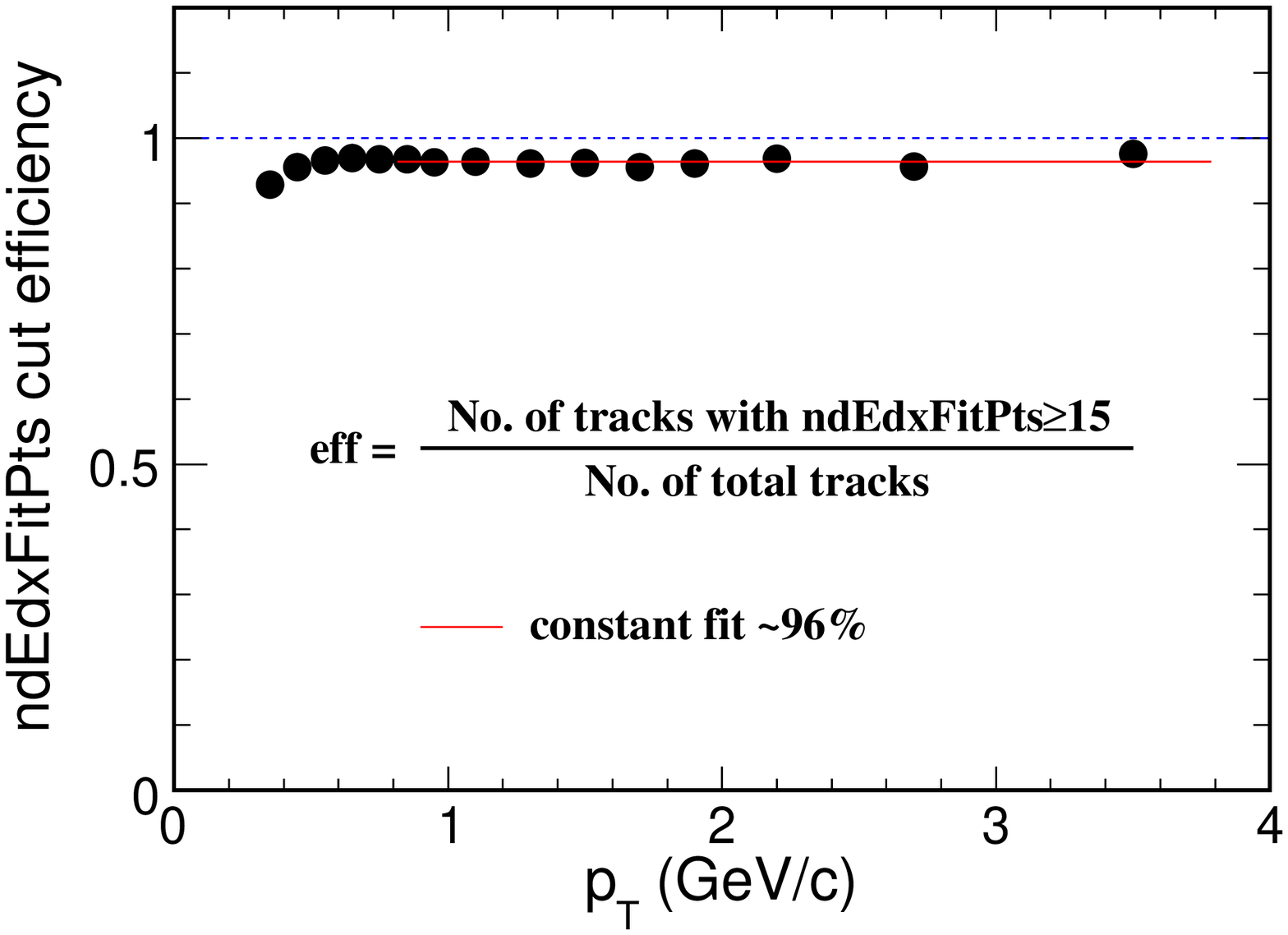}
\emn\\[10pt]
\caption[ndEdxFitPts cut efficiency]{Efficiency of cutting on
ndEdxFitPts $\geq$ 15 as a function of \pt.} \label{eff2} \ef

\subsection{Photonic background}

In this analysis, the dominant sources of photonic electron
background are considered as gamma conversions and scalar meson
Dalitz decay:

\begin{itemize}
\item $\gamma\rightarrow e^{+}e^{-}$ photon conversions in the
material in STAR detector. \item $\pi^0\rightarrow\gamma
e^{+}e^{-}$ ($1.198\pm0.032$)\%. \item $\eta\rightarrow\gamma
e^{+}e^{-}$ ($0.60\pm0.08$)\%.
\end{itemize}

The huge photonic background electrons come from conversions,
especially at low \pt ($<1$ \gevc), due to the amount of materials
in STAR detector. Fig.~\ref{convr} shows the conversion point and
conversion radius distributions from the simulation data with
HIJING generator + STAR GEANT detector configuration in 200 GeV
\AuAu\ collisions. It shows most of converted electrons are from
the huge material of 3-layer SVT ($R \sim 7, 11, 16cm$) and its
supporting structures.

\bf \centering \bmn[b]{0.5\textwidth} \centering
\includegraphics[width=1.0\textwidth]{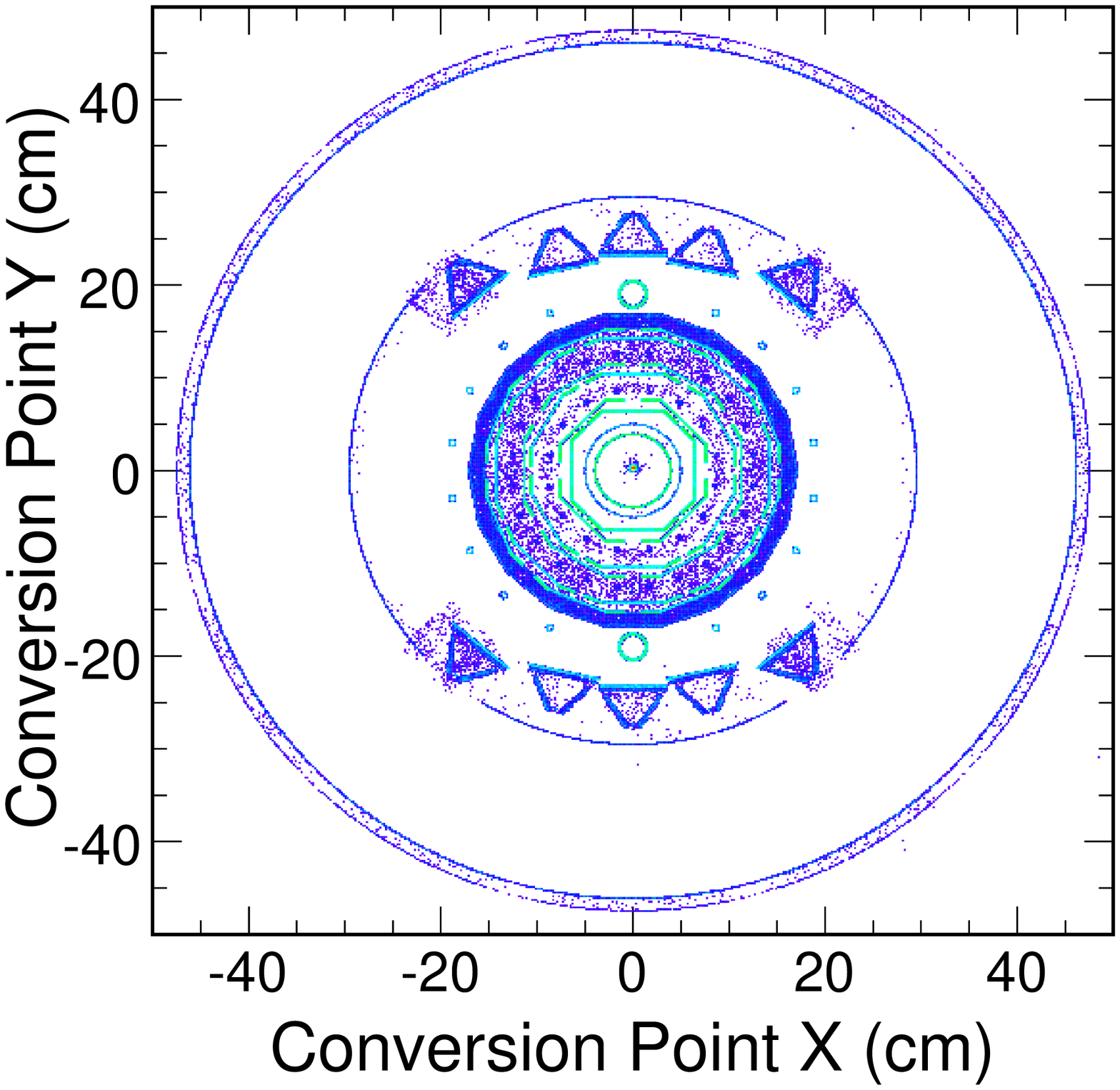}
\emn%
\bmn[b]{0.5\textwidth} \centering
\includegraphics[width=1.0\textwidth]{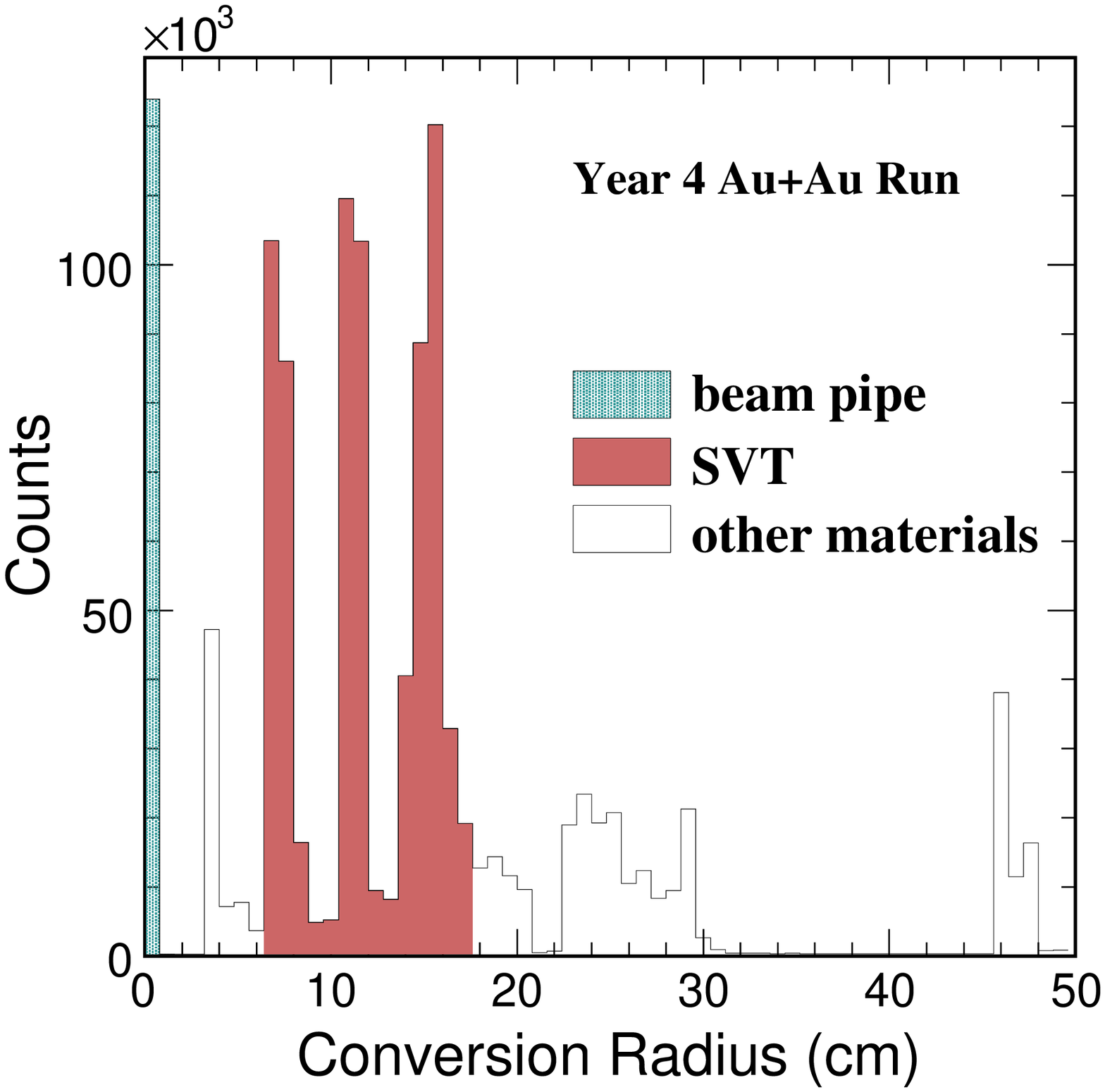}
\emn\\[10pt]
\caption[Fit \dedx\ tail with different width]{Left panel: The 2-D
scattering plot for conversion point, where the photonic electron
come from. It reflects the STAR material structure. Right panel:
The conversion radius distributions.} \label{convr} \ef

Photonic electrons dominate the total yield, especially at low
\pt. At \pt $>1$ \gevc, The electrons from heavy flavor hadron
decays, which is considered as the signal, may become visible due
to the enhancement of the total spectrum w.r.t background
spectrum. From the previous measurements, background from photon
conversion {\em etc.} was reconstructed experimentally using
topological method~\cite{IanPRC,IanThesis}. The $e^+e^-$ from
photonic background have the characteristic feature of low
invariant mass. The $e^+e^-$ invariant mass were reconstructed
from a tagged electron (positron) from TOF at low \pt\ (we use TPC
\dedx\ to identify electron at high \pt, due to the poor
statistics from TOF events at high \pt) combined with every other
global positron (electron) candidate, whose helixes have a dca
($dca_{e^{+}e^{-}}$) to the tagged tracks' less than 1 cm, to find
the other partner track reconstructed in the TPC. Because the TPC
acceptance is large enough, the pair reconstruction efficiency is
reasonable. Two methods to reconstruct the invariant mass for
$e^+e^-$ pairs were used in this analysis: One is to reconstruct
the invariant mass in the Cartesian coordinate. In the other
method, the invariant mass of the pair is constructed in the r-z
plane with an opening angle cut of ($\phi_{e^{+}e^{-}}<\pi/10$) in
the azimuthal plane, here the z-axis is the beam direction. The
random combinatorial background is constructed by rotating the
partner candidate lepton by $180^{o}$. Raw yields of the photonic
electrons are the excess over the combinatorial background in the
low mass region ($M_{e^{+}e^{-}}<150$ \mev). Comparing these two
methods, the second one with an opening angle cut has the ability
to reduce the combinatorial background and the effect of the
secondary vertex resolution, which shift the mass to higher region
(the second bump observed in the first method), see
Fig.~\ref{phbgmass}.

\bf \centering \bmn[b]{0.5\textwidth} \centering
\includegraphics[width=1.0\textwidth]{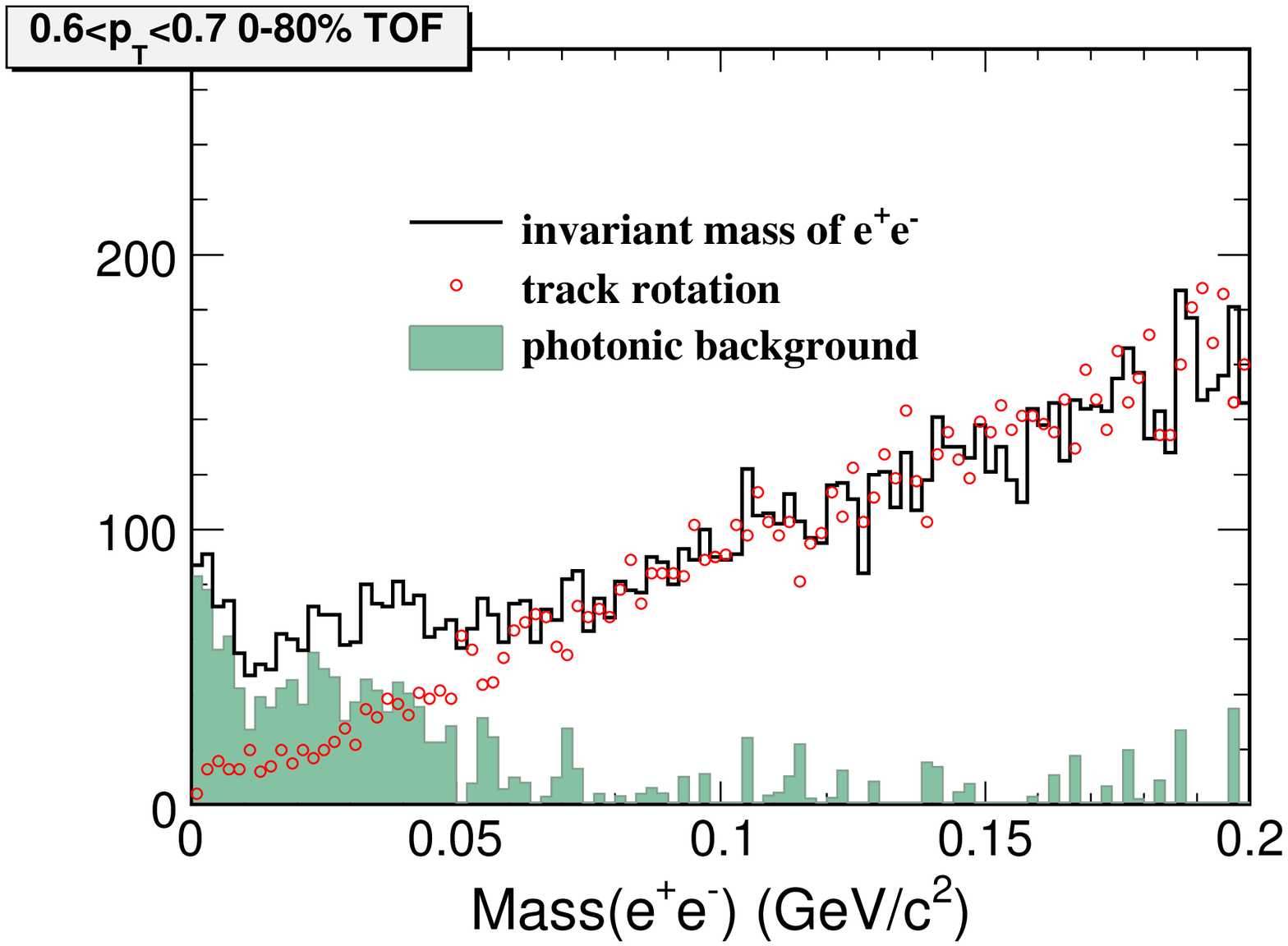}
\emn%
\bmn[b]{0.5\textwidth} \centering
\includegraphics[width=1.0\textwidth]{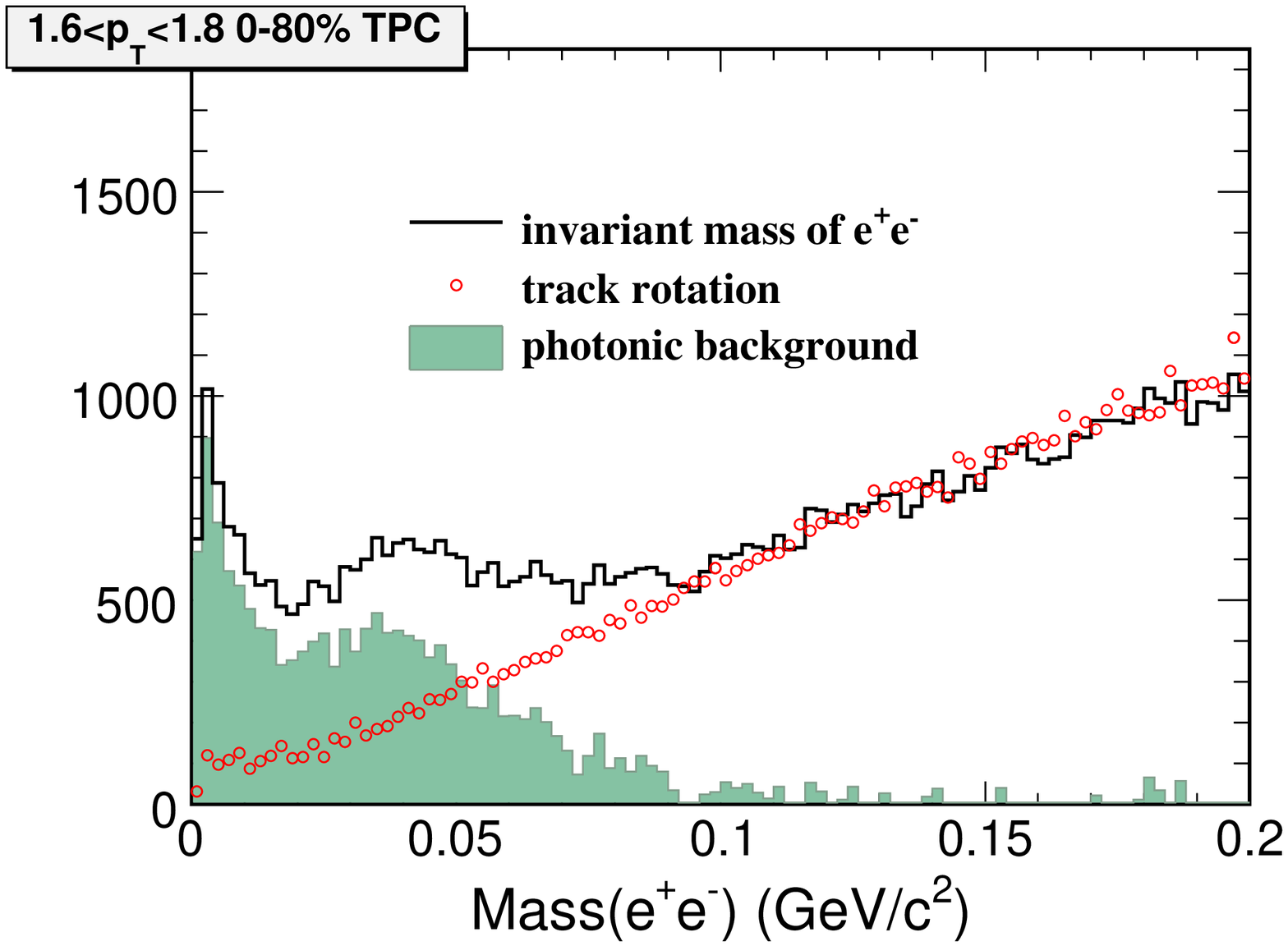}
\emn\\[10pt]
\bmn[b]{0.5\textwidth} \centering
\includegraphics[width=1.0\textwidth]{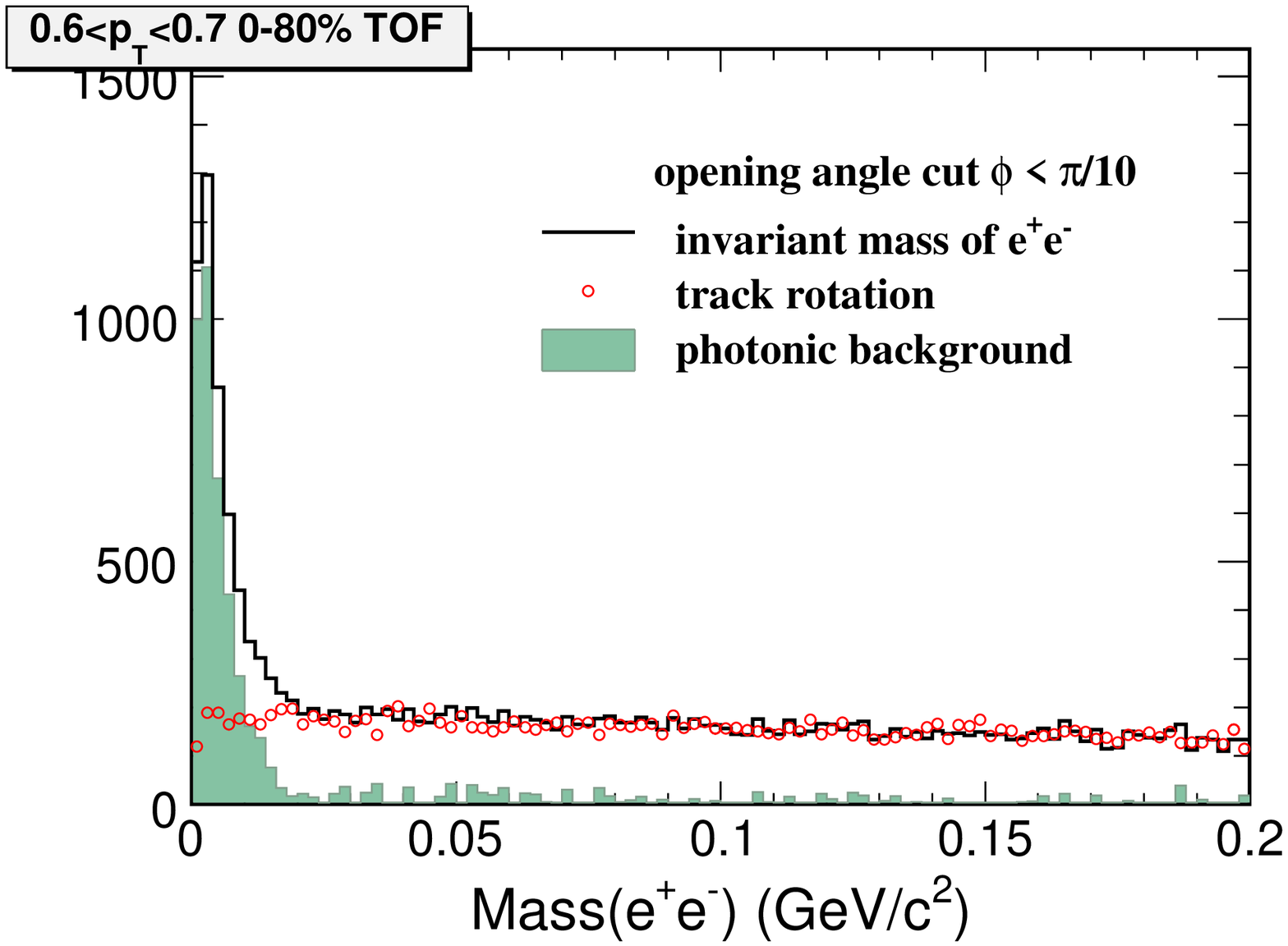}
\emn%
\bmn[b]{0.5\textwidth} \centering
\includegraphics[width=1.0\textwidth]{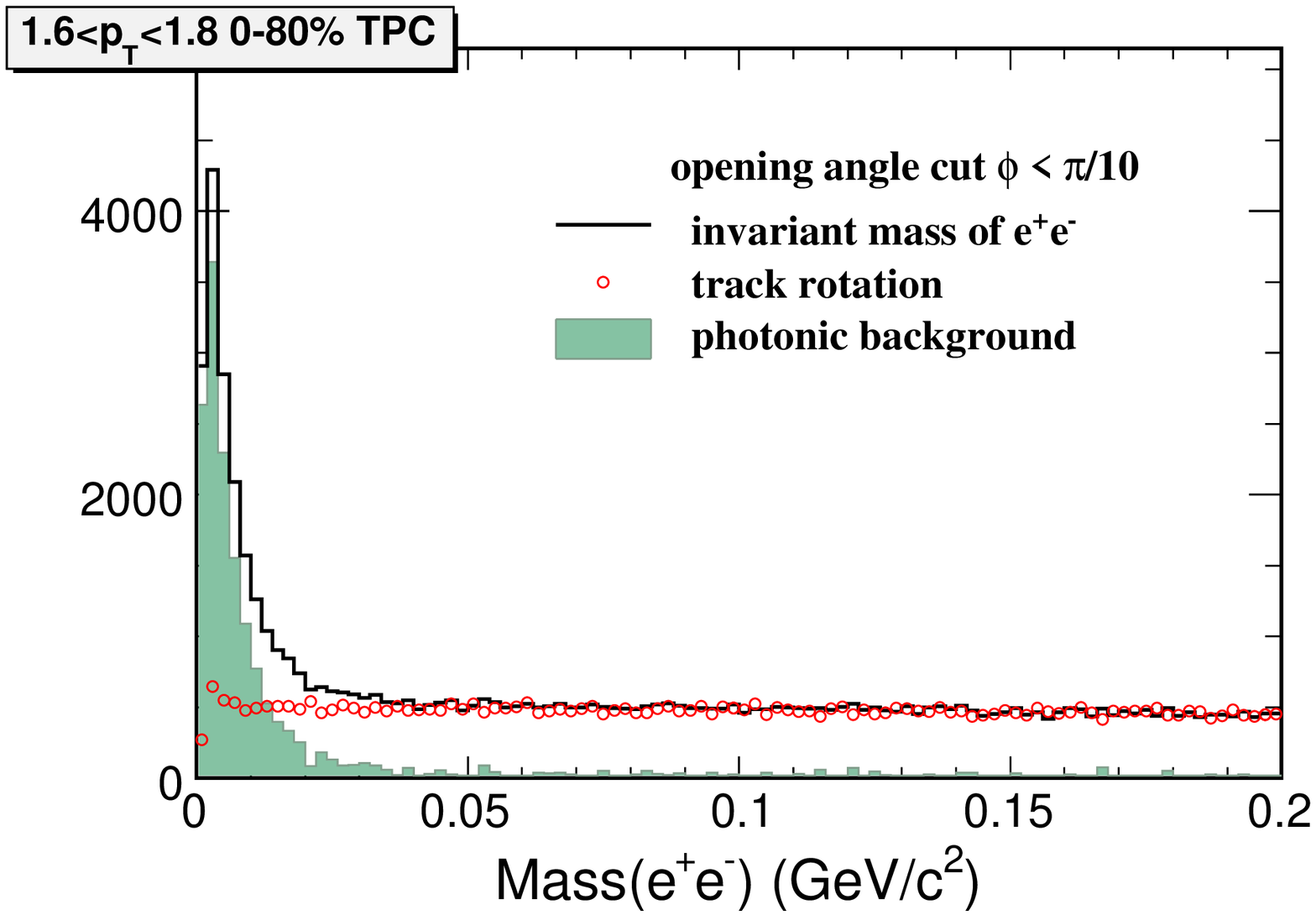}
\emn\\[10pt]
\caption[Photonic background invariant mass] {Left Column:
Photonic invariant mass from TOF at low \pt. Right Column:
Photonic invariant mass from TPC at high \pt. Upper Raw: Photonic
invariant mass reconstructed in the Cartesian coordinate. Bottom
Raw: Photonic invariant mass reconstructed in the r-z plane with
an opening angle cut of ($\phi<\pi/10$) in the azimuthal plane.}
\label{phbgmass} \ef

From the measurement of the photonic invariant mass with the
combinatorial background subtracted in each \pt\ bin, the photonic
electron raw yield is obtained. Electrons could be randomly
combined with hadrons due to hadron contamination from the \dedx\
selections. The different \dedx\ cuts were tried to estimate the
systematic uncertainties. The uncertainties of the combinatorial
background ($5-15$\%) dominate the systematic error for photonic
electron spectra. The photonic background raw yields need
correction by the efficiencies and detector acceptance as what we
did for inclusive electron analysis. In addition, the partner
finding efficiency, which was calculated by taking the ratio of
$e^+$($e^-$) with partner $e^-$($e^+$) found to total converted
electrons, was done from the same reconstruction process in
embedding. Since the embedded $\pi^{0}$ was flat in \pt, the
charged $\pi$ \pt\ distribution measured by STAR~\cite{lqeloss}
was used to weight the efficiency, shown in Fig.~\ref{pi0emb}.

\bf \centering\mbox{
\includegraphics[width=1.0\textwidth]{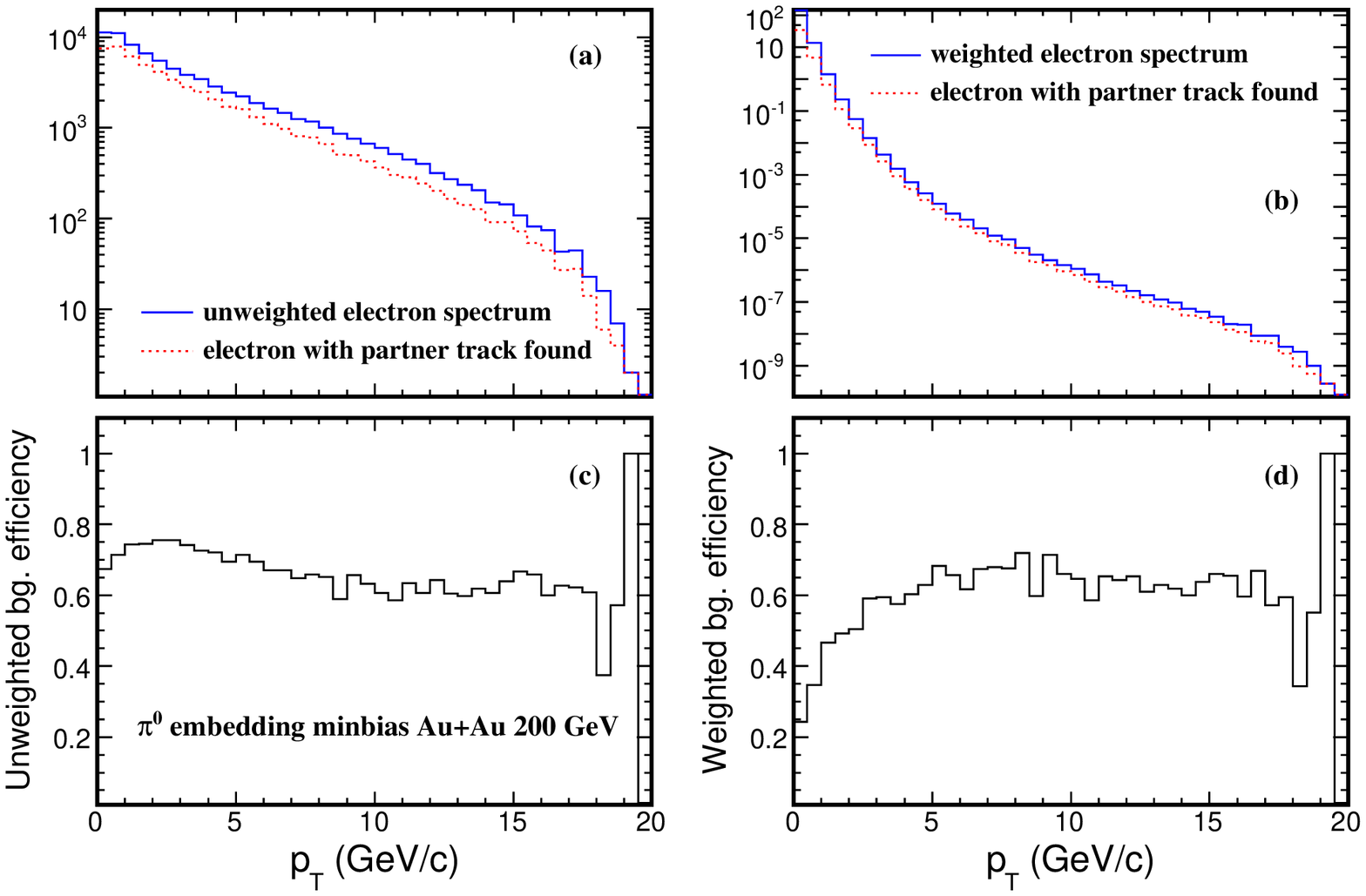}}
\caption[$\pi^{0}$ embedding and background efficiencies]{Panel
(a): Converted electron \pt\ distribution (blue solid line) and
electron \pt\ distributions with partner track found (red dashed
line). Panel (b): Weighted electron \pt\ distributions. Panel (c):
Background reconstruction efficiency. Panel (d): Weighted
background reconstruction efficiency.} \label{pi0emb} \ef

From the same embedding data, EMC electron analysis gives the
consistent background efficiency, shown in the right panel of
Fig.~\ref{bgeffcom1}. At TOF electron \pt\ region, the effect of
the momentum resolution to the background efficiency is small. A
random sampling method was also used to cross check the normal
weighted method. The normalized \pt\ spectra shape of $\pi^{+}$
was used to sample the electron candidates randomly in each \pt\
bin. Both the two methods give consistent efficiencies, shown in
the left panel of Fig.~\ref{bgeffcom1}.

\bf \centering \bmn[b]{0.5\textwidth} \centering
\includegraphics[width=1.0\textwidth]{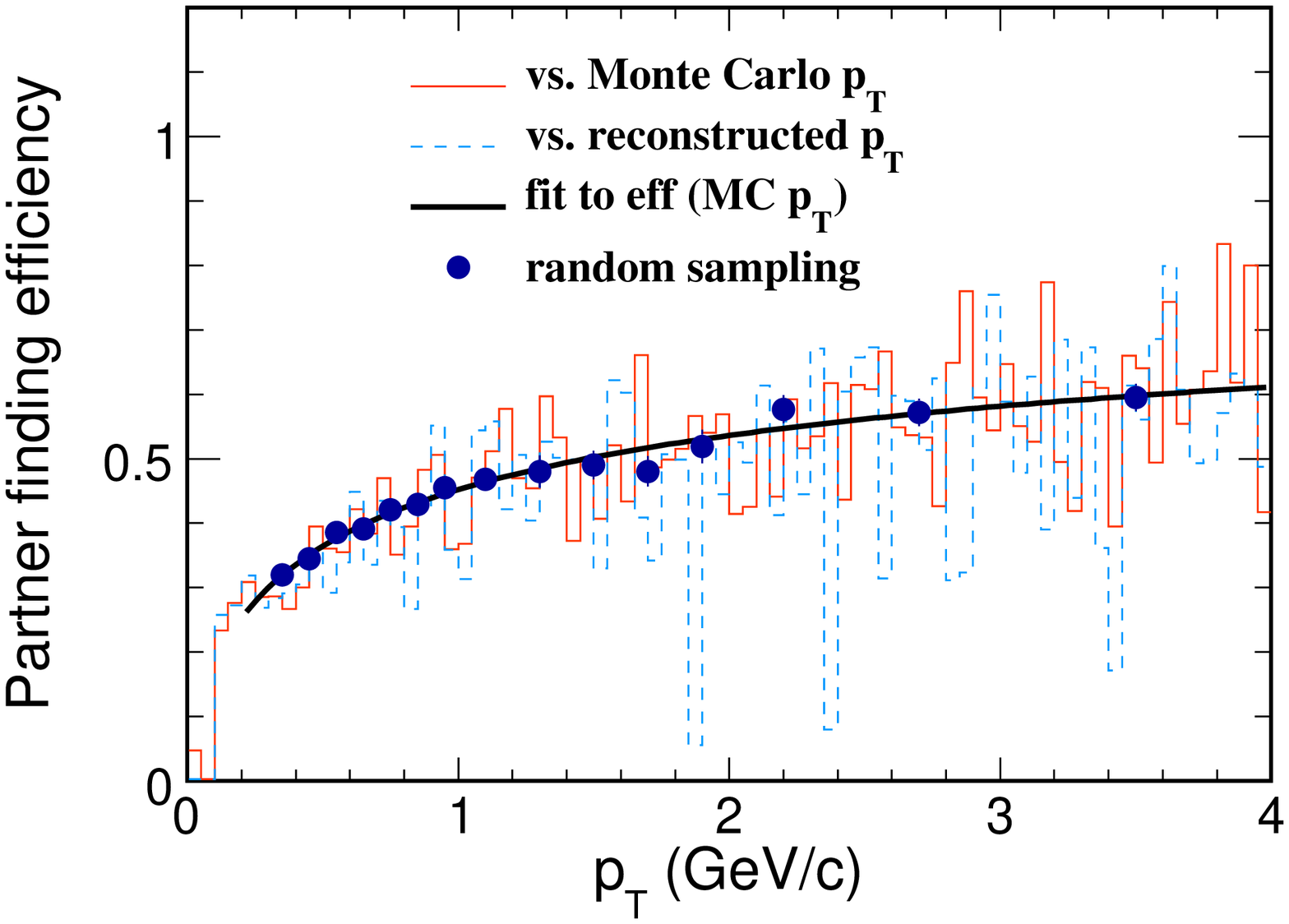}
\emn%
\bmn[b]{0.5\textwidth} \centering
\includegraphics[width=1.0\textwidth]{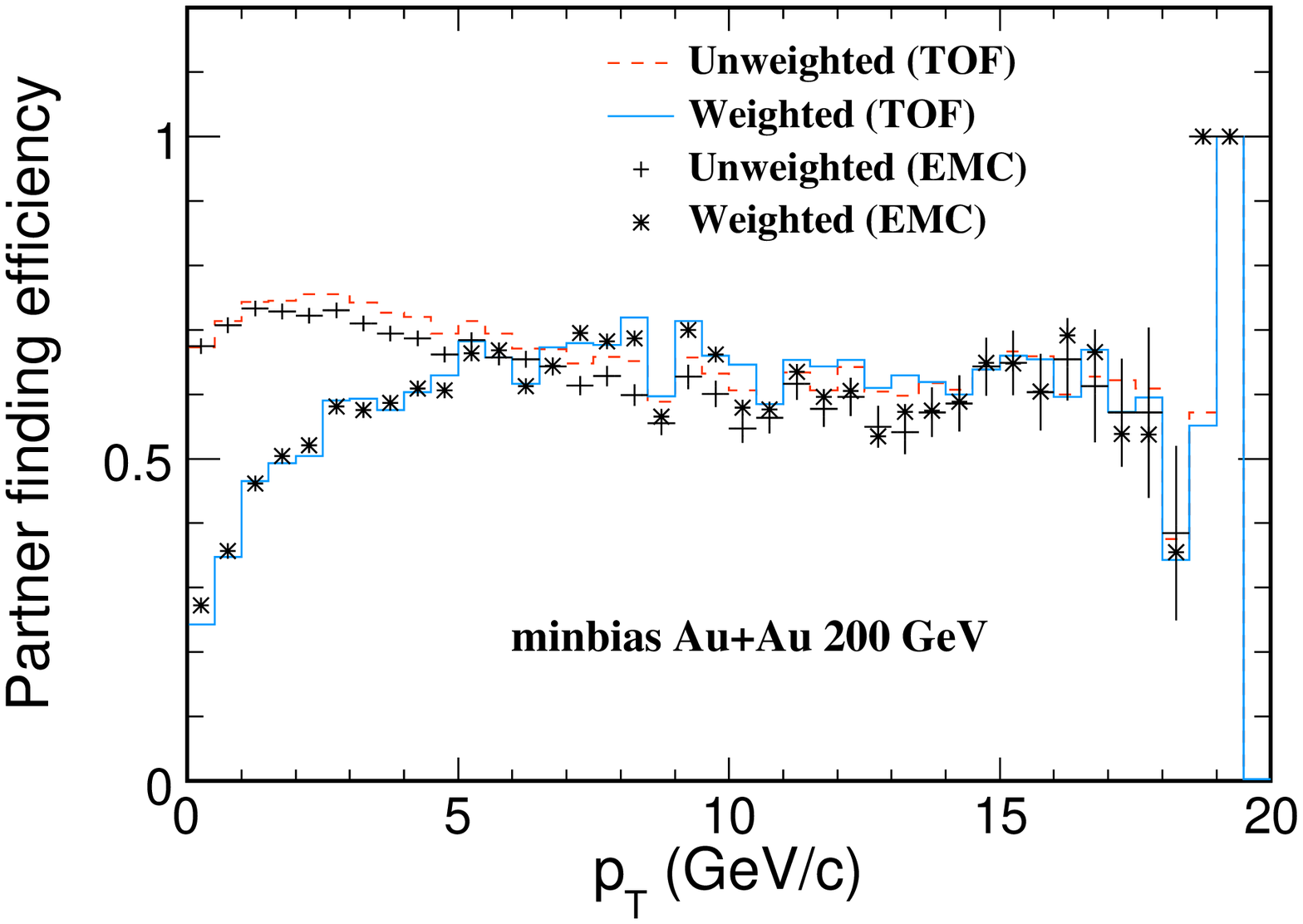}
\emn\\[10pt]
\caption[Comparisons of background efficiencies]{Left panel:
Comparison of different ways to derive background efficiency at
TOF electron \pt\ region from embedding. Right panel: Comparison
of the background efficiency from different electron analysis.}
\label{bgeffcom1} \ef

To systematically estimate the uncertainties of different $\pi$
spectra for the background efficiency weighting, the PHENIX
$\pi^{0}$ spectrum~\cite{pi0PHENIX} was also used to compare with
STAR $\pi^{+}$ spectrum, shown in the Panel (c) of
Fig.~\ref{bgeffcom2}. The difference of the efficiencies weighted
by STAR $\pi^{+}$ and PHENIX $\pi^{0}$ \pt\ distributions is
small. At \pt\ around 1-5 \gevc, the efficiencies are nearly the
same. Below 1 \gevc, the difference is up to 20\%. At high \pt\
($>5$ \gevc), it is $\sim$5\%, see the Panel (a),(b) of
Fig.~\ref{bgeffcom2}.

\bf \centering \bmn[c]{0.5\textwidth} \centering
\includegraphics[width=1.1\textwidth]{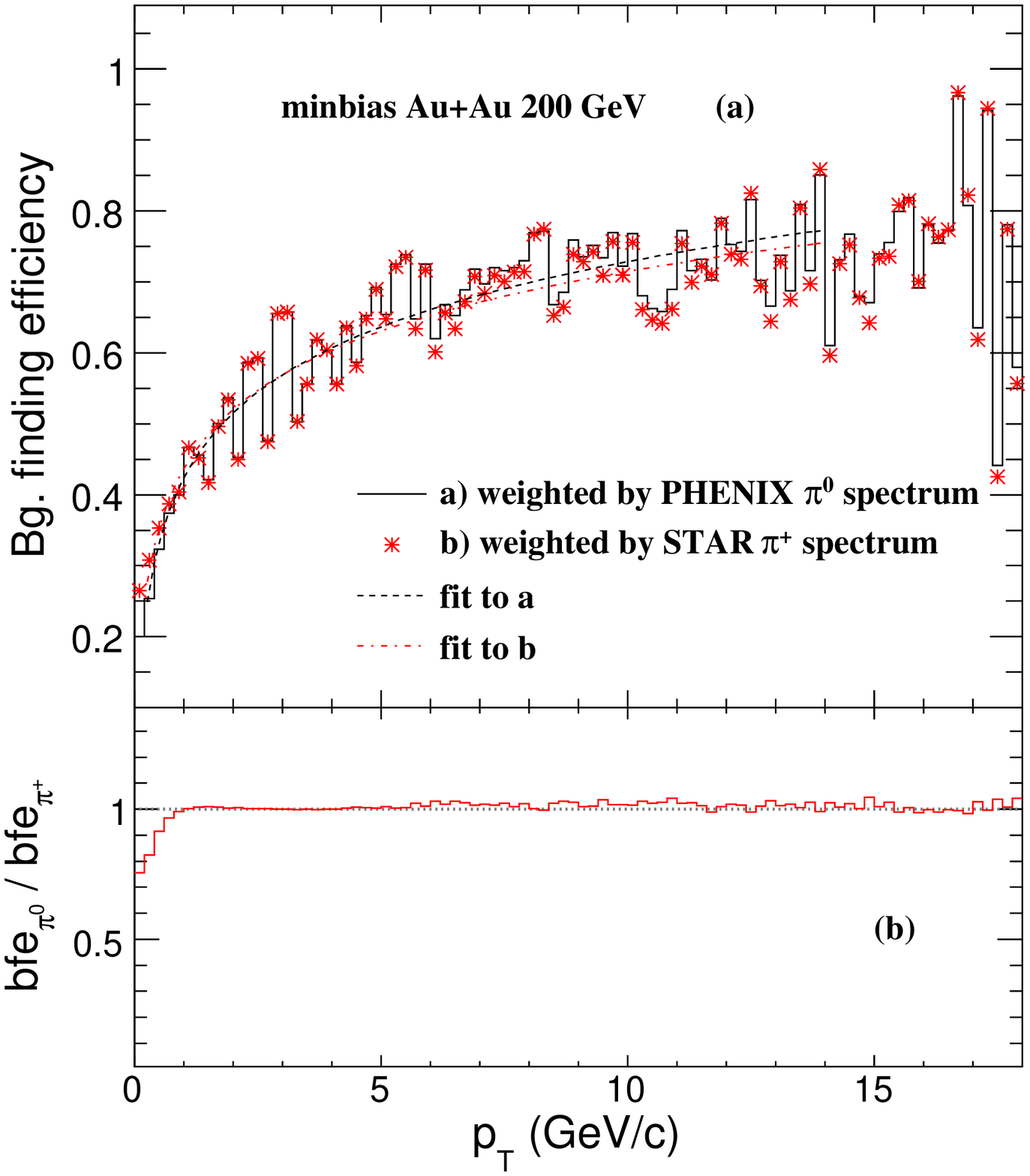}
\emn%
\bmn[c]{0.5\textwidth} \centering
\includegraphics[width=0.9\textwidth]{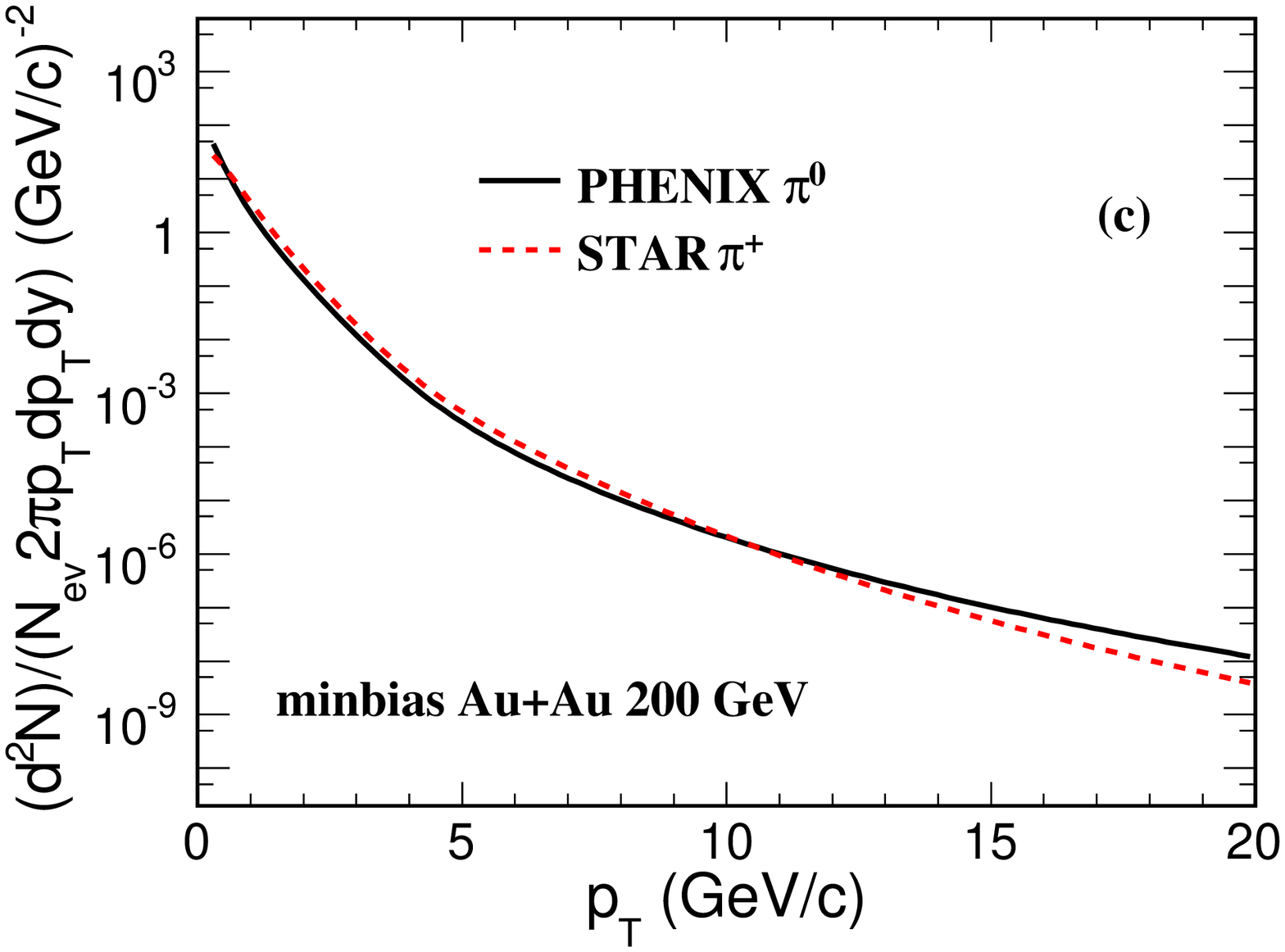}
\emn%
\caption[Comparisons of background efficiencies from different
weighting]{Panel (a): Comparison of the background efficiency from
different weighting. Panel (b): The ratio of the efficiencies.
Panel (c): Comparison of PHENIX $\pi^{0}$ spectrum and STAR
$\pi^{+}$ spectrum.} \label{bgeffcom2} \ef

Simply, the techniques are just shown in detail for minbias 0-80\%
and central 12\% collisions, since good statistics can be
collected in these two data sets. The techniques for other
sub-centralities from minbias events are the same.

After corrected by detector acceptance and efficiencies, the
inclusive and photonic electron spectra were shown in the Panel
(a) of Fig.~\ref{espectra1}. The ratios of the inclusive spectra
and the photonic electron spectra are consistent with unit at low
\pt\ ($<0.9$ \gevc). The increasing ratios above unit at $p_T>0.9$
\gevc\ indicate that the observable signal, shown in the Panel
(b). Due to poor statistics of the invariant mass by selecting
tagged electrons from TOF, we use TPC \dedx\ instead. Panel (c)
shows the comparison of the photonic background spectra from TOF
and TPC. Within errors, they are consistent.

\bf \centering \bmn[c]{0.5\textwidth} \centering
\includegraphics[width=1.0\textwidth]{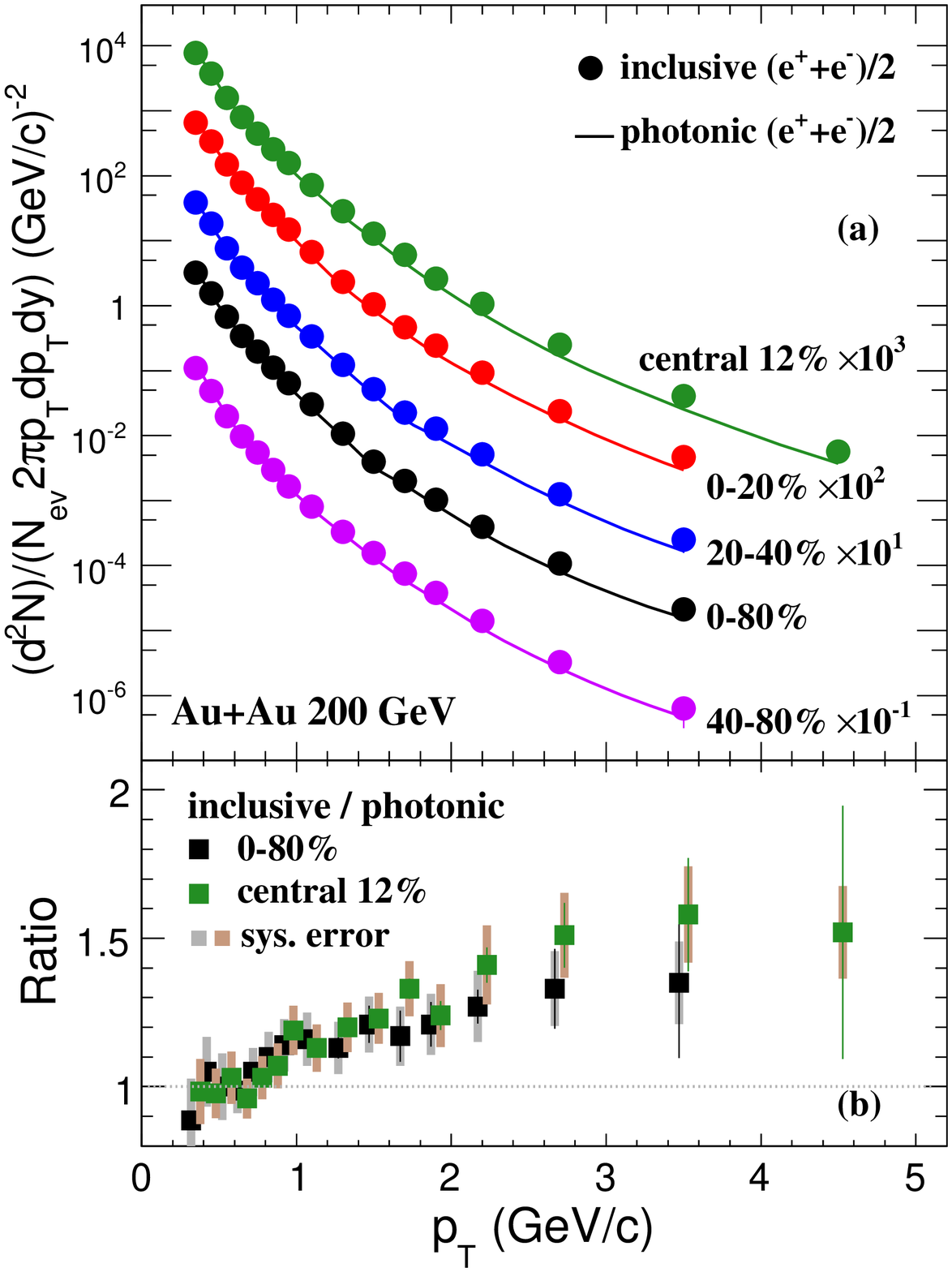}
\emn%
\bmn[c]{0.5\textwidth} \centering
\includegraphics[width=0.95\textwidth]{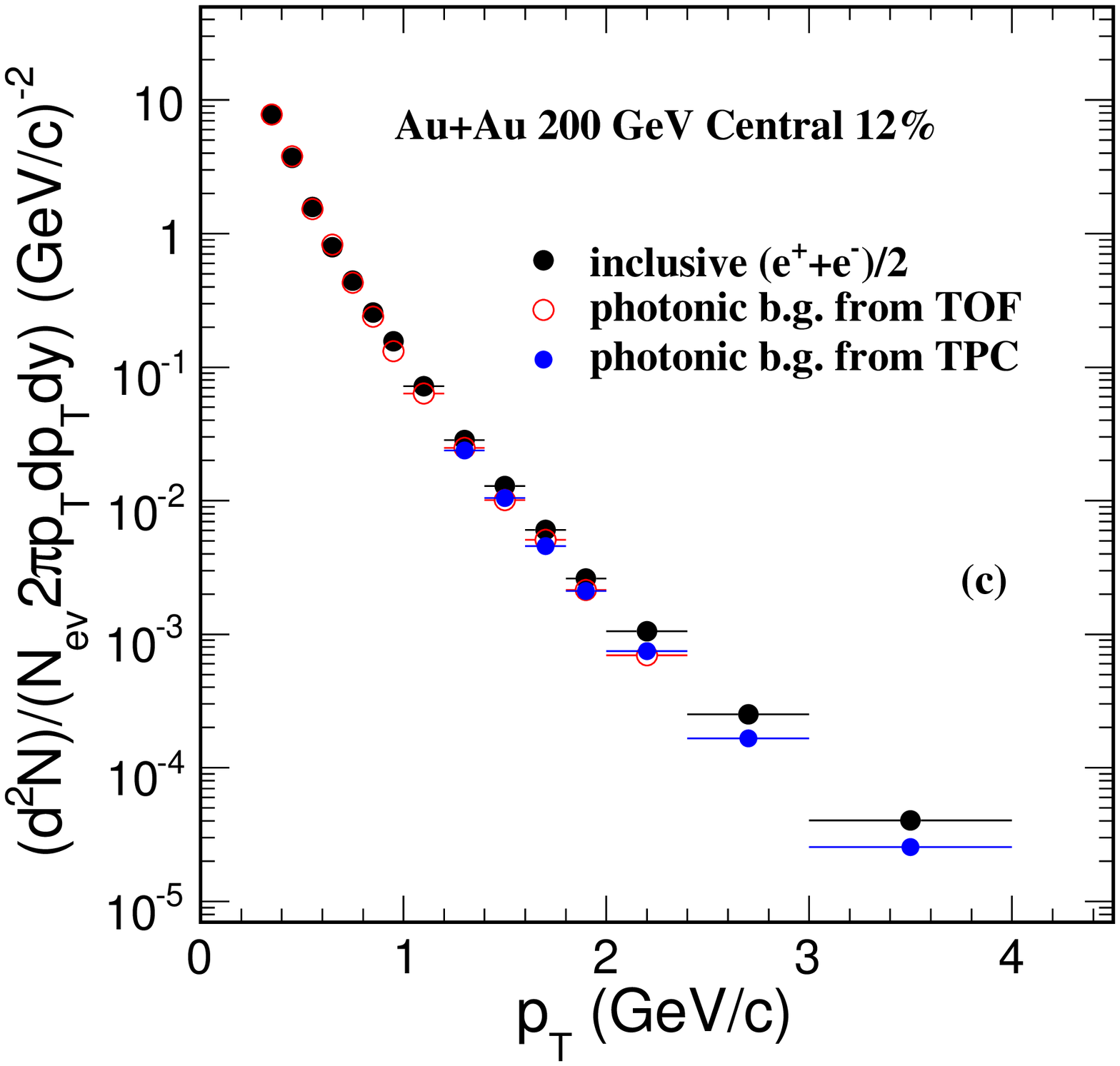}
\emn%
\caption[Centrality dependence of inclusive electron and photonic
electron spectra]{Panel (a): Centrality dependence of inclusive
electron and photonic electron spectra. Panel (b): The ratio of
inclusive electron over photonic electron spectra. Panel (c):
Comparison of photonic spectrum from TOF and that from TPC in
central 12\% collisions.} \label{espectra1} \ef

Table~\ref{phoecut} summarizes the cuts for tagged electron and
its partner electron for the conversion pair selection.

\begin{table}[hbt]
\caption[Electron cuts for photonic background]{Electron cuts for
photonic background analysis} \label{phoecut} \vskip 0.1 in
\centering\begin{tabular}{|c|c|} \hline \hline
Tagged electron & partner electron\\
\hline inclusive electron cuts required & nFitPts $\geq$ 15\\
& nFitPts/nMax $>0.52$\\
\hline $-1<n\sigma_{e}<3$ (TOF) & $-1<n\sigma_{e}<3$\\
\cline{1-1} $0<n\sigma_{e}<2$ (TPC) &\\
\hline
\hline \multicolumn{2}{|c|}{$dca_{e^{+}e^{-}}$ $<1$ cm}\\
\hline \multicolumn{2}{|c|}{$M_{e^{+}e^{-}}<150$ MeV}\\
\hline \multicolumn{2}{|c|}{$M_{e^{+}e^{-}(\theta)}<150$ MeV}\\
\hline \multicolumn{2}{|c|}{Opening angle ($\phi$) $<\pi/10$}\\
\hline \hline
\end{tabular}
\end{table}

\bf \centering\mbox{
\includegraphics[width=0.55\textwidth]{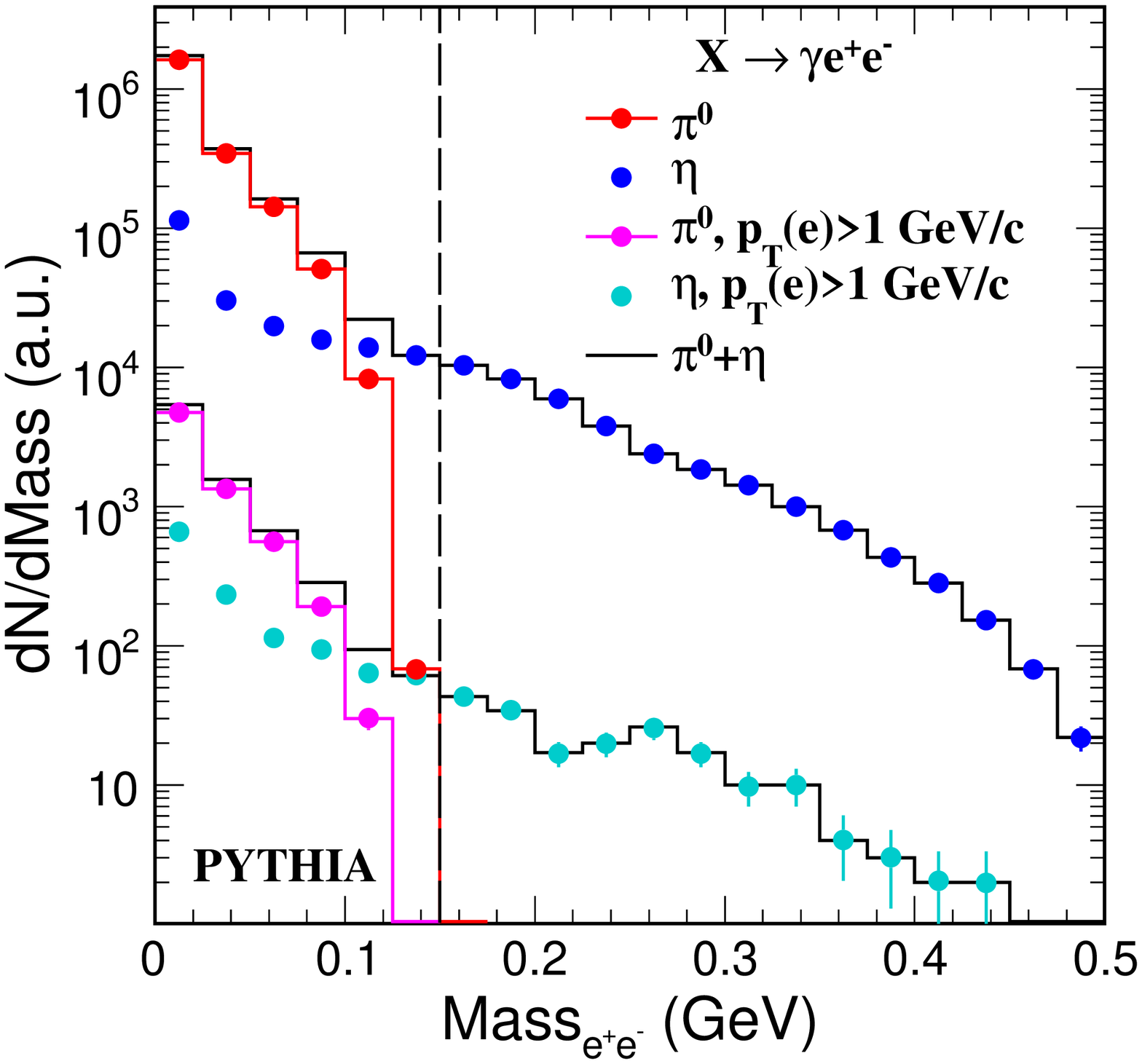}}
\caption[$\pi^{0}$ and $\eta$ Dalitz decays from PYTHIA]{Invariant
mass distributions of $\pi^{0}$ and $\eta$ Dalitz decays from
PYTHIA. The lost fraction above the mass cut is $\sim$1.4\% for
all electron \pt. For electron $p_T>1$ \gevc, the lost fraction is
$\sim$2\%.} \label{dalitz} \centering\mbox{
\includegraphics[width=0.55\textwidth]{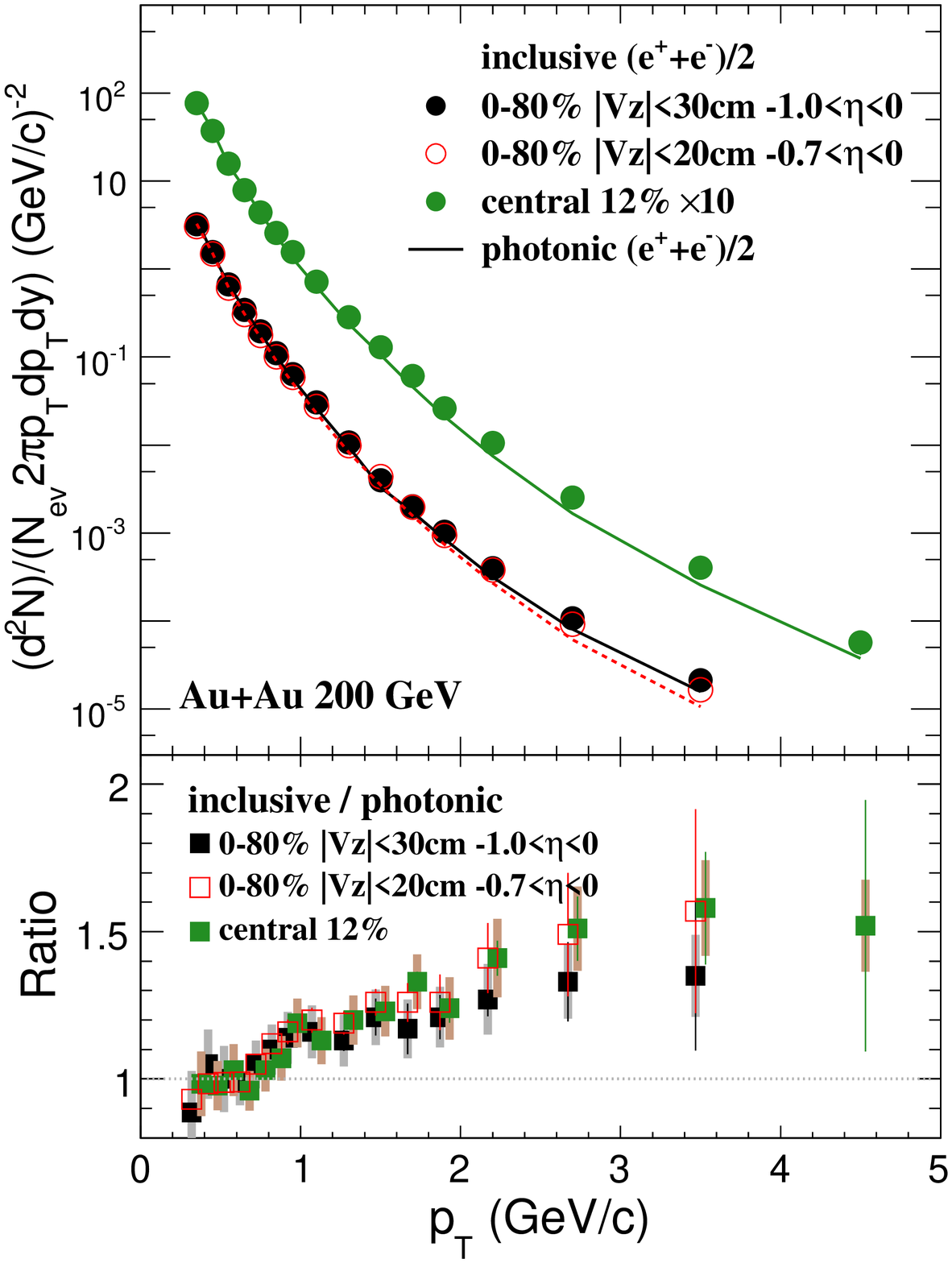}}
\caption[Inclusive and photonic electron spectra comparison with
different Vz and $eta$ cuts]{Comparison of the inclusive and
photonic electron spectra and their ratios with different Vz and
$\eta$ cuts.} \label{vzetacut}
\ef

The final spectra were also corrected by the efficiencies of the
$n\sigma_{e}$ cuts. The efficiencies of $dca_{e^{+}e^{-}}$,
$M_{e^{+}e^{-}}$ and opening angle cuts were included in the
background reconstruction efficiencies from embedding. From PYTHIA
study, only $\sim$2\% electrons from $\pi^{0}$ and $\eta$ Dalitz
decay are lost by cutting on $M_{e^{+}e^{-}}<150$ MeV, shown in
Fig.~\ref{dalitz}. Including vector meson decays, a 2-3\%
contribution from other sources to the total background was taken
into account~\cite{stardAucharm}.


Since the tighter cuts for vertex-Z and eta could reduce the
material influence, which is sensitive to the ratio of signal to
photonic background, another cuts of $|Vz|<20$ cm and
$-0.7<|\eta|<0$ were applied to cross check the final inclusive
and photonic electron spectra, shown in Fig.~\ref{vzetacut}. The
inclusive and photonic electron spectra are both little lower than
those from former cuts, due to the reduced yields of converted
electrons. The inclusive/photonic ratio is similar as central top
12\%, whose Vz distributions is narrow from -20 cm to 20 cm. At
higher \pt\ ($>2$ \gevc), the ratio from former cuts is a little
lower than the central top 12\% and the minbias result from new
cuts, which is probably because the decrease of material by
applying the new cuts.

\subsection{Centrality dependence of non-photonic electron}

After the photonic background subtraction from inclusive
electrons, the non-photonic electrons, which are the signal from
heavy flavor decay, are extracted. Due to low \pt\ huge photonic
background and high \pt\ low statistics, the non-photonic
electrons were measured covering \pt\ from 0.9 \gevc\ to 4 \gevc\
in minbias \AuAu\ collisions and \pt\ from 0.9 \gevc\ to 5 \gevc\
in central 12\% \AuAu\ collisions.

\bf \centering\mbox{
\includegraphics[width=0.65\textwidth]{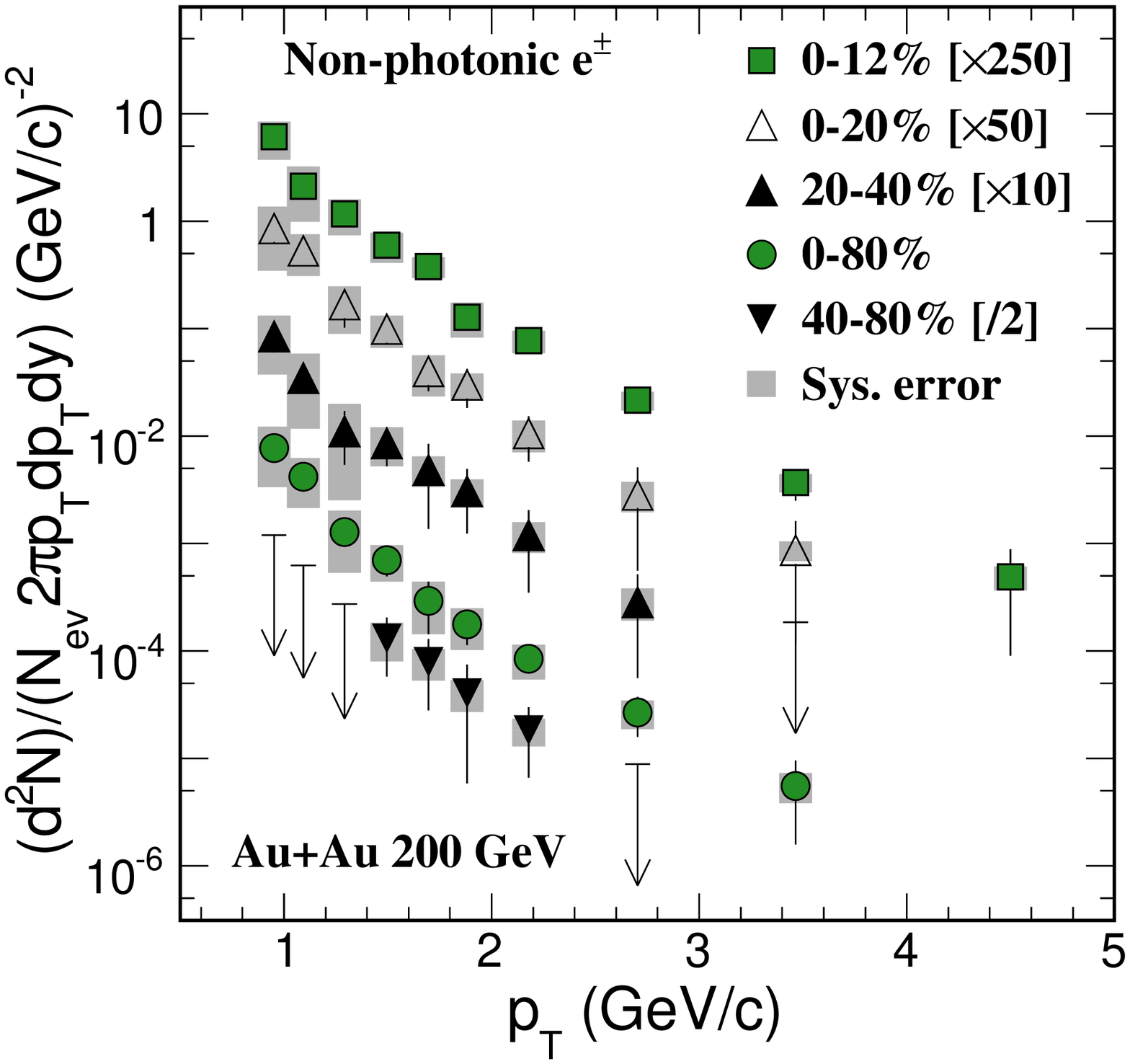}}
\caption[Centrality dependence of the non-photonic electron
spectra]{Centrality dependence of the non-photonic electron
spectra.} \label{nonphespec} \ef

Fig.~\ref{nonphespec} shows the centrality dependence of the
non-photonic electron spectra. The bin-by-bin errors dominated by
the uncertainties of inclusive electron yield extraction, photonic
combinatorial background reconstruction and photonic background
efficiency {\em etc.}, were propagated to the final non-photoinc
electron spectra. The overall systematic errors, like the tracking
efficiency uncertainty {\em etc.}, were counted in the total
systematic errors, which are shown as the shaded boxes.

Table~\ref{syserr} lists the systematic error contributions.

\begin{table}[hbt]
\caption[Systematic errors list]{Systematic errors contributions
to the electron spectra.} \label{syserr} \vskip 0.1 in
\centering\begin{tabular}{|c|c|} \hline \hline
Inclusive electron yield extraction & $5-15$\% (point-to-point)\\
\hline Combinatorial background uncertainties & minbias:
$6-15$\% (point-to-point)\\
 & Central $5-10$\% (point-to-point)\\
\hline $n\sigma_{e}$ cut: (0,3)(-1,3) - (-1,3)(-1,3), (0.2)(-1,3) & $3-5$\%\\
\hline Background reconstruction efficiency &  $0.3<p_T<0.9$
\gevc\ up to 20\%
(weight)\\
 & $p_T>0.9$ \gevc\ $3-5$\% (weight, EMC) \\
\hline $\eta$ Dalitz, Kaon decay, vector meson decay & $2-3$\%\\
\hline TPC tracking efficiency & 5\%\\
\hline Central 10\% $\rightarrow$ 12\% & 5\%\\
\hline \hline
\end{tabular}
\end{table}

\section{Non-photonic electron azimuthal anisotropy distributions}

\subsection{Heavy flavor anisotropic flow}

In non-central nucleus-nucleus collisions, the anisotropy in the
initial coordinate space is transferred into final momentum
anisotropy. The anisotropic flow is strongly sensitive to the
partonic rescattering. The elliptic flow parameter \vv\ from
Fourier expansion of the azimuthal distributions is considered as
good probe to extrapolate early pressure and
density~\cite{flowth0}. The mass dependence of \vv\ for identified
light hadrons can be well reproduced by a hydrodynamical model in
the low \pt\ region ($p_T<2$ \gevc), which indicates that the
collective motion evolves in early partonic phase of the
collisions~\cite{hydro1}. The \pt\ dependence of \vv\ scaled by
the number of constituent quarks is observed to be universal. The
scaling behavior can be illuminated by a quark coalescence
model~\cite{coalMolnar}. But for heavy flavor quark, due to its
extremely heavy mass, it can acquire flow only when light quarks
punch it very frequently in a very dense medium. The thermal
equilibrium is approached through sufficient interactions in the
bulk matter. So the measurement of \vv\ for heavy quarks is vital
to test the light flavor thermalization and partonic density in
the early stage of heavy ion collisions.

The single electron from heavy flavor hadrons semileptonic decay
was used to infer the production of heavy quarks. The angular
correlation between charmed meson and electron from its
semileptonic decay is significantly strong. That indicates charmed
meson \vv\ can be indirectly measured through single electron
\vv~\cite{kocharmflow,minepiv2}. The non-zero single electron \vv\
was observed by PHENIX detector at RHIC~\cite{Phenixv2}. STAR
detector provides a large acceptance with $|\phi|<2\pi$,
$|\eta|<1.5$. The measurement of single electron \vv\ from STAR
will systematically enhance our understanding of the partonic
thermalization in the high dense matter. But it is still a
challenge due to the huge photon conversion from the virulent
material in the STAR detector.

\subsection{Inclusive and photonic electron elliptic flow}

In this analysis, we try to develop a method to measure single
electron \vv\ from the data taken with the STAR experiment during
the \sNN$=$200 GeV Au+Au run in 2004. A total of 12 and 19 million
0-80\% minimum bias Au+Au events were used for the inclusive
electron identification from the TOF and photonic electron
reconstruction from the TPC, respectively. The vertex Z is
required to be less than 20 cm in order to reduce the material
influence as well as the pseudorapidity ($-0.7<\eta<0$).

Inclusive electrons are separated from hadrons with more than 90\%
purity after applying a cut of their $dE/dx$ ($0<n\sigma_e<3$) in
the TPC and a cut of $|1/\beta-1|<0.03$ in the TOF.

As the photonic electron identification in the previous analysis
for electron spectra, in the cylinder with beam axis and azimuthal
plane, the invariant mass of the $e^+e^-$ pairs is reconstructed
with an opening angle cut less than $\pi/10$. The photonic
electrons are identified from the previous invariant mass
subtracted by the combinatorial background in a very low mass
region ($M_{e^{+}e^{-}}<15$ MeV). Other cuts such as
$M_{e^{+}e^{-}}<10$ MeV, $M_{e^{+}e^{-}}<20$ MeV are also tried to
systematically study the photonic electron \vv.

\bf \centering \bmn[b]{0.5\textwidth} \centering
\includegraphics[width=1.0\textwidth]{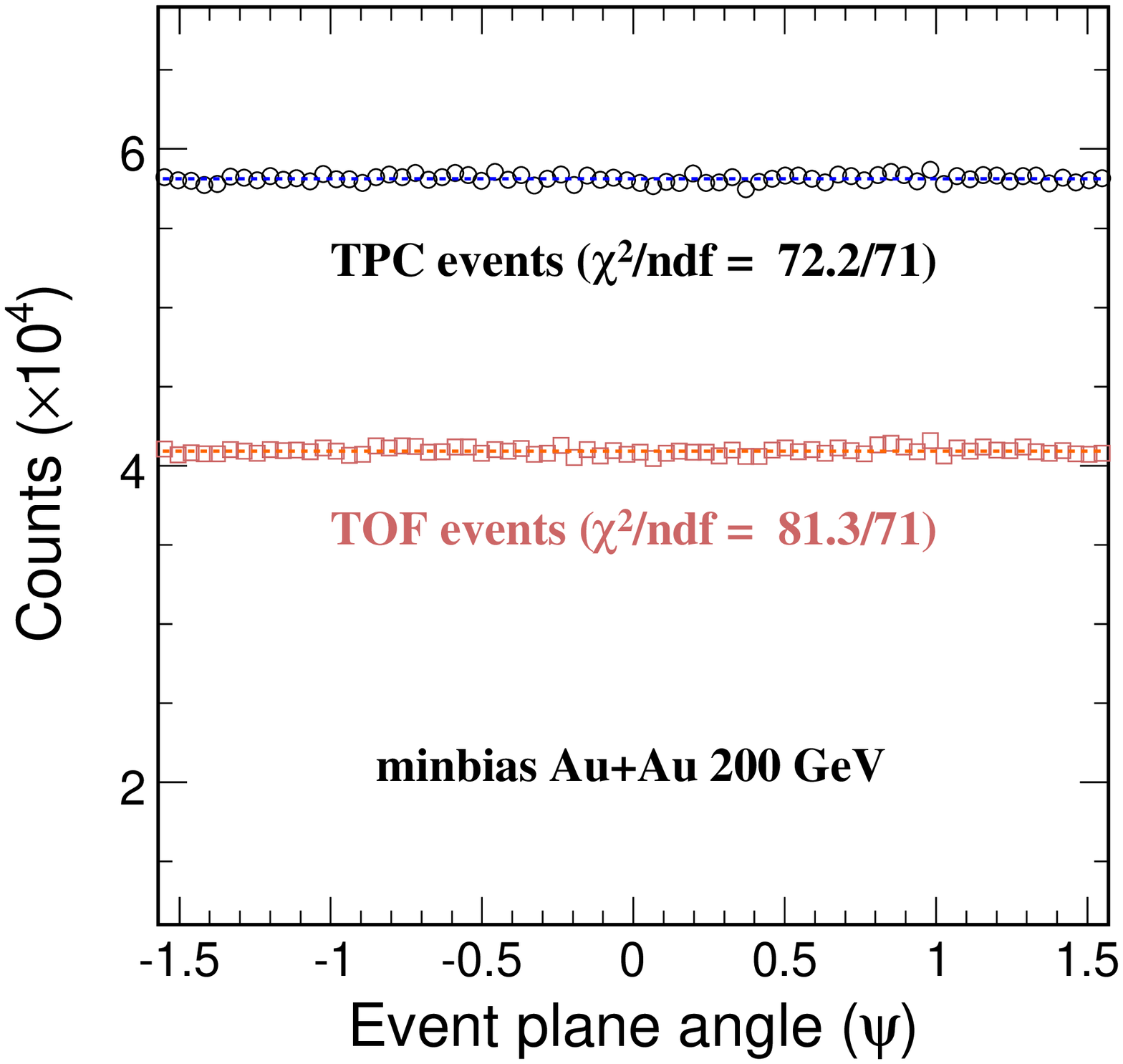}
\emn%
\bmn[b]{0.5\textwidth} \centering
\includegraphics[width=1.0\textwidth]{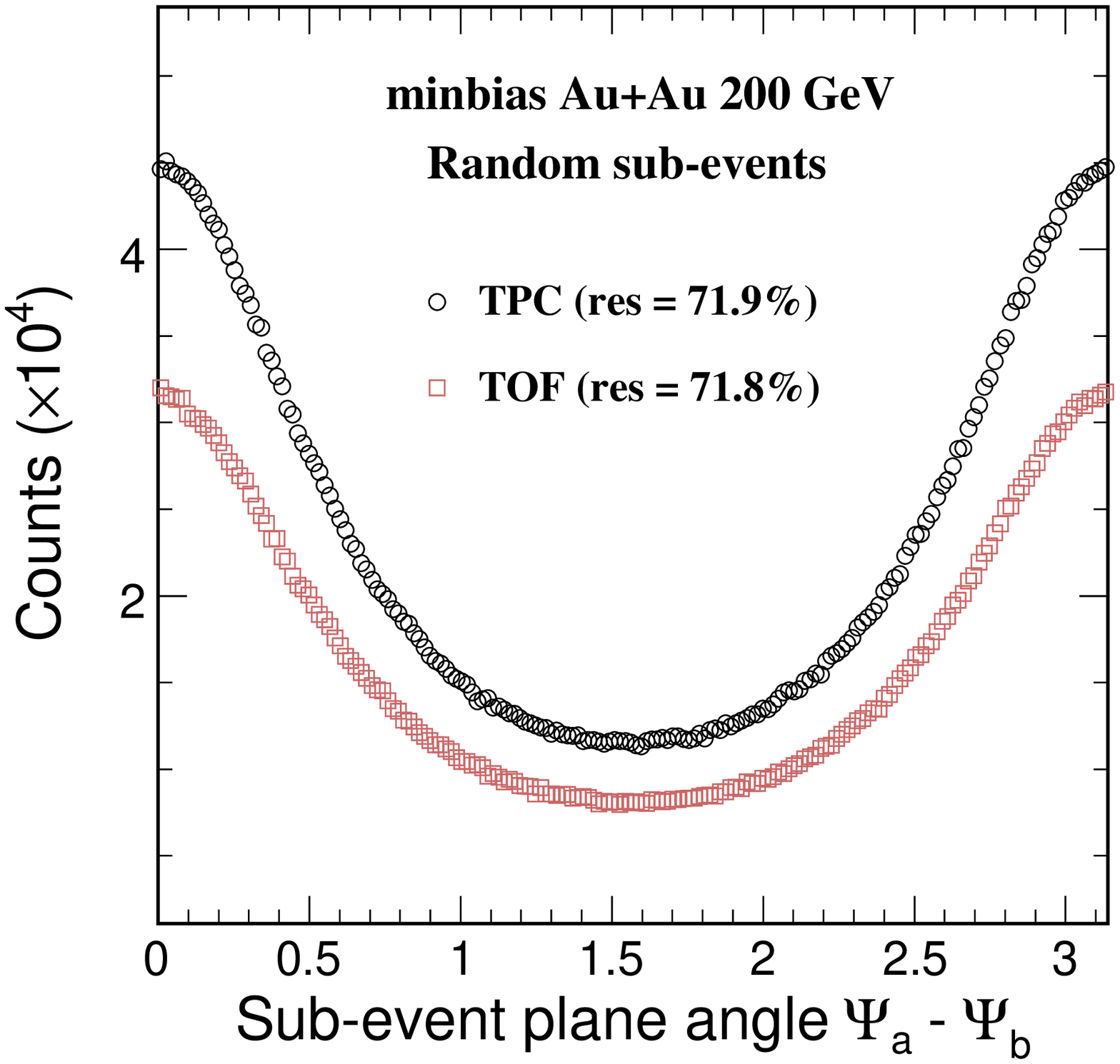}
\emn\\[10pt]
\caption[Event plane azimuth and resolution]{Left panel: The event
plane azimuthal angle distributions from TPC and TOF events. Right
panel: Event plane resolution from random sub-event method.}
\label{eventplane} \ef

An event plane method was used for electron \vv~\cite{v2EPmethod}.
The event plane was reconstructed from the final particles'
azimuths excluding electron candidates ($n\sigma_e<0$ or
$n\sigma_e>3$) to remove the auto-correlations. The acceptance and
efficiency of the detectors in azimuth was corrected by
compensating the azimuth to a flat distribution with $\phi$
weights from day-by-day run. Additional \pt\ weights were also
applied to improve the event plane resolution. The second order
harmonic azimuth angle $\Psi_2$ for event plane can be calculated
from the $\overrightarrow{Q}$ vector, as
Eq.~\ref{eventP},\ref{Qvector}:

\be \Psi_2 = \left(\arctan\frac{Q_y}{Q_x}\right)/2, \hskip 1 in
0<\Psi_2<\pi \label{eventP} \ee \be \overrightarrow{Q}=(Q_x,
Q_y)=\left(\sum_{i}w_i\cdot \cos(2\phi_i), \hskip 0.2 in
\sum_{i}w_i\cdot \sin(2\phi_i)\right) \label{Qvector} \ee Here,
$w_i$ is the weight for each track with both the track azimuthal
compensation and \pt\ weight contributions.

\bf \centering \bmn[b]{0.5\textwidth} \centering
\includegraphics[width=1.0\textwidth]{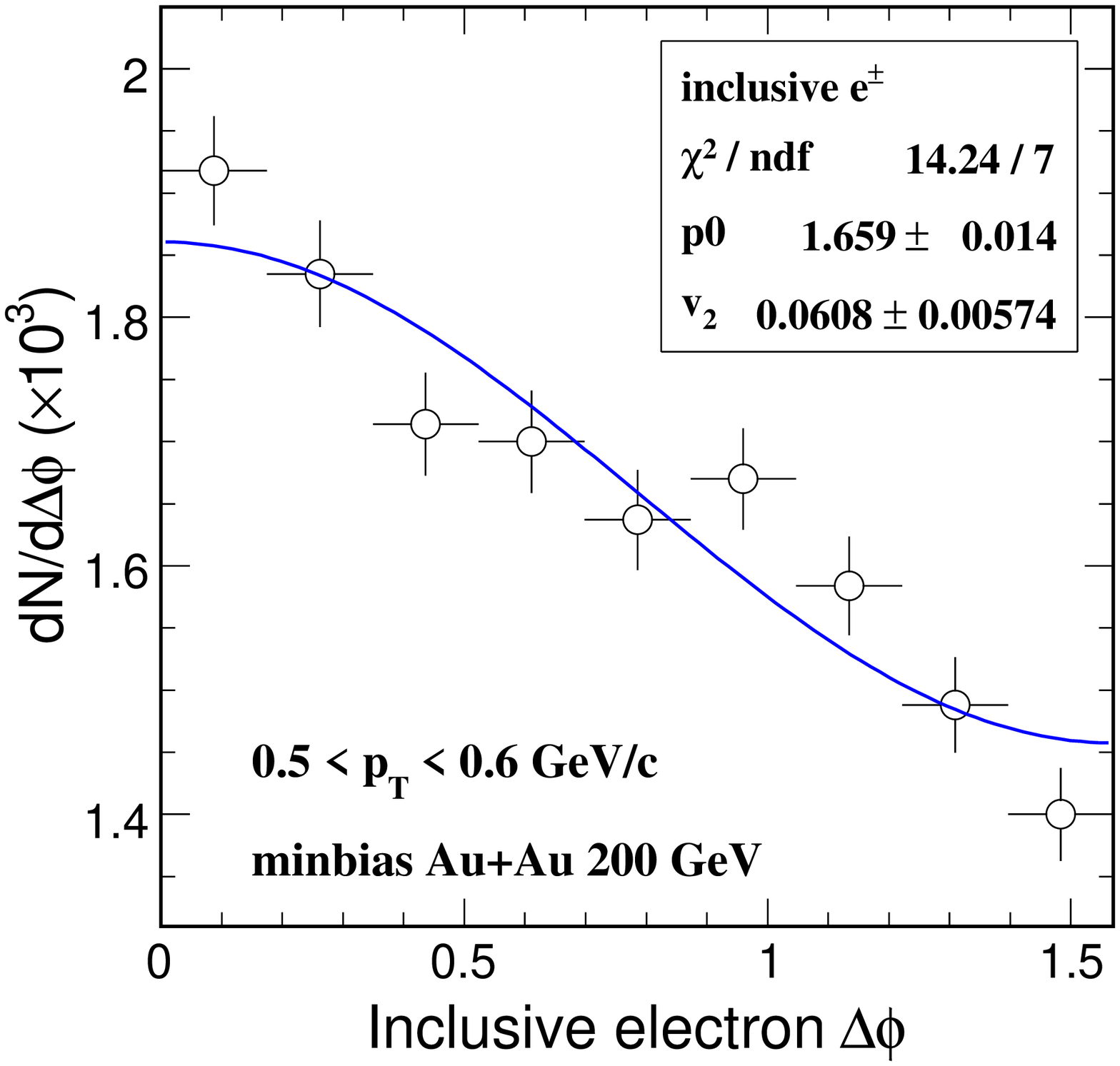}
\emn%
\bmn[b]{0.5\textwidth} \centering
\includegraphics[width=1.0\textwidth]{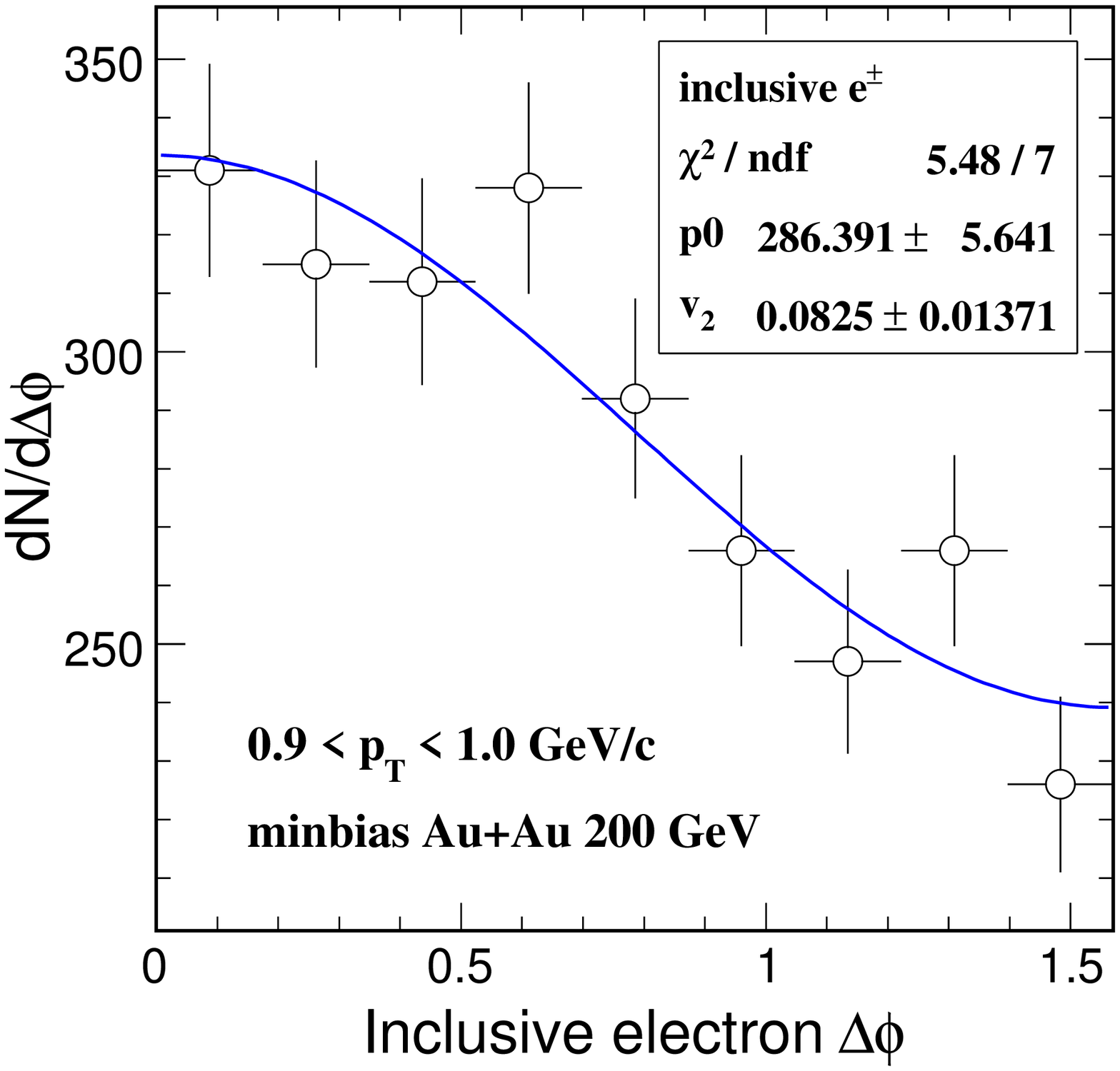}
\emn\\[10pt]
\caption[Inclusive electron $dN/d\Delta\phi$
distributions]{Inclusive electron $\Delta\phi$ distributions from
TOF.} \label{incluephi} \ef

The left panel of Fig.~\ref{eventplane} shows the event plane
azimuthal angle distributions from both TPC and TOF events. A
constant function fit to the distributions gives reasonable
$\chi^{2}/ndf$, which indicates that the event plane was
reconstructed successfully.

The event plane resolution was calculated using the random
sub-event method~\cite{v2EPmethod}. Each event was divided into
two sub-events "a" and "b" randomly. Two event plane azimuthal
angles $\Psi_2^{a}$ and $\Psi_2^{b}$ were reconstructed
correspondingly. Then the event plane resolution $r=\la
cos[2(\Psi_2-\Psi_{rp})]\ra$ can be calculated from Eq.(14) and
(11) of ~\cite{v2EPmethod}:

\be \la \cos[2(\Psi_2-\Psi_{rp})]\ra =
\frac{\sqrt{\pi}}{2\sqrt{2}}\chi_{2}^{}
\exp(-\chi_2^2/4)\times[I_0(\chi_2^2/4)+I_1(\chi_2^2/4)]
\label{evtResFunc} \ee \be \la \cos[2(\Psi_2^{a}-\Psi_{rp})]\ra =
\sqrt{\la \cos[2(\Psi_2^{a}-\Psi_2^{b})]\ra} \label{subEvtResFunc}
\ee \be \chi_2^{}=v_2^{}/\sigma=v_2^{}\sqrt{2N} \label{chim} \ee

$\Psi_{rp}$ is the event plane angle calculated from the event
plane $\overrightarrow{Q}$ vector. $I_0$ and $I_1$ are the Bessel
functions. The event plane resolution is around 72\% from random
sub-event method, shown in the right panel of
Fig.~\ref{eventplane}. The event plane angle $\Psi_{rp}$ is used
as reference to obtain the azimuthal distributions
($dN/d\Delta\phi$) of inclusive and photonic electrons.
Fig.~\ref{incluephi} shows the $dN/d\Delta\phi$ distributions for
inclusive electron in two typical \pt\ bins. The \vv\ can be
extracted from fitting the $dN/d\Delta\phi$ distributions in each
\pt\ bin corrected by the event resolution.

\bf \centering \bmn[b]{0.5\textwidth} \centering
\includegraphics[width=1.0\textwidth]{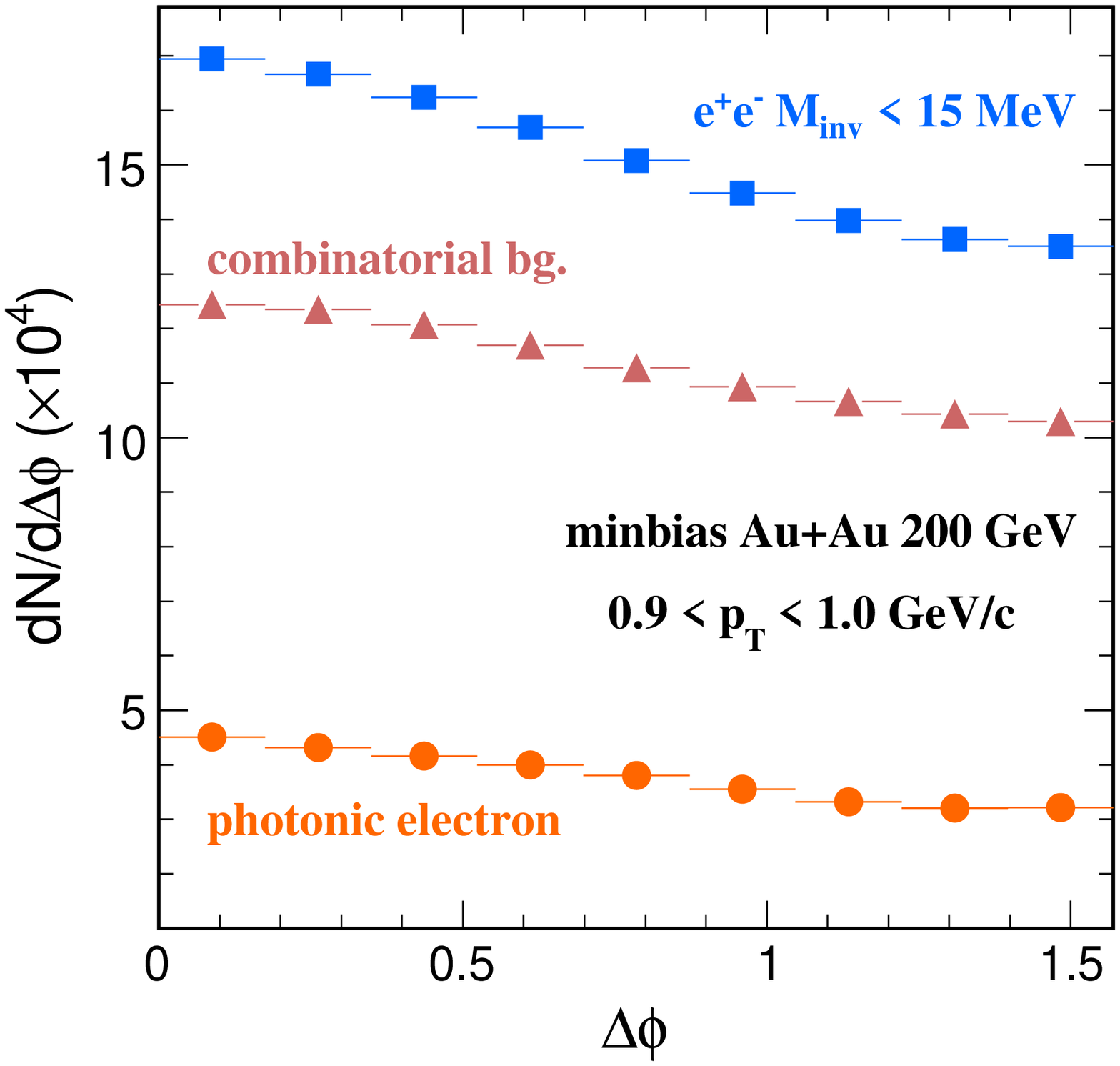}
\emn%
\bmn[b]{0.5\textwidth} \centering
\includegraphics[width=1.0\textwidth]{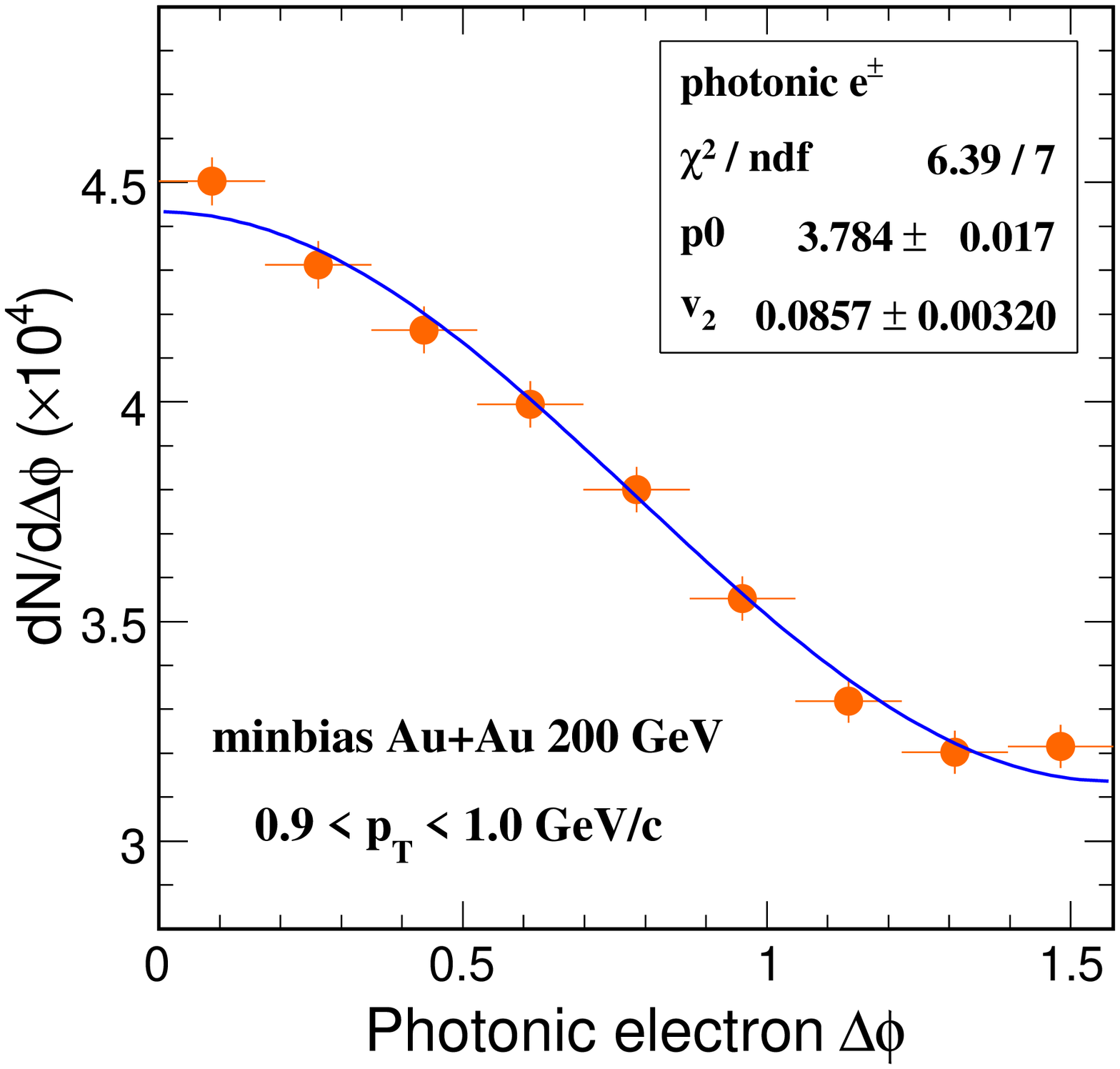}
\emn\\[10pt]
\caption[Photonic electron $dN/d\Delta\phi$ extraction]{Left
panel: Extraction of photonic electron $dN/d\Delta\phi$
distribution from TPC. Right panel: Photonic electron
$dN/d\Delta\phi$ distribution (zoom in the photonic electron
$\Delta\phi$ distribution in the left panel).} \label{phbgephi}
\ef

The electron pairs in the mass window $M_{e^{+}e^{-}}<15$ MeV are
filled into $dN/d\Delta\phi_{all}$ distributions as well as the
combinatorial background electrons $dN/d\Delta\phi_{com}$. Then
the photonic electron $dN/d\Delta\phi_{ph}$ distributions were
derived as:

\be dN/d\Delta\phi_{ph} =
dN/d\Delta\phi_{all}-dN/d\Delta\phi_{com}  \label{phdphi} \ee

Fig.~\ref{phbgephi} shows the extraction of the $dN/d\Delta\phi$
distributions for photonic electrons. Photonic electron \vv\ was
also obtained from fitting to the $dN/d\Delta\phi$ distributions
in each \pt\ bin corrected by the event resolution.

\bf \centering\mbox{
\includegraphics[width=0.62\textwidth]{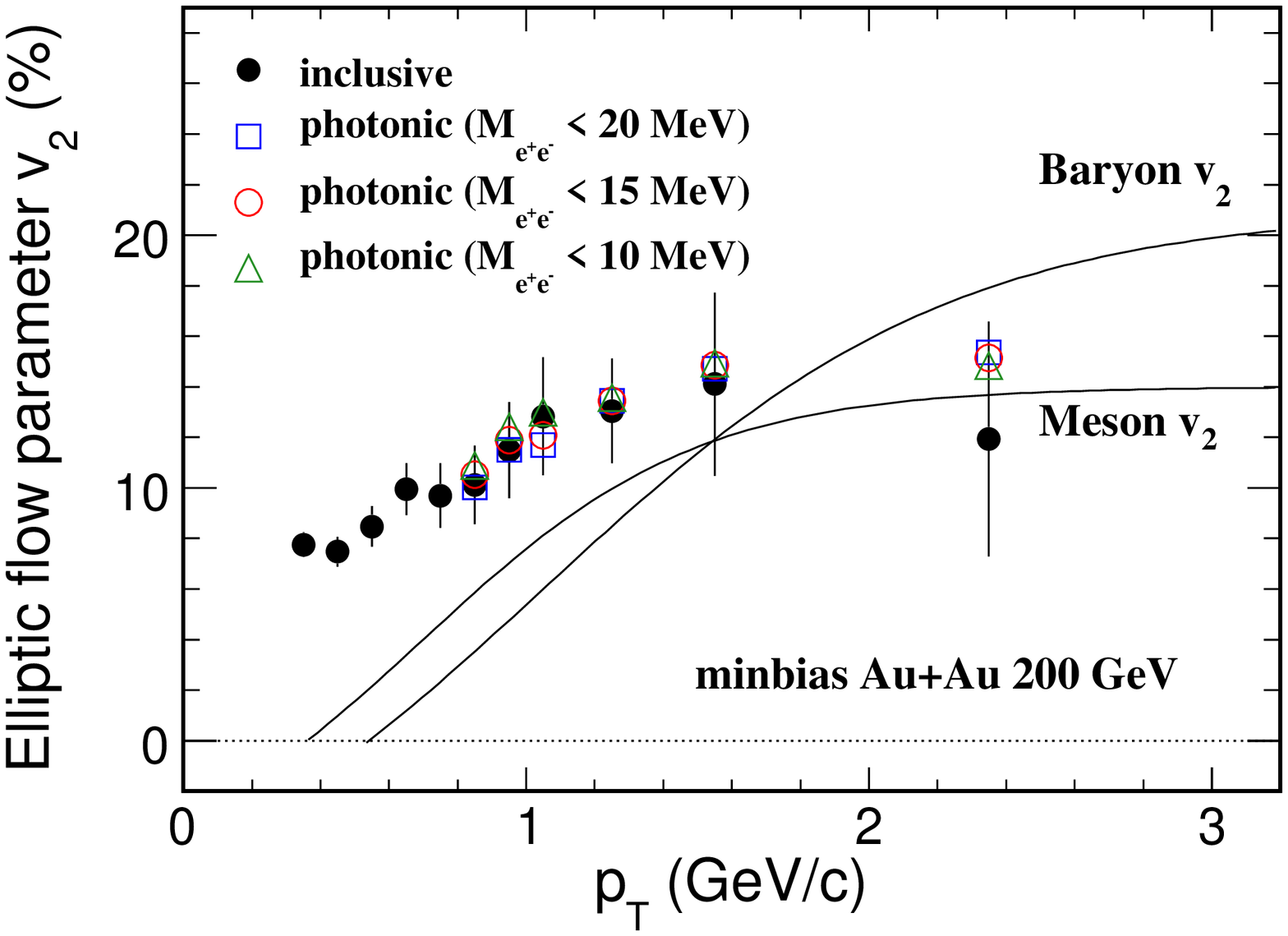}}
\caption[Inclusive and photonic electron \vv]{Inclusive electron
\vv\ from TOF and photoinic electron \vv\ from TPC comparing with
meson \vv\ and baryon \vv. Error bars are only statistical.}
\label{ev2} \ef

The inclusive and photonic electron \vv\ are shown in
Fig.~\ref{ev2}. They are higher than the meson and baryon \vv\
shown in lines~\cite{minepiv2} in the low \pt\ due to the decay
kinematics. The different invariant mass cuts are systematically
tried for the photonic electron \vv. The statistical errors are
limited by inclusive electron yield.

\subsection{non-photonic electron elliptic flow}

The non-photonic electron \vv\ and its propagated error can be
derived from the formulae: \vspace{-0.1in}
$$
{v_2^{non}}={{rv_2^{inc}-v_2^{pho}}\over{r-1}},
{\sigma_{v_2^{non}}}={{r\sigma_{v_2^{inc}}}\over{r-1}},
\vspace{-0.1in}
$$where r is the yield ratio of inclusive over photonic electrons,
which is from the measurement of their spectra. The statistic
error of photonic electron is extremely small to be neglected
comparing to that of inclusive electron. The non-photonic electron
$v_2$ is shown in the bottom panel of Fig.~\ref{nonphev2}. The
wild error bars are due to the very small r ($\sim1.3-1.5$ at
$2-3$ GeV/$c$) and large errors of inclusive electron $v_2$. The
small r is due to the huge photon conversion from the material in
the STAR detector. The poor statistics of inclusive electron is
due to the small acceptance of current TOF tray.

\bf \centering\mbox{
\includegraphics[width=0.62\textwidth]{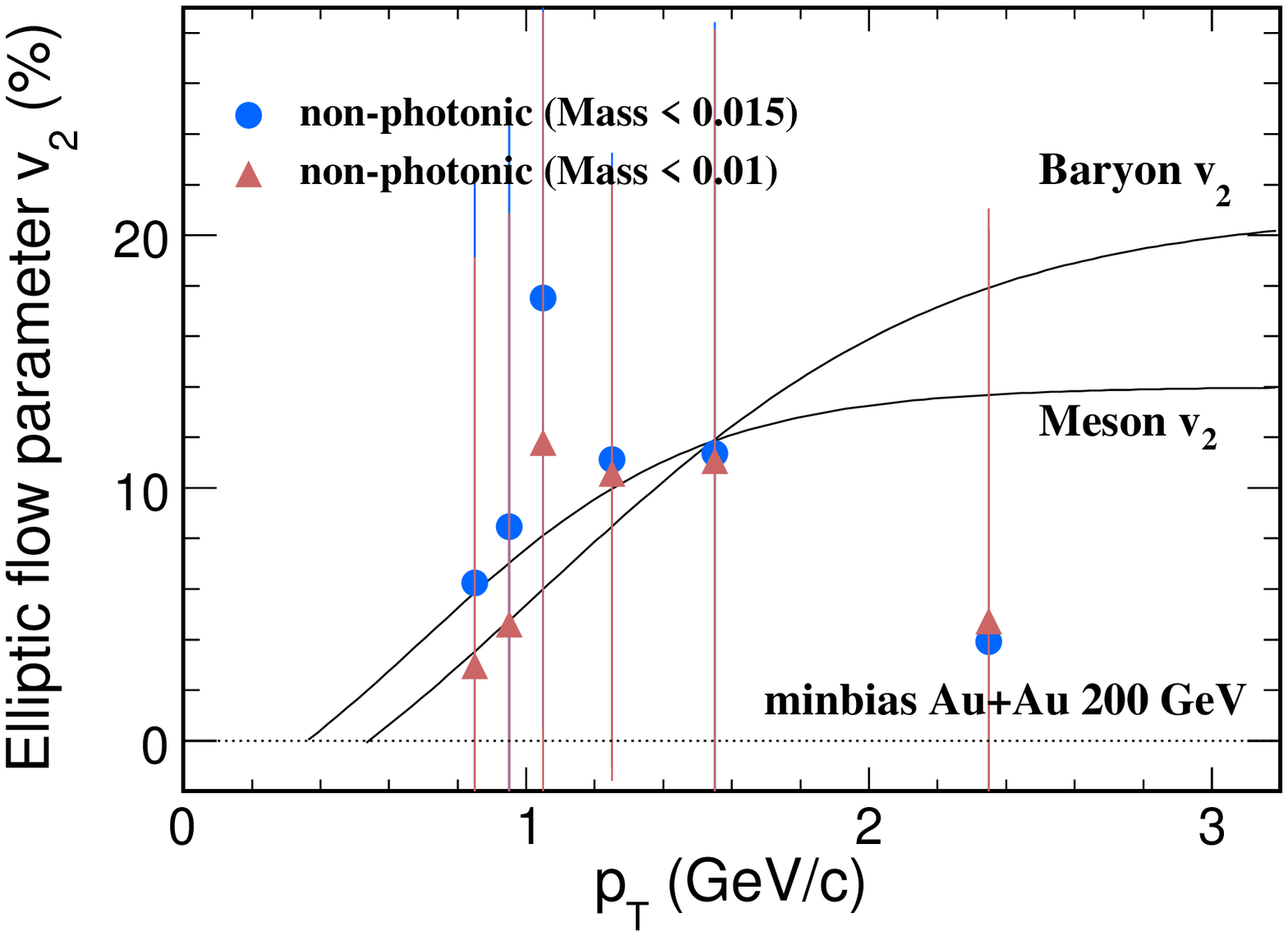}}
\caption[Non-photonic electron \vv]{Current status of TOF
non-photoinic electron \vv\ from inclusive electron \vv\ with the
photonic background subtracted.} \label{nonphev2} \ef

The coming update for STAR detector will give us a very good
chance to measure non-photonic electron $v_2$ precisely. The
hydrodynamic model and quark coalescence model for heavy flavors
will be tested from this future experiment at STAR.

\chapter{Charm energy-loss, freeze-out, flow properties and cross-section}

In this chapter, the measurements of $D^{0}\rightarrow K\pi$ at
low \pt\ ($<2$ \gevc) in 200 \gev\ \AuAu\ collisions will be
introduced. In addition, we use a newly proposed technique to
identify muons from charm decays at low \pt~\cite{ffcharm}.
Combined with non-photonic electron measurement, all three
measurements together stringently constrain the total charm
production cross-section at mid-rapidity covering $\sim90\%$ of
the kinematics. They allow us to extract the charmed hadron
spectral shape and to study the charm energy-loss and radial flow
properties.

\section{$D^0$ reconstruction in \AuAu\ collisions}

A total of 13.3 million 0-80\% minbias triggered \AuAu\ events in
year 2004 Run IV were used for the \dzero\ ($\bar{D^{0}}$) direct
reconstruction through hadronic channel ($D^{0}\rightarrow
K^{-}\pi^{+}$, $\bar{D^{0}}\rightarrow K^{+}\pi^{-}$). The decay
branching ratio is 3.83\%. The collision vertex Z was required
from $-30$ cm to 30 cm. In what follows, we imply
($D^0+\bar{D^{0}}$)/2 when using the term $D^{0}$ unless otherwise
specified.

Without the inner tracker devices, the exact $D^{0}$ decay vertex
cannot be reconstructed due to insufficient track projection
resolution close to the collision vertex. The invariant mass
spectrum of $D^{0}$ mesons was obtained by pairing each oppositely
charged kaon and pion candidate in the same event. The kaon and
pion daughter tracks were identified by the \dedx\ measured in the
TPC. The cuts for selection of the $D^{0}$ daughter candidates are
listed in Table~\ref{kpicut}.

\begin{table}[hbt]
\caption[$D^{0}$ daughter candidates cuts list]{Cuts for $D^{0}$
daughter candidates.} \label{kpicut} \vskip 0.1 in
\centering\begin{tabular}{|c|c|} \hline \hline
momentum $p$ & $>0.3$ \gevc\\
\hline \pt\ & $>0.2$ \gevc\\
\hline nFitPts & $>15$\\
\hline Global dca & $<1.5$ cm\\
\hline pseudorapidity $\eta$ & (-1,1)\\
\hline pair rapidity & (-1,1)\\
\hline $n\sigma_{\pi}$ & (-3,3)\\
\hline $n\sigma_{K}$ & $p<0.7$ \gevc\ (-2,2)\\
 & $p>0.7$ \gevc\ (-1,1)\\
\hline \hline
\end{tabular}
\end{table}

Table~\ref{kpicut} lists the cuts for the kaon and pion candidate
tracks.

\bf \centering \bmn[b]{0.5\textwidth} \centering
\includegraphics[width=1.0\textwidth]{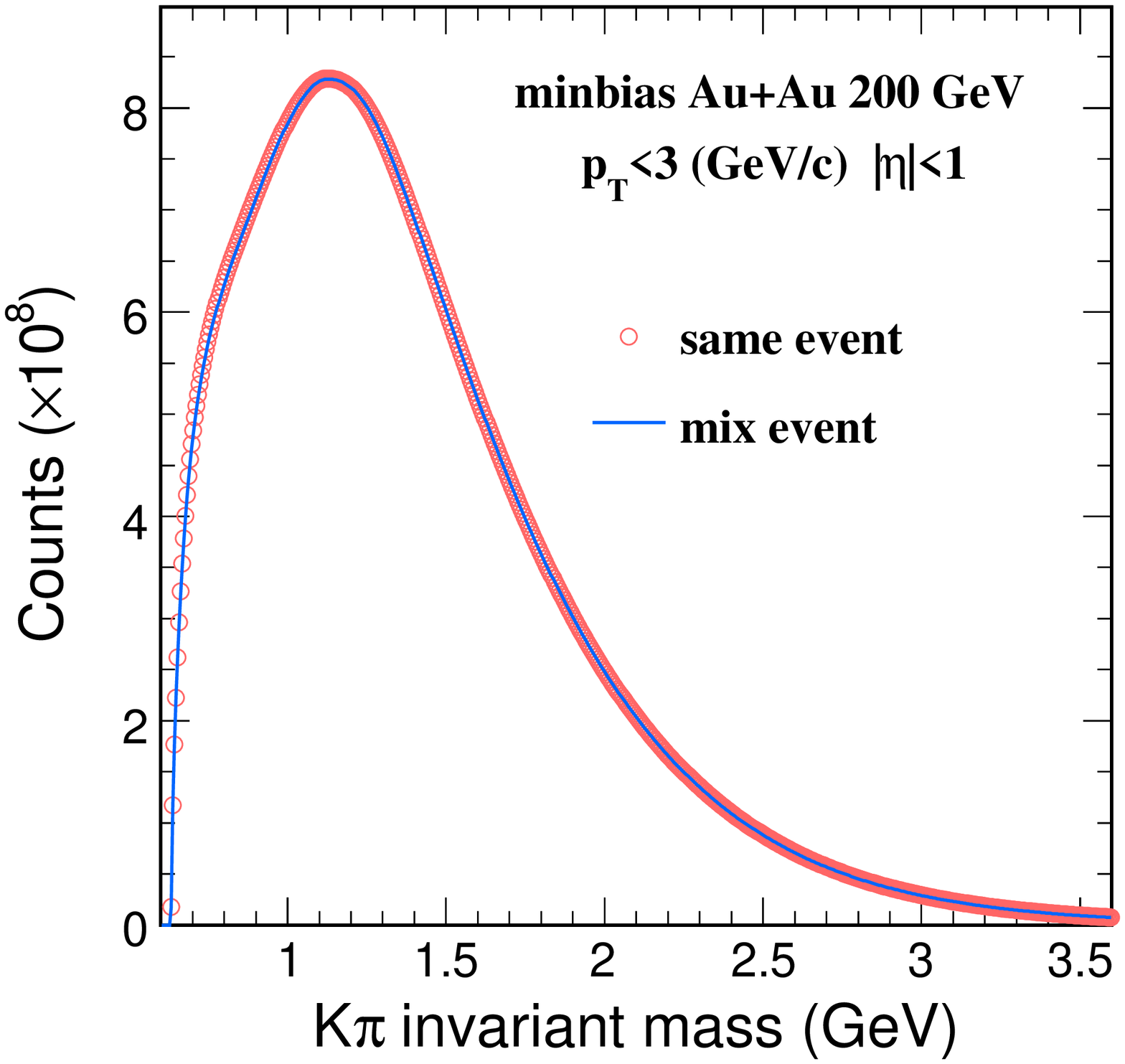}
\emn%
\bmn[b]{0.5\textwidth} \centering
\includegraphics[width=1.0\textwidth]{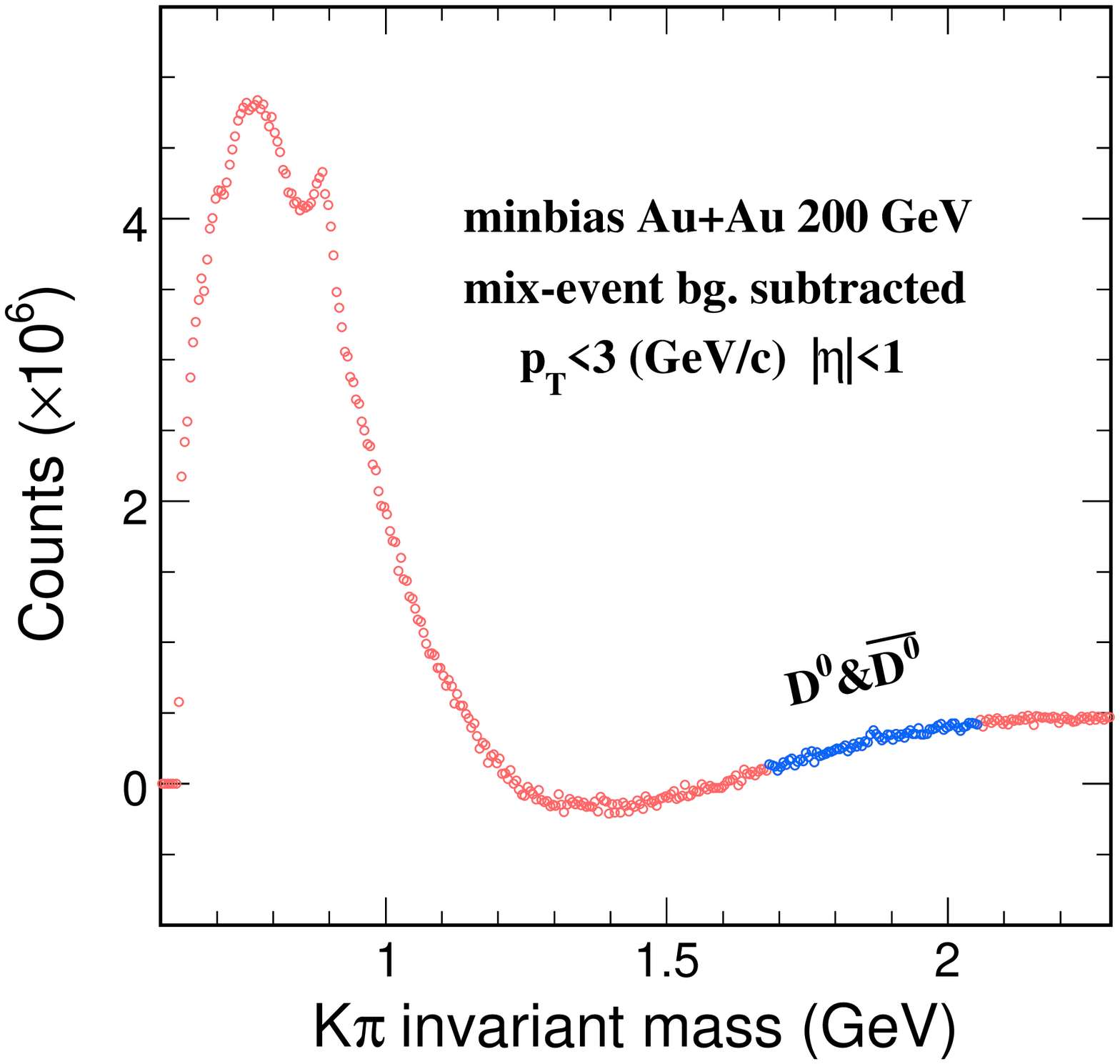}
\emn\\[10pt]
\caption[Kaon/pion invariant mass reconstruction from event-mixing
method]{Left panel: Kaon/pion invariant mass distributions from
same event (red open circle) and invariant mass distributions from
mixed event (blue curve). Right panel: Kaon/pion invariant mass
distributions from same event after the combinatorial background
subtracted.} \label{eventmix} \ef

The random combination of kaon and pion pairs contribute to a huge
combinatorial background. An event-mixing
method~\cite{HaibinThesis,stardAucharm} was provided to represent
the random combinatorial background. In this method, the kaon in
one event and the pion in another event from the event buffer were
selected to generate a reference event-mixing invariant mass
distribution. The buffered event candidates were required to have
similar event environment. The mixing-buffer is divided into 10
RefMult bins and 10 Vertex Z bins. Each RefMult bin and Vertex Z
bin has roughly the same number of events. Each event is mixed
with two other events in the same RefMult bin and Vertex Z bin.
Due to the multi-combination with different event buffers, the
statistics of the combinatorial background can be increased
$\sim4$ times. Then the event-mixing spectrum was normalized with
a factor obtained by comparing the entries in the two spectra with
invariant mass $>2.5$ \gevcc. In order to increase statistics,
$D^0$ and $bar{D^0}$ signals are added together. Left panel of
Fig.~\ref{eventmix} shows kaon/pion invariant mass distributions
from same events and mixed events after normalization. Due to the
huge magnitude of the combinatorial background, the signal is
invisible. But if the invariant mass distributions from same
events are subtracted by the event-mixing combinatorial
background, the signal can be seen within a mass region of
1.68$-$2.05 \gev, shown in the right panel of Fig.~\ref{eventmix}.
The small peak near mass $\sim0.96$ \gev\ is known as $K^*$.

\bf \centering \bmn[b]{0.33\textwidth} \centering
\includegraphics[width=1.0\textwidth]{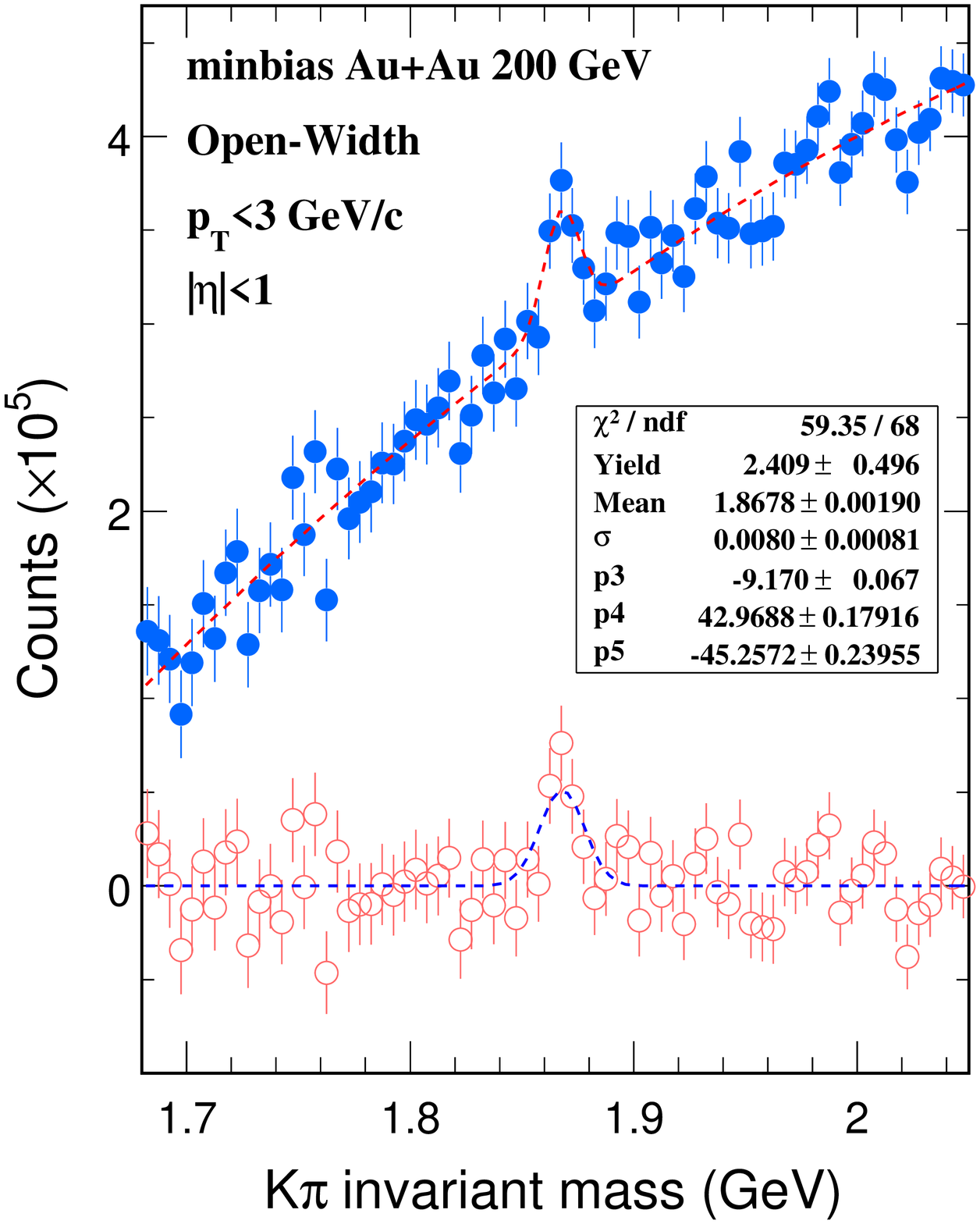}
\emn%
\bmn[b]{0.33\textwidth} \centering
\includegraphics[width=1.0\textwidth]{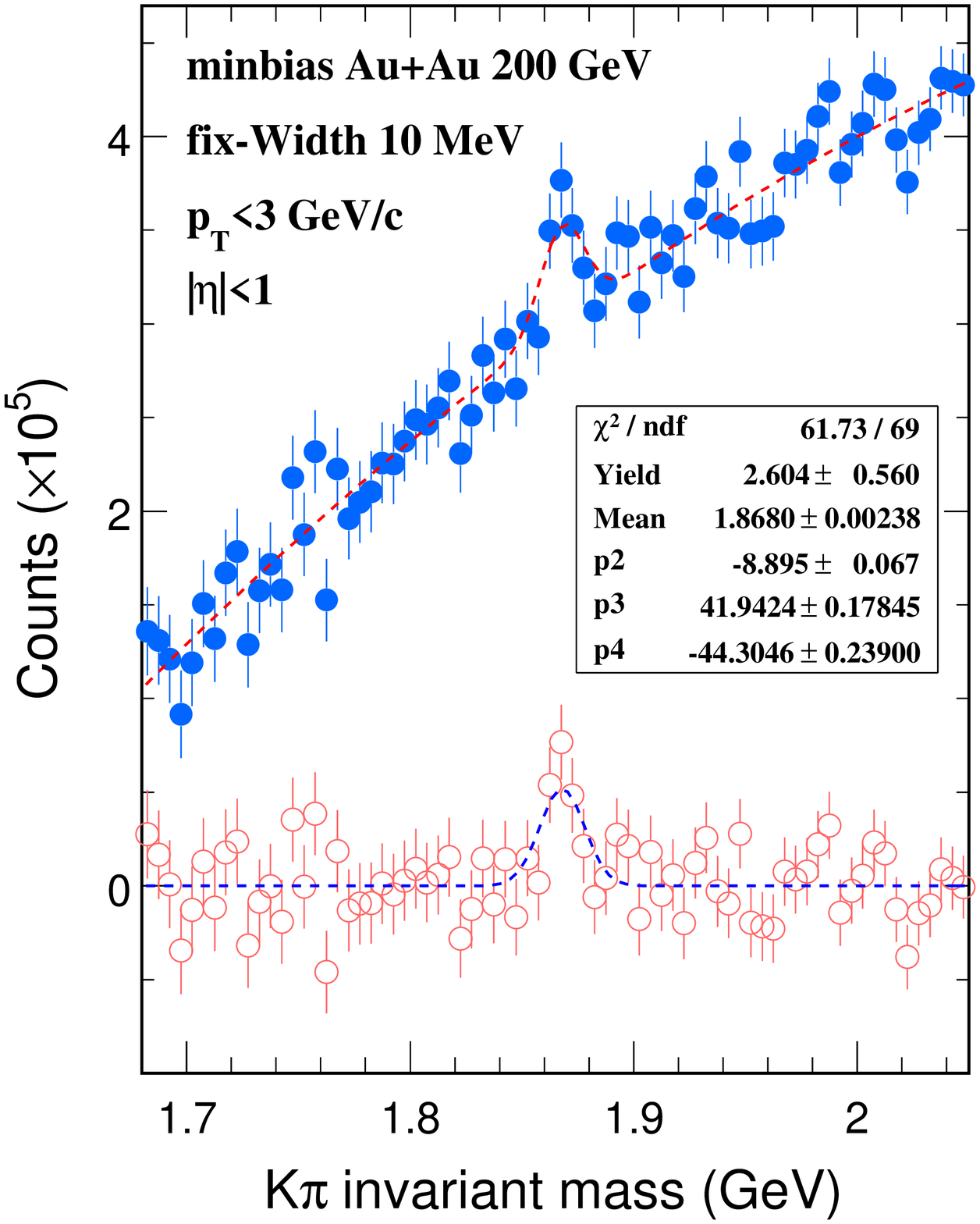}
\emn%
\bmn[b]{0.33\textwidth} \centering
\includegraphics[width=1.0\textwidth]{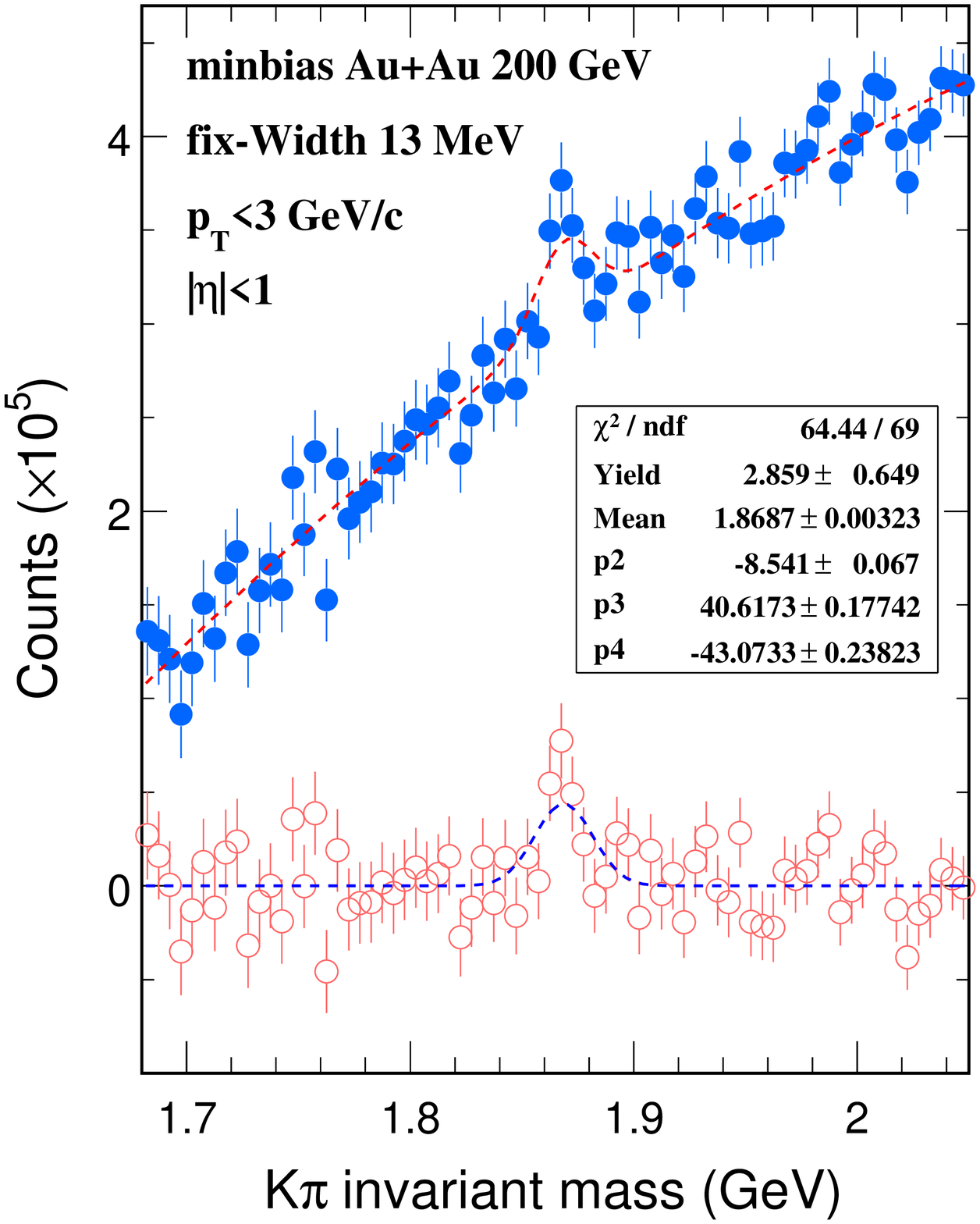}
\emn\\[10pt]
\caption[Kaon/pion invariant mass distributions by fitting with
open-width and fix-width]{Extraction of the signal raw yields by
fitting with open-width and fix-width from kaon/pion invariant
mass distributions.} \label{d0width} \ef

After the event-mixing background subtraction, the kaon/pion
invariant mass distributions within the mass window (1.68$-$2.05
\gev) were shown as the blue solid circles in Fig.~\ref{d0width}.
The red open circle is the invariant mass subtracted by the
residual background. A gauss+linear or second order polynomial
(Pol-2) function as \be f(m)={W\times A_{0}\over{\sqrt{2\pi}\times
\sigma}} \times e^{-(m-\bar{m})^2\over{2\sigma^2}}+f',
\label{kpimassfit} \ee

was used to fit the data to extract the signal, where $W=0.005 $
\gev\ is the bin width. $A_{0}$ is the signal raw yield. $f'$ is
the linear or the Pol-2 function. $\sigma$ is the width of the
gaussian mass distribution. From embedding, the width of the mass
is around 13 MeV, but in real data, the width turns out to be
smaller. Since in sub-\pt\ bins, due to statistical limit, the
mass distributions can not be fitted by open width. So the width
was fixed as 10 MeV, and the overall uncertainty of fix-width and
open-width fits, which is around 10\%, was done at $p_T<3$ \gevc\
for good statistics, shown as Fig.~\ref{d0width}.

To estimate the uncertainties of the combinatorial background, we
use linear and Pol-2 function to describe the background by
varying several fit mass regions. The average yield from linear
and Pol-2 fit with best $\chi^{2}$ was extracted as the final raw
yield. The dominant bin-by-bin systematic error was estimated as
the largest deviation from all the fits with reasonable
$\chi^{2}$, since all the fit results are considered as a uniform
distribution.

Fig.~\ref{d02fun} gives the fits for three sub-\pt\ bins using
linear function (the first row) and Pol-2 function (the second
row) with best $\chi^{2}$.

\bf \centering \bmn[b]{0.33\textwidth} \centering
\includegraphics[width=1.0\textwidth]{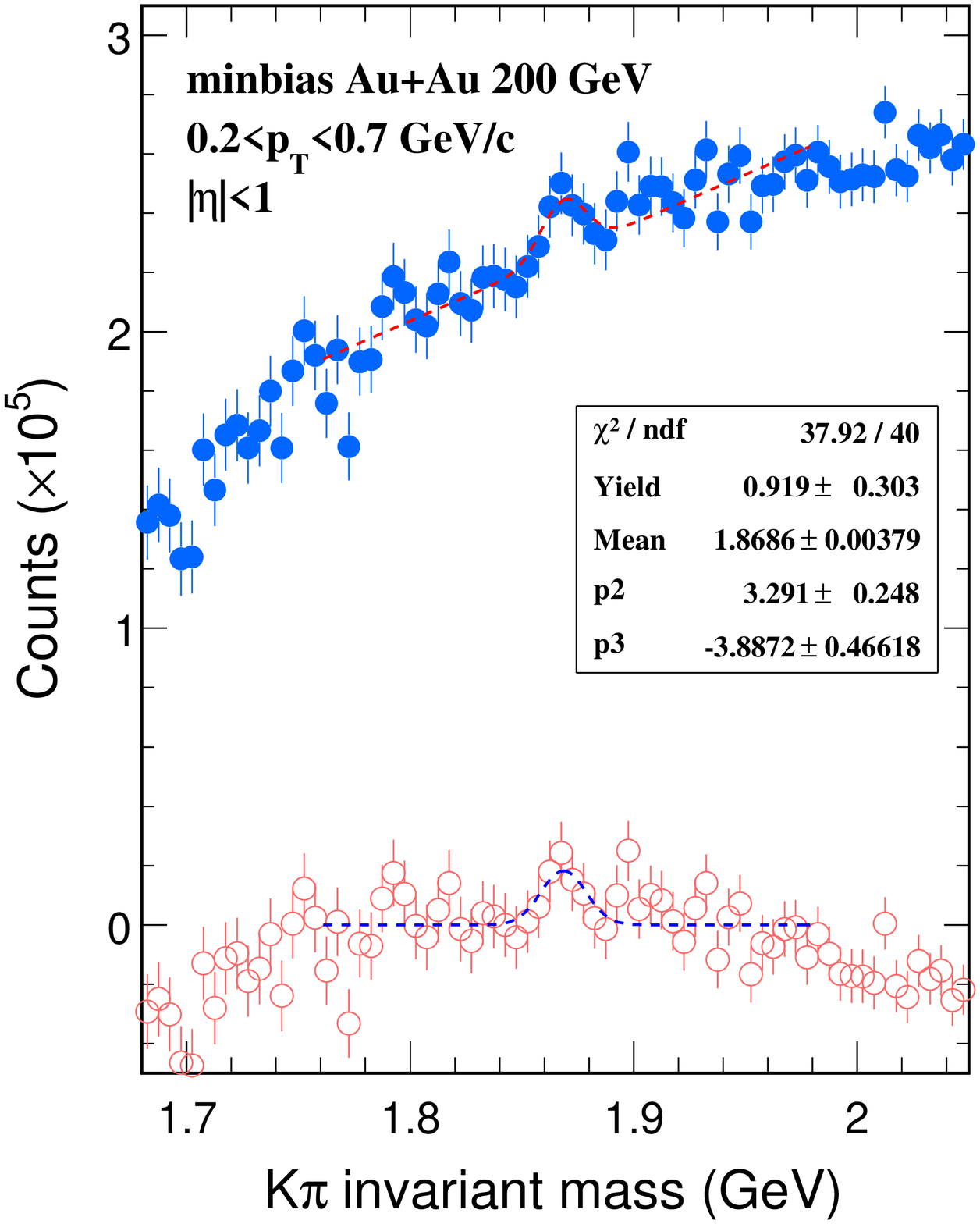}
\emn%
\bmn[b]{0.33\textwidth} \centering
\includegraphics[width=1.0\textwidth]{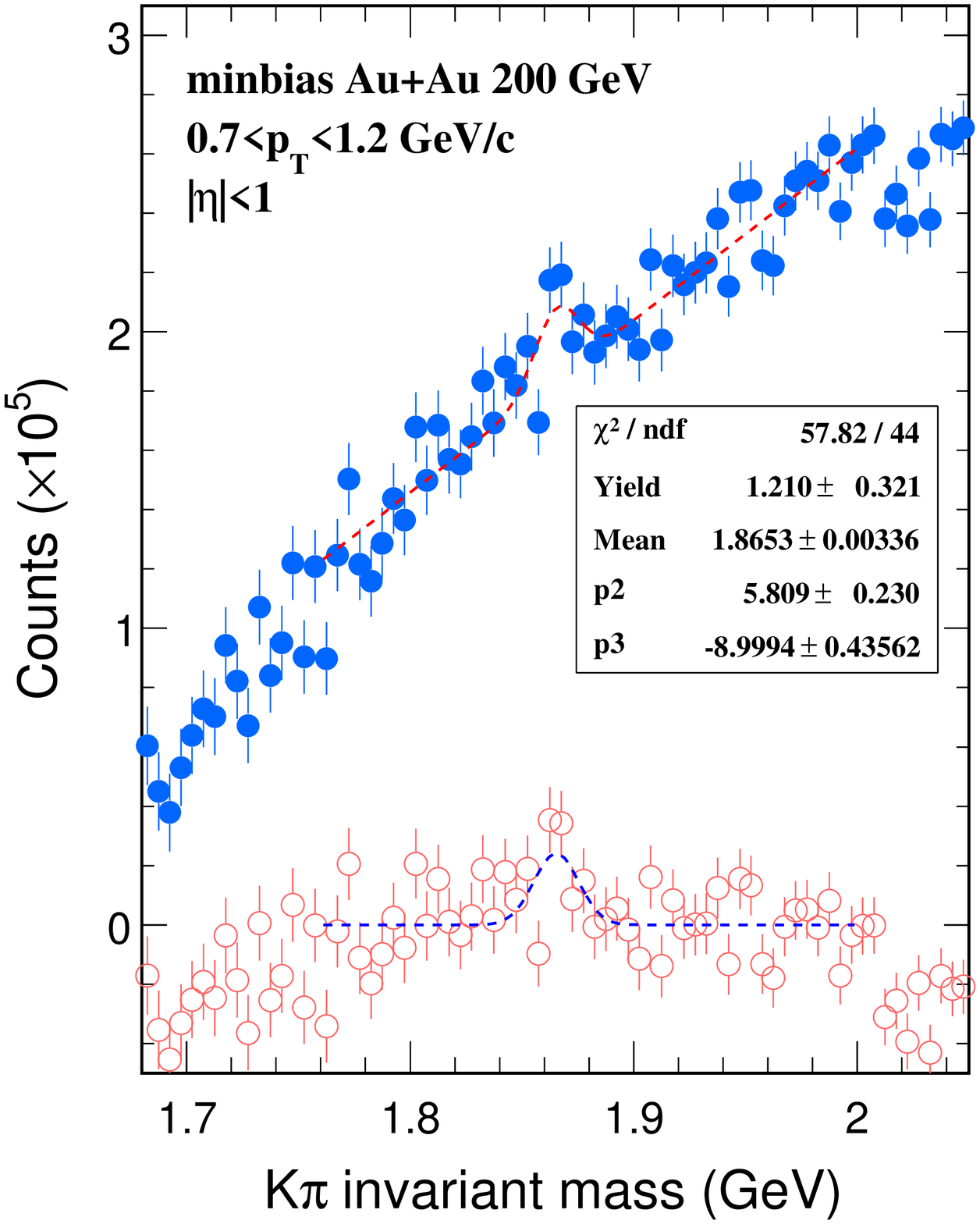}
\emn%
\bmn[b]{0.33\textwidth} \centering
\includegraphics[width=1.0\textwidth]{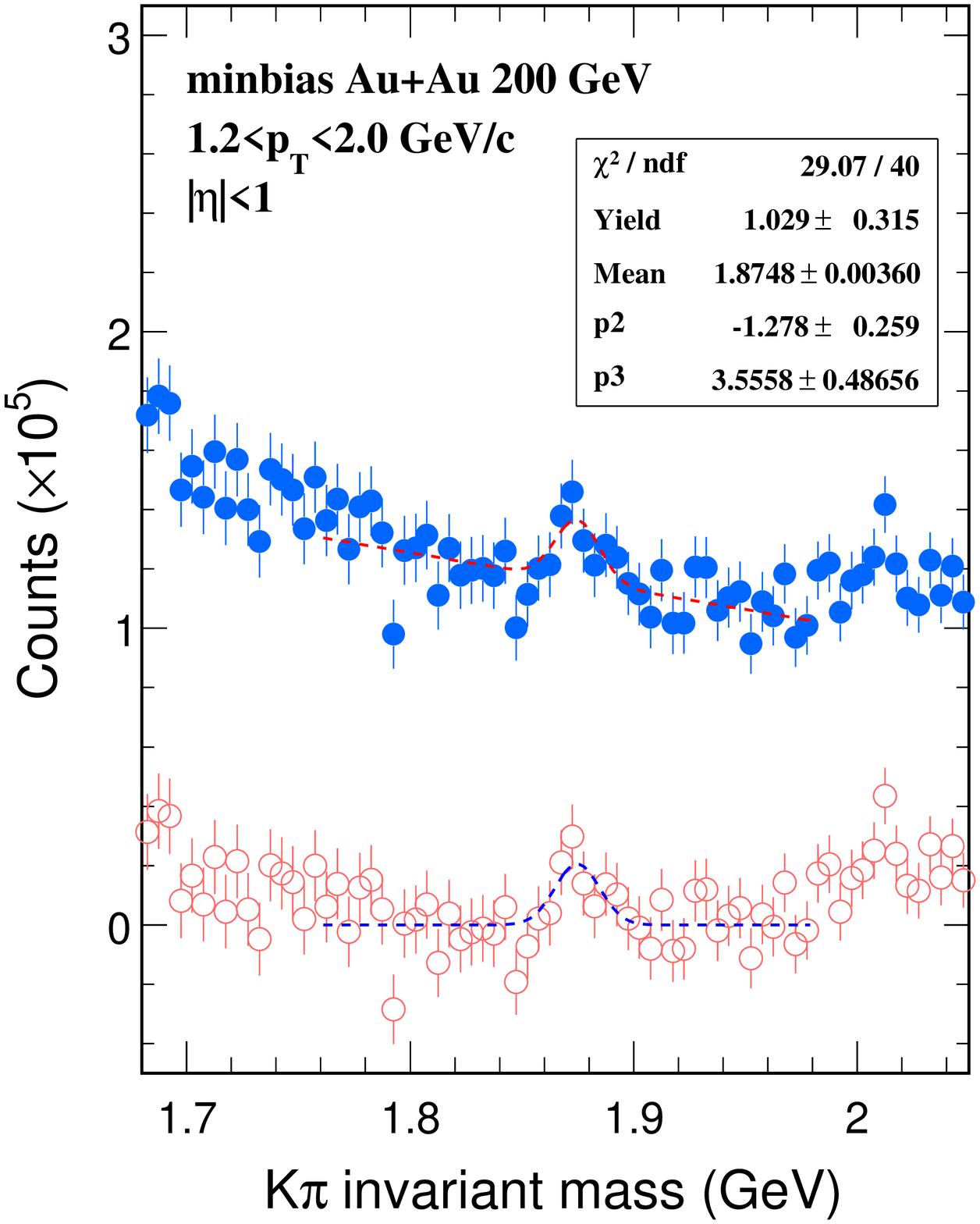}
\emn\\[10pt]
\bmn[b]{0.33\textwidth} \centering
\includegraphics[width=1.0\textwidth]{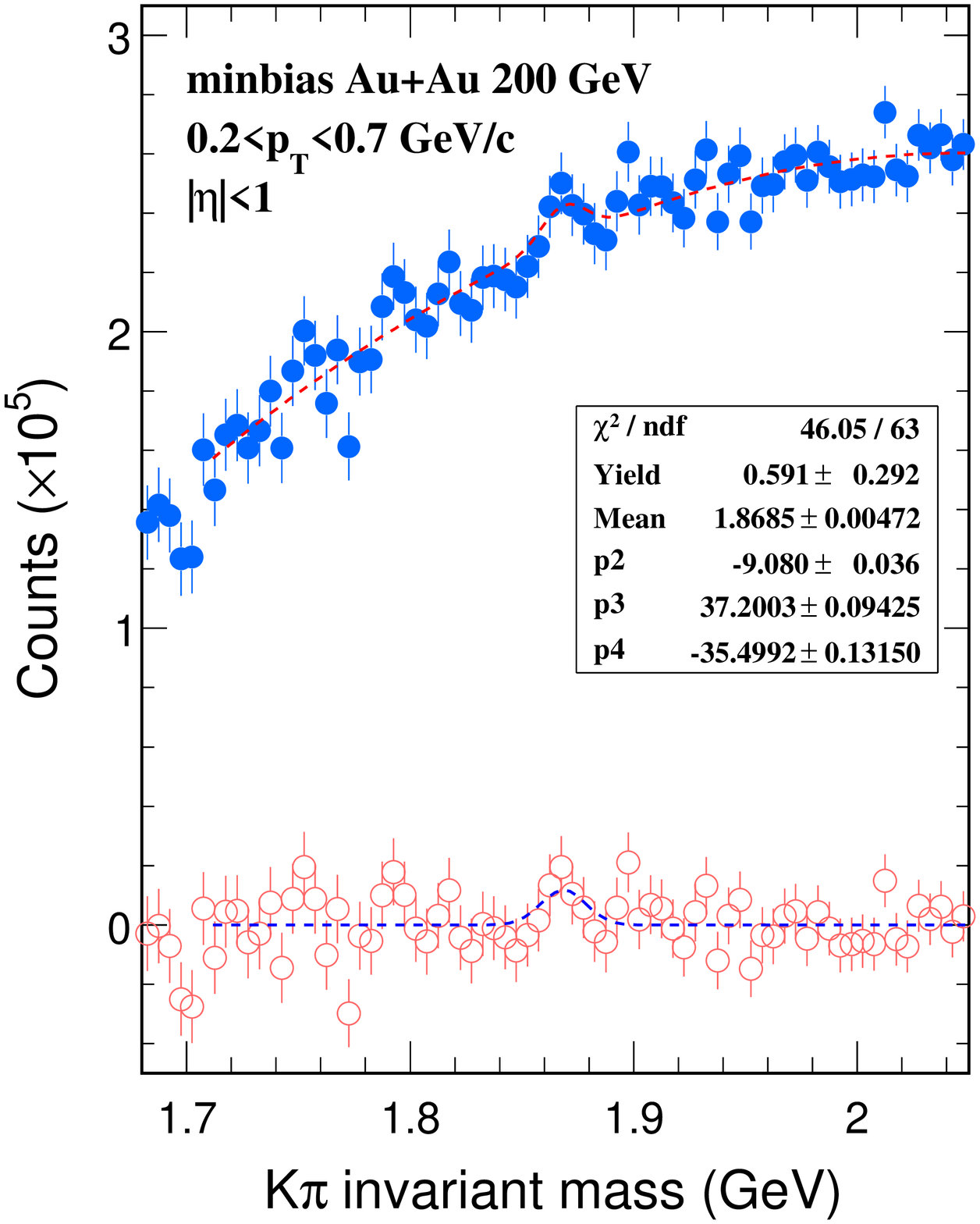}
\emn%
\bmn[b]{0.33\textwidth} \centering
\includegraphics[width=1.0\textwidth]{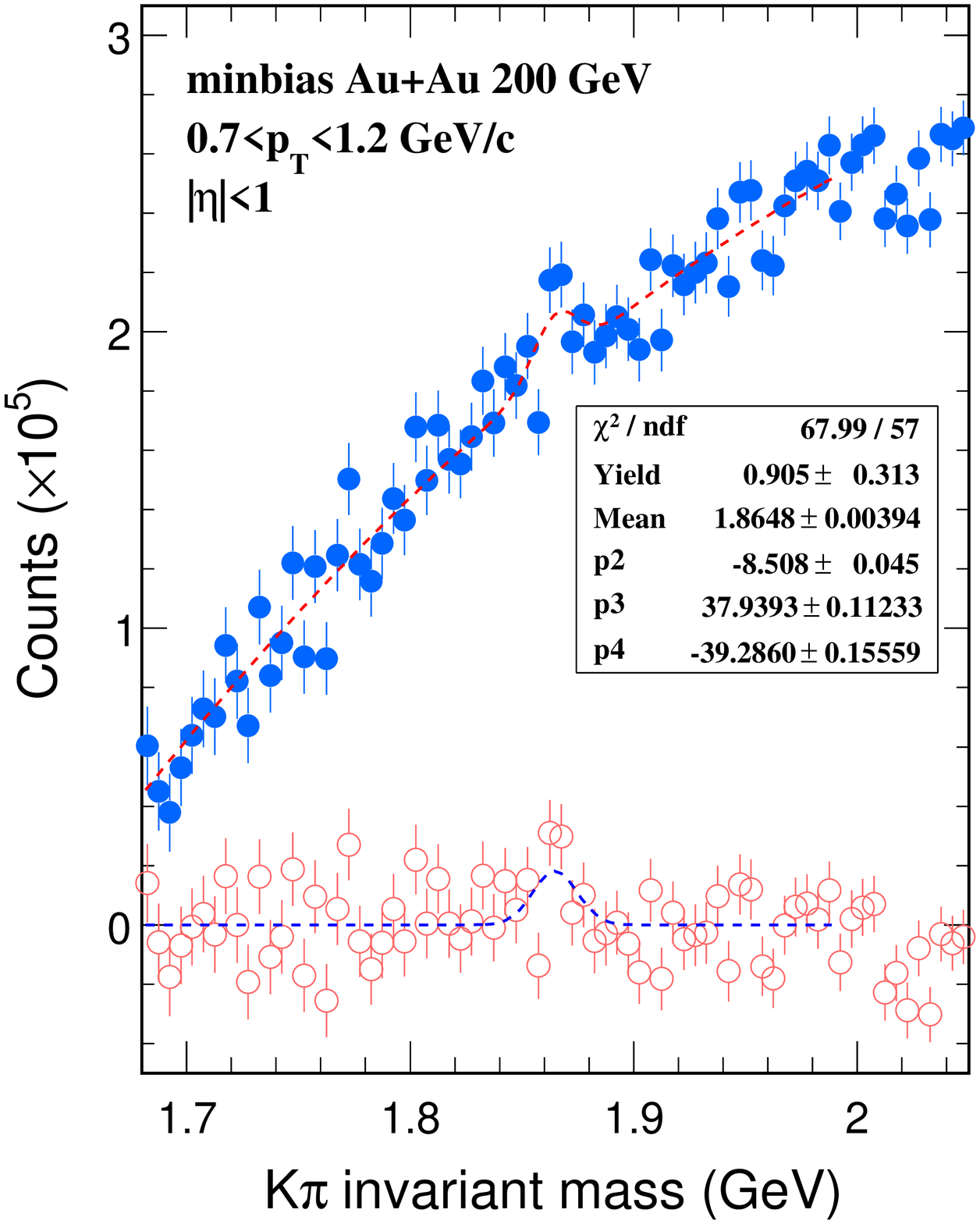}
\emn%
\bmn[b]{0.33\textwidth} \centering
\includegraphics[width=1.0\textwidth]{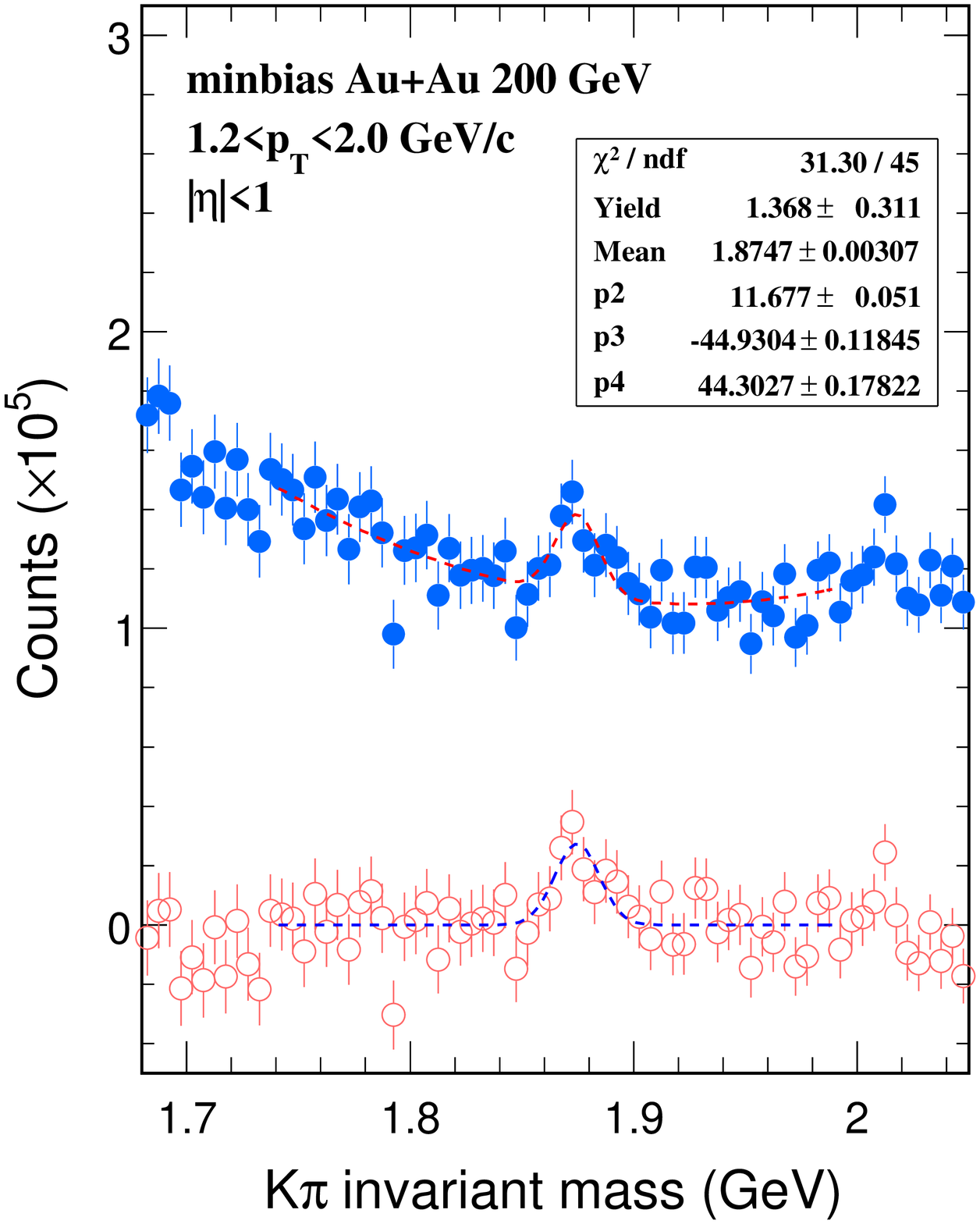}
\emn\\[10pt]
\caption[Kaon/pion invariant mass distributions by fitting with
linear and Pol-2 function in each \pt\ bin] {Kaon/pion invariant
mass distributions by fitting with linear and Pol-2 function in
each \pt\ bin.} \label{d02fun} \ef


Table~\ref{d0sys} lists the systematical uncertainties for \dzero\
spectrum in minimum bias 200 \gev\ \AuAu\ collisions.

\begin{table}[hbt]
\caption[\dzero\ systematical uncertainties]{\dzero\ systematical
uncertainties in minimum bias 200 \gev\ \AuAu\ collisions.}
\label{d0sys} \vskip 0.1 in
\centering\begin{tabular}{|c|c|c|c|c|c|c|} \hline \hline
\pt\ & bin-by-bin & open/fix- & Full Field / & Tracking & Track cuts & Total \\
(\gevc) & fit & width & Reversed FF & eff. & (\dedx) & sys.\\
\hline
0.2-0.7 & 42.9\% & & & & & 49.5\% \\
\cline{1-2} \cline{7-7}
0.7-1.2 & 26.8\% & 10\% & 9\% & 10\% & 18\% & 36.4\% \\
\cline{1-2} \cline{7-7}
1.2-2.0 & 27.0\% & & & & & 36.6\% \\
\cline{1-2} \cline{7-7}
\hline \hline
\end{tabular}
\end{table}

The final $D^0$ spectrum in 0-80\% minbias AuAu collisions at 200
GeV are shown in Fig.~\ref{spectrafit}.

\section{Muon from charm decay at low transverse momentum}

In recent experiments in STAR, due to large random combinatorial
background, it is difficult to reconstruction charmed hadrons
directly. And single electron measurement is impossible due to
overwhelming photon conversions in the detector material and
$\pi^{0}$ Dalitz decay at low \pt, where the yield accounts for a
large fraction of the total cross-section. Nevertheless, the charm
total cross-sections have been measured in d+Au collisions at RHIC
by a combination of the directly reconstructed low \pt\
$D^0\rightarrow K\pi$ and the non-photonic electron
spectra~\cite{stardAucharm}, and by electron spectra
alone~\cite{phenixAuAu,phenixpp}. Although the systematic and
statistical errors are large, the result indicates a much larger
charm yield than predicted by pQCD
calculations~\cite{stardAucharm,cacciari}. Since most of the
measurements to date at RHIC are from indirect heavy-flavor
semileptonic decays, it is therefore important to find novel
approaches to improve the measurements and also study in detail
how to extract the maximum information about the heavy-flavor
spectrum from its lepton spectrum.

In this section, we propose a new method to extract the charm
total cross-section by measuring muons from charmed hadron
semileptonic decay at low \pt\ (e.g. $0.17\leq p_{T} \leq0.25$
\gevc). Since muons in this \pt\ range are a very uniform sample
of the whole charmed hadron spectrum, the inferred charm total
cross-section is insensitive to the detail of the charm spectrum.
Once the cross-section is determined, the electron spectrum at
higher \pt\ can be used to sensitively infer the charmed hadron
spectral shape. Left panel of Fig.~\ref{emuyields} shows the
charmed hadron \pt\ spectrum before and after requiring its
decayed muons at $0.16<p_T<0.26$ \gevc. The similarity of the
spectral shape shows that the muon selection reasonably uniformly
samples the entire charmed hadron spectrum. The muons in this \pt\
range sample 14\% of the charmed hadron spectrum. The muon yield
is about $1/70$ of the charmed yield due to an additional 9.6\%
($c\rightarrow l+anything$) decay branching ratio. Right panel of
Fig.~\ref{emuyields} shows the dependence of the muon yield on
$\langle$\pt$\rangle$ for a fixed total charm yield. The yield is
normalized to yields at $n=10$ and $\langle$\pt$\rangle=1.3$
\gevc, see the power-law function Eq.~\ref{plfun}. We also note
that the muon yield has a very weak dependence on $n$ which
demonstrates that over a wide range in $\langle$\pt$\rangle$, the
muon yield is within $\pm15\%$. This is in contrast to the large
variation of the electron yield integrated above \pt\ of 1.0
\gevc, where a factor of 8 variation is seen in
Fig.~\ref{emuyields}.

\bf \centering \bmn[b]{0.51\textwidth} \centering
\includegraphics[width=1.0\textwidth]{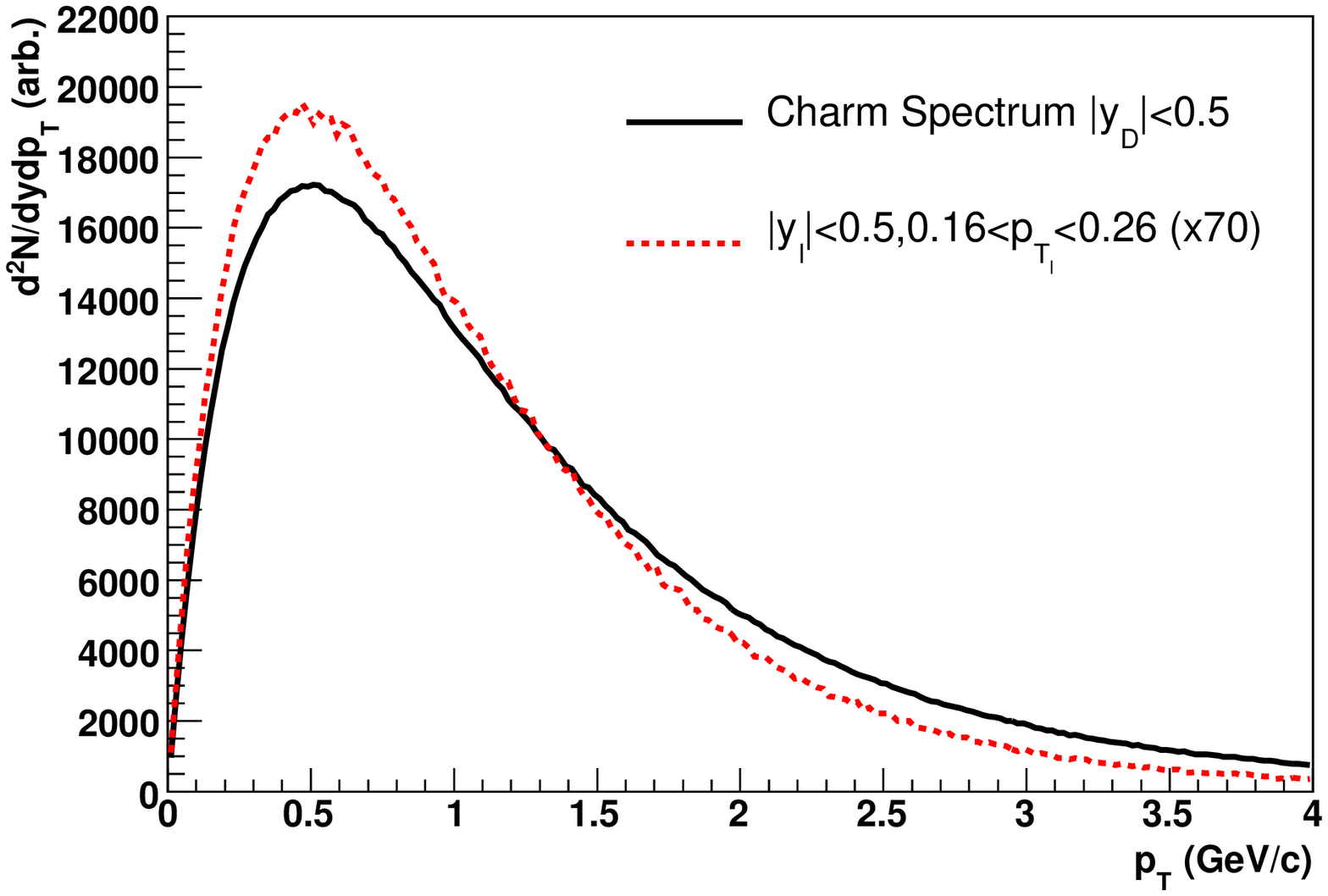}
\emn%
\bmn[b]{0.49\textwidth} \centering
\includegraphics[width=1.0\textwidth]{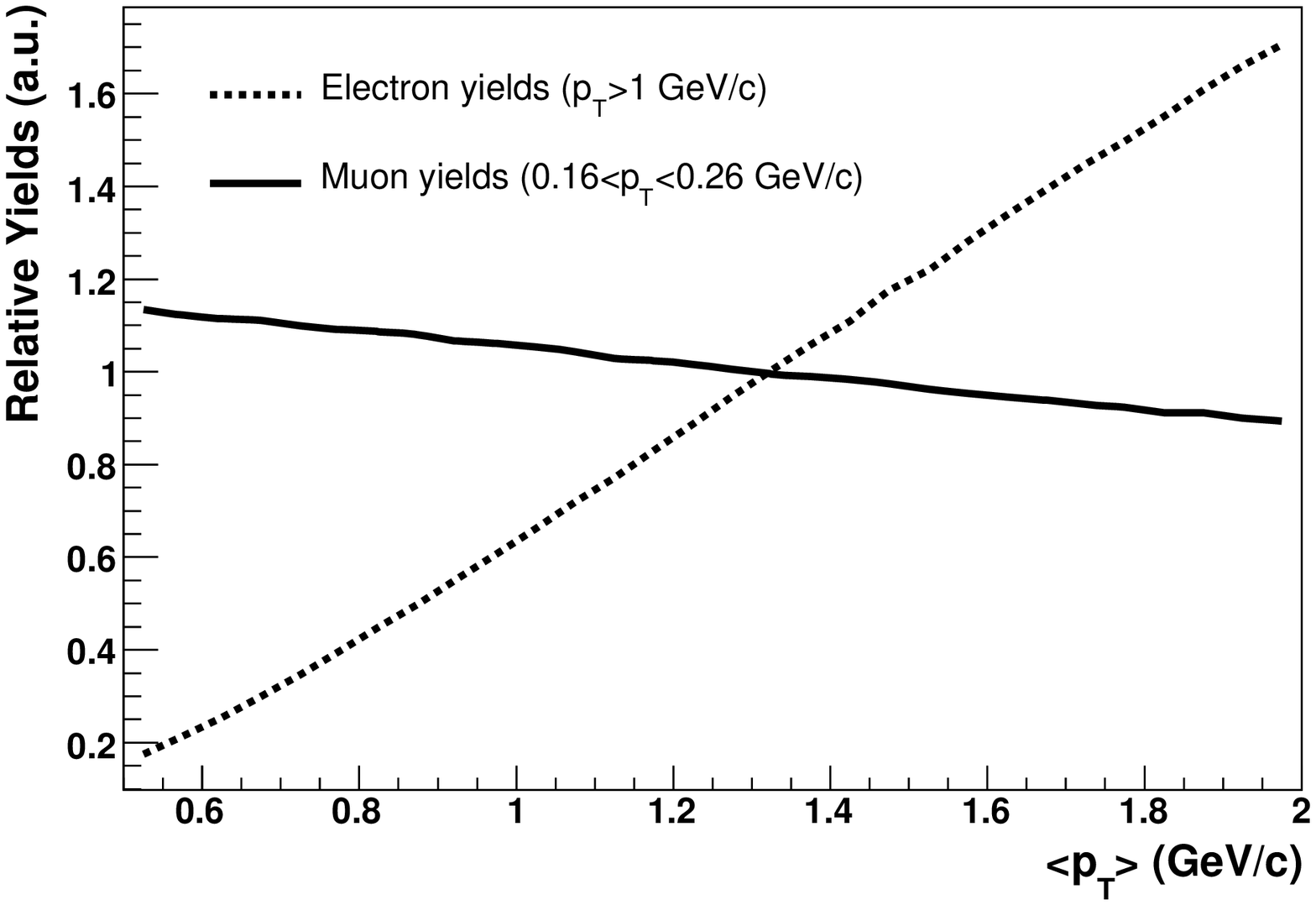}
\emn\\[10pt]
\caption[Muon uniformly sample charm cross-section]{Left panel:
Charmed hadron ($D$) spectra ($dN/dydp_T$) as a function of \pt\
at midrapidity before (solid line) and after (dashed line) a muon
selection of $0.16<p_T<0.26$ \gevc\ and $|y_l|<0.5$. The later was
scaled up by a factor of 70. Right panel: Lepton yields relative
to the fixed total charm cross- section as a function of
$\langle$\pt$\rangle$ for a charmed hadron transverse momentum
spectrum. Solid line shows muon yields with a kinematics selection
$0.16<p_T<0.26$ \gevc\ and $|y_l|<0.5$. Dashed line shows electron
yields with $p_T>1.0$ \gevc.} \label{emuyields} \ef

A total of 7.8 million 0-80\% minbias and 15 million 0-12\%
central triggered \AuAu\ events in year 2004 Run IV were used for
the charm-decayed prompt muons analysis covering a \pt\ range of
0.17-0.25 \gevc\ at mid-rapidity ($-1<\eta<0$). The collision
vertex Z was required from $-30$ cm to 30 cm. The number of fit
points was required above 24.

The low \pt\ single muons were analyzed by combining the \dedx\
measured in the TPC and $m^{2}=(p/\beta/\gamma)^{2}$ from the TOF
at STAR \cite{pidNIMA}. A variable called $n\sigma_{\mu}$, which
is calculated as Eq.~\ref{nsigmamu}, was used for the muon \dedx\
selection. \be n\sigma_{\pi}={{A\over{\sqrt{N_{dE/dx}}}}\times
\ln{dE/dx_{TPC}^{\pi}\over{dE/dx_{Bichsel}^{\pi}}}},
\label{nsigmapi} \ee \be
n\sigma_{e}={{A\over{\sqrt{N_{dE/dx}}}}\times
\ln{dE/dx_{TPC}^{e}\over{dE/dx_{Bichsel}^{e}}}}, \label{nsigmae}
\ee \be n\sigma_{\mu}={{A\over{\sqrt{N_{dE/dx}}}}\times
\ln{dE/dx_{TPC}^{\mu}\over{dE/dx_{Bichsel}^{\mu}}}}.
\label{nsigmamu} \ee

where $A$ is the calibration factor which only basically depends
on the detector intrinsic character, $N_{dE/dx}$ is the number of
\dedx\ fit points for the track reconstruction, $dE/dx_{TPC}^{x}$
is the energy-loss of track $x$ measured in TPC,
$dE/dx_{Bichsel}^{x}$ is the predicted value for track $x$ from
Bichsel calculation.

The muons were selected as $-3<n\sigma_{\mu}<0$,
$-3<n\sigma_{\mu}<-0.5$ for the two \pt\ bins: $0.17<p_T<0.21$ and
$0.21<p_T<0.25$ \gevc, respectively.

\bf \centering \bmn[b]{0.33\textwidth} \centering
\includegraphics[width=1.0\textwidth]{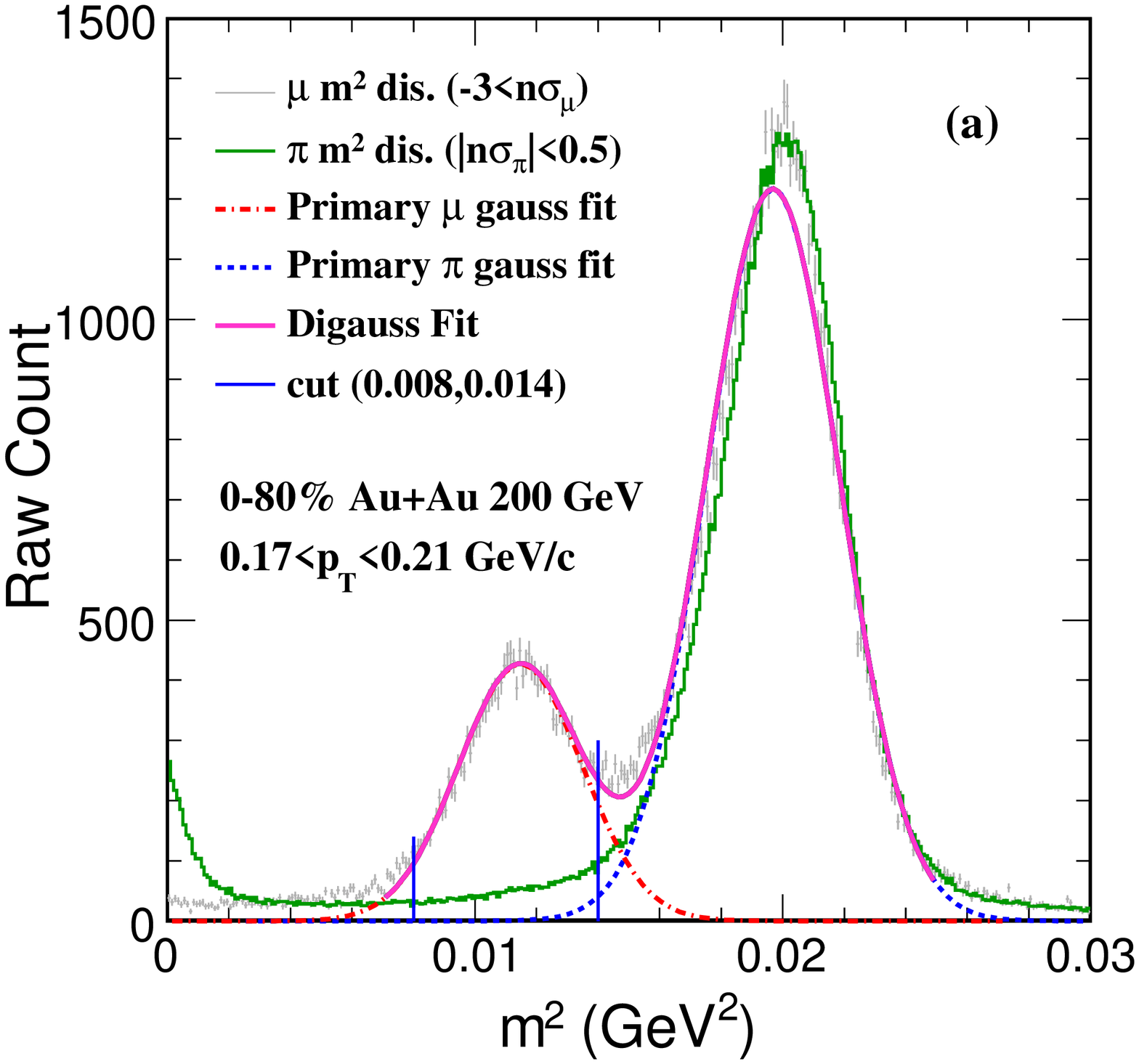}
\emn%
\bmn[b]{0.33\textwidth} \centering
\includegraphics[width=1.0\textwidth]{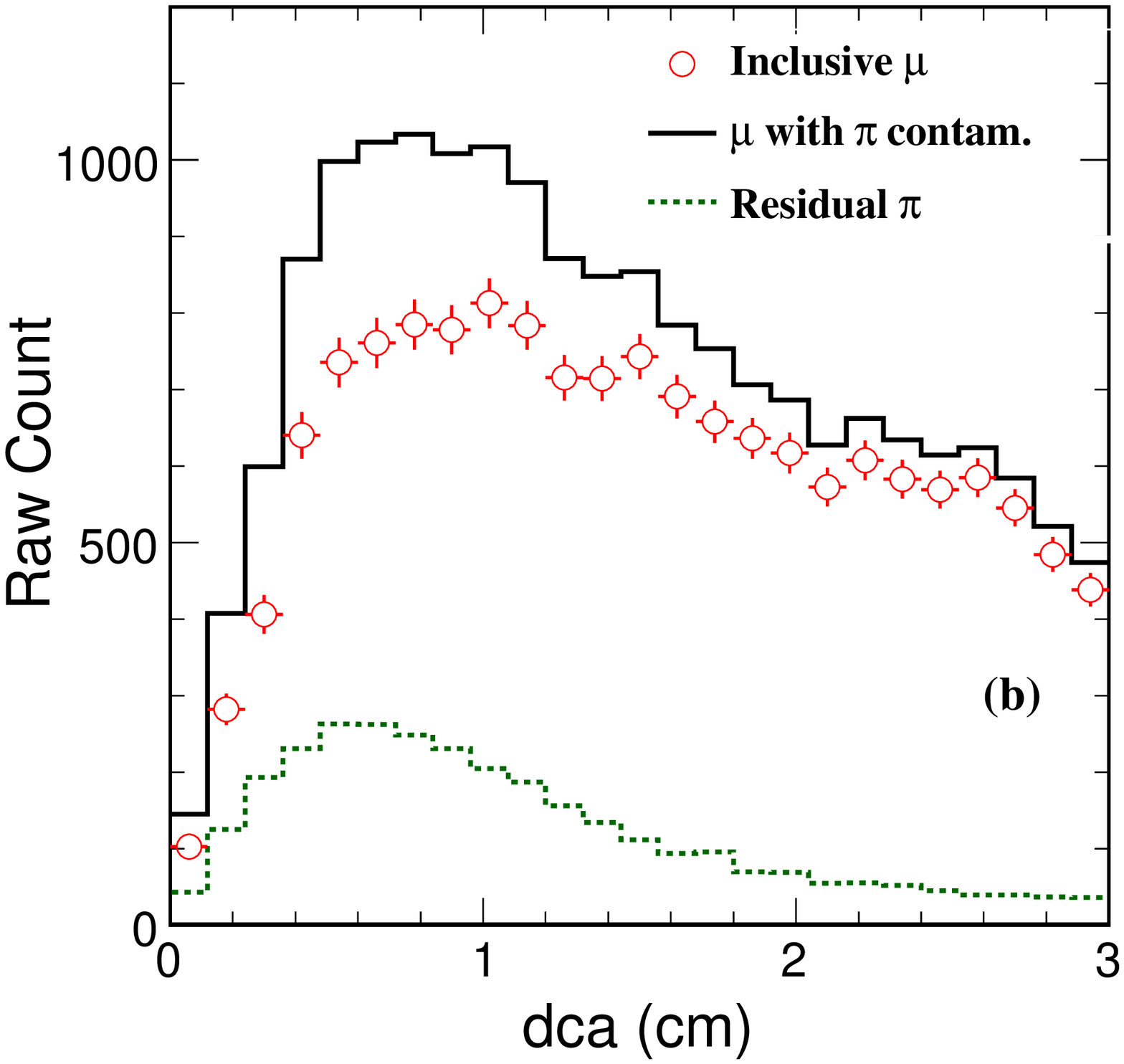}
\emn%
\bmn[b]{0.33\textwidth} \centering
\includegraphics[width=1.0\textwidth]{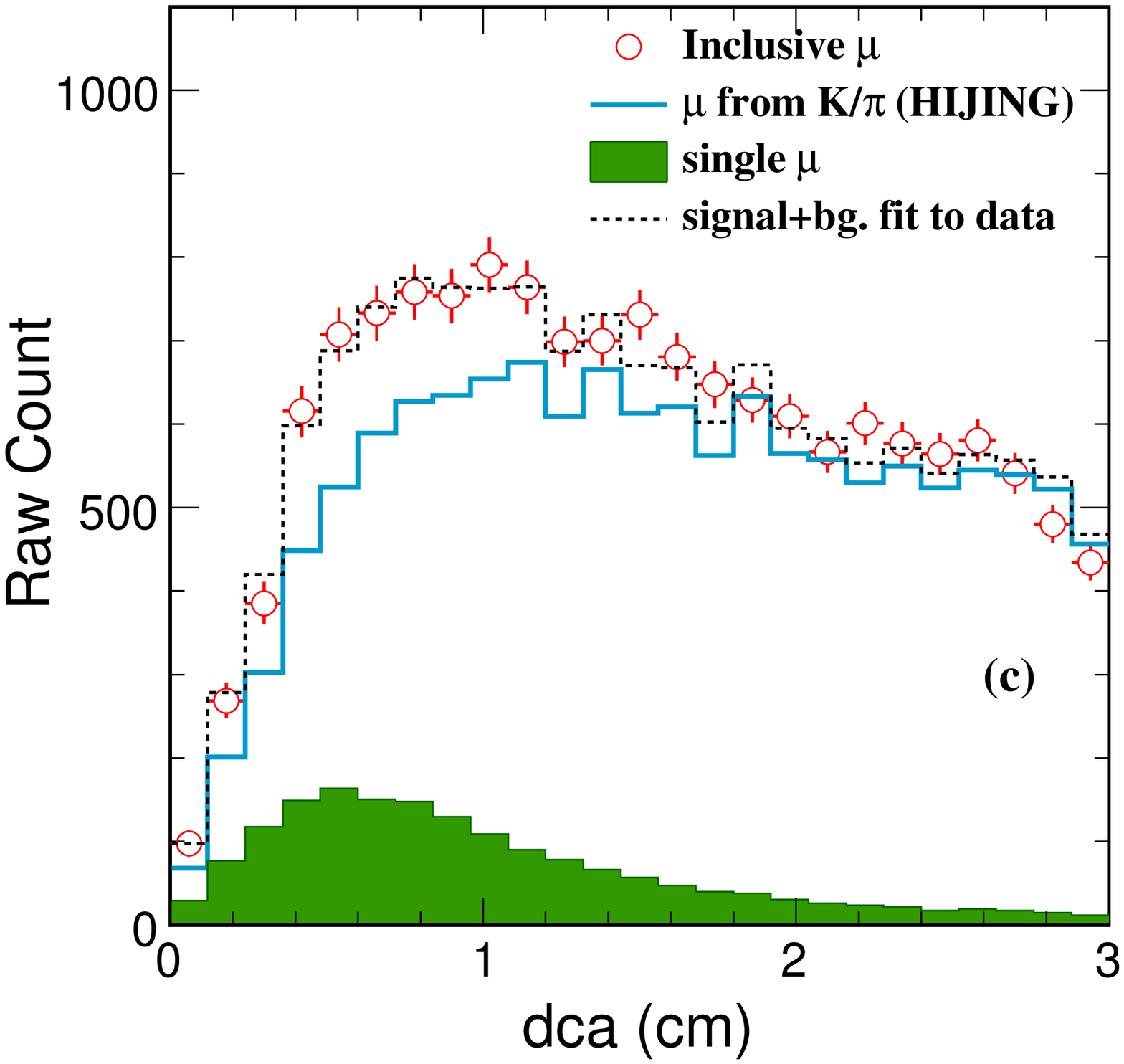}
\emn\\[10pt]
\caption[Muon identification and single muon raw yields
extraction]{Panel (a): Particle mass squared distribution
$m^{2}=(p/\beta/\gamma)^{2}$ from the TOF measurements. A clear
muon mass peak is shown on the left and the primary pion
candidates are shown at the right. The pion contamination was
estimated by applying pion $m^2$ samples using \dedx\ selection.
Panel (b): Inclusive muon DCA distributions were obtained by
subtracting the residual pion DCA in the muon mass window. Panel
(c): The single muon raw yield was obtained from a fit to muon DCA
distributions with the background DCA distributions combining the
primary particle DCA distributions.} \label{muextract} \ef

After the muon \dedx\ selection, a clean muon peak can be
identified within a mass window of $0.008<m^{2}<0.014$
GeV$^{2}/c^{4}$ measured from the TOF, see Fig.~\ref{muextract}
(a). A di-gaussian function was used to fit the muon $m^2$
distributions (purple curve). The residual pions contributing to
the muon mass window were estimated by a pion $m^2$ distributions
with $|n\sigma_{\pi}|<0.5$ (green histogram). After \dedx\ and
$m^2$ selections, the muon yields in the mass window were counted
in each bin of the DCA distributions, see Fig.~\ref{muextract}
(b). The residual pions (green dotted histogram) were subtracted
statistically from the DCA distribution within the muon mass
window applied to the \dedx\ selected pion sample \cite{ffcharm}.
The systematic uncertainties ($18-25$\%) are dominated by the pion
contaminations, which were estimated using different \dedx\ cuts
for the residual pions. After the residual pion contamination
subtracted, the inclusive muon DCA distributions were obtained
(red open circle).

The dominant background muons are from pion/kaon weak decays at
low \pt. Other sources of background
($\rho\rightarrow\mu^+\mu^-$,$\eta\rightarrow\gamma\mu^+\mu^-$,
etc.) are found to be negligible from simulations, due to their
very low yield at low \pt. The measurement of $\eta/\pi$ ratio
shows around 0.5 at high \pt\ ($>3$ \gevc), but drops very fast at
low \pt, its contribution to the low \pt\ muons is very small. The
background muon from pion/kaon weak decays were subtracted using
the DCA distribution from HIJING simulation. The single muon raw
yield was obtained from a fit to muon DCA distributions with the
background DCA distributions combining the primary particle DCA
distributions \cite{ffcharm}, see Fig.~\ref{muextract} (c). The
\pt\ distributions for muon invariant yields in 0-12\% central and
minbias \AuAu\ collisions are shown as open crosses and diamonds
in Fig.~\ref{spectrafit}, respectively.

To understand the background shape, some relevant checks were
performed. The muon DCA distribution from pion decay and that from
kaon decay have similar shape, shown in the left panel of
Fig.~\ref{mucheck}, the relative yield of pion and kaon in HIJING
dose not affect the background DCA shape. Embedding results give
consistent shape of the background DCA, see right panel of
Fig.~\ref{mucheck}.

\bf \centering \bmn[b]{0.5\textwidth} \centering
\includegraphics[width=1.0\textwidth]{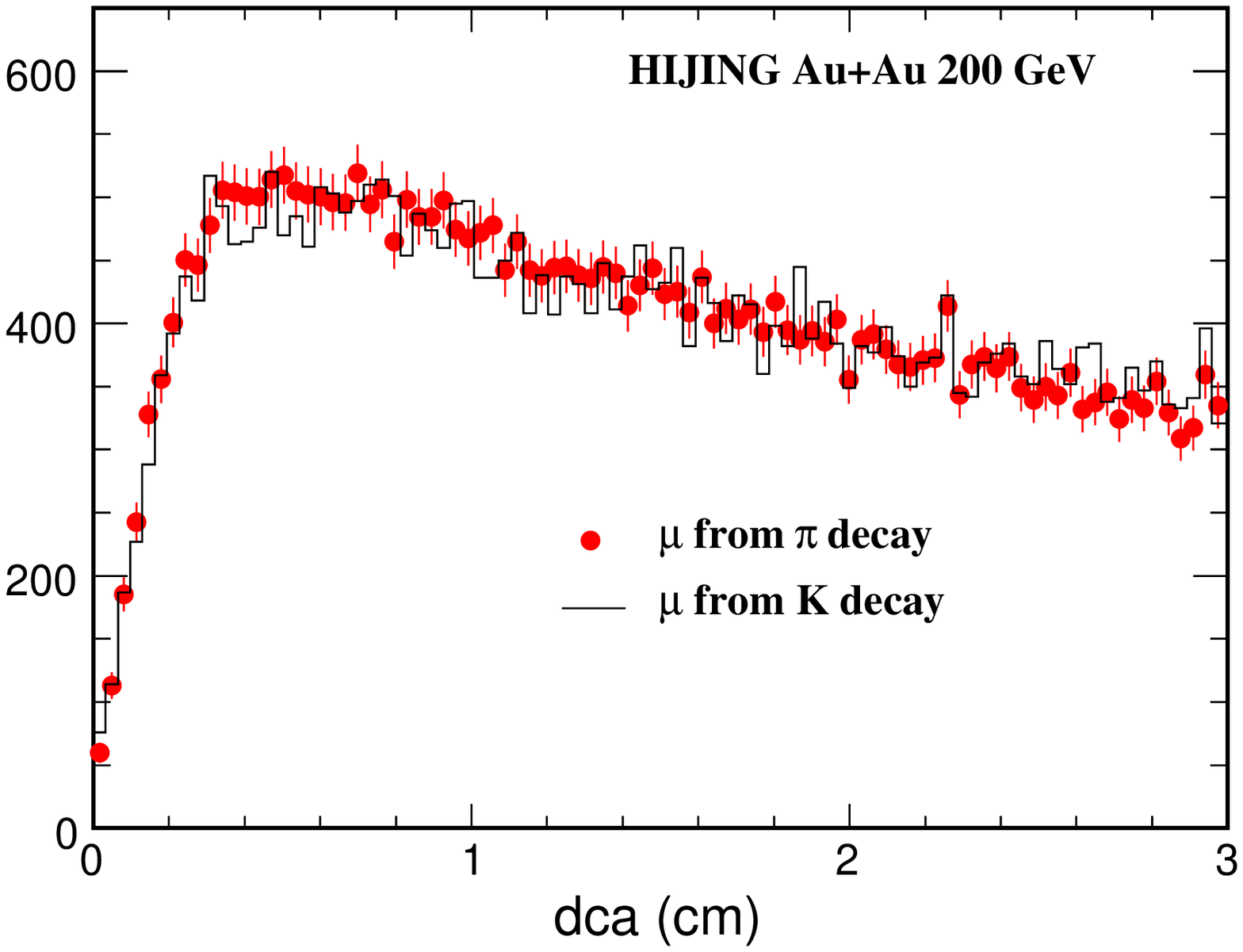}
\emn%
\bmn[b]{0.5\textwidth} \centering
\includegraphics[width=1.0\textwidth]{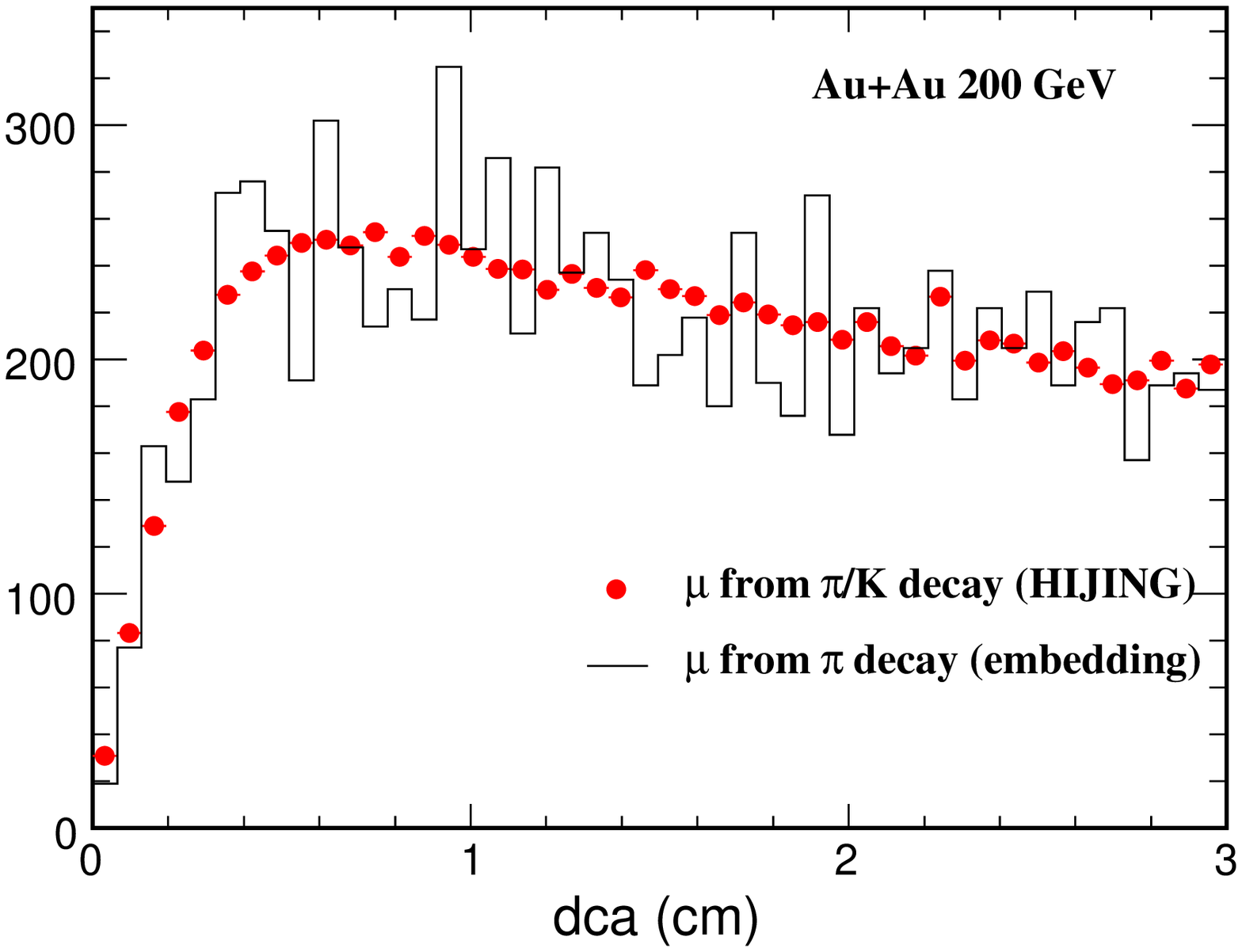}
\emn\\[10pt]
\caption[Muon background check]{Left panel: Background muon DCA
distributions from pion decay (circle) and kaon decay (histogram).
Right panel: Background muon DCA distributions from HIJING
simulation (circle) and embedding (histogram).} \label{mucheck}
\ef

\section{Decay form factors}

In this section, we discuss the charmed hadron semileptonic decay
form factor and its effect on the lepton spectrum.
Fig.~\ref{ffactorfig} shows the electron momentum spectra from
charmed meson decays at rest generated using the Particle Data
Table~\cite{pdgcharmff},
PYTHIA~\cite{stardAucharm,phenixAuAu,phenixpp,pythia}, pQCD
calculations~\cite{cacciari} and from the CLEO preliminary
inclusive measurement~\cite{cleocetalk}. The spectrum generated by
the PDG is according to the form factor of charmed meson decays to
pseudoscalar $K+l+\nu$, vector meson $K^*+l+\nu$ and non-resonance
$(K\pi)+l+\nu$ where the $K^*$ mass is used for the $(K\pi)$
system.  The decay partial widths ($\Gamma$) of the three dominant
decay channels are:
\begin{enumerate}
\item $K+l+\nu$ with pseudo scalar meson in final state ($D^{\pm}$ B.R.=7.8\%)\\
$${{d\Gamma}\over{dq^{2}}} \propto
{{p_{k}^{3}}\over{(1-q^{2}/M^{*2})^2}}$$ where $q^{2}$ is the
invariant mass of the virtual $W\rightarrow\nu l$, $p_K$ is the
momentum of the kaon, and $M^{*}=0.189$ GeV/$c^2$ is a
parameterization of the effective pole mass in the decay.
\item $(K\pi)+l+\nu$ with non-resonant $K\pi$ in final state($D^{\pm}$ 4.0\%)\\
We use the $K^*$ mass (0.892 GeV/$c^2$) for the $K\pi$ invariant
mass and the form factor is the same as in the decay to the pseudo
scalar meson.
\item $K^*+l+\nu$ with vector meson in final state($D^{\pm}$ 5.5\%)\\
$${{d\Gamma}\over{dq^{2}d\cos{\theta_{l}}}} \propto {{p_Vq^{2}}\over{M^2}}
[(1-\cos{\theta_l})^2|H_{+}(q^{2})|^2+
{4\over3}(1+\cos{\theta_l})^2|H_{-}(q^{2})|^2+
{8\over3}\sin^2{\theta_l}|H_{0}(q^{2})|^2]$$\\

where $\theta_l$ is the decay angle between the lepton and the
vector meson, $p_V$ is the vector meson momentum,

$$H_{\pm}(q^{2})= (M+m)A_{1}(q^{2})\mp{{2Mp_V}\over{M+m}}V(q^{2})$$

and

$$H_{0}(q^{2}) = {1\over{2mq}}[(M^2-m^2-q^{2})(M+m)A_{1}(q^{2})-
{{4M^{2}p_{V}^{2}}\over{M+m}}A_{2}(q^{2})]$$

where $A_{1,2},V$ take the form of $1/(1-q^{2}/M^{*2}_{A,V})$ with
$M^*_{A}=2.5$ GeV/$c^2$, $M^*_{V}=2.1$ GeV/$c^2$ and $r_V =
V(0)/A_{1}(0)=1.62\pm0.08$, $r_2 =
A_{2}(0)/A_{1}(0)=0.83\pm0.05$~\cite{pdgcharmff,polemass}.
\end{enumerate}
For different charmed hadrons, we assume that the relative
branching ratios among these three channels are the same, and
their decay electron spectra are the same. The overall charmed
hadron to electron branching ratio $\Gamma(c\rightarrow
e)/\Gamma(c\rightarrow anything)$ is 10.3\%~\cite{pdgcharmff}.
There is a possible $\sim5\%$ difference between electron and muon
decays due to phase space which was not taken into account in this
analysis. Electrons at high momentum are mainly from decay channel
(1) $K+l+\nu$ because the kaon is lighter than the $K^*$ and the
form factor of the decay channel to a vector meson ($K^*+l+\nu$)
favors a low momentum lepton and higher momentum neutrino. Since
PYTHIA uses a simplified vector meson decay form
factor~\cite{pythia}, it tends to produce a softer electron
spectrum. Both the parameterization by Cacciari~\cite{cacciari}
and formulae from the PDG agree with CLEO's preliminary electron
spectrum. In addition, we also find that although the charmed
mesons ($D^{\pm}$ and $D^{0}$) from $\Psi$(3770) decay have a
momentum of 244 MeV/c only and without correction of final state
radiation~\cite{cleocetalk}, it affects slightly its subsequent
electron spectrum.

\bf \centering \bmn[b]{0.49\textwidth} \centering
\includegraphics[width=1.0\textwidth]{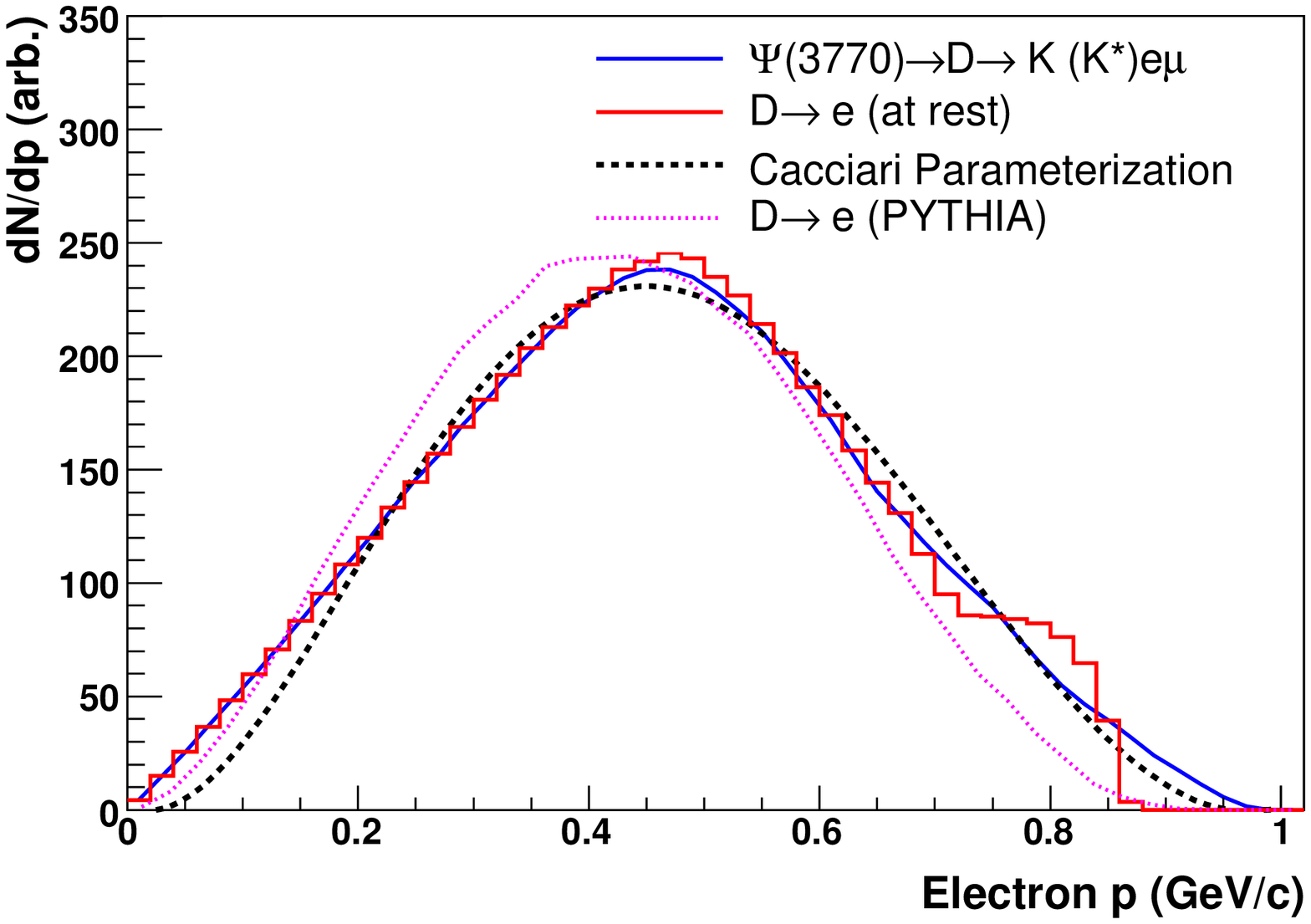}
\emn%
\bmn[b]{0.51\textwidth} \centering
\includegraphics[width=1.0\textwidth]{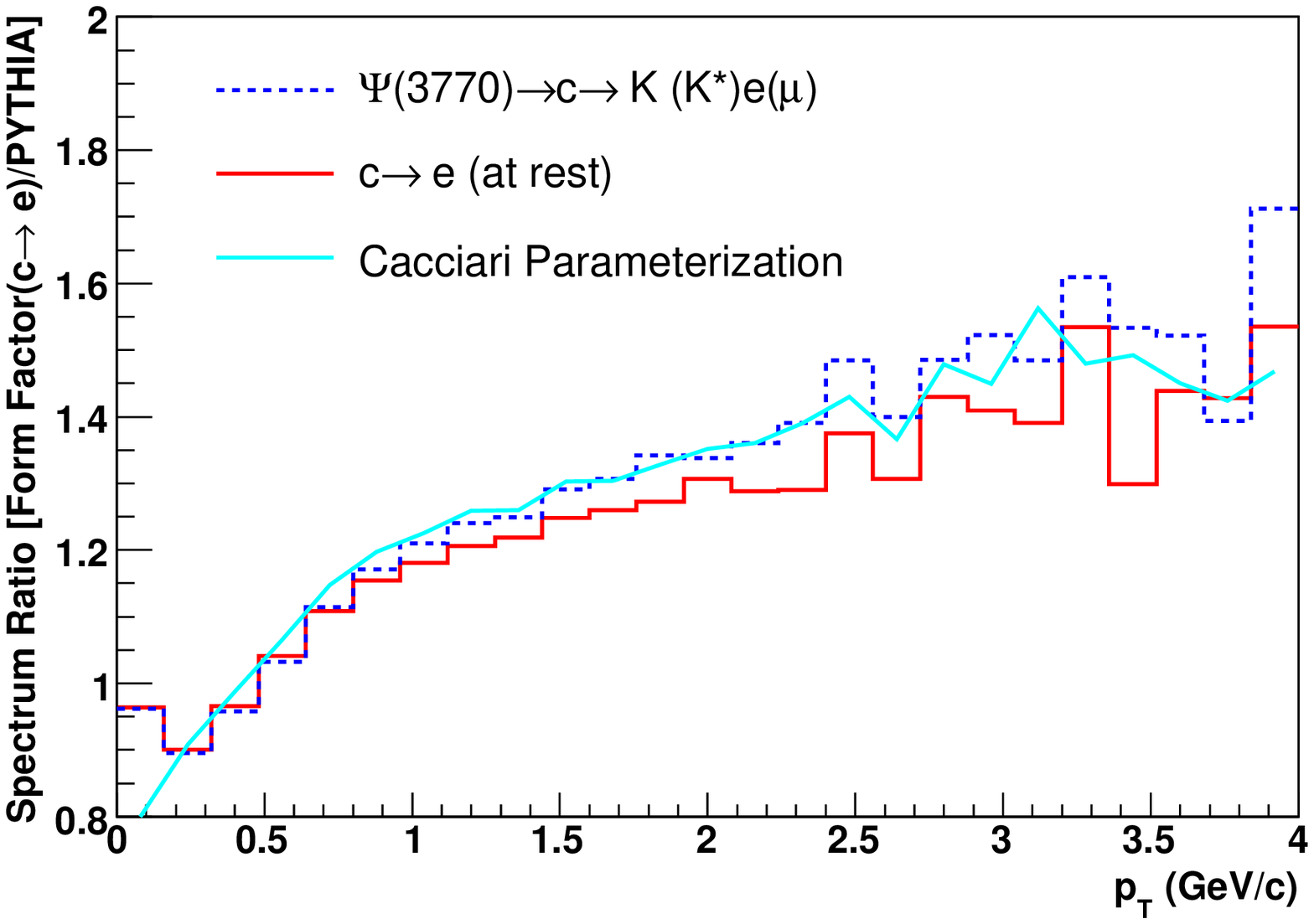}
\emn\\[10pt]
\caption[Muon background check]{Left panel: Electron momentum
spectra from charmed meson decays at rest.  Histogram is the one
with form factor from PDG. The dashed line is that of Cacciari's
parameterization.  The dotted line is from the simplified
vector-meson form factor in PYTHIA. The solid line is that from
the PDG which takes a form factor from $\Psi$(3770)$\rightarrow
D\rightarrow e$. Right panel: Charm-decay electron spectra for
three different form factors divided by the spectrum using the
PYTHIA form factor. The histogram represents the ratio using the
PDG form factor. The solid line is Cacciari's parameterization,
and the dashed line is from PDG and takes a form factor from
$\Psi$(3770)$\rightarrow D\rightarrow e$. See text for detail.}
\label{ffactorfig} \ef

We use our electron momentum spectra and that of Cacciari to
generate electron spectra from charmed decay at RHIC. A power-law
function of the charmed hadron transverse momentum spectrum was
obtained from minimum-bias Au+Au collisions~\cite{starcharmQM05}.
Fig.~\ref{ffactorfig} shows the ratios of those electron spectra
divided by the spectrum using PYTHIA decay form
factors~\cite{pythia,stardAucharm}. The slightly soft form factor
of the charm semileptonic decay in PYTHIA convoluted with a
steeply falling charm spectrum produces an electron $p_T$ spectrum
in Au+Au collisions at RHIC, which can be lower than the correct
one by up to a factor of 1.5 at high \pt. Part of the discrepancy
between experimental results and PYTHIA in electron
spectra~\cite{stardAucharm,phenixAuAu} can be explained by the
decay form factor. Taking $D^{\pm}$ and $D^{0}$ from $\Psi$(3770)
decay as if it were at rest, results in slight change on the
electron spectrum.

In the following, if we say 'form factor decay' that means we let
the charmed hadrons decay using the form factor discussed above.

\section{Combined fit: Charm cross-section, freeze-out and flow, energy-loss}

At RHIC energy, charm total cross-section have been measured from
single electron spectra alone at $p_T>0.8$ \gevc\ by PHENIX. But
the charmed hadron \pt\ spectrum is unknown and the electron \pt\
region is corresponding to higher \pt\ of the charmed hadron
spectrum. Then the charmed hadron at low \pt\, where the yield
accounts for a large fraction of the total cross-section, is
missing. The method to estimate the background from cocktail is
also strongly model dependent. The direct extraction from electron
spectrum will have large systematic uncertainties. STAR provided a
method to extract the charm production cross-section by a
combination of the directly reconstructed low \pt\ $D^0\rightarrow
K\pi$ and the non-photonic electron spectra~\cite{stardAucharm}.
The $D^0$ measurement constrains most of the cross-section. But
due to large combinatorial background, the systematic
uncertainties are still large.

On the other hand, as discussed in the previous sections, STAR
collaboration has measured $D^0$ hadronic decay channel
($0.2<p_T<2$ \gevc), single muon with high precision at low \pt\
($0.17<p_T<0.25$ \gevc) and single electron ($0.9<p_T<5$ \gevc) at
200 GeV \AuAu\ collisions. Charm cross-sections at mid-rapidity
($d\sigma_{c\bar{c}}^{NN}/dy$) were extracted from a combination
of the three measurements covering $\sim90\%$ of the kinematics.
In this section, the detail of how we perform the fit combining
all the data points from these three measurements to extract the
charm cross-section will be presented. In addition, the extraction
of charmed hadron freeze-out temperature and flow velocity based
on blast-wave model will be discussed.

Firstly, the decay kinematics of charmed hadron to electrons were
studied in previous section. The same improved charm semileptonic
decay form factors were used for all charmed hadrons. Assume
similar \pt\ spectrum shape between different charmed hadrons. And
assume their decay electron spectra are similar. We applied the
$D^0$ mass (1.863 GeV) for the form factor decay. The decay
electron spectra were normalized by the $D^0$ fraction in total
charmed hadrons from $e^+e^-$ collisions at \sqrts = 91 GeV from
PDG~\cite{PDG}: $R\equiv N_{D^0}/N_{c\bar{c}}= 0.54\pm0.05$, and
the charm branching ratio to electron ($c\rightarrow e$ $B.R.=
10.3\%$). There is $\sim$15\% systematical difference between
charm decay to electron and muon. In the following combined-fit,
the single electrons are all from charm decays (Bottom
contributions will be discussed in next chapter).

A power-law function is used to create charmed hadron \pt\
spectra. The function takes the form: \be {{dN}\over{2\pi
dyp_Tdp_T}}={{dN}\over{dy}}
{{2(n-1)(n-2)}\over{\pi(n-3)^{2}\langle p_{T}
\rangle^{2}}}(1+{{2p_T}\over{\langle p_{T} \rangle(n-3)}})^{-n},
\label{plfun} \ee where {$dN/dy$} is the yield and $n$ and
$\langle p_{T} \rangle$ are the parameters controlling the shape
of the spectrum. A $D^{0}$ \pt\ distribution with a set of these
parameters was used as input for form factor decay, and the decay
electron spectrum was obtained. A 3-dimensional scan on the
($dN/dy$, $\la p_T\ra$, $n$) ``plane" was done to fit $D^{0}$ and
muon/electron data points simultaneously. The point with the
smallest $\chi^2$ value was set to be the fit result. When we
calculate the $\chi^2$, we try to exclude the correlations between
different measurements~\cite{pdgerr}. Correlated errors are,
however, treated explicitly when there are a number of results of
the form $A_{i}\pm\sigma_{i}\pm\Delta$ that have identical
systematic errors $\Delta$, where $\sigma_{i}$ includes the
statistical error and uncorrelated (bin-to-bin) systematical
error. In this case, we use the quadratic sum of these errors
$(\sigma_{i}^{2}\pm\Delta_{i}^{2})^{1/2}$, where the modified
systematic error \be
\Delta_{i}=\sigma_{i}\Delta[\Sigma(1/\sigma_{j}^{2})]^{1/2}. \ee
Then $\chi^2$ was calculated from the following equation. \be
\chi^2 = \sum_{D} \bigl(\frac{y_D-f_D}{\sigma_D}\bigr)^2 +
         \sum_{\mu} \bigl(\frac{y_{\mu}-f_{\mu}}{\sigma_{\mu}}\bigr)^2 +
         \sum_{e} \bigl(\frac{y_e-f_e}{\sigma_e}\bigr)^2,
\label{chi2} \ee where $y_D$, $y_{\mu}$, $y_e$ denote the measured
yields of $D^{0}$, muons and electrons. $\sigma_D$,
$\sigma_{\mu}$, $\sigma_e$ denote the measured errors, where we
use statistical errors only to calculate the statistical errors
for the fit results and use total errors calculated as above to
estimate the systematical errors for the fit results. $f_D$,
$f_{\mu}$, $f_e$ denote the expected values from input power law
function for $D^0$ and its decay curve respectively. To avoid the
\pt\ position issue in large \pt\ bins, we used the integral yield
$dN$ instead of $dN/p_Tdp_T$ in each \pt\ bin.

In addition, since the \pt\ distributions of $D^{0}$ is unknown,
we also tried a blast-wave function as the input $D^{0}$ \pt\
distribution. The blast-wave function is written as the
following:
\be
{{dN}\over{m_{T}dm_{T}}}\propto\int^{R}_{0}rdrm_{T}K_{1}({{m_{T}cosh\rho}\over{T_{fo}}})I_{0}({{p_{T}sinh\rho}\over{T_{fo}}}),
\label{bwfun} \ee where $\rho=tanh^{-1}\beta_{t}$. The 3 free
parameters are: $dN/dy$, freeze-out temperature $T_{fo}$ and
collective velocity $\beta_{m}$, where $\la \beta_{t}\ra =
\beta_{m}\times {2\over3}$. A 3-dimensional scan on the ($dN/dy$,
$T_{fo}$, $\beta_{m}$) ``plane" was also performed to fit $D^{0}$
and muon/electron data points simultaneously. The way to calculate
errors and $\chi^2$ is the same as above.

The error estimation was through the contour scan in the 3-D
"plane" with the $\chi^2 =\chi^2_{min}+1$. The error of $dN/dy$
was then obtained by projecting this 3-D contour into $dN/dy$
axis.

Once the $dN/dy$ was extracted, the charm production cross-section
per nucleon-nucleon interaction at mid-rapidity can be calculated
from Eq.~\ref{Xsec}: \be
\frac{d\sigma_{c\bar{c}}^{NN}}{dy}\Bigr\rvert_{y=0} =
\frac{dN_{D^0}}{dy}\rvert_{y=0}\times R
\times\frac{\sigma_{inel}^{pp}}{\la N_{bin}\ra}. \label{Xsec} \ee
In this equation, the factor $R$ is the $D^0$ fraction in total
charmed hadrons, as mentioned before. The number of binary
collisions \nbin, which is from Glauber calculations, is
$293\pm35$ for 0-80\% minbias \AuAu\ collisions and is $900\pm71$
for 0-12\% central \AuAu\ collisions.

Table~\ref{combinefit} lists all the fitting results.

\begin{table}
\caption[Combined fit]{The combined fit results for $D^0$, muons
and electrons in \AuAu\ collisions.} \label{combinefit} \vskip 0.1
in \centering
\begin{tabular}{|c|c|c|c|c|c|} \hline \hline
power- & measurements & $d\sigma_{c\bar{c}}^{NN}/dy$ ($\mu$b) &
$\la p_T\ra$ (GeV) & n & $\chi^2$/ndf
\\ \cline{2-6}
law & mb ($D^0$+$\mu$+e) & $277\pm26\pm63$ & $0.92\pm0.06\pm0.12$
& $11.5\pm6.5\pm7$ & 18.6(4.2)/11
\\ \cline{2-6}
fit & 0-12\% ($\mu$+e) & $311\pm26\pm64$ & $0.95\pm0.04\pm0.16$ &
$13\pm1.5\pm6.5$ & 18.5(0.6)/8
\\ \hline
blast- & measurements & $d\sigma_{c\bar{c}}^{NN}/dy$ ($\mu b$) &
$T_{fo}$ (MeV) & $\la \beta_{t}\ra$ & $\chi^2$/ndf
\\ \cline{2-6}
wave & mb ($D^0$+$\mu$+e) & $271\pm24\pm57$ & 222 & 0.27 ($<0.60$)
& 13.9(3.2)/8
\\ \cline{2-6}
fit & 0-12\% ($\mu$+e) & $283\pm21\pm61$ & 220 & 0.35 ($<0.63$) &
45.2(2.3)/5
\\ \hline \hline
\end{tabular}
\end{table}

Fig.~\ref{spectrafit} shows $D^0$ , muons spectra and centrality
dependence of non-photonic electrons spectra and the combining fit
results for $D^0$, muons and electrons spectra in \AuAu\
collisions.

\bf \centering\mbox{
\includegraphics[width=0.6\textwidth]{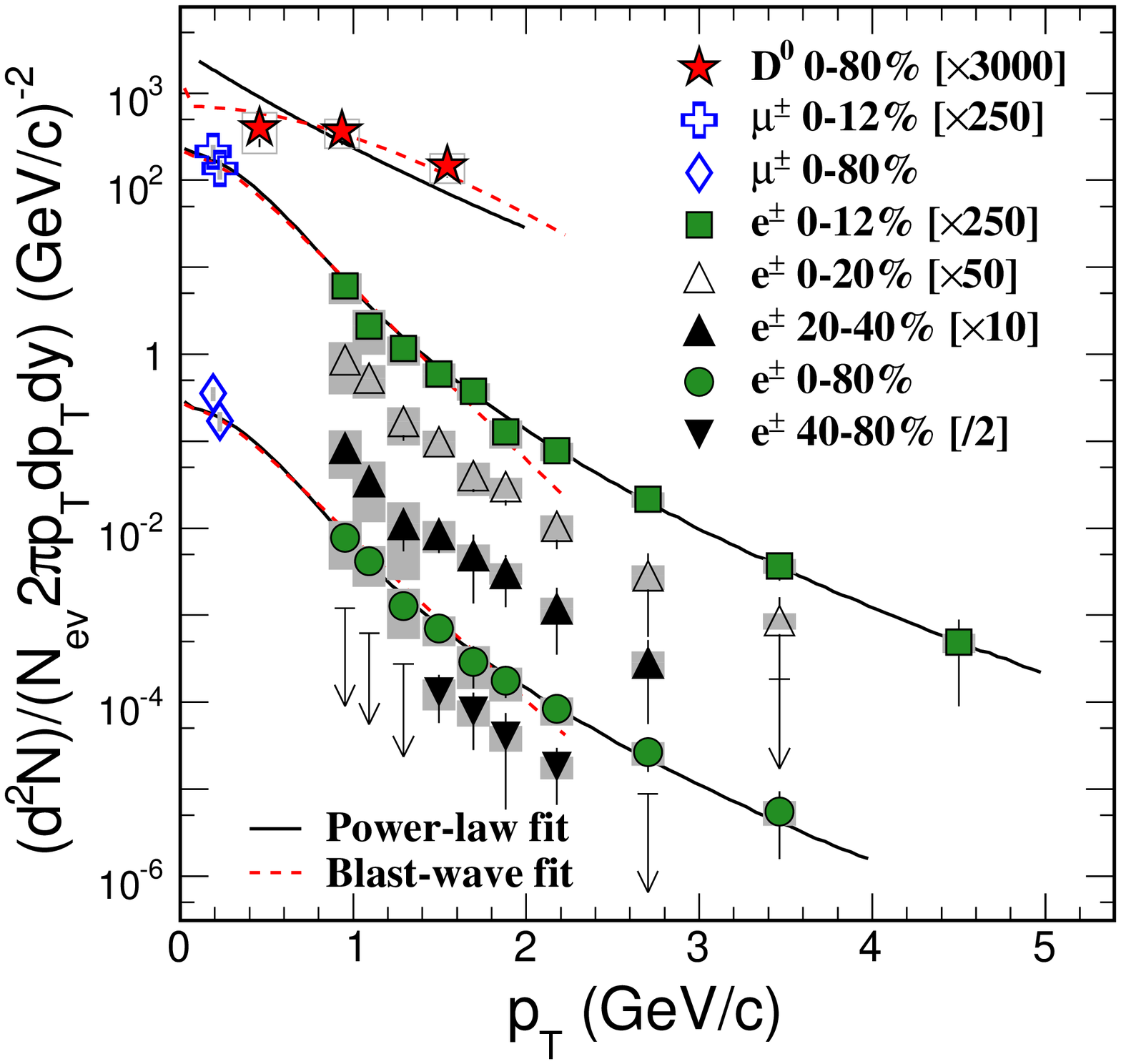}}
\caption[$D^0$, muon spectra and centrality dependence of
non-photonic electron spectra]{$D^0$ spectrum in minbias \AuAu\
collisions (star), muons spectra in minbias (diamond) and 0-12\%
(open cross) \AuAu\ collisions. Centrality dependence of
non-photonic electron spectra. Solid curves show the power-law
combined fit results. Dashed curves are from the blast-wave
combined fit.} \label{spectrafit} \ef

Within errors, both power-law fit and blast-wave fit give the
similar $d\sigma_{c\bar{c}}^{NN}/dy$. Therefore, by averaged the
two fits, $d\sigma_{c\bar{c}}^{NN}/dy$ is presented to be
274$\pm$25(stat.)$\pm$60(sys.) $\mu$b in minbias \AuAu\ and
297$\pm$24$\pm$63 $\mu$b in 0-12\% central \AuAu\ collisions at
$\sqrt{s_{NN}}$=200 \gev. The total charm cross-section per
nucleon-nucleon collision ($\sigma_{c\bar{c}}^{NN}$) following the
method addressed in Ref. \cite{stardAucharm} is presented to be
1.40$\pm$0.11(stat.)$\pm$0.39(sys.) mb in 0-12\% central \AuAu\
and 1.29$\pm$0.12$\pm$0.36 mb in minbias \AuAu\ collisions at
$\sqrt{s_{NN}}$=200 \gev. Fig.~\ref{xsecnbin} shows the
$d\sigma_{c\bar{c}}^{NN}/dy$ as a function of \nbin\ for minbias
\dAu, minbias \AuAu\ and 0-12\% central \AuAu\ collisions. It can
be observed that the charm cross-section seems to follow \nbin\
scaling from \dAu~\cite{stardAucharm} to \AuAu\ collisions which
supports the conjecture that charm quarks are produced at early
stages in relativistic heavy-ion collisions. However, the recent
cross-section result from PHENIX is a factor of 2 lower than STAR
and the FONLL (NLO) calculations
\cite{cacciari,vogtXsec,vogtCronin} shown as the band, which
under-predicts the minbias data by a factor of
$5.1\pm0.48{(stat.)}\pm1.2(syst.)_{-3.1}^{+6.8}(theory)$.

\bf \centering\mbox{
\includegraphics[width=0.6\textwidth]{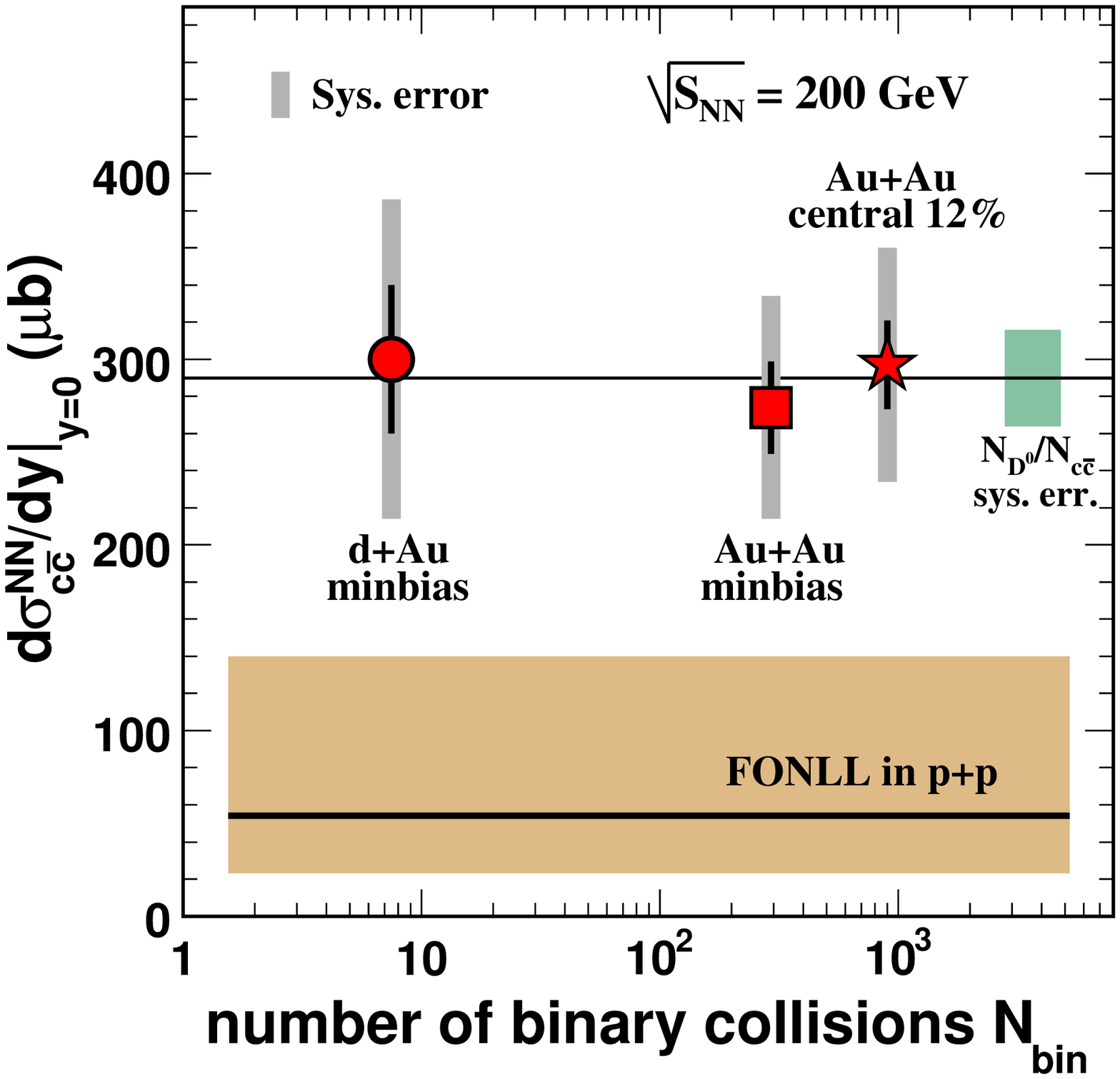}}
\caption[\nbin\ scaling of charm cross-section at
mid-rapidity]{Mid-rapidity charm cross-section per nucleon-nucleon
collision as a function of number of binary collisions (\nbin) in
\dAu, minbias and 0-12\% central \AuAu\ collisions. R factor is
the $D^0$ fraction in total charmed hadrons. The solid line is
from the average of the three values. Within the errors, the
measured cross-sections are consistent with the number of binary
collisions scaling. FONLL prediction, shown as the band,
under-predicts the charm cross-section for collisions at RHIC.}
\label{xsecnbin} \ef

The nuclear modification factor (\RAA)~\cite{highpt130} for single
muon (open crosses) and non-photonic electron \RAA\ (solid
squares) are shown in Fig.~\ref{raa}. The \RAA\ are obtained
taking the ratio of the \pt\ spectra in 0-12\% central \AuAu\
collisions and the \nbin\ scaled decay electron spectra in \dAu\
collisions. The decay \dAu\ electron spectra curve is from the
combined fit with both TOF and EMC
data~\cite{starcraa,phenixcraa}, see Fig.~\ref{dAucombinefit}. The
bin-by-bin errors from the fit are propagated in the systematic
errors of the \RAA.

\bf \centering\mbox{
\includegraphics[width=0.6\textwidth]{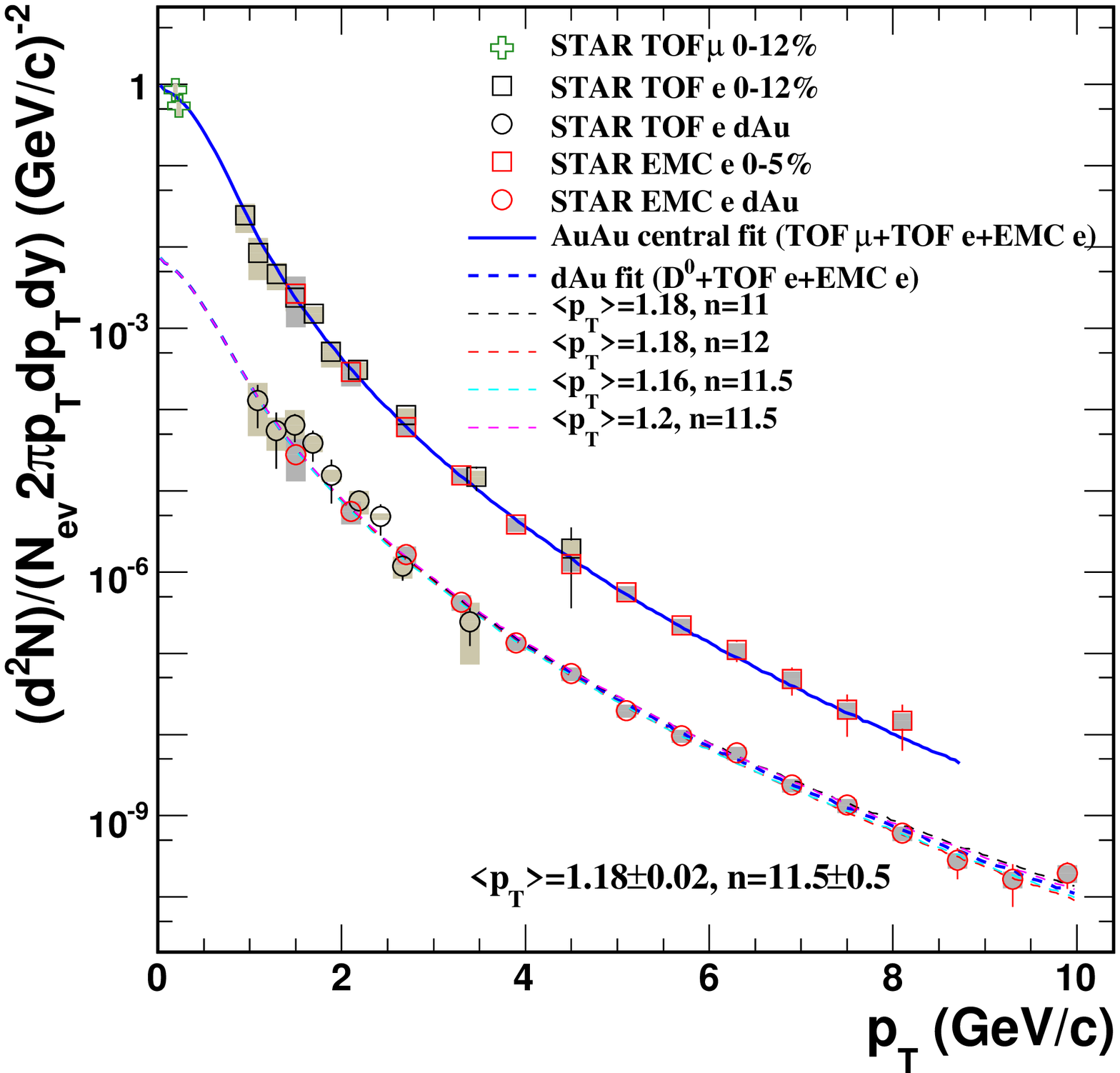}}
\caption[Combined fit of single electron spectra from TOF and EMC
in \dAu\ collisions]{Single electron spectra measured from TOF are
consistent with EMC results. In order to reduce the errors of the
reference spectrum in \dAu\ collisions, a combined fit was
performed to fit both TOF data points and EMC data points. The
variation of the power-law parameters in $1-\sigma$ was applied to
propagate the uncertainties of the spectra shape.}
\label{dAucombinefit} \ef

The muon \RAA\ at low \pt\ is consistent with unity considering
uncertainties. The non-photonic electron \RAA\ in 0-12\% central
\AuAu\ collisions is observed to be significantly below unity at
$1<p_T<5$ \gevc\ and is suppressed as strongly as that of light
hadrons \cite{lqeloss}, which indicates a large amount of
energy-loss for heavy quarks in central \AuAu\ collisions. The
measurement of non-photonic electron at high \pt\ from STAR EMC
also shows strong suppression \cite{starcraa,phenixcraa}.
Theoretical calculations \cite{armesto,heavyDMPRL} considering
only the charm contributions to the non-photonic electrons agree
with the measured non-photonic electron \RAA, while calculations
with single electrons decayed from both bottom and charm quarks
give larger \RAA\ values. The discrepancy from pQCD calculation
assuming only gluon radiative energy-loss challenges our
understanding of the detailed mechanisms of quark and gluon
energy-loss in strongly interacting matter. Model calculations
incorporating in-medium charm resonances/diffusion or collisional
dissociation can reasonably describe the non-photonic electron
spectra~\cite{teaney,vanHCharmflow,RappRaa,Ivancoll}, see
Fig.~\ref{raa}.

\bf \centering\mbox{
\includegraphics[width=0.6\textwidth]{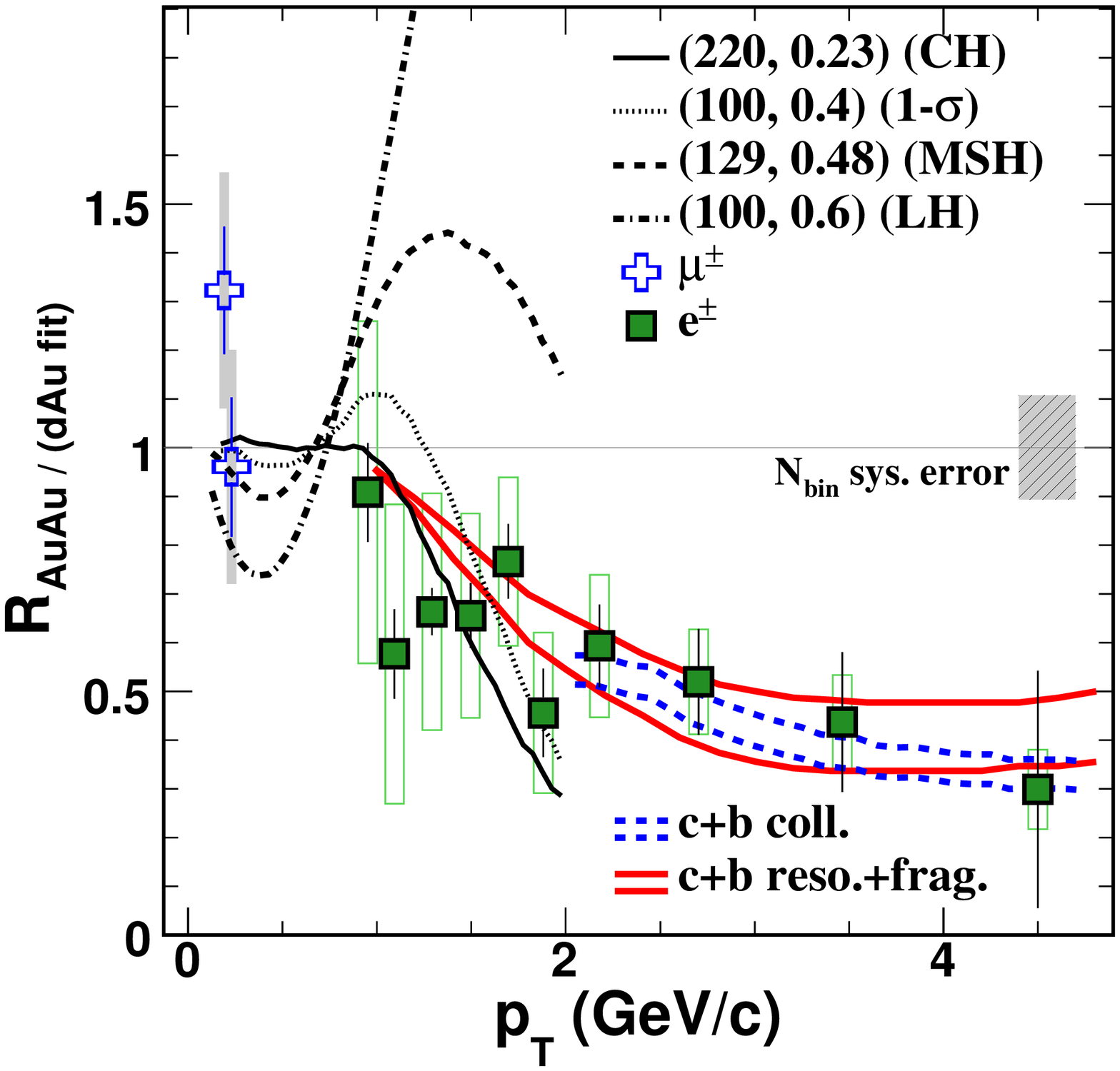}}
\caption[Nuclear modification factor of single muons and
electrons]{Nuclear modification factor (\RAA) of the spectra in
$0-12$\% \AuAu\ collisions divided by corresponding spectra in
\dAu\ collisions. Bin to bin systematic errors are represented by
the open boxes. Box on the right at unity shows the common
normalization uncertainty in \nbin. Model calculations are
presented: coalescence and fragmentation~\cite{RappRaa}
(double-solid lines), and collisional dissociation of heavy
meson~\cite{Ivancoll} (double-dashed lines). The solid, dashed and
dot-dashed lines are Blast-Wave calculations with different
freeze-out parameters of ($T_{fo}$ in MeV,
$\langle\beta_t\rangle$) for charmed hadrons (CH,$1-\sigma$),
multi-strange hadrons (MSH) and light hadrons (LH), respectively.}
\label{raa} \ef

The precise measurement of charmed hadron flow properties are
expected to be a good test for partonic thermalization. The solid
curve in Fig.~\ref{raa} from bast-wave model with the parameters
($T_{fo}$=129 \mev, $\langle \beta_t \rangle$)= 0.477) for
multi-strange hadrons cannot describe the data. The parameters
obtained from pion, kaon and proton spectra ($T_{fo}$=100 \mev,
$\langle \beta_t \rangle$)= 0.6, dot-dashed curve) show even
larger discrepancy~\cite{ffcharm,pikpspectra}. Due to the smearing
of the charm semileptonic decay kinematics, only qualitative
conclusion can be reached, that the charm spectra are not
consistent with large flow and late freeze-out. Bast-wave
parameters with low temperature and moderate radial flow(dotted
line), or with high temperature and low radial flow (dashed line)
can describe our results. The data are also consistent with the
dynamical models~\cite{teaney,vanHCharmflow,RappRaa,Ivancoll}
using finite charm interaction cross-section in a strongly
interacting medium. This may connect the freeze-out parameters
(temperature and flow velocity) to the drag constant in those
dynamical models. Future upgrades with a direct reconstruction of
charmed hadrons are crucial for more quantitative
answers~\cite{hft}.

In summary, we report measurements of charmed hadron production at
mid-rapidity from analysis of $D\rightarrow K\pi$ , muons and
electrons from charm semileptonic decays in minbias and central
\AuAu\ collisions at RHIC. The transverse momentum spectra from
non-photonic electrons are strongly suppressed at $1<p_T<5$ \gevc\
in \AuAu\ collisions relative to those in \dAu\ collisions. For
electrons with \pt$\sim$2 \gevc, corresponding to charmed hadrons
with \pt$\sim3-5$ \gevc, the suppression is similar to that of
light baryons and mesons. Detailed model-dependent analysis of the
electron spectra with $p_T<2$ \gevc\ indicates that charmed
hadrons have a different freeze-out pattern than the more
copiously produced light hadrons. Charm differential
cross-sections at mid-rapidity ($d\sigma_{c\bar{c}}^{NN}/dy$) are
extracted from a combination of the three measurements covering
$\sim90\%$ of the kinematics. The cross-sections are found to
follow binary scaling as a signature of charm production
exclusively at the initial impact. This supports the assumption
that hard processes scale with binary interactions among initial
nucleons and charm quarks can be used as a probe sensitive to the
early dynamical stage of the system.

\chapter{Discussion}

\section{Heavy flavor energy loss}

As we discussed in the introduction chapter, the energy loss of
heavy quarks is considered as a unique tool to study the
interactions between heavy quarks and the medium created in the
heavy-ion collisions, and provide us important information of the
medium properties. Experimentally, the high \pt\ modifications of
heavy flavor hadron (D-mesons, B-mesons, {\em etc.}) yields are
expected to be the direct variables to reveal heavy quark energy
loss. But in current STAR experiment, it is very difficult to
topologically reconstruct D-mesons or B-mesons. And due to large
random combinatorial background, the same- and mixing-event method
can only provide the $D^0$ \pt\ spectrum below 2 \gevc\ with large
systematical uncertainties in minimum bias \AuAu\ collisions. In
central \AuAu\ collisions, the combinatorial background becomes
even larger, the signals are not good enough to extract useful
information. Due to small acceptance, direct measurement of heavy
flavor hadrons in PHENIX experiment becomes even harder.
Nevertheless, the measurements of single electron from heavy quark
decays were used as the indirect substitute.

The single electron strong suppression, similar to light hadrons,
was observed in recent measurements~\cite{starcraa,phenixcraa}.
Fig.~\ref{eraa} shows the nuclear modification factor of single
muons (open crosses) and electrons (solid squares), \RAA\, as a
function of \pt\ measured from STAR TPC+TOF detector. The \RAA\
are obtained using \dAu\ fit curve as reference, which has been
discussed in previous section. The PHENIX result~\cite{phenixcraa}
and STAR EMC result~\cite{starcraa}, \raa\, which use electron
spectra measured in \pp\ collisions as references, are shown as
open circles and solid circles for comparison. Within errors,
these results are consistent in different experiments. The \RAA\
may show some Cronin effect~\cite{cronin,LijuanThesis,Xinthesis}
due to enhancement in \dAu\ collisions relative to \pp\ collisions
$R_{dAu}>1$. But this effect is not significant compared to huge
error bars.

\bf \centering\mbox{
\includegraphics[width=0.6\textwidth]{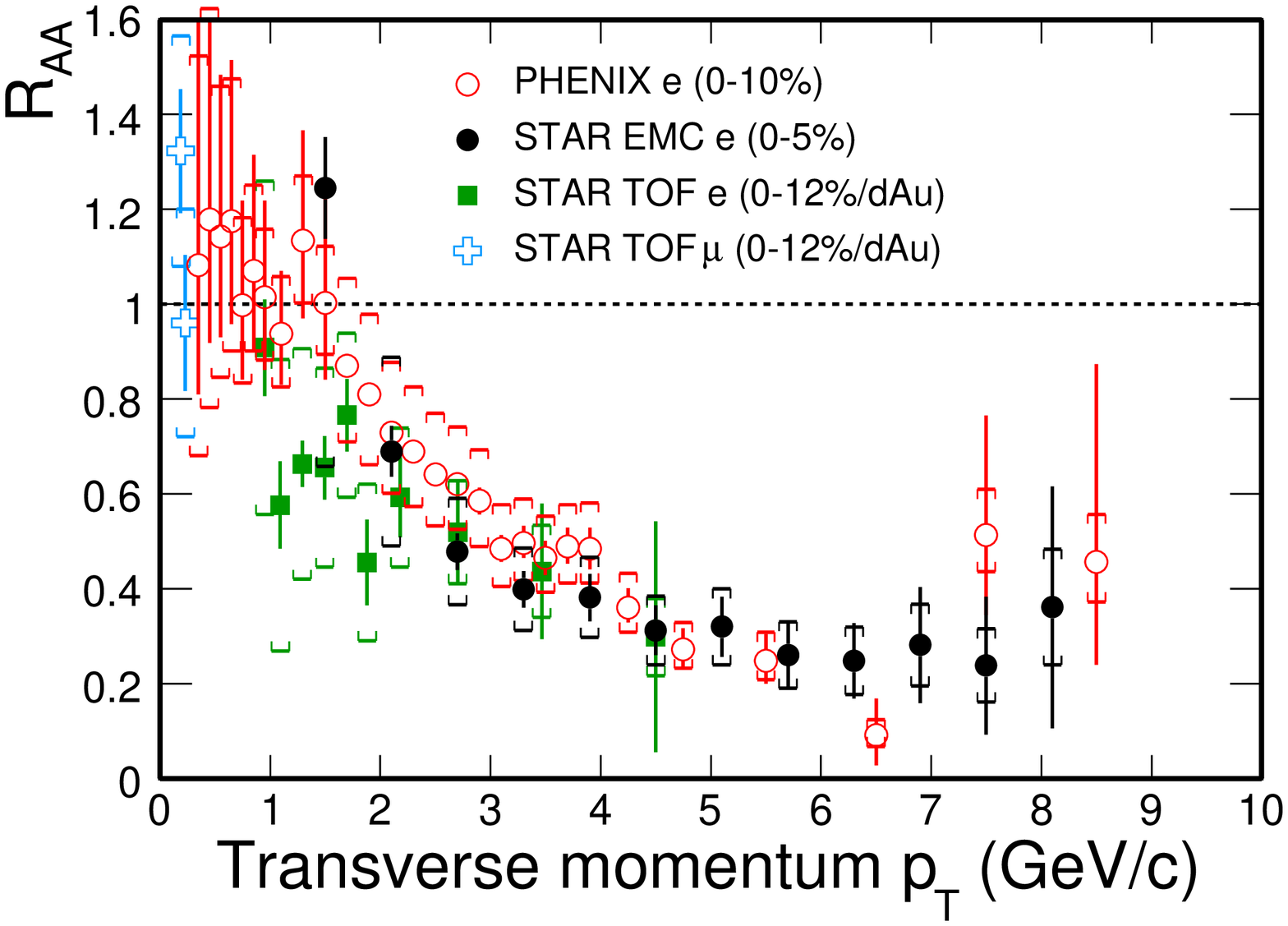}}
\caption[Single electron nuclear modification factors]{Nuclear
modification factors of single electrons as a function of \pt.}
\label{eraa} \ef

There are still some outstanding issues: how can the behavior of
heavy flavor decay electrons reflect heavy quarks? Is the medium
also opaque to heavy quarks? What is the contribution of bottom in
single electron measurements?

Fig.~\ref{Dptvsept} shows the 2D-scattering plot of D-meson \pt(D)
versus decay electron \pt(e) from form factor decay. The inserted
small panel shows the D-meson \pt(D) distributions when select
electron at $1.45<p_T(e)<1.55$ \gevc. $\sim$ 92\% yields of
D-meson are from $1<p_T(D)<4$ \gevc. This indicates that the \pt\
correlation between single electrons and their parent D-mesons is
weak.

\bf \centering\mbox{
\includegraphics[width=0.6\textwidth]{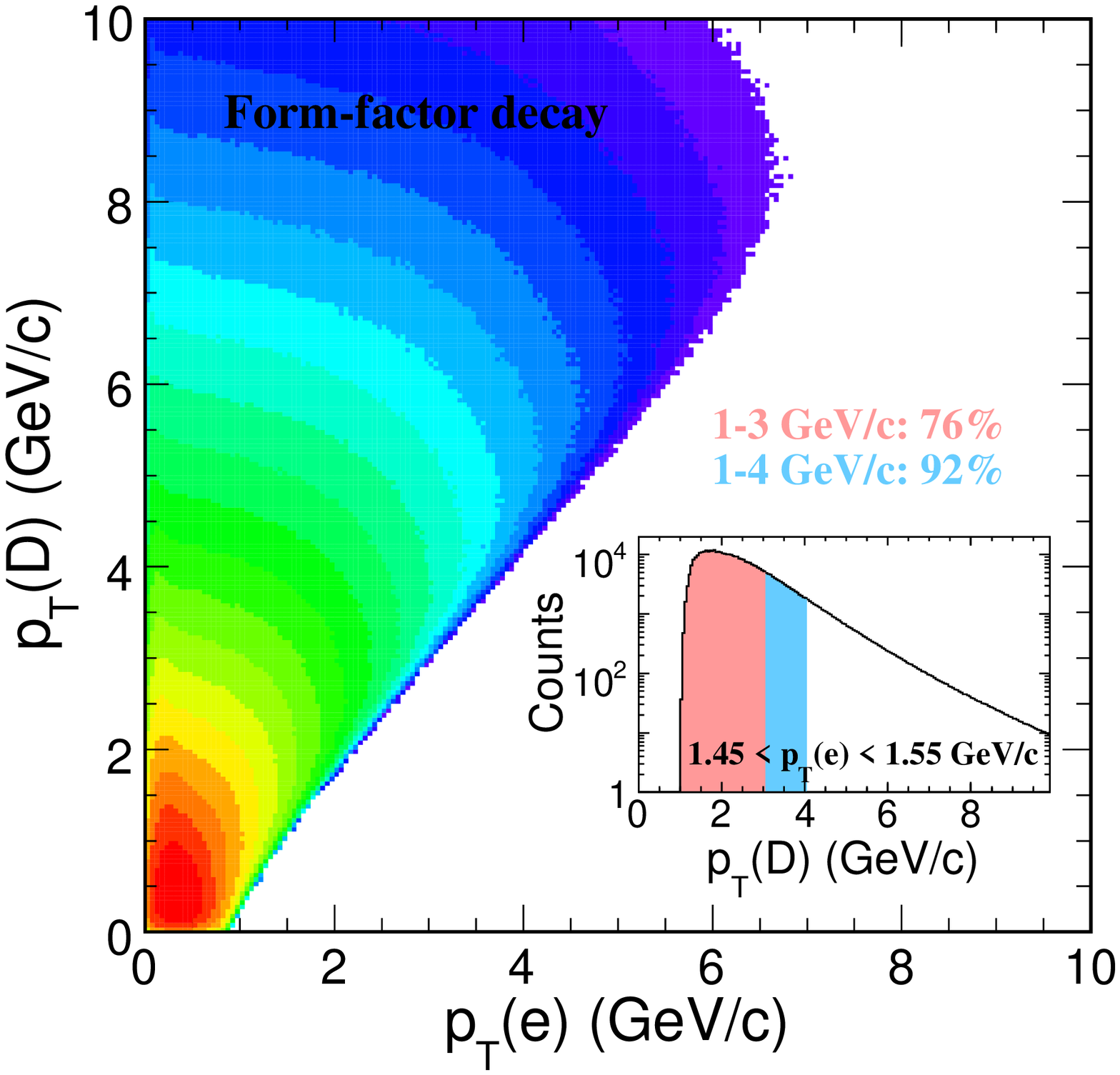}}
\caption[2D-scattering plot of D-meson \pt\ versus electron
\pt]{2D-scattering plot of D-meson \pt(D) versus electron \pt(e)
from form factor decay.} \label{Dptvsept} \ef

On the other hand, due to large mass and small radiative angle,
heavy quarks are predicted to lose less energy than light quarks
via only gluon radiation -- the "dead cone"
effect~\cite{deadcone}. Many theoretical calculations tried to
explain the single electron strong suppression observed in
experiments. A collisional (elastic) energy loss has been proposed
to be taken into account for heavy quark energy loss calculations.

\bf \centering\mbox{
\includegraphics[width=0.6\textwidth]{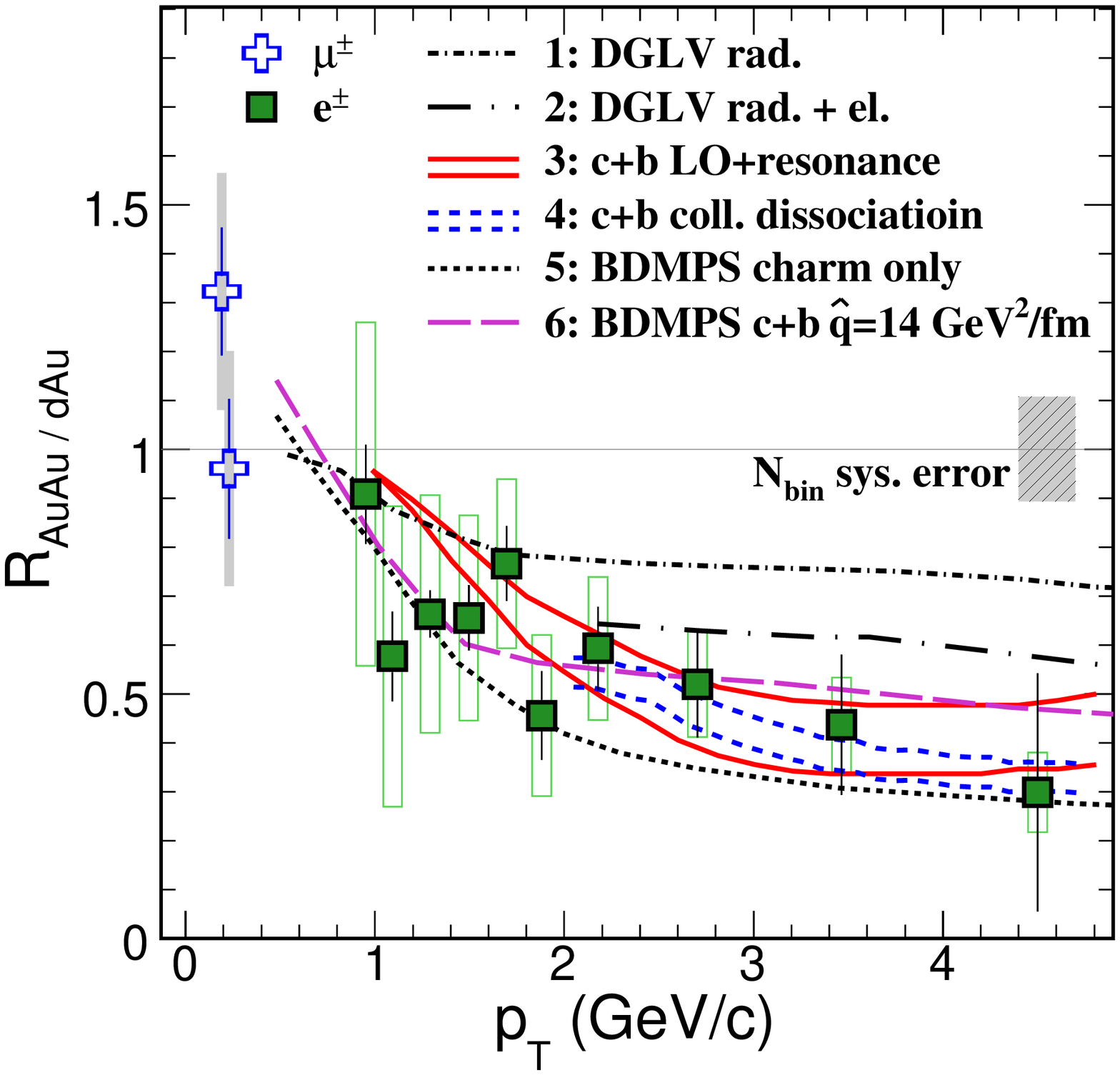}}
\caption[Theoretical calculations for single electron \raa\
compared to data]{Theoretical calculations for single electron
\raa\ compared to data.} \label{raa1} \ef

Fig.~\ref{raa1} shows some recent theoretical calculations for
single electron \raa\ from heavy flavor semi-leptonic decays. The
DGLV radiative energy loss via few hard scatterings with initial
gluon density $dN_g/dy=1000$~\cite{DGLV06} predicts significantly
less suppression than data, shown as curve 1. The curve drops
dramatically after including collisional energy
loss~\cite{Wicks05}, shown as curve 2, but still predicts less
suppression than observed. Curve 3, which has a good agreement
with data, is from the calculation of heavy quark energy loss via
elastic scattering mediated by resonance excitations of D- and B-
mesons and LO pQCD gluon exchange, and including heavy-light
quark-coalescence at hadronization~\cite{RappRaa}. The double
curves indicate the uncertainties from the calculation assuming
the D- and B- resonance width $\Gamma=0.4-0.75$ \gev. Curve 4
shows the calculation of single electron suppression from
collisional dissociation of heavy mesons in QGP by deriving heavy
meson survival and dissociation probability from the collisional
broadening of their light cone wave function~\cite{Ivancoll}. The
double-curve indicate the uncertainties from tuning the typical
value of $\xi\sim2-3$. This calculation also shows consistent
suppression with data. Curve 5 is for single electron only from
D-meson decays with only radiative energy loss from BDMPS
calculations via multiple soft collisions~\cite{armeloss}. Curve 6
is the same calculation as curve 5 except that it includes
electron from B-meson decays. And assumes the transport
coefficient $\hat{q}=14$ GeV${}^2$/fm~\cite{armeloss}. Both BDMPS
calculations agree with data.

There are still a few assumptions and uncertainties in theory.
Several different processes can describe data within the large
experimental uncertainties. The exact mechanism of heavy quark
energy loss is still under intense theoretical and experimental
investigations.

\section{Heavy flavor collectivity}

Theoretical calculations have shown that interactions between the
surrounding partons in the medium and heavy quarks could change
the measurable kinematics~\cite{teaney,vanHCharmflow,dd1}, and
could boost the radial and elliptic flow resulting in a different
heavy quark \pt\ spectrum shape. Fig.~\ref{hfcollectivity} shows
the picture that heavy quarks participate in collective motion. We
can simply treat heavy quark as an intruder, put into the hot
medium with relatively very high density of light quarks. Due to
their large mass and partonic density gradient, such a heavy quark
may acquire flow from the sufficient interactions with the
constituents of a dense medium in analog to Brownian motion. We
expect that bottom quark has small collective velocity due to its
extremely large mass and small interaction cross section.

\bf \centering \bmn[c]{0.47\textwidth} \centering
\includegraphics[width=0.9\textwidth]{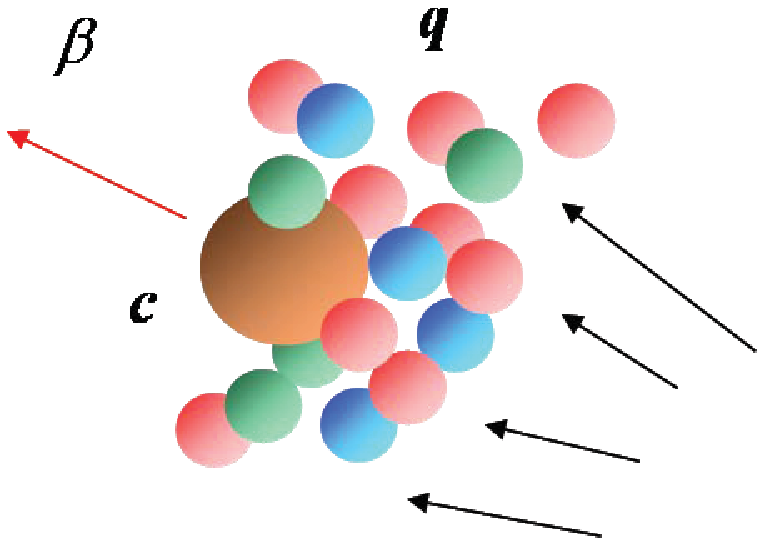}
\emn%
\bmn[c]{0.5\textwidth} \centering
\includegraphics[width=0.9\textwidth]{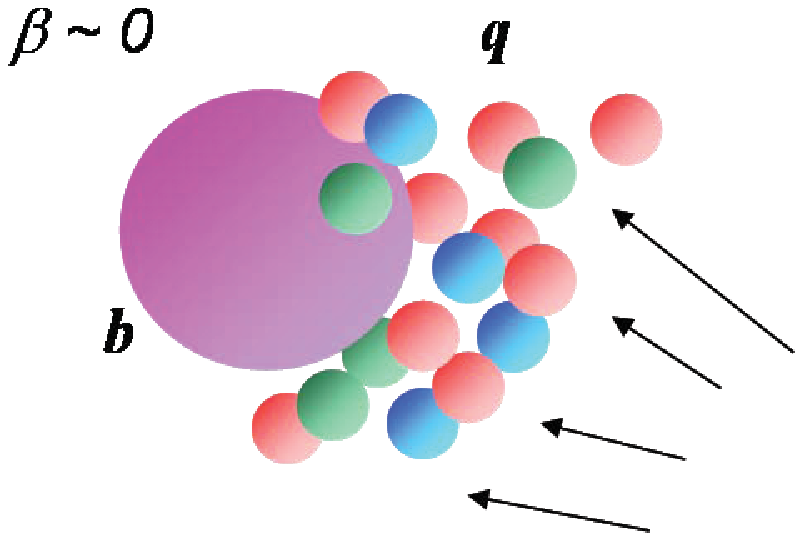}
\emn%
\caption[heavy quark collective motion]{Heavy quarks (c, b)
participate in the collective motion.} \label{hfcollectivity} \ef

Base on blast wave model~\cite{blastwave} discussed in previous
section, a combined fit was performed to describe the spectral
shape and to provide the freeze-out temperature $T_{fo}$ and
collective velocity $\la\beta\ra$. Fig.~\ref{tfovsbeta} shows the
2D distribution of $T_{fo}$ versus $\beta_m$
($\la\beta\ra=\frac{2}{3}\beta_m$) extracted from the combined fit
to muon and electron spectra measured in $0-12$\% central \AuAu\
collisions. The best fit with minimum $\chi^2$ gives the
parameters as: $T_{fo}=220$ \mev\ and $\la\beta_t\ra=0.23$. The
1-$\sigma$ (dotted curve) and 2-$\sigma$ (white solid curve)
contours are from combined fits with statistical errors only.
Black solid curve is for the 1-$\sigma$ contour from combined fit
with both statistical errors and systematical errors. The fit
results are not sensitive to $T_{fo}$, and $\la\beta\ra$ is
smaller than that of light hadrons. One can also see that $T_{fo}$
and $\beta_m$ are strongly correlated.

\bf \centering\mbox{
\includegraphics[width=0.7\textwidth]{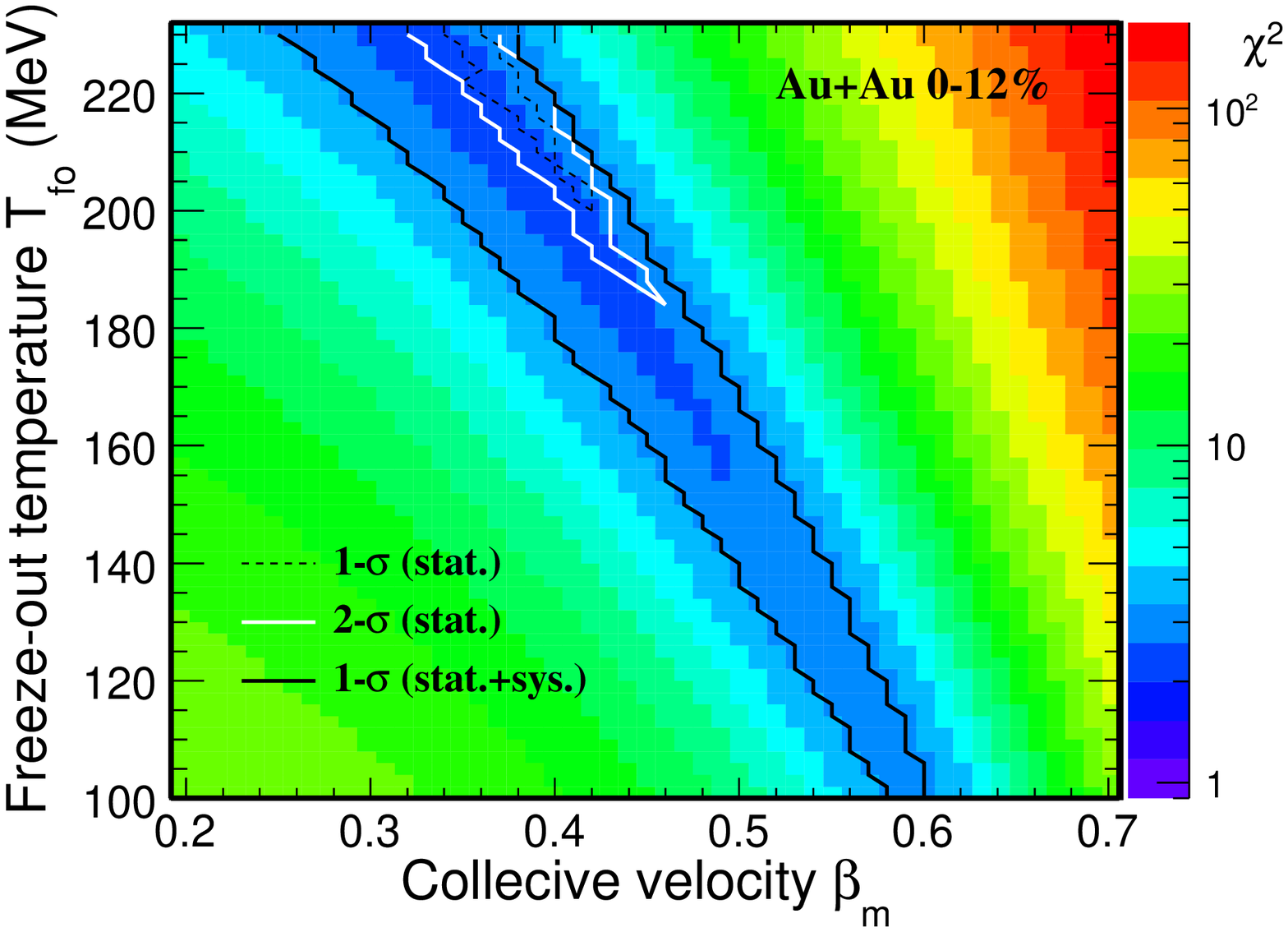}}
\caption[Blast wave parameters $T_{fo}$ versus $\beta_m$ and
$1-\sigma$ contours]{Blast wave parameters $T_{fo}$ versus
$\beta_m$ and 1-$\sigma$ contours from the combined fit to muon
and electron spectra in 0-12\% central \AuAu\ collisions.}
\label{tfovsbeta} \ef

To understand whether charmed hadrons may have similar radial flow
as lighter hadrons, the blast wave fit result of charmed hadrons
was compared to multi-strange hadrons and light hadrons
($\pi$,K,p). Fig.~\ref{bwcurve} (a) shows these comparisons. Solid
curve shows the blast wave fit for charmed hadrons ($T_{fo}=220$
\mev\ and $\la\beta_t\ra=0.23$). The blast wave fit parameters for
multi-strange hadrons (short-dashed curve) are $T_{fo}=129$ \mev\
and $\la\beta_t\ra=0.48$, and for light hadrons (short-dot-dashed
curve) are $T_{fo}=100$ \mev\ and $\la\beta_t\ra=0.6$. The
long-dot-dashed curve, as a reference, is from a power law
combined fit ($\langle p_T\rangle=1.18\pm0.02$, $n=11.5\pm0.5$) to
both TOF and EMC data in \dAu\ collisions, also shown in
Fig.~\ref{dAucombinefit}. All these curves are scaled to match the
measured cross sections. One may consider these comparisons highly
model dependent. In principle, we can directly compare the nuclear
modification function between Omega and charmed hadrons since
their masses are very similar. In practice, we take the blast-wave
fit to Omega, which describes the Omega data well, and apply the
charmed hadron semileptonic decay to obtain lepton spectra if the
Omega and charmed hadrons would have the same freeze-out
properties. This comparison is practically independent of model.

\bf \centering \bmn[c]{0.5\textwidth} \centering
\includegraphics[width=1.\textwidth]{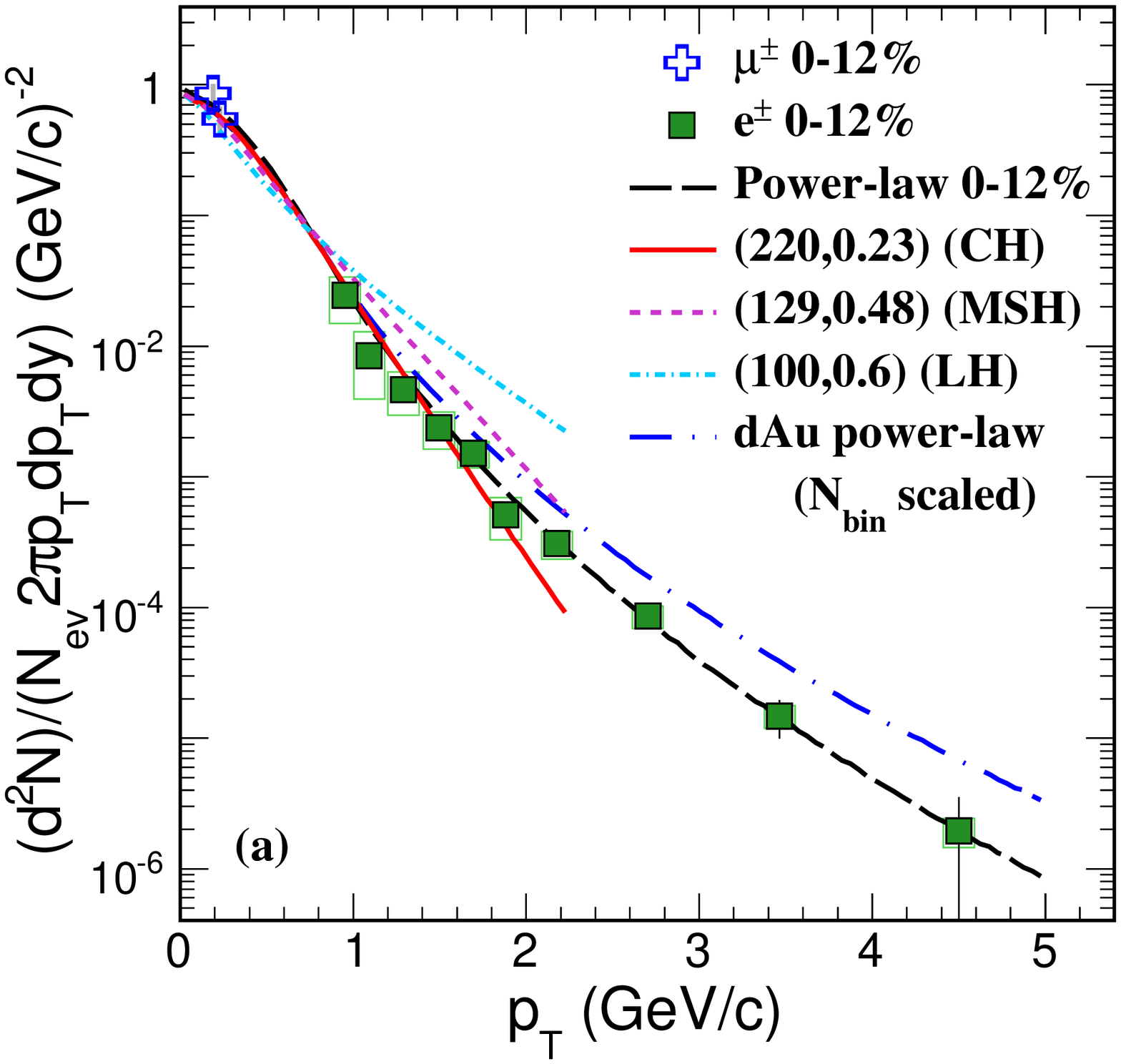}
\emn%
\bmn[c]{0.5\textwidth} \centering
\includegraphics[width=1.\textwidth]{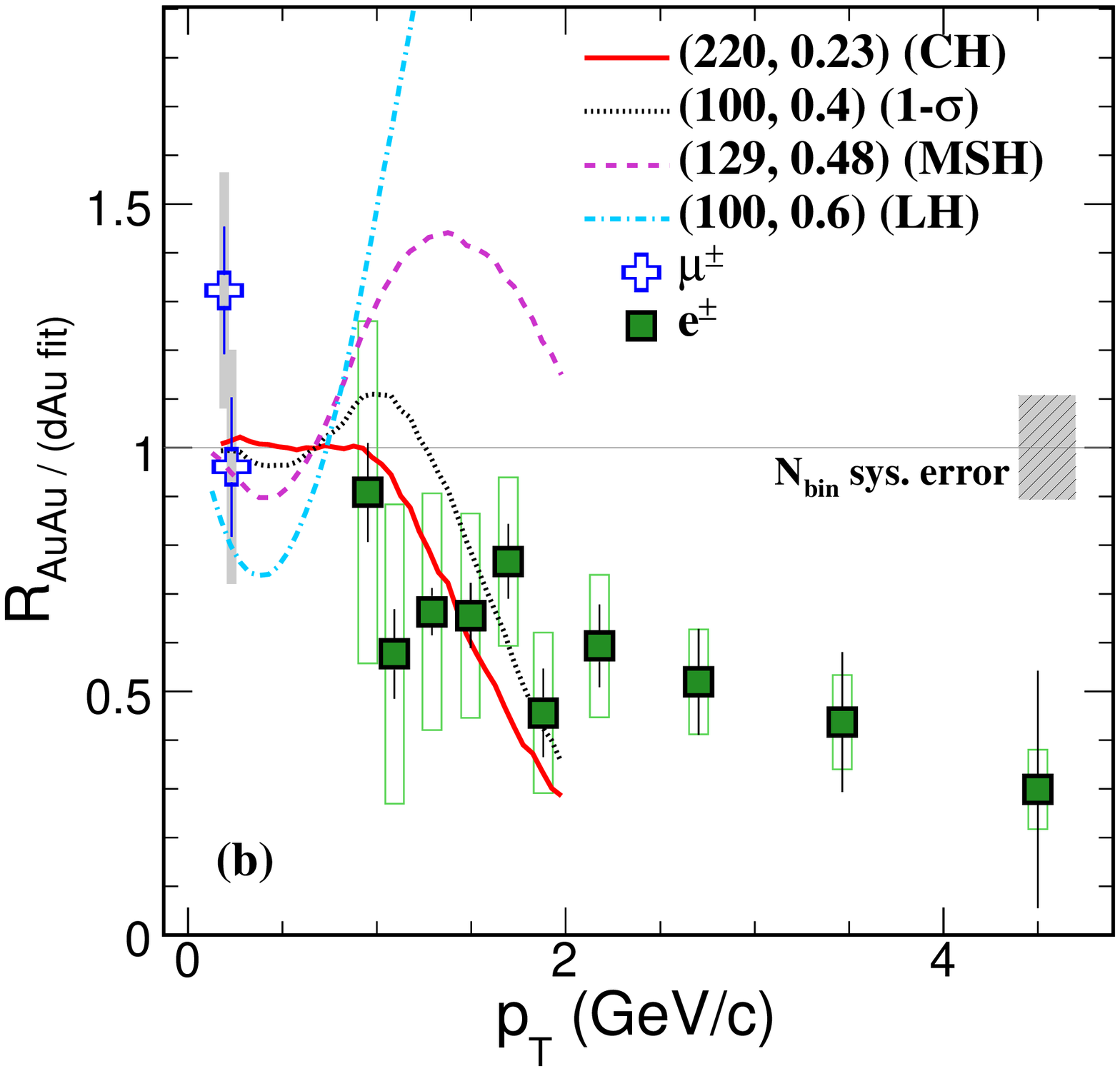}
\emn%
\caption[Blast wave fit results for charmed hadrons compared to
multi-strange hadrons and light hadrons]{Panel (a): Different sets
of blast wave parameters compared to measured lepton \pt\ spectra.
The solid, short-dashed and short-dot-dashed curves are blast-wave
calculations with different freeze-out parameters of ($T_{fo}$ in
MeV, $\langle\beta_t\rangle$) for charmed hadrons (CH),
multi-strange hadrons (MSH) and light hadrons (LH), respectively.
The long-dot-dashed curve is the power law decay curve fit to
\dAu\ data. Bin to bin systematic errors are represented by the
gray bands for muons and open boxes for electrons. Panel (b):
Nuclear modification factor (\RAA) of the spectra in $0-12$\%
\AuAu\ collisions divided by corresponding power law decay curve
fit to \dAu\ data. Box on the right at unity shows the common
normalization uncertainty in \nbin. The curves show the ratio of
blast wave fit results and \dAu\ power law, corresponding to panel
(a).} \label{bwcurve} \ef

The short-dashed curve in Fig.~\ref{bwcurve} (b) from blast-wave
model with the parameters ($T_{fo}=129$ \mev,
$\langle\beta_t\rangle=0.48$) for multi-strange hadrons cannot
describe the data~\cite{hyperon}. The parameters obtained from
pion, kaon and proton spectra ($T_{fo}=100$ \mev,
$\langle\beta_t\rangle=0.6$, short-dot-dashed curve) show even
larger discrepancy~\cite{highpt130}. Due to the smearing of the
charm semileptonic decay kinematics, only the qualitative
conclusion that the charm spectra are not consistent with large
flow and late freeze-out can be drawn. Bast-wave parameters with
low temperature and moderate radial flow(dotted curve), or with
high temperature and low radial flow (solid curve) can describe
our results. This may indicate that charmed hadrons interact and
decouple from the system differently from lighter hadrons. Future
upgrades with a direct reconstruction of charmed hadrons are
crucial for more quantitative answers~\cite{hft}.

\section{Charm production cross-section -- consistency and discrepancy}

We reported the charm cross sections from a combination of three
independent measurements: $D\rightarrow K\pi$ , muons and
electrons from charm semileptonic decays in minbias and central
\AuAu\ collisions at RHIC. While the discrepancy between STAR and
PHENIX also exists. We perform a detailed comparison in this
section.

Charm total cross sections have been extracted from single
electron spectra alone at $p_T>0.8$ \gevc\ by PHENIX. The single
electron spectrum from charmed hadron decay peaks at $p_T\sim0.5$
\gevc\ and therefore a large fraction of the total cross section
is missing from the measurement. Recently, PHENIX reported new
charm total cross section in \pp\ collisions from single electron
spectrum at $p_T>0.3$ \gevc. In all the cases, the inclusive
electron spectrum contains photonic background. The photonic
background level in this \pt\ range is comparable to STAR electron
photonic background level at high \pt. Cocktail and converter
methods have been used by PHENIX to obtain the background.

In this thesis, the charm cross sections are reported from
combined three independent measurements: $D^0\rightarrow K\pi$,
low \pt\ muons and single electrons in STAR experiments.
Currently, the consistencies within STAR experiments are observed,
while the discrepancy between STAR and PHENIX still exists.

The comparison of the single muon/electron spectra measured from
STAR TOF detector and the single electron spectrum measured from
EMC detector in 0-12\% and 0-5\% central \AuAu\ collisions,
respectively, is shown in Fig.~\ref{dAucombinefit}. These data
points are scaled by \nbin\ since they are from different
centralities. In order to see the comparison more clearly, we take
the ratio of the data and the fit curve, see
Fig.~\ref{comEspecCen}. Since EMC results have smaller errors, so
it constrains the fit curve, thus the EMC data points are around
unity and TOF result is a little higher, probably due to 0-5\% has
more suppression than 0-12\%. But within errors, the results from
two independent measurements are consistent.

\bf \centering\mbox{
\includegraphics[width=0.45\textwidth]{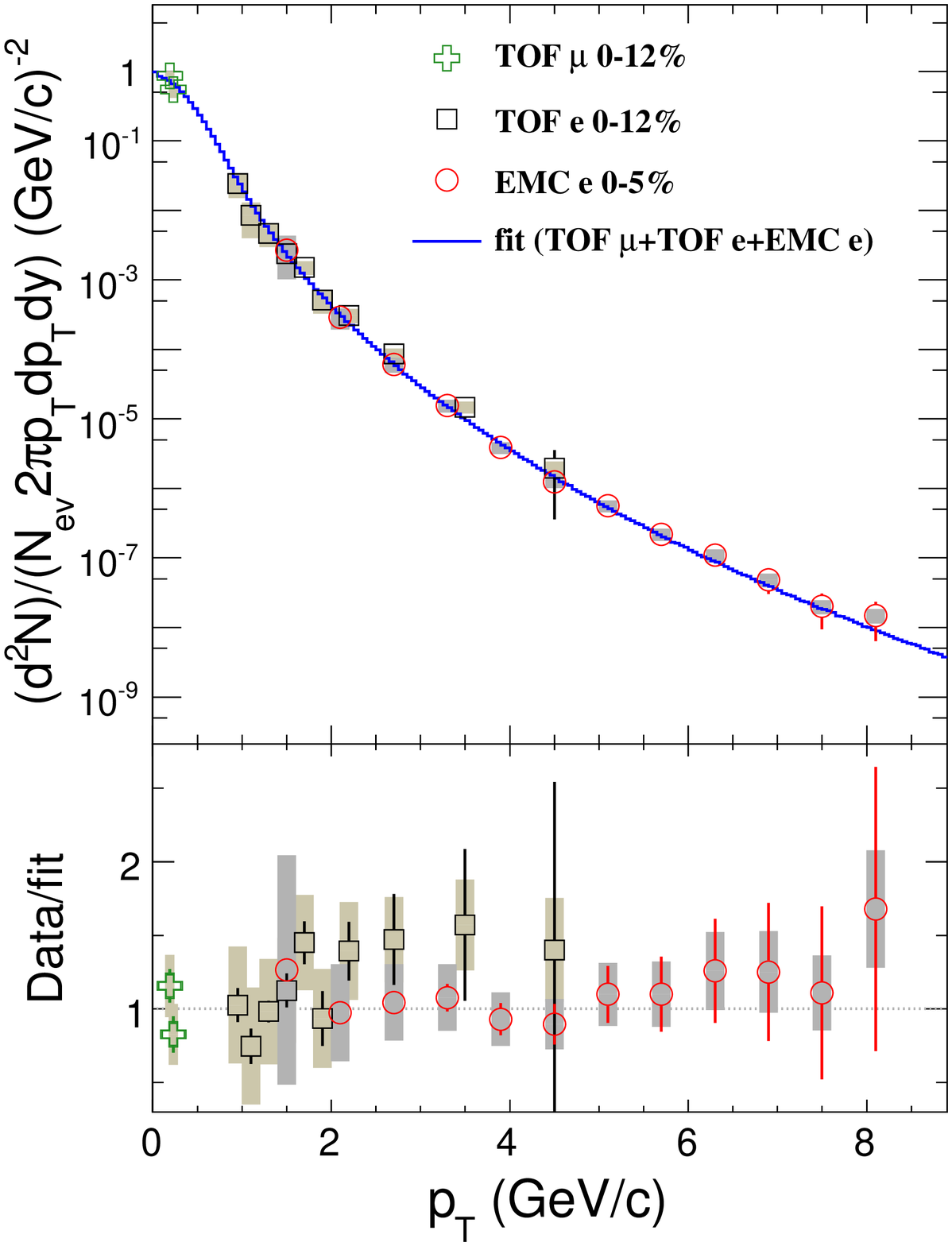}}
\caption[The comparison of single lepton spectra measured from
STAR TOF detector and EMC detector]{The comparison of the single
muon (open crosses) and electron (open squares) spectra measured
from STAR TOF detector and the single electron spectrum (open
circles) measured from EMC detector in 0-12\% and 0-5\% central
\AuAu\ collisions, respectively. The curve shows the fit to all
the data points. Bottom panel shows the ratio of data and the fit
curve.} \label{comEspecCen} \ef

In Fig.~\ref{comESTARPHENIX} (a), the single electron spectrum
from PHENIX measurement in \pp\ collisions, shown as triangles,
has similar shape but systematically lower than STAR measurements,
while STAR TOF result (circles) is consistent with STAR EMC result
(diamonds) within errors. STAR TOF muon measurement (crosses) has
also good agreement with STAR TOF electron measurement (squares)
in 0-12\% central \AuAu\ collisions. A form factor decay combined
fit (solid curve) describes both low \pt\ muons and higher \pt\
electrons and gives the charm cross section at mid-rapidity as 297
mb. In addition, STAR muons can not be fitted together with PHENIX
electrons (scaled by \nbin), see the dashed curve, which is mostly
constrained by PHENIX electron data points with small errors.
PHENIX published charm cross section is 123 mb. The cross section
from fit to PHENIX data only is 131 mb. And the fit combining STAR
\AuAu\ muons and PHENIX electrons gives 137 mb with large
$\chi^2$. The PHENIX cross sections are systematically a factor of
2 lower than STAR. But both STAR and PHENIX are self-consistent in
different collisions. Both observed the number of binary
collisions scaling behavior of charm cross section.
Fig.~\ref{comESTARPHENIX} (b), which shows the charm cross section
at mid-rapidity as a function of number of binary collisions for
STAR (circles), PHENIX (triangles) and FONULL calculations (solid
line and dashed box), clearly summarizes the consistency and
discrepancy among different data sets and theories.

\bf \centering \bmn[c]{0.5\textwidth} \centering
\includegraphics[width=1.\textwidth]{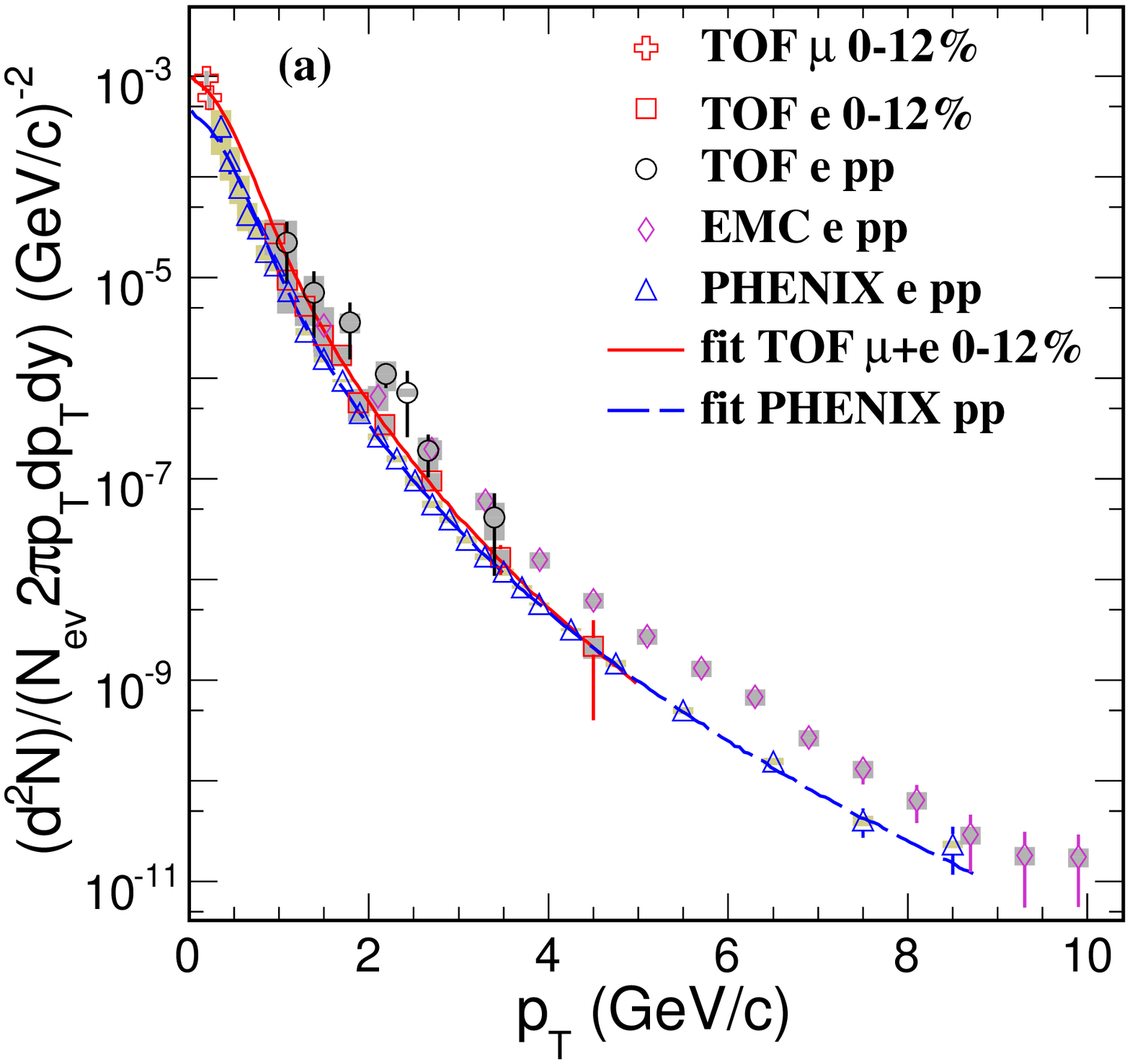}
\emn%
\bmn[c]{0.5\textwidth} \centering
\includegraphics[width=1.\textwidth]{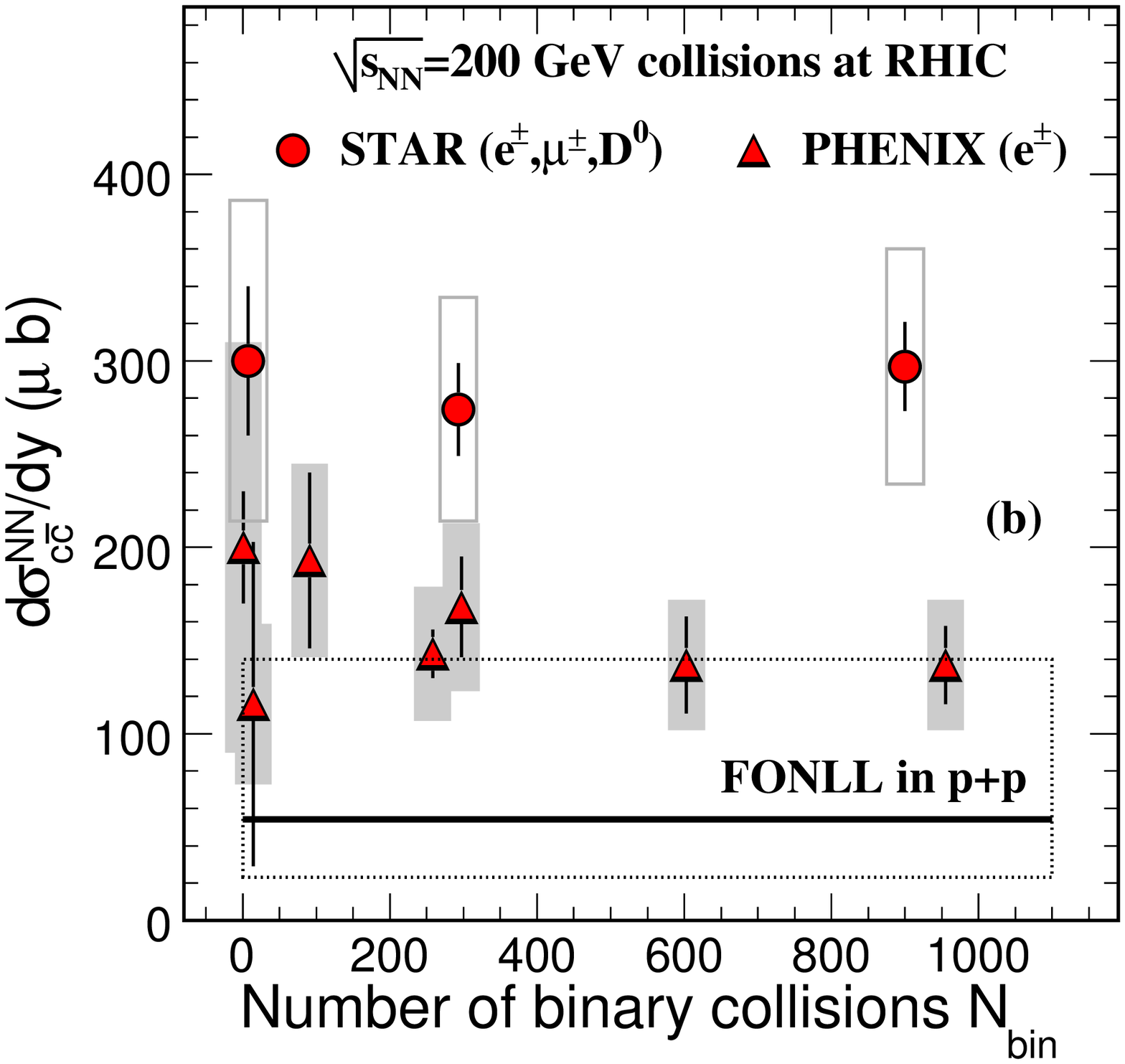}
\emn%
\caption[Spectra and cross section comparisons between STAR and
PHENIX]{Panel (a) Spectra comparisons between STAR and PHENIX.
Spectra measured in \pp\ collisions are scaled by \nbin\ to
compare with central results. Curves, from power law charmed
hadron distributions via form factor decay, are used to fit data.
Panel (b): The comparisons of charm cross section at mid-rapidity
in different centralities and different measurements. FONULL
calculation is shown as the solid line for comparison, and the
dashed box stands for the uncertainties of theory.}
\label{comESTARPHENIX} \ef

The FONLL and NLO pQCD
calculations~\cite{cacciari,vogtXsec,vogtCronin} underpredict the
STAR minbias data by a factor of
$5.1\pm0.48{(stat.)}\pm1.2(syst.)_{-3.1}^{+6.8}(theory)$. As
discussed previously~\cite{Xinthesis}, theoretical calculations
tuned parameters to match the low energy measurements and then
extrapolated to high energies, and the parameters are probably
energy dependent but not understood well yet. The charm cross
section will be precisely measured with a direct reconstruction of
charmed hadrons in the future, more detail will be discussed in
the outlook chapter.

The mid-rapidity charm cross-section per nucleon-nucleon collision
($d\sigma_{c\bar{c}}^{NN}/dy$) can be converted to the total charm
cross-section per nucleon-nucleon collision
($\sigma_{c\bar{c}}^{NN}$) following the method addressed in
Ref.~\cite{dAuCharm}. The $\sigma_{c\bar{c}}^{NN}$ is presented to
be $1.40\pm0.11(stat.)\pm0.39(sys.)$ mb in $0-12$\% central \AuAu\
and $1.29\pm0.12\pm0.36$ mb in minbias \AuAu\ collisions at
$\sqrt{s_{NN}}=200$ \gev. But the different widths of the rapidity
distributions from theoretical models will lead to substantial
systematical uncertainties. Fig.~\ref{xsecrapidity} shows the
charm cross sections as a function of rapidity distributions
measured from STAR and PHENIX, compared to theoretical
models~\cite{vogtPrivate,raufeisenXsec,HSD}. Here the systematical
uncertainties are dominant. PHENIX forward muon
measurement~\cite{phenixmu} with large errors, shown as the solid
triangle at rapidity $\sim1.6$, gives consistent result with STAR,
which is also significant higher than theoretical predictions.

\bf \centering\mbox{
\includegraphics[width=0.6\textwidth]{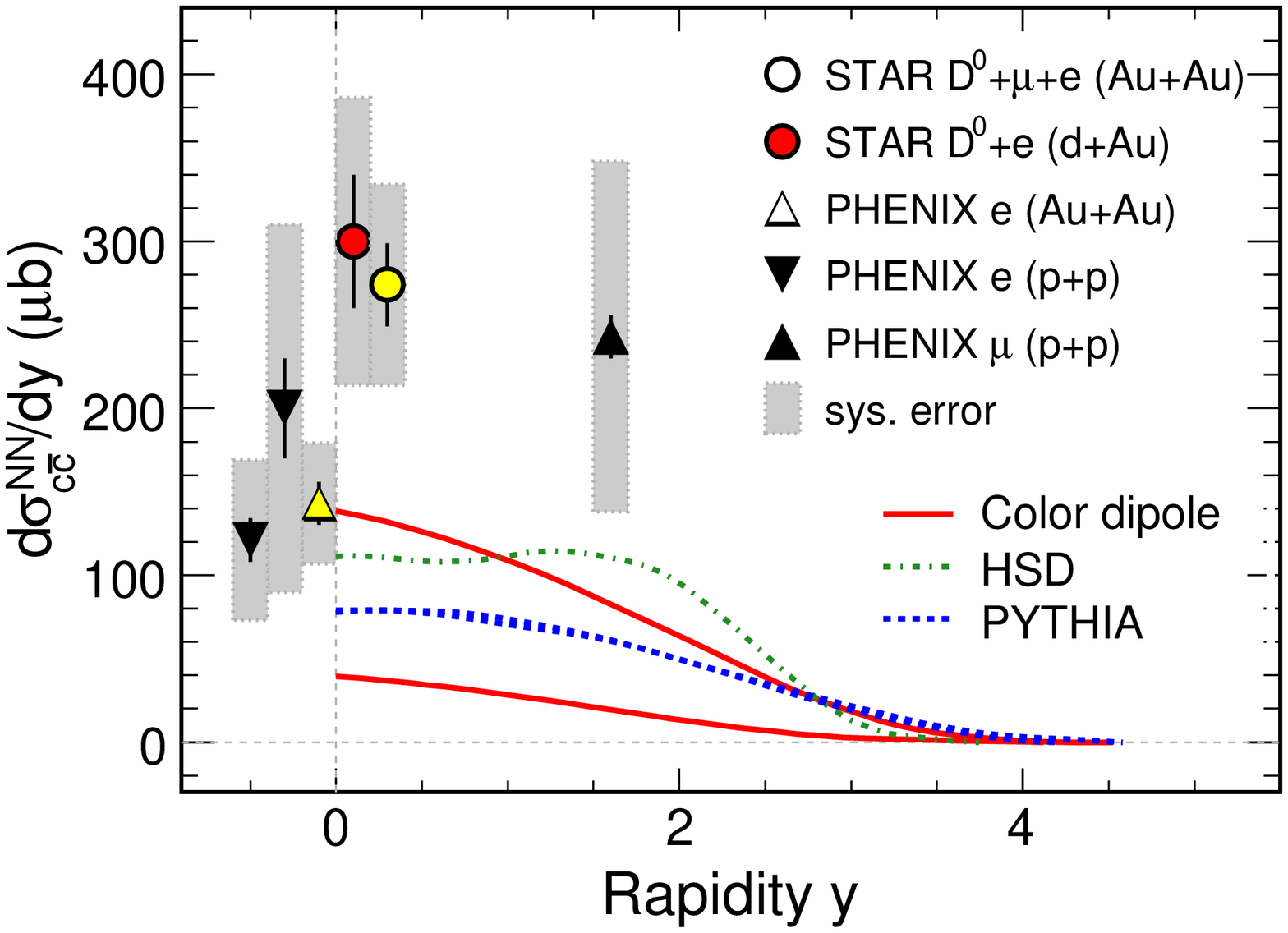}}
\caption[Charm cross sections as a function of rapidity
distributions]{Charm cross sections as a function of rapidity
distributions measured from STAR and PHENIX, compared to
theoretical models.} \label{xsecrapidity} \ef

\section{Identify bottom contribution in non-photonic electron
spectra and $v_2$ from \AuAu\ collisions at RHIC}

In the above discussions, the bottom contribution to the
non-photonic electron spectrum is neglected. We have discussed in
the introduction chapter that the separation of bottom and charm
contributions in current non-photonic electron measurements is
very difficult. There are large uncertainties in the model
predictions for charm and bottom production in high-energy nuclear
collisions. And many theoretical calculations with charm or with
both charm and bottom contributions describe the non-photonic
electron
suppression~\cite{DGLV06,Wicks05,RappRaa,Ivancoll,armeloss} and
$v_2$~\cite{kocharmflow,AmptCharmflow,vanHCharmflow}. Thus
identification of bottom from the non-photonic electron
measurements is crucial to better understand charm physics. In
this section, we try a fit to non-photonic electron spectrum and
estimate the bottom contributions. We also compare the $v_2$
distribution from simulation to the experimental data and estimate
the possible charm $v_2$.

\subsection{Fit to non-photonic electron spectrum and relative cross section ratio}

The non-photonic electron spectrum up to 10 \gevc\ has been
measured by STAR experiment in 200 \gev\ \pp\ collisions. The idea
is that we use the sum of electron spectra from both charm and
bottom decays in PYTHIA model~\cite{pythia} to fit the STAR \pp\
data~\cite{starcraa} to extract the fraction of the bottom
contribution. As we discussed previously, the D-mesons and their
decay electrons spectra from default PYTHIA parameters are
soft~\cite{ffcharm}, see Fig.~\ref{ffactorfig}. Thus a modified
Peterson Fragment Function (FF) and the high \pt\ tuned parameter
are used to make spectra harder to be comparable with the form
factor decays~\cite{XYLin04}.

Table~\ref{PYpar} lists the parameter initialization for PYTHIA 6.131:

\begin{table}[hbt]
\caption[PYTHIA parameters for heavy flavor decays]{PYTHIA
parameters for heavy flavor decays.} \label{PYpar} \vskip 0.1 in
\centering\begin{tabular}{|c|c|} \hline \hline
Parameter & Value \\
\hline
MSEL    & 4 (charm), 5 (bottom) \\
\hline
quark mass & $m_c=1.25$, $m_b=4.8$ (\gev) \\
\hline
parton dist. function & CTEQ5L \\
\hline
$Q^2$ scale & 4 \\
\hline
K factor & 3.5 \\
\hline
$\la K_t\ra$ & 1.5 \\
\hline
Peterson Frag. function & $\varepsilon=10^{-5}$ \\
\hline
high \pt\ tuned PARP(67) & 4 \\
     \hline \hline
\end{tabular}
\end{table}

Fig.~\ref{pyspecratio} (a) shows the \pt\ distributions of the
heavy flavor hadrons and their decay electrons from PYTHIA with
above parameters. The D-meson spectrum, shown as the hatched band,
is normalized to \be dN/dy = dN/dy(D^0)/\la N_{bin}\ra/R_{dAu}/R,
\ee where $dN/dy(D^0)=0.028\pm0.004\pm0.008$ measured in \dAu\
collisions~\cite{stardAucharm}. $\la N_{bin}\ra=7.5\pm0.4$ in
\dAu\ collisions. $R_{dAu}=1.3\pm0.3$~\cite{Xinthesis}. $R$ factor
stands for $D^0$ fraction in total charmed hadrons, the
fragmentation ratio $R(c\rightarrow D^0)\equiv
N_{D^0}/N_{c\bar{c}}=0.54\pm0.05$~\cite{PDG}. All these
normalization errors are propagated into the uncertainty band of
the D-meson spectrum. The curve in this band is the lower limit of
the D-meson spectrum in our simulation. Correspondingly, its decay
electron spectrum is shown as the solid band. The non-photonic
electron spectrum measured in \pp\ collisions at
STAR~\cite{starcraa} is shown as the open squares. The decay
electron band alone can describe the data, indicating that the
contribution of electrons from bottom decay could be very small.
In order to estimate the upper limit of bottom contribution, we
use the lower limit of the decay electron spectrum, shown as the
open circles. B-meson spectrum (solid curve) and its decay
electron spectrum (open triangles) are normalized by varying the
ratio of $\sigma_{b\bar{b}}/\sigma_{c\bar{c}}$. The summed
spectrum (solid circles) by combining the lower limit of
$D\rightarrow e$ and $B\rightarrow e$ is used to fit STAR data in
\pp\ collisions, and then the upper limit of $B\rightarrow e$
contribution will be extracted.

\bf \centering \bmn[c]{0.5\textwidth} \centering
\includegraphics[width=1.\textwidth]{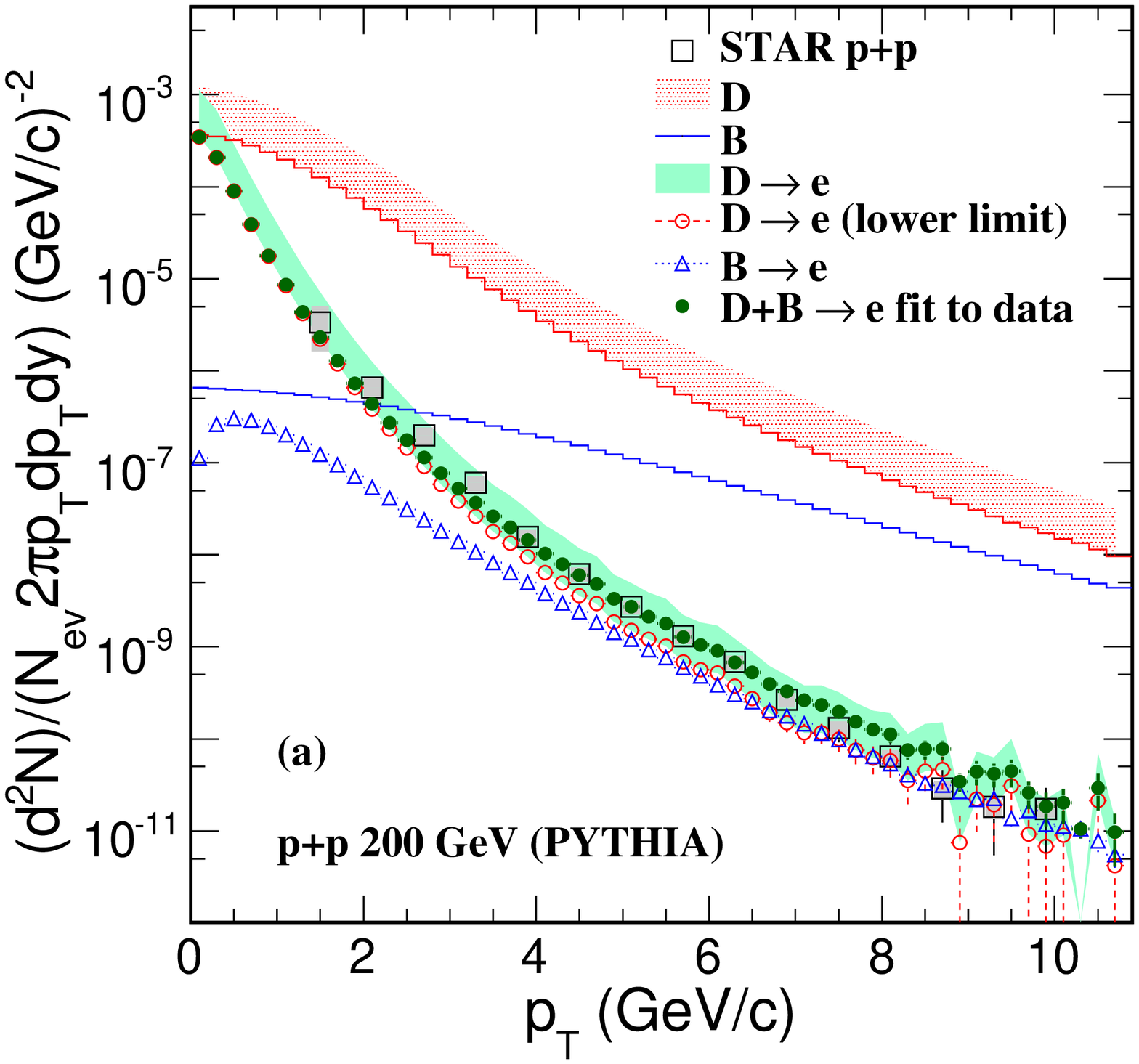}
\emn%
\bmn[c]{0.5\textwidth} \centering
\includegraphics[width=1.\textwidth]{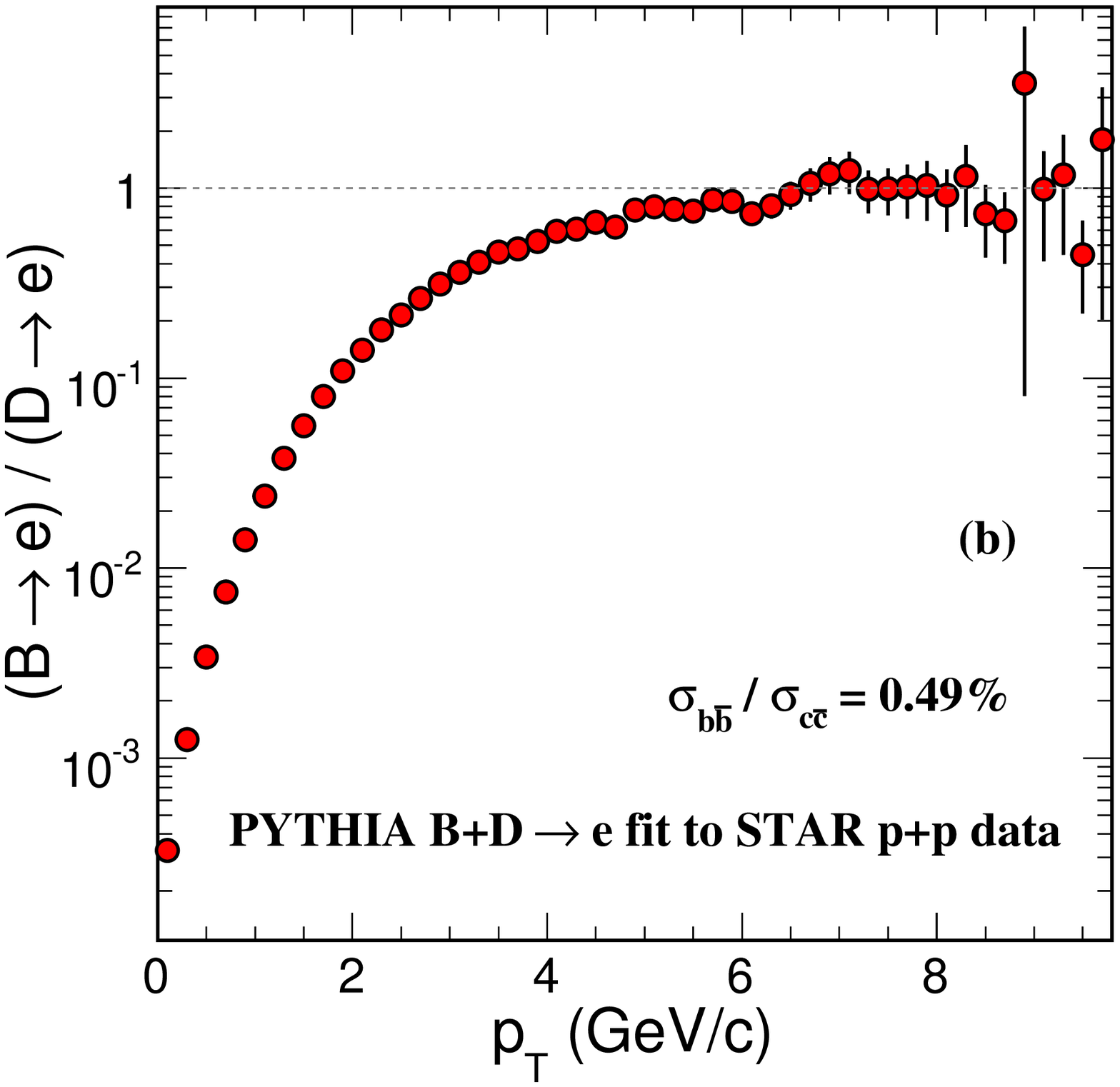}
\emn%
\caption[D-/B- mesons and their decay electron spectra from PYTHIA
and the relative spectra ratio]{Panel (a): D-/B- mesons and their
decay electron spectra from PYTHIA. The $B+D\rightarrow e$ fit to
STAR non-photonic electron data in \pp\ collisions. Panel (b): The
relative spectra ratio, upper limit of $B\rightarrow e$
contributions as a function of \pt.} \label{pyspecratio} \ef

Fig.~\ref{pyfitchi2} (a) shows the fit $\chi^2$ as a function of
the unique variable $\sigma_{b\bar{b}}/\sigma_{c\bar{c}}$. The
best fit with a minimum $\chi^2/ndf=16.6/14$ gives the upper limit
of the total cross section ratio as
$\sigma_{b\bar{b}}/\sigma_{c\bar{c}}\leq(0.49\pm0.09\pm0.09)\%$.
The first term of the errors is calculated from
$\chi^2=\chi_{min}^2+1$. The second term is from the 15\%
normalization error of the $dN/dy$ converted to total cross
sections due to the uncertainties of the model dependent rapidity
distributions~\cite{Xinthesis}. Fig.~\ref{pyfitchi2} (b) shows the
B-/D- mesons rapidity distributions from PYTHIA. The cross section
ratio from FONLL calculation is 0.18\%-2.6\%~\cite{cacciari}. The
upper limit is consistent with theory prediction.

\bf \centering \bmn[c]{0.5\textwidth} \centering
\includegraphics[width=1.\textwidth]{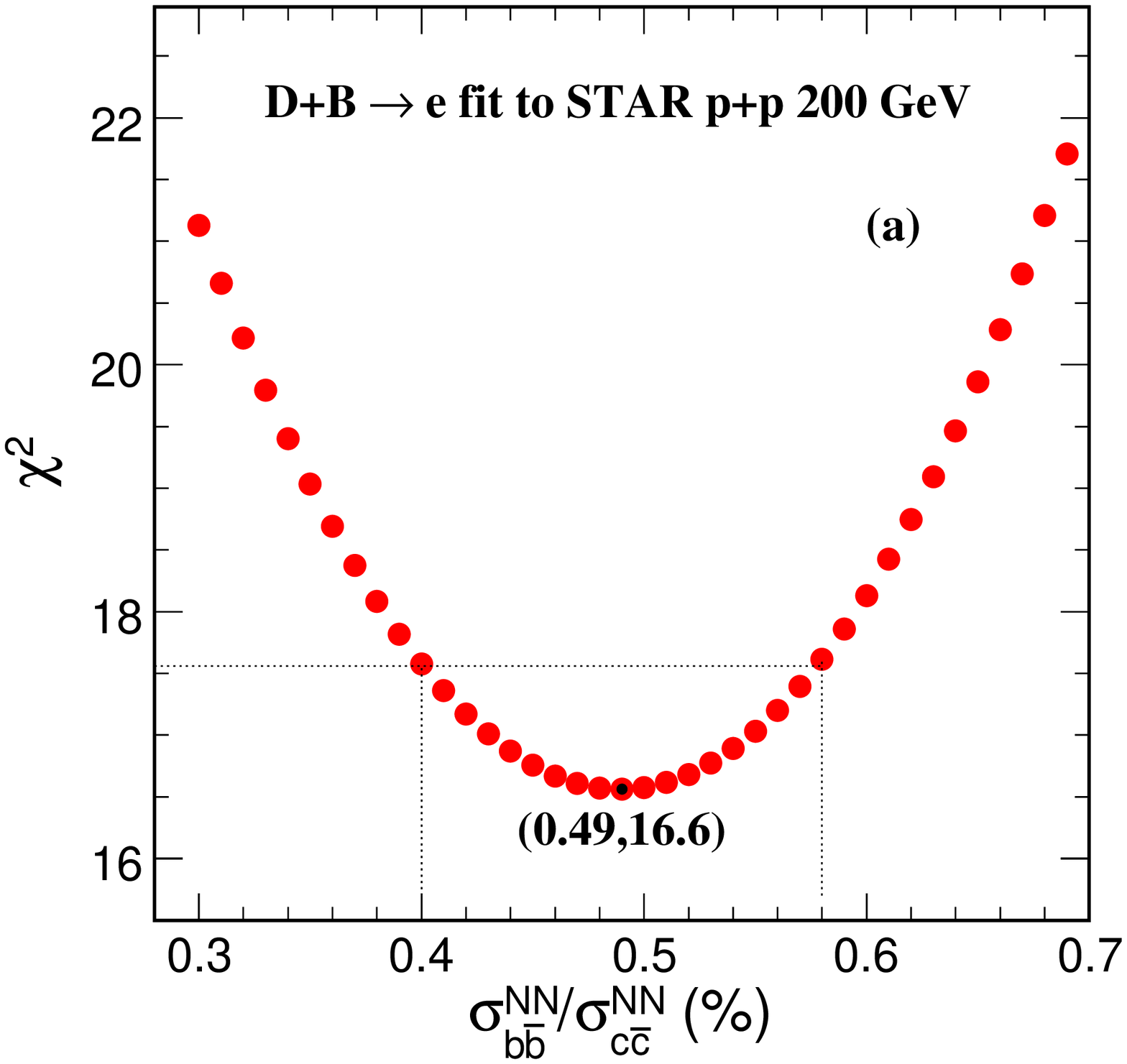}
\emn%
\bmn[c]{0.5\textwidth} \centering
\includegraphics[width=1.\textwidth]{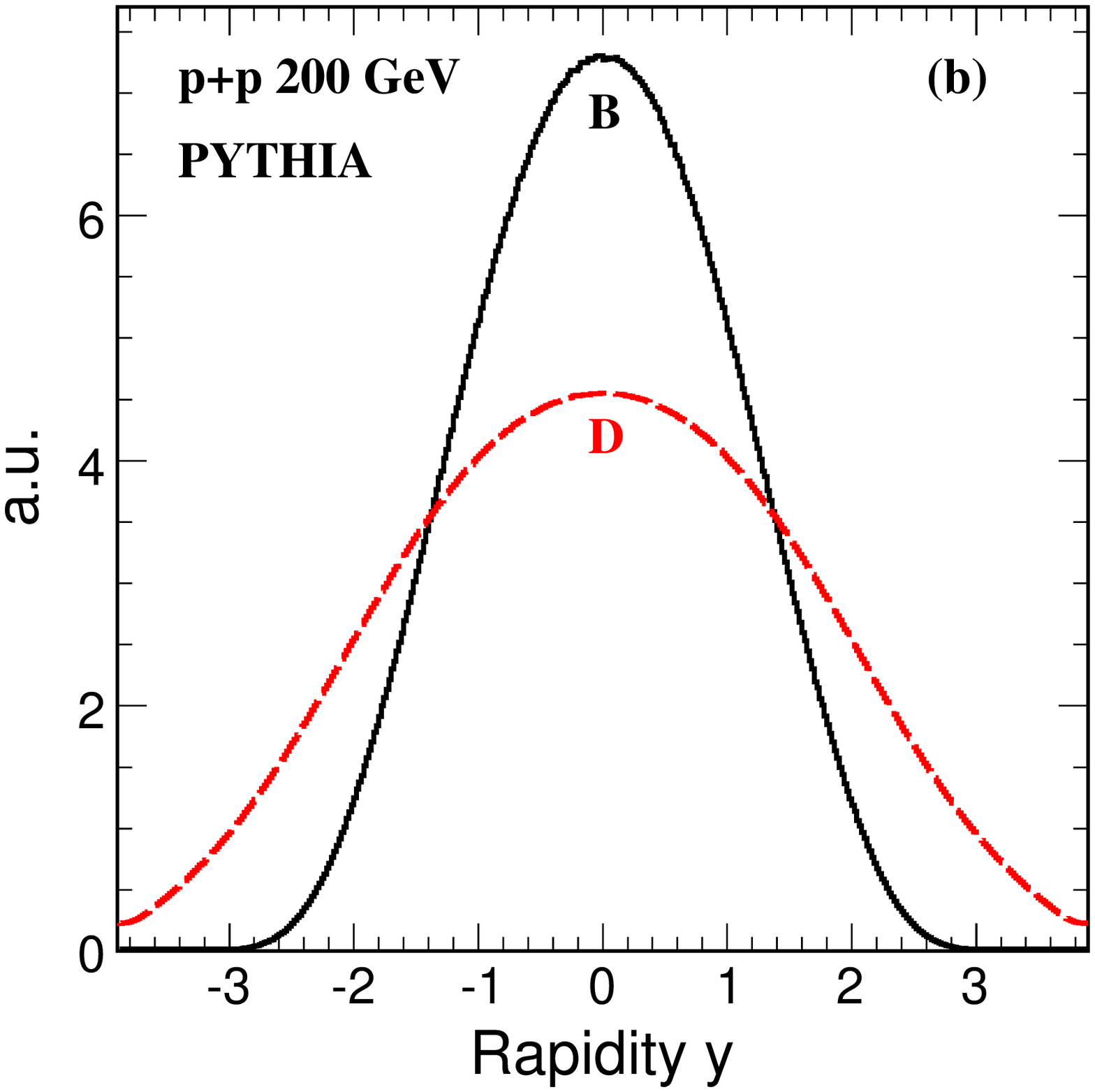}
\emn%
\caption[$\chi^2$ distribution from fitting to non-photonic
electron spectrum and rapidity distributions from PYTHIA]{Panel
(a): Fit $\chi^2$ as a function of
$\sigma_{b\bar{b}}/\sigma_{c\bar{c}}$. Straight lines is for the
$\chi^2=\chi_{min}^2+1$. Panel (b): B- (solid curve) /D- (dashed
curve) mesons rapidity distributions from PYTHIA.}
\label{pyfitchi2} \ef

The upper limit of $B\rightarrow e$ contributions as a function of
\pt\ is shown in Fig.~\ref{pyspecratio} (b). It is increasing and
becomes flat around 7 \gevc. The \pt\ crossing point, where the
bottom contribution is equal to charm, of electron spectra from
B,D decay is very sensitive to the cross section ratio, since at
high \pt, these electron spectra shapes are similar. From the
$B+D\rightarrow e$ fit to STAR \pp\ data, we estimate the crossing
point $p_T^c\geq7$ \gevc.

Table~\ref{crosspt} lists the crossing points of heavy flavor
decay electrons in several \pt\ bins.

\begin{table}[hbt]
\caption[Crossing points of heavy flavor decay electrons]{Crossing
points of heavy flavor decay electrons as a function of $p_T$.}
\label{crosspt} \vskip 0.1 in
\centering\begin{tabular}{|c|c|c|c|c|c|c|c|} \hline \hline
\pt\ (\gevc) & 2 & 3 & 4 & 5 & 6 & 7 ($p_T^c$) & 8 \\
\hline
$(B\rightarrow e)/(D\rightarrow e)\leq$ & 0.11 & 0.31 & 0.53 & 0.77 & 0.85 & 1.2 & 1.1 \\
\hline \hline
\end{tabular}
\end{table}

\subsection{Fit to non-photonic electron $v_2$}

Besides the non-photonic electron spectrum, the non-photonic
electron $v_2$ has also been measured in 200 \gev\ \AuAu\
collisions at RHIC~\cite{phenixv2}. In this measurement, bottom
contribution has not been separated, which can be studied by
comparing simulations and data. Since heavy flavor hadrons \pt\
distributions and $v_2$ are unknown, our simulations have to base
on the following assumptions:

\begin{description}
\item[--] The same relative $(B\rightarrow e)/(D\rightarrow e)$
ratio from \pp\ to \AuAu.
\item[--] Assume the B-/D- meson $v_2$
as the inputs for the simulation, here we assume three aspects:
\begin{itemize}
\item \uppercase \expandafter {\romannumeral 1}: B-/D- meson $v_2$
are similar as light meson $v_2$.
\item \uppercase \expandafter
{\romannumeral 2}: D-meson $v_2$ as light meson $v_2$ but B-meson
does not flow.
\item \uppercase \expandafter {\romannumeral 3}:
$B\rightarrow e$ contribution is neglected and D-meson $v_2$
decreases at $p_T>2$ \gevc.
\end{itemize}
\end{description}

Here heavy flavor baryons, $\Lambda_c$, $\Lambda_b$ are taken into
account as 10\% of total heavy flavor hadrons~\cite{PDG,pdgerr}.
Their $v_2$ are assumed to follow light baryon $v_2$. This baryon
contribution effect in this simulation is small.

We use the light meson $v_2$ curve from fitting experimental
data~\cite{minepiv2} as the input B/D $v_2$ distributions
(Assumption \uppercase \expandafter {\romannumeral 1}), see
Fig.~\ref{cbev2} (a). That means in each \pt\ bin, the B/D
$\Delta\phi$ distribution is initialized. The electron
$\Delta\phi$ distributions in each \pt\ bin will be obtained via
B/D decays in PYTHIA model. Then the electron $v_2$, shown in
Fig.~\ref{cbev2} (b), will be extracted by fitting the
$\Delta\phi$ distributions in each \pt\ bin.

\bf \centering\mbox{
\includegraphics[width=1.0\textwidth]{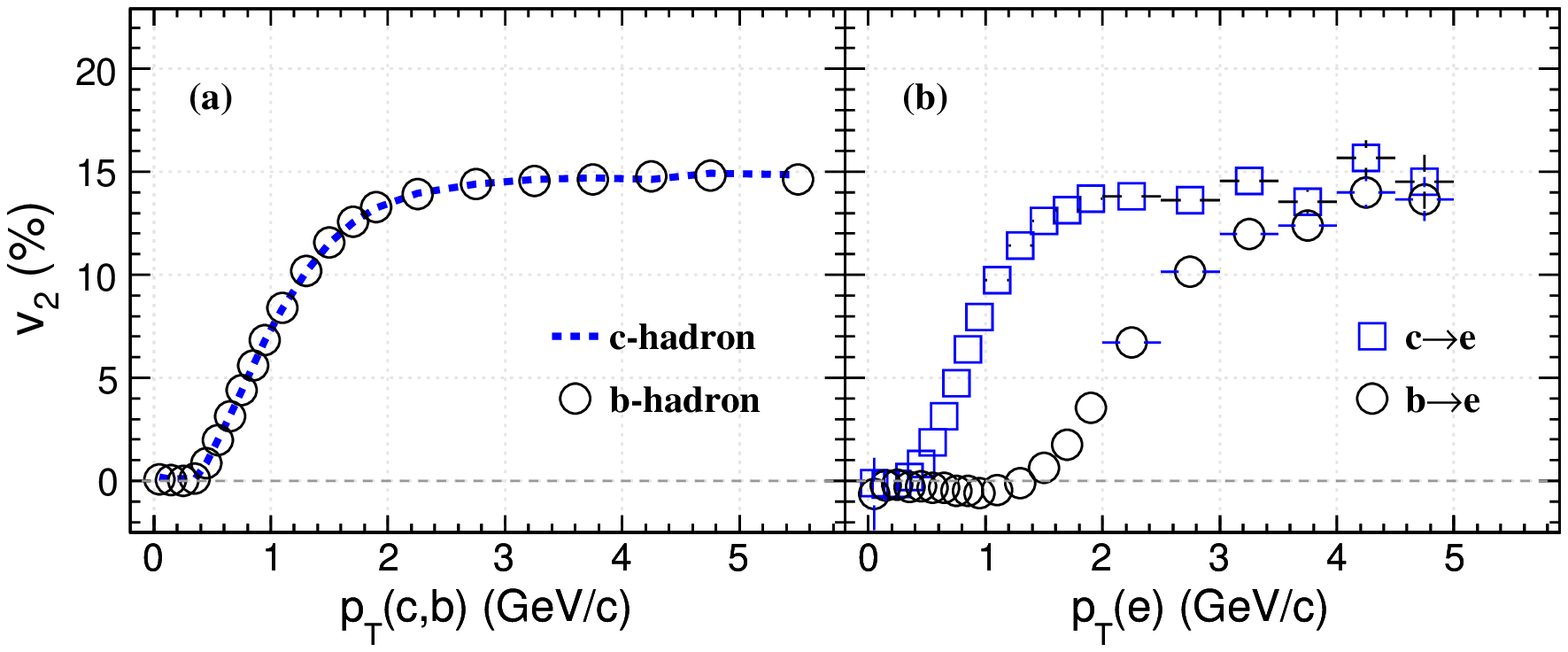}}
\caption[Decay electron $v_2$ from assumed B-/D- mesons
$v_2$]{Panel (a): Assumed B-meson $v_2$ (open circles) and D-meson
$v_2$ (dashed curve) as light meson $v^2$. Panel (b): Electron
$v_2$ from B-meson decays (open circles) and D-meson decays (open
squares).} \label{cbev2} \ef

Fig.~\ref{cbev2} shows the obvious mass effect: The B/D $v_2$ are
assumed as the same, but the decay electron $v_2$ can be very
different due to decay kinematics~\cite{YFHP06}. This is not
surprising, since we know B-meson is much heavier than D-meson and
light hadrons. The decay electrons can only have a small momentum
fraction of B-mesons. The momentum and angular correlations
between decay electrons and B-mesons are weak, especially at low
\pt. Therefore, at low \pt\ the decay electron $\phi$ angle will
almost randomly distribute. So we see the zero or negative $v_2$
for the electron from B-meson decays. But from previous study, we
know that bottom contribution below 3 \gevc\ is small, thus the
mass effect to the total electron $v_2$ is not significant.

Fig.~\ref{v2bnob} (a) shows the total electron $v_2$ from PYTHIA
simulation compared to data. The measured non-photonic electron
$v_2$ from PHENIX is shown as the triangles. The solid curve
(Assumption \uppercase \expandafter {\romannumeral 1}) is the sum
$v_2$ of the two decay electron $v_2$ distributions in
Fig.~\ref{cbev2} (b) by taking the relative ratio of
$(B\rightarrow e)/(D\rightarrow e)$ into account. It can not
describe the data. If we assume B-meson does not flow (Assumption
\uppercase \expandafter {\romannumeral 2}), the total decay
electron $v_2$ will become decreasing, shown as the band. The band
is corresponding to the
$\sigma_{b\bar{b}}/\sigma_{c\bar{c}}=(0.3-0.7)\%$ (The upper
limit, 0.49\%, is in between). It has better agreement with data,
but still higher. The decreasing of non-photonic electron $v_2$
could be due to $B\rightarrow e$ contribution and B-meson $v_2$
could be very small. But below 3 \gevc, $B\rightarrow e$
contribution is not significant. That indicates D-meson $v_2$
should be smaller than light meson $v_2$ and start decreasing at
higher \pt\ ($>2$ \gevc).

\bf \centering\mbox{
\includegraphics[width=1.0\textwidth]{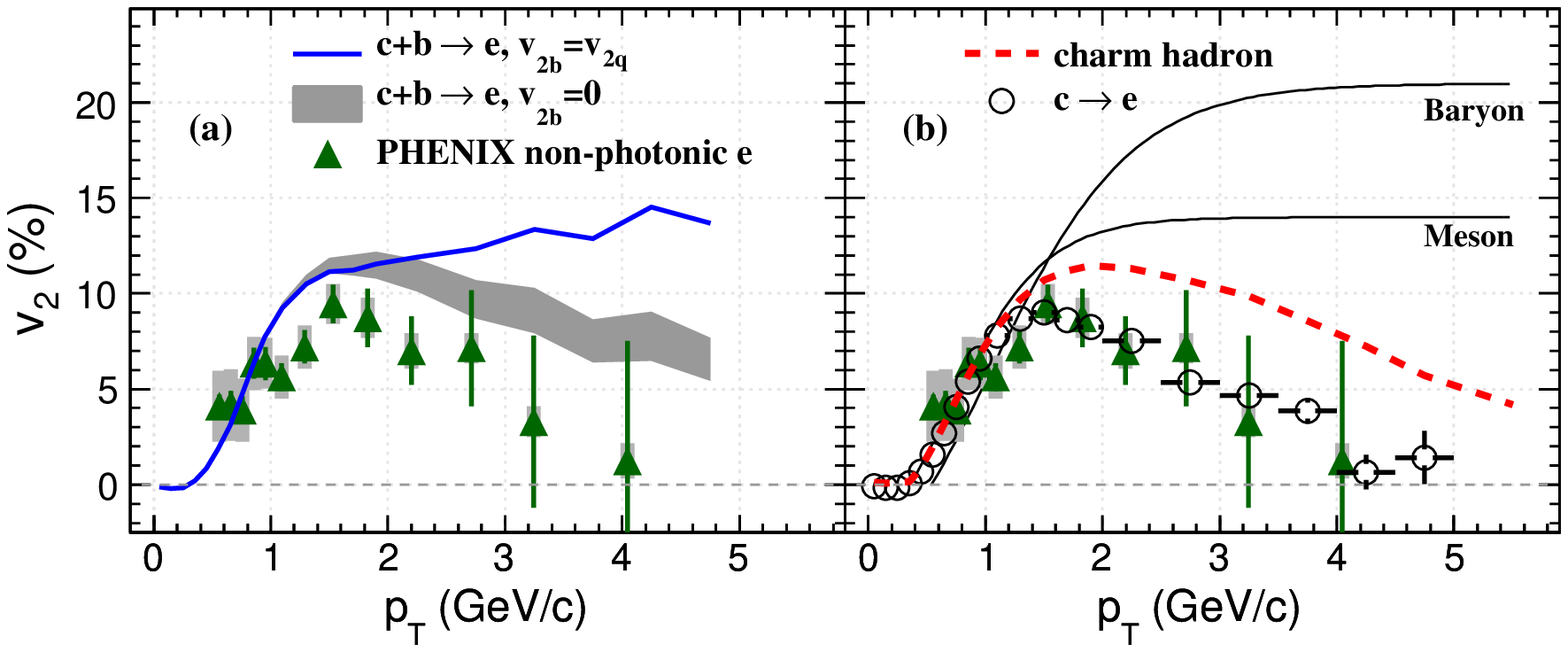}}
\caption[The total electron $v_2$ from PYTHIA simulation compared
to data]{Panel (a): The total electron $v_2$ from PYTHIA
simulation assuming that bottom flows (solid curve) and bottom
does not flow (band) compared to data. Panel (b): The total
electron $v_2$ from PYTHIA simulation fit to data and the
estimated D-meson $v_2$.} \label{v2bnob} \ef

So ignoring $B\rightarrow e$ contribution, we try to speculate the
D-meson $v_2$ by fitting the data using decay electron $v_2$
(Assumption \uppercase \expandafter {\romannumeral 3}). In
Fig.~\ref{v2bnob} (b), the best fit of the decay electron $v_2$ is
shown as the open circles. The estimated D-meson $v_2$ is shown as
the dashed curve, which is smaller than light meson $v_2$ above 1
\gevc\ and start decreasing above 2 \gevc. This most possible
D-meson $v_2$ distribution shows that at $p_T<3$ \gevc, where the
bottom contribution is negligible, D-meson has large $v_2$,
indicating that charm strongly flows in high dense medium, which
could be the evidence of light flavor thermalization in QGP
created at RHIC energy.

\chapter{Detector upgrades and outlook}

\section{STAR detector upgrades: TOF and HFT}

The experimental physics results are inevitably limited by the
detector resolutions, luminosities, acceptances and electronics
{\em etc.} Physics developments require the upgrade of the
detectors and electronics. The STAR Collaboration has proposed
some important sub-detector upgrades. The full barrel
Time-Of-Flight (TOF) detector~\cite{tofproposal} and the Heavy
Flavor Tracker (HFT)~\cite{hft} are two of them. The other
detector upgrades like the Intermediate Silicon Tracker (IST) and
the Forward GEM Tracker (FGT) {\em etc}, will not be introduced in
this thesis.

The barrel TOF detector based on recently developed Multi-gap
Resistive Plate Chamber (MRPC) technology will surround the outer
edge of the TPC, covering $-1<\eta<1$ and $\sim2\pi$ in azimuth.
The TOF system can achieve the required timing resolution $<100$
ps and the particle detecting efficiency $>95\%$. This detector
will significantly extend the reach of the STAR scientific
program, doubling the percentage of kaons and protons for which
particle identification is possible to more than 95\% of all those
produced within the MRPC-TOF acceptance. Combined with existing
STAR detectors, the barrel TOF detector will allow STAR to extract
the maximum amount of information and statistics for heavy flavor,
dilepton and resonances physics, such as the single electron, muon
\pt\ spectra, the single electron \vv, $J/\Psi$, $\rho$, $\omega$,
vector mesons, {\em etc}.

When combined with a possible future vertex detector upgrade, the
proposed TOF detector will also reduce the integrated luminosity
needed to measure a statistically robust sample of $D^0$, $D^+$,
$D_s^+$ mesons by approximately an order of magnitude, enabling
STAR to make systematic studies of charm thermalization and $D^0$
meson flow.

In addition, with the barrel TOF detector, it becomes possible to
separate the bottom contribution in non-photonic electron
measurement experimentally and charm physics will be more clearly
understood.

The expectant vertex detector HFT brings extremely high precision
tracking capabilities to STAR with a resolution of 10 $\mu m$ at
the first layer of the detector, over a large pseudo-rapidity
range, and with complete azimuthal angular coverage $\Phi=2\pi$.
The two layers of the Heavy Flavor Tracker (HFT), which are
composed of monolithic CMOS pixel detectors using 30 $\mu
m\times30\mu m$ square pixels, will be placed closed to beam pipe
in STAR at radii of 2.5 cm and 7 cm, respectively.

The HFT is designed to measure displaced vertices that are
displaced 100 micros, or less, from the primary vertex. Therefore,
the neutral and charged particles with very short lifetimes can be
distinguished from the primary particles which originate at the
collision vertex. The addition of the HFT will extend STAR's
unique capabilities even further by providing particle
identification for hadrons containing charm and beauty and
electrons decaying from charm and beauty hadrons. Thus, the HFT is
the enabling technology for making direct charm and beauty
measurements at STAR.

\section{Direct measurements of charmed hadrons with HFT}

The Monte Carlo events in \AuAu\ collisions at $\sNN=200$ GeV are
generated by using HIJING model. The standard GSTAR simulation
package with modifications of the new detector design is used for
the material geometry configurations. The new STAR tracking
software package ITTF with HFT layers added is used for tracking.

The physical interactiona between particles and the material of
the detector are simulated using the STAR implementation of the
GEANT simulation package. This package is used in STAR and is a
standard analysis tool which includes a detailed understanding of
the TPC response function; including dead areas and realistic
detector resolutions and responses. Realistic detector resolutions
are used to smear the perfect position information, and the
resulting simulated hits are used in tracking. For the SSD, the
hits were smeared by 20 $\mu m$ in and 750 $\mu m$ in z, in
agreement with the SSD specifications. For the HFT, the hits were
smeared by 6 $\mu m$ in both $\Phi$ and z.

In this section, we will focus on the charmed hadron
reconstruction. The optimization of the HFT hit finding accuracy,
vertex reconstruction performance, single track efficiency and
ghost rate {\em etc}, will not be discussed. The details can be
found in Ref.~\cite{hft}.

To demonstrate the power of the HFT, we have simulated several
specific charm hadron decay channels, including $D^0\rightarrow
K^-\pi^+$ and $D^+\rightarrow K^-\pi^+\pi^+$. Table~\ref{Dbr}
displays some of the properties of these channels.

\begin{table}[hbt]
\caption[Open charm hadron properties]{Open charm hadron
properties.} \label{Dbr} \vskip 0.1 in
\centering\begin{tabular}{|c|c|c|c|} \hline
\hline Particle & Daughters & c$_{\tau} ~(\mu m)$ & Mass (GeV) \\
\hline $D^0$ & $K^-\pi^+$ (3.8\%) & 122.9 & 1.8645 \\
\hline $D^+$ & $K^-\pi^+\pi^+$ (9.5\%) & 311.8 & 1.8693 \\
\hline $D_s^+$ & $\Phi\pi^+$ (4.4\%) & 149.9 & 1.9682 \\
       & $\pi^+\pi^+\pi^-$ (1.2\%) & & \\
\hline $\Lambda_c$ & $pK^-\pi^+$ (5.0\%) & 59.9 & 2.2865 \\
\hline \hline
\end{tabular}
\end{table}

Signal and background events are generated separately. The signal
consists of one $D^0$ or $D^+$ per event. The transverse momentum
distribution of the charmed hadrons follows a Boltzman
distribution which reproduces the $\la p_T\ra$ of D-mesons as
measured by STAR in \dAu\ collisions at $\sNN=200$
GeV~\cite{stardAucharm} and the rapidity distribution suggested by
perturbative QCD calculations applying the program code
Pythia~\cite{pythia}. The background is simulated using the MevSim
event generator parameterized to reproduce the experimentally
measured particle multiplicities in \AuAu\ collisions at
$\sNN=200$ GeV. Our parameterization is accurate for particles
below 3 \gevc. It may underestimate the background above this
momentum. The distributions of reconstructed D-meson signal and
background were scaled to match the expected D-meson production
per central \AuAu\ collision. Also, the higher track
reconstruction efficiency in single (signal) events compared to
central \AuAu\ collisions (background) was taken into account.

The heavy flavor tracker is designed to allow us to directly
reconstruct mesons containing charm quarks. If this is true, we
should see differences between the charm meson daughter tracks and
the background and the primary tracks in several important
variables. The DCA between the tracks and the vertex is an
important example of this; see Fig.~\ref{DCAd0dp}. The
distributions of reconstructed $D^0$ and $^+$ daughters is
compared to the primary track background. The charm meson tracks
clearly have a broader distribution, driven by the decay length of
the charm mesons. Cutting on the track DCA, then, will improve the
signal to noise in the analyses.

\bf \centering \bmn[b]{0.6\textwidth} \centering
\includegraphics[width=1.0\textwidth]{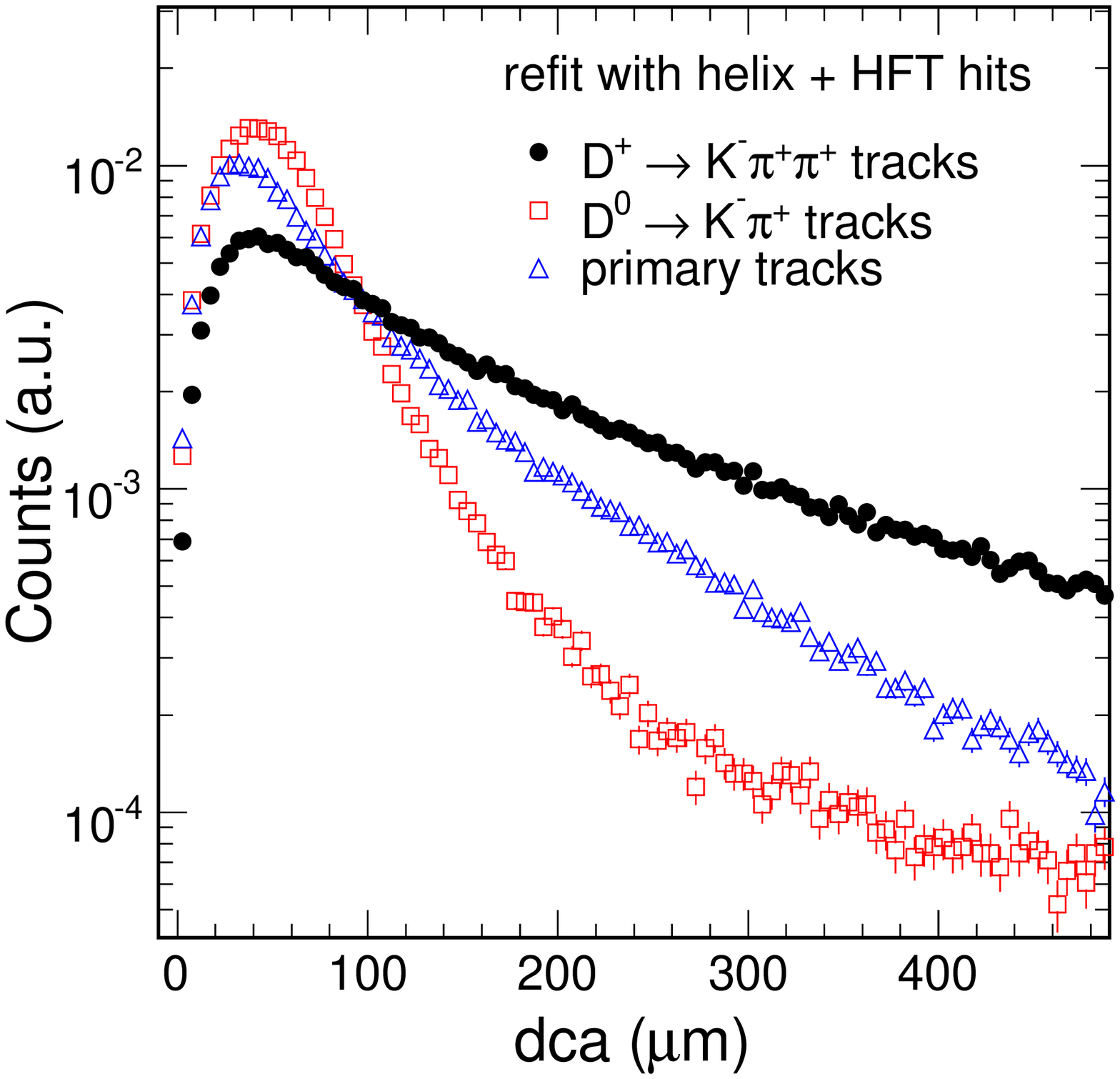}
\emn\\[10pt]
\caption[The DCA distributions for $D^0$ and $D^+$ reconstructed
by HFT]{The DCA distributions for $D^0$ (triangles) and $D^+$
(circles) reconstructed by HFT. The results of the dca
distribution for primary tracks are shown in squares for
comparison.} \label{DCAd0dp} \ef

The following selection criteria were also used to separate the
D-meson signal from background.

\begin{itemize}
\item The decay length l: the distance between the primary vertex
and the D-meson decay vertex (V0). \item The DCA between the
daughter tracks and the reconstructed decay vertex. \item Momentum
angle correlation cos($\theta$): $\theta$ is defined as the angle
between the D-meson momentum (vector sum of the two daughter
momenta) and the vector joining the primary vertex to the D-meson
decay vertex.
\end{itemize}

The topological reconstruction of $D^+\rightarrow
K^-\pi_1^+\pi_2^+$ is introduced here as an example. The better
signal to background can be obtained for the reconstruction of
$D^0\rightarrow K^-\pi^+$, since it is two-body decay. The HFT is
designed as close to collision vertex as possible, so that it
gives good primary vertex performance ($\sim10\mu m$), which is
good enough for the secondary vertex reconstruction. And the
reconstructed TPC helixes are required to hit on the two layers of
HFT. After refit the helixes with two HFT hits, the new helixes
were obtained. The new helixes were used to calculate those
variables mentioned above. Since it is three-body decay, $D^+$
decay vertex is estimated by calculating the average of three
displaced points of $K^-\pi_1^+$, $K^-\pi_2^+$ and
$\pi_1^+\pi_2^+$.

The reconstruction of charmed baryon $\Lambda_c$ is similar as
$D^+$. But due to the smaller c$_{\tau}$ and different charm
fragmentation ratios, the cuts for $\Lambda_c$, $D^0$ and $D^+$
should be different. The cuts are optimized to obtain good signal
to background ratio. Table~\ref{chadroncut} lists the cuts for the
charmed hadrons.

\begin{table}[hbt]
\caption[Topological cuts for open charm hadrons]{Topological cuts
for open charm hadrons.} \label{chadroncut} \vskip 0.1 in
\centering\begin{tabular}{|c|c|c|c|} \hline
\hline Cuts & $D^0$ & $D^+$ & $\Lambda_c$ \\
\hline nFitPts $>$ & 15 & 15 & 15 \\
\hline $|\eta|$ $<$ & 1.0 & 1.0 & 1.0 \\
\hline DCA (global) $\geq$ &  & 100 $\mu m$ & 35 $\mu m$ \\
\hline DCA (V0) $\leq$ & 35 $\mu m$ & 100 $\mu m$ & 40 $\mu m$ \\
\hline decay length $\geq$ & 150 $\mu m$ & 150 $\mu m$ & 50 $\mu m$ \\
\hline cos($\theta$) $>$ & 0.996 & 0.85 & 0.92 \\
\hline \hline
\end{tabular}
\end{table}

\bf \centering \bmn[b]{1.0\textwidth} \centering
\includegraphics[width=1.0\textwidth]{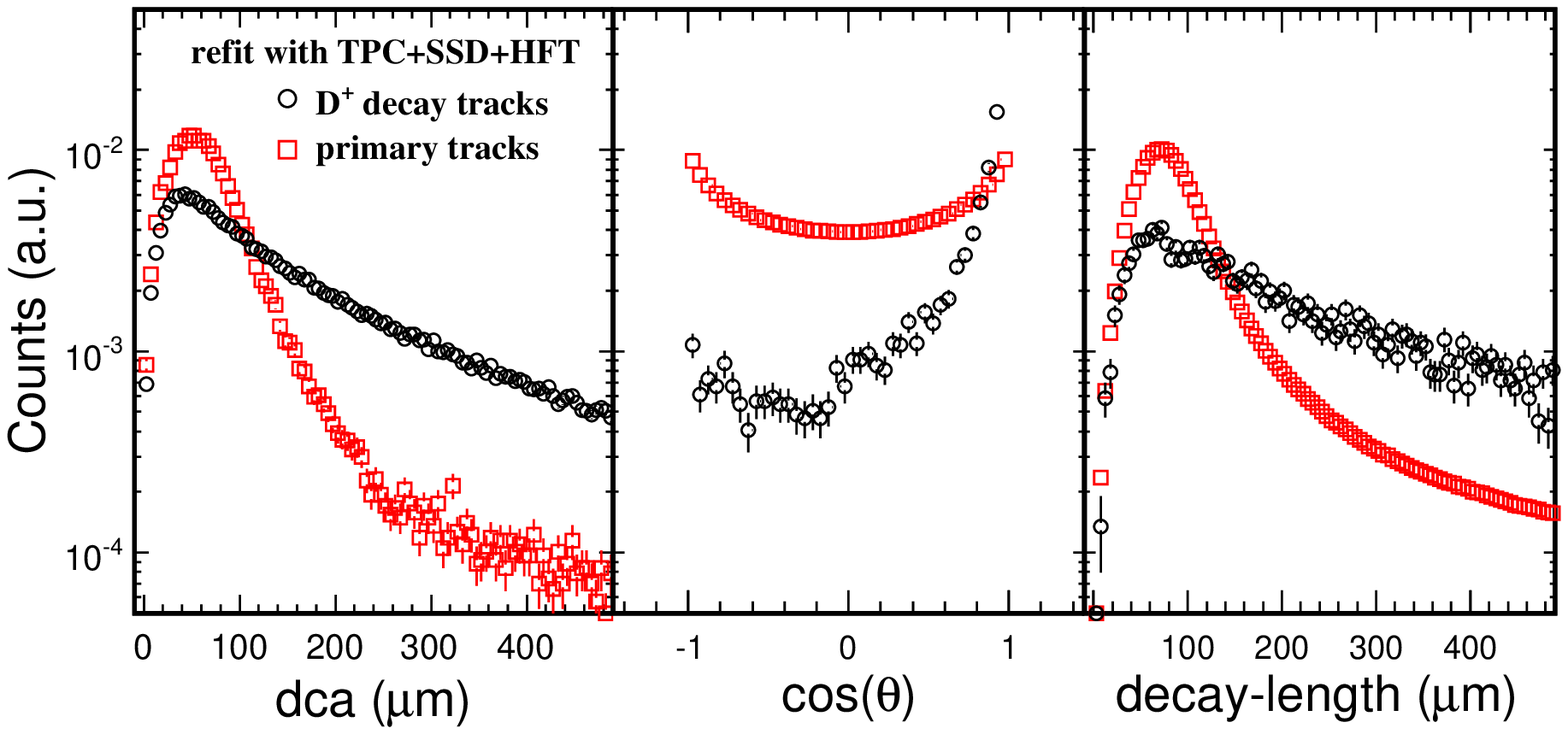}
\emn\\[10pt]
\caption[The decay-length, cos($\theta$) and DCA distributions for
$D^+$ reconstructed by HFT]{The $D^+$ daughter tracks
decay-length, dca, and cos($\theta$) distributions (circles). The
same distributions for primary tracks are shown as squares.}
\label{dpcut} \ef

The dca, cos($\theta$), and decay-length distributions for both
primary tracks (open-squares) and $D^+$ decayed tracks are shown
in Fig.~\ref{chadroncut}. Clearly the decayed tracks are well
separated from the primary track. For $D^+$ reconstruction, a
slightly different method was used compared to that of the $D^0$.
The 'signal event' and 'background event' were mixed together. The
distribution of the invariant mass from a K and two $\pi$ tracks
is then formed. The number of tracks used in the background events
is consistent with the top 10\% central \AuAu\ collisions at RHIC.
The resulting invariant mass distributions for several \pt\ bins
are shown in Fig.~\ref{dpmass}. For the \pt\ bins studied so far,
the significance $S/\sqrt{S+B}$ is better than 3.

\bf \centering \bmn[b]{1.0\textwidth} \centering
\includegraphics[width=1.0\textwidth]{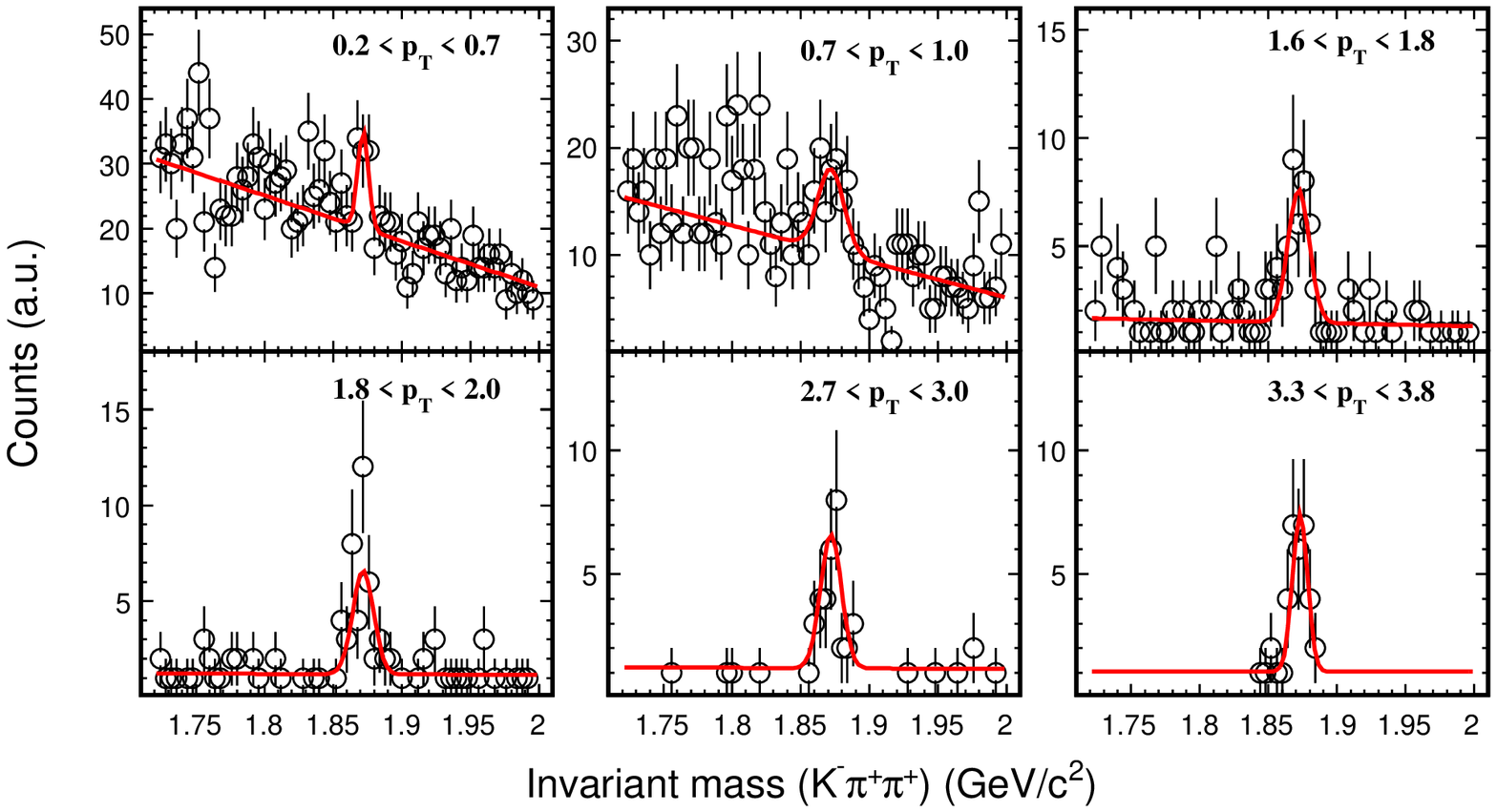}
\emn\\[10pt]
\caption[Invariant mass distributions for $D^+$ reconstructed by
HFT]{Invariant mass distributions for $D^+$ for several \pt\
windows. The lines are a polynomial (up to 2nd order) + Gaussian
fit.} \label{dpmass} \ef

\section{Bottom separation in heavy flavor measurements}

With the barrel TOF detector and vertex detector HFT, the D-meson
\pt\ spectrum and non-photonic electron \pt\ spectrum will be
measured precisely. The D-meson semileptonic decayed electron
spectrum will be known based on models via decay kinematics.
Comparison between the D-meson decayed electron and non-photonic
electron \pt\ distributions will tell us the information of the
bottom contribution in non-photonic electron measurements. This
will also be tested from the measurements of D-meson \vv\ and
non-photonic electron \vv\ when the TOF detector and HFT are
available.

But before the barrel TOF detector and HFT installation in STAR,
the separation of B-meson and D-meson due to their different life
time (decay length) is proposed to be a good try.
Fig.~\ref{hfdecl} shows the different decay length distributions
of D-meson and B-meson.

\bf \centering \bmn[c]{0.5\textwidth} \centering
\includegraphics[width=1.\textwidth]{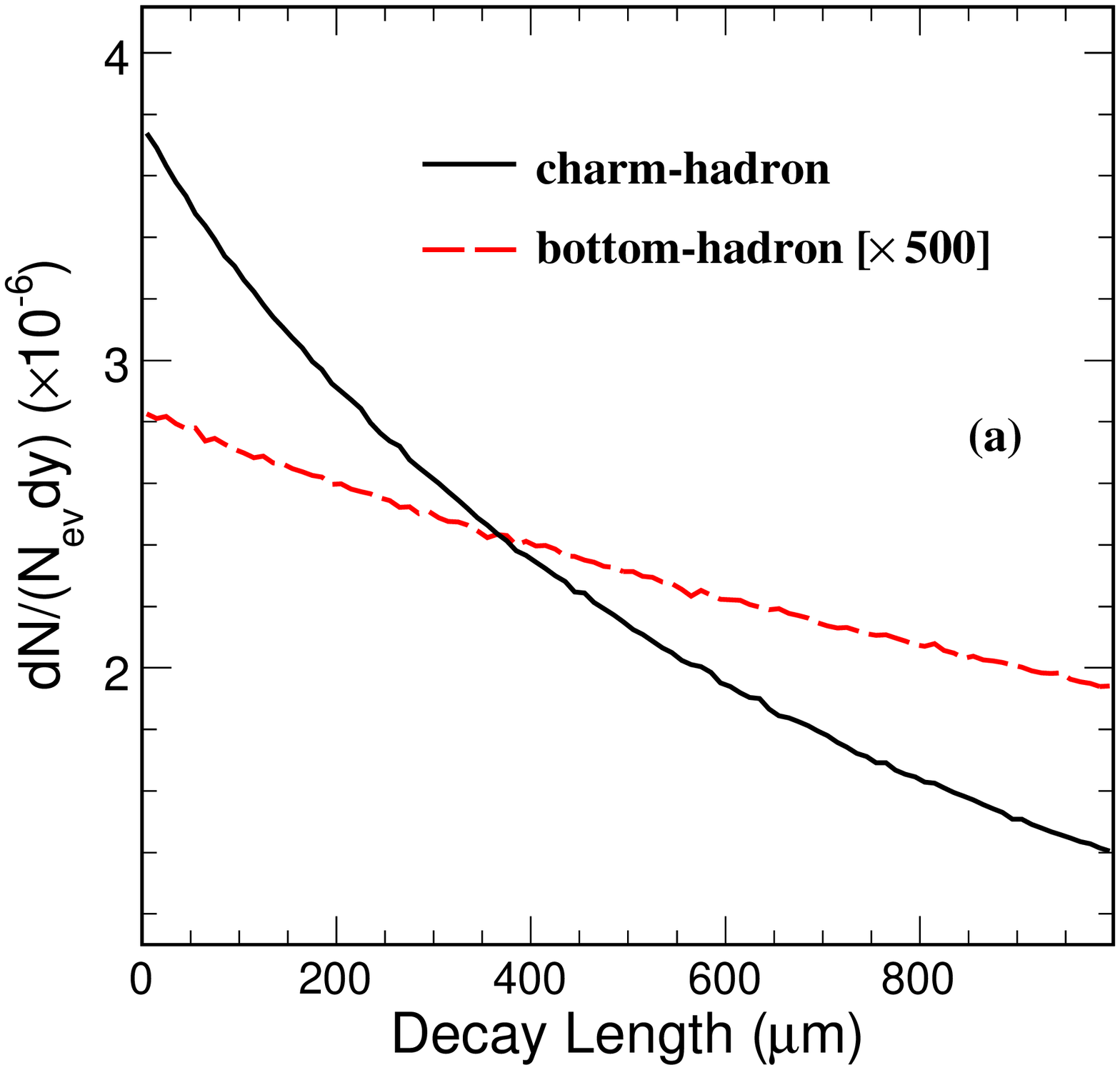}
\emn%
\bmn[c]{0.5\textwidth} \centering
\includegraphics[width=1.\textwidth]{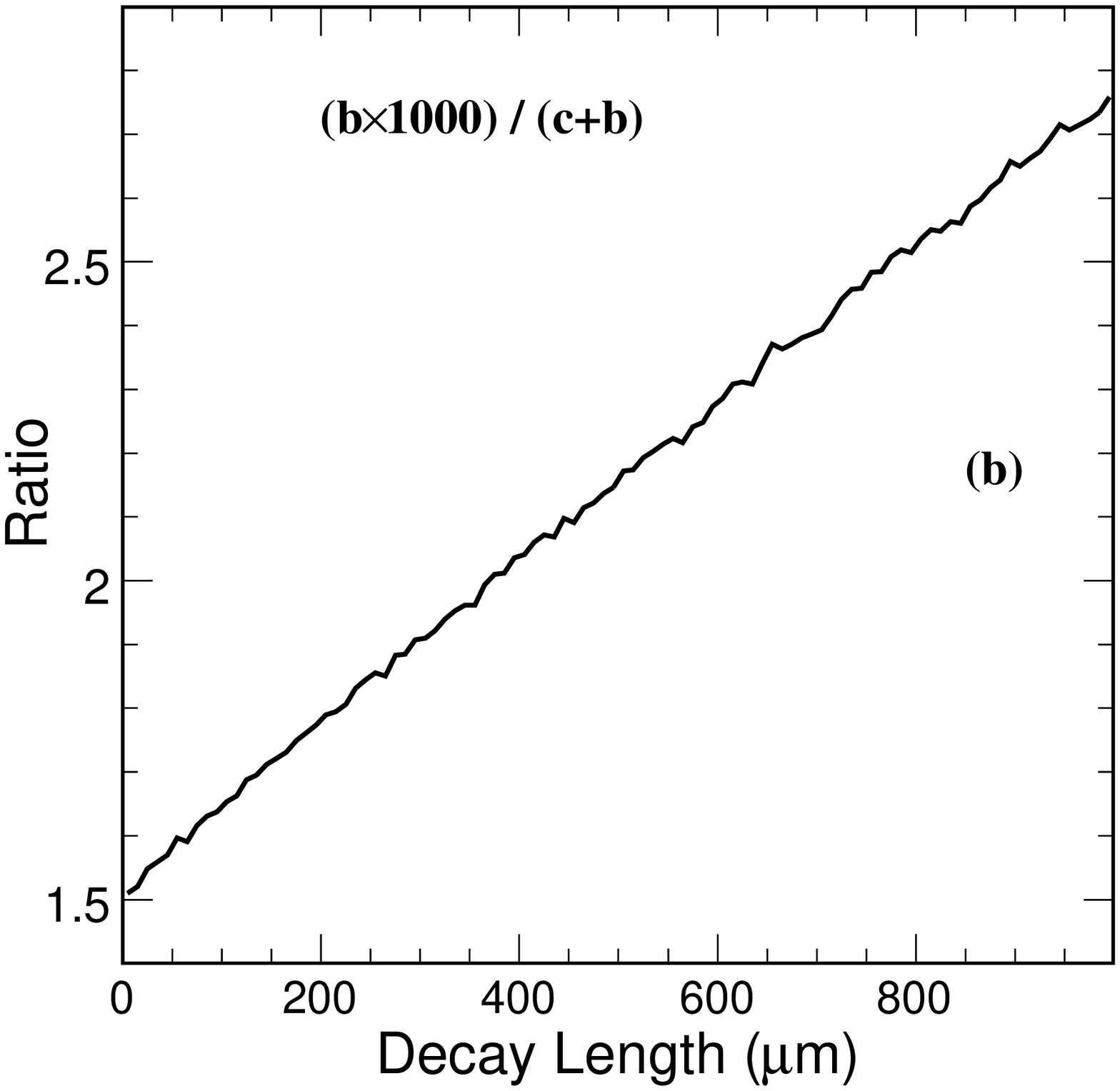}
\emn%
\caption[D-/B- mesons decay length distributions]{Panel (a): D-/B-
mesons decay length distributions. Panel (b): The fraction of
B-meson in the total decay length of (B+D)-meson.} \label{hfdecl}
\ef

D-meson decay vertex will be closer to the collision vertex than
B-meson. Currently, the existing vertex detector SVT and
intermediate detector SSD will be used to help the reconstruction
of the B-meson decay vertex.

\appendix

\chapter{Presentations and publication list}
\vskip 0.1 in \Large{$\underline{\textbf{Presentations}}$}
\normalsize
\begin{itemize}
\item {\em Electron Spectra in Au+Au Collisions at 200 GeV} (poster)\\
2005 Gordon Research Conference on Nuclear Chemistry, Bates
College, Lewiston, ME, USA, 07/10/2005 - 07/15/2005.

\item {\em Charm Hadron Reconstructions with HFT} \\
STAR Upgrades Workshop, Brookhaven, New York, USA, 12/01/2005 -
12/07/2005.

\item {\em Scaling of Charm Integrated Cross Section and
Modification of its $p_T$ Spectra in 200 GeV Au+Au Collisions at
STAR} \\
2006 International Conference of Strangeness in Quark Matter,
UCLA, California, USA, 03/25/2006 - 04/01/2006.

\item {\em Charm Cross Sections and Collectivity from its
Semileptonic Decay at RHIC} \\
2nd International Conference on Hard and Electromagnetic Probes of
High-Energy Nuclear Collisions, Asilomar, California, USA,
06/09/2006 - 06/16/2006.

\item {\em Identify Bottom Contribution in Non-photonic Electron
Spectra and $v_2$ from p+p and Au+Au Collisions at RHIC} \\
23rd Winter Workshop on nuclear Dynamics, Big Sky, Montana, USA,
02/11/2007 - 02/18/2007.

\end{itemize}

\vskip 2.0 in \Large{$\underline{\textbf{Publication List}}$}
\normalsize
\begin{itemize}

\item Y.~F. Zhang (for STAR collaboration), ``Open charm
production in $\sqrt(s_{NN})$ = 200 GeV Au+Au collisions.", {\em
J. Phys. G: Nucl. Part. Phys.}, {\bfseries 32}, S529-S532, 2006.

\item Z. Xu {\em et al.}, ``Measurement of charm flow with the
STAR Heavy Flavor Tracker.", {\em J. Phys. G: Nucl. Part. Phys.},
{\bfseries 32}, S571-S574, 2006.

\item Y.~F. Zhang, S.Esumi, H.Huang, Y.Miake, S.Sakai and N.Xu,
``Identifying bottom contribution in non-photonic electron spectra
and $v_2$ from Au+Au collisions at RHIC.", {\em Nucl. Phys. A},
{\bfseries 783}, 489-492, 2007.

\item H.~D. Liu, Y.~F. Zhang, C. Zhong and Z. Xu, ``Extracting the
charm cross-section from its semileptonic decay at RHIC." , {\em
Phys. Lett. B}, {\bfseries 639}, 441, 2006.

\item Y.~F. Zhang, S.~L. Huang, Z.~P. Zhang, J. Wu, ``Study on
"soft" and "hard" interactions in $pp (\bar{p}p)$ collisions using
HIJING and PYTHIA.", {\em Chinese Physics}, {\bfseries 16}, 58-61,
2007.

\item Y.~F. Zhang, S.~L. Huang, Z.~P. Zhang, J. Wu, ``Simulation
on "soft" and "hard" process in $pp (\bar{p}p)$ collisions using
HIJING and PYTHIA.", {\em Journal of University of Science and
Technology of China}, {\bfseries 35(6)}, 821-824, 2005 (in
Chinese).

\item Q. Shan, J. Wu, X.~L. Wang, H.~F. Chen, Y.~F. Zhang, Z.~P.
Zhang, M. Shao, ``Upper limit of the yield of di-omega in central
Au+Au collision at $\sqrt(s_{NN})$ = 200 GeV with HIJING.", {\em
High Energy Phys. and Nucl. Phys.}, {\bfseries 29}, 1146-1149,
2005.


\item J. Adams {\em et al.}, STAR Collaboration. ``Pion, Kaon,
Proton and Anti-proton Transverse Momentum Distributions from p +
p and d + Au Collisions at \sNN = 200 GeV.", {\em Phys. Lett. B},
{\bfseries 616} 8, 2005.

\item J. Adams {\em et al.}, STAR Collaboration. ``Hadronization
Geometry and Charge-dependent Number Autocorrelations on Axial
Momentum Space in \AuAu~Collisions at \sNN = 130 GeV. ", {\em
Phys. Lett. B}, {\bfseries 634} 347, 2006.

\item J. Adams {\em et al.}, STAR Collaboration. ``Azimuthal
Anisotropy in \AuAu~Collisions at \sNN = 200 GeV. ", {\em Phys.
Rev. C}, {\bfseries 72} 014904, 2005.

\item J. Adams {\em et al.}, STAR Collaboration. ``Minijet
Deformation and Charge-independent Angular Correlations on
Momentum Subspace ($\eta$, $\phi$) in \AuAu~Collisions at \sNN =
130 GeV. ", {\em Phys. Rev. C}, {\bfseries 73} 064907, 2006.

\item J. Adams {\em et al.}, STAR Collaboration. ``Experimental
and Theoretical Challenges in the Search for the Quark Gluon
Plasma: The STAR Collaboration's Critical Assessment of the
Evidence from RHIC Collisions. ", {\em Nucl. Phys. A}, {\bfseries
757} 102, 2005.

\item J. Adams {\em et al.}, STAR Collaboration. ``Multiplicity
and Pseudorapidity Distributions of Photons in \AuAu~Collisions at
\sNN = 62.4 GeV. ", {\em Phys. Rev. Lett}, {\bfseries 95} 062301,
2005.

\item J. Adams {\em et al.}, STAR Collaboration. ``Multi-strange
Baryon Elliptic Flow in \AuAu~collisions at \sNN = 200 GeV. ",
{\em Phys. Rev. Lett.}, {\bfseries 95} 122301, 2005.

\item J. Adams {\em et al.}, STAR Collaboration. ``Incident Energy
Dependence of $p_{T}$ Correlations at RHIC. ", {\em Phys. Rev. C},
{\bfseries 72} 044902, 2005.

\item J. Adams {\em et al.}, STAR Collaboration.
``Transverse-momentum $p_{T}$ Correlations on ($\eta$, $\phi$)
from mean $p_{T}$ Fluctuations in \AuAu~Collisions at \sNN = 200
GeV. ", {\em J. Phys. G}, {\bfseries 32} L37, 2006.

\item J. Adams {\em et al.}, STAR Collaboration. ``Directed Flow
in \AuAu~Collisions at \sNN = 62 GeV. ", {\em Phys. Rev. C},
{\bfseries 73} 034903, 2006.

\item J. Adams {\em et al.}, STAR Collaboration. ``Proton-Lambda
Correlations in Central \AuAu~Collisions at \sNN = 200 GeV. ",
{\em Phys. Rev. C}, {\bfseries 74} 064906, 2006.

\item J. Adams {\em et al.}, STAR Collaboration. ``Multiplicity
and Pseudorapidity Distributions of Charged Particles and Photons
at Forward Pseudorapidity in \AuAu~Collisions at \sNN = 62.4 GeV.
", {\em Phys. Rev. C}, {\bfseries 73} 034906, 2006.

\item J. Adams {\em et al.}, STAR Collaboration. ``Strangelet
Search at RHIC. ", {\em arXiv: nucl-ex/0511047 }.

\item M. Calderon de la Barca Sanchez {\em et al.}, STAR
Collaboration. ``Open Charm Production from $d$ + Au Collisions in
STAR. ", {\em Eur. Phys. J. C}, {\bfseries 43} 187, 2005.

\item A. A. P. Suaide  {\em et al.}, STAR Collaboration. ``Charm
Production in the STAR Experiment at RHIC. ", {\em Eur. Phys. J.
C}, {\bfseries 43} 193, 2005.

\item C. A. Gagliardi {\em et al.}, STAR Collaboration. ``Recent
high-$p_{T}$ Results from STAR. ", {\em Eur. Phys. J. C},
{\bfseries 43} 263, 2005.

\item J. Adams {\em et al.}, STAR Collaboration. ``Identified
Hadron Spectra at Large Transverse Momentum in $p$ + $p$ and $d$ +
Au Collisions at \sNN = 200 GeV. ", {\em Phys. Lett. B},
{\bfseries 637} 161, 2006.

\item J. Adams {\em et al.}, STAR Collaboration. ``Measurements of
Identified Particles at Intermediate Transverse Momentum in the
STAR Experiment from \AuAu~Collisions at \sNN =200 GeV. ", {\em
arXiv: nucl-ex/0601042 }.

\item J. Adams {\em et al.}, STAR Collaboration. ``Forward Neutral
Pion Production in $p$ + $p$ and $d$ + Au Collisions at \sNN = 200
GeV. ", {\em Phys. Rev. Lett.}, {\bfseries 97} 152302, 2006.

\item J. Adams {\em et al.}, STAR Collaboration. ``Direct
Observation of Dijets in Central \AuAu~Collisions at \sNN = 200
GeV. ", {\em Phys. Rev. Lett.}, {\bfseries 97} 162301, 2006.

\item J. Adams {\em et al.}, STAR Collaboration. ``Strange Baryon
Resonance Production in \sNN = 200 GeV $p$ + $p$ and
\AuAu~Collisions. ", {\em Phys. Rev. Lett}, {\bfseries 97} 132301,
2006.

\item J. Adams {\em et al.}, STAR Collaboration. ``The Energy
Dependence of $p_{T}$ Angular Correlations Inferred from mean
$p_{T}$ Fluctuation Scale Dependence in Heavy Ion Collisions at
the SPS and RHIC. ", {\em J. Phys. G}, {\bfseries 33} 451, 2007.

\item J. Adams {\em et al.}, STAR Collaboration. ``Identified
Baryon and Meson Distributions at Large Transverse Momenta from
\AuAu~Collisions at \sNN = 200 GeV. ", {\em Phys. Rev. Lett},
{\bfseries 97} 152301, 2006.

\item J. Adams {\em et al.}, STAR Collaboration. ``Scaling
Properties of Hyperon Production in \AuAu~Collisions at \sNN = 200
GeV. ", {\em arXiv: nucl-ex/0606014}.

\item J. Adams {\em et al.}, STAR Collaboration. ``The
Multiplicity Dependence of Inclusive $p_{T}$ Spectra from $p$ +
$p$ Collisions at \sNN = 200 GeV. ", {\em Phys. Rev. D},
{\bfseries 74} 032006, 2006.

\item J. Adams {\em et al.}, STAR Collaboration. ``Delta phi Delta
eta Correlations in Central \AuAu~Collisions at \sNN = 200 GeV. ",
{\em Phys. Rev. C}, {\bfseries 75} 034901, 2007.

\item B. I. Abelev {\em et al.}, STAR Collaboration. ``Transverse
Momentum and Centrality Dependence of high-$p_{T}$ Non-photonic
Electron Suppression in \AuAu~Collisions at \sNN = 200 GeV. ",
{\em arXiv: nucl-ex/0607012}.

\item B. I. Abelev {\em et al.}, STAR Collaboration. ``Strange
Particle Production in $p$ + $p$ Collisions at \sNN = 200 GeV. ",
{\em arXiv: nucl-ex/0607033}.

\item B. I. Abelev {\em et al.}, STAR Collaboration. ``Neutral
Kaon Interferometry in \AuAu~Collisions at \sNN = 200 GeV. ", {\em
Phys. Rev. C}, {\bfseries 74} 054902, 2006.

\item B. I. Abelev {\em et al.}, STAR Collaboration.
``Longitudinal Double-spin Asymmetry and Cross Section for
Inclusive Jet Production in Polarized Proton Collisions at \sNN =
200 GeV. ", {\em Phys. Rev. Lett}, {\bfseries 97} 252001, 2006.

\item B. I. Abelev {\em et al.}, STAR Collaboration. ``Rapidity
and Species Dependence of Particle Production at Large Transverse
Momentum for $d$ + Au Collisions at \sNN = 200 GeV. ", {\em arXiv:
nucl-ex/0609021}.

\item B. I. Abelev {\em et al.}, STAR Collaboration. ``Partonic
Flow and Phi-meson Production in Au + Au Collisions at \sNN = 200
GeV. ", {\em arXiv: nucl-ex/0703033}.

\item B. I. Abelev {\em et al.}, STAR Collaboration. ``Charged
Particle Distributions and Nuclear Modification at High Rapidities
in $d$ + Au Collisions at \sNN = 200 GeV. ", {\em arXiv:
nucl-ex/0703033}.

\item B. I. Abelev {\em et al.}, STAR Collaboration. ``Mass,
Quark-number, and \sNN~Dependence of the Second and Fourth Flow
Harmonics in Ultra-Relativistic Nucleus-Nucleus Collisions. ",
{\em arXiv: nucl-ex/0701010}.

\end{itemize}

\bibliography{PhDthesis}
\bibliographystyle{uclathes}

\end{document}